\documentclass[a4paper,14pt,openany,epsfig]{extbook}
\usepackage[cp1251]{inputenc}
\usepackage{amsmath,euscript,amssymb}
\usepackage[T2A]{fontenc}
\usepackage{epsfig}
\usepackage{graphicx}
\usepackage{bm}
\usepackage{color}
\newcommand{\be}{\begin{equation}}
\newcommand{\ee}{\end{equation}}
\newcommand{\bea}{\begin{eqnarray}}
\newcommand{\eea}{\end{eqnarray}}

\begin{document}

\vskip 6cm
\title
{\bf\Huge Principles \\ of Quantum Universe} \vskip 4cm
\author
{\bf V.N. Pervushin,  A.E. Pavlov\\
Laboratory of Theoretical Physics\\
Joint Institute for Nuclear Research, Dubna} \maketitle

\tableofcontents

\newpage
\makeatletter
\renewcommand{\@evenhead}{}
\renewcommand{\@oddhead}{}
\makeatother

\newpage

\makeatletter
\renewcommand{\@evenhead}{\raisebox{0pt}[\headheight][0pt]
{\vbox{\hbox to\textwidth{\hfil \strut  Table of contents  \quad\rm\thepage}\hrule}}}
\renewcommand{\@oddhead}{\raisebox{0pt}[\headheight][0pt]
{\vbox{\hbox to\textwidth{\strut Table of contents \hfil \rm\thepage}\hrule}}}
\makeatother

\newpage

\makeatletter
\renewcommand{\@evenhead}{}
\renewcommand{\@oddhead}{}
\makeatother

\begin{flushright}
\section*{Preface to\\
the first English edition}
\end{flushright}
\vspace{.31cm}
The Russian edition of the present book was published in June 2013.  It just happened that it was the time
between two significant dates:
in 2011 the Nobel Prize was awarded ``{\it for the discovery of the accelerated expansion of the Universe through
observations of distant Supernovae}’’
and in 2013 the Nobel Prize was awarded for ``{\it the theoretical discovery of a mechanism that contributes
to our understanding of the origin
of the mass of subatomic particles}’’.
Both these formulations left the questions about the explanations of these phenomena in the framework
of the fundamental principles open.

Our book is devoted to attempts to explain the observed {\it long distances to the Supernovae} and the
{\it small value of
the Higgs particle mass}
by the principles of affine and conformal symmetries
and the vacuum postulate.
Both these phenomena are described by quantum gravity in the form
 of joint irreducible unitary representations
of the affine and conformal symmetry groups.
These representations were used in our book to classify physical processes in the Universe,
including its origin from the vacuum.
The representations of the
Poincar\'e group were used in the same way by Wigner to classify particles and their bound states.

We are far from considering our understanding of the ``distant  Supernovae'' and the ``origin
of the mass of subatomic particles''
 to be conclusive, but we do not abandon hope that the present revised
and enlarged English edition encourages a deeper and worthier
investigation of these open questions in the future.

The authors express their appreciation and gratitude to the coauthors of the papers
on which this book is based. The authors are grateful to
I.V. Kronshtadtova and G.G. Sandukovskaya for proofreading the text of the book.
The authors are grateful
to Academician V.A. Matveev, Professors V.V. Voronov and M.G. Itkis for the support.

\vspace{.99cm}

\begin{flushright}
V.N. Pervushin\\
A.E. Pavlov
\end{flushright}
\vspace{.31cm}

\begin{flushleft}
\vspace{.3cm}
December, 2013\\
Dubna
\end{flushleft}
\newpage

\begin{flushright}
\section*{Preface to\\
the first Russian edition\footnote{Victor Pervushin, Alexander Pavlov:
{\it Principles of Quantum Universe.}
LAP LAMBERT Academic Publishing. 420 pp. (2013).
Saarbr\"ucken, Deutschland (in Russian)
}}
\end{flushright}
This monograph is based on papers published during last 25 years by the authors and lectures
delivered by one of the authors (V.P.) at the universities of Graz (Austria), Berlin, Heidelberg, Rostock (Germany),
New Delhi (India), Fairfield, the Argonne National Laboratory (USA), the physical faculty of Moscow State University and in
the Joint Institute for Nuclear Research. The main goal of the authors is to bring
readers into the interesting and intriguing problem of description of modern experimental and observational data in the framework of
ideas and methods elaborated until 1974 by the founders of the modern relativistic classical and quantum physics.
The distinction of our approach from the standard ones consists in using conformal symmetry:
everywhere, from the horizon of the Universe to quarks,
we use scale-invariant versions of modern theories on the classical level with dimensionless coupling constants, breaking
scale invariance only at the quantum level by normal ordering of products of field operators. The method of classification of
novae data, obtained in the last fifteen years in cosmology and high--energy physics, essentially uses quantum theories and representations.
From here the title of our book originated: ``\textit{Principles of Quantum Universe}''. Let us briefly present the
content of the book.

In Introduction (Chapter 1) we discuss the evolution of ideas and mathematical methods of theoretical physics
during last five centuries of its development from Copernicus to Wheeler, focusing on the problem of classification
of physical measurements and astrophysical observations. In Chapter 2 we present the problems of choosing initial data and
frames of reference in Newton's mechanics, relativistic theory of a particle, cosmological standard models of a
miniuniverse. Chapter 3 is devoted to principles of symmetries, widely used in modern theoretical physics.
In Chapter 4 we acquaint readers with the method of nonlinear realizations of groups of symmetries developed at the end of the
sixtieth of the last century, which applied for derivation of the theory of gravitation by joint nonlinear realization
of affine and conformal symmetries.
In Chapter 5 the generally accepted Dirac -- Bargmann's Hamiltonian formulation is presented; it is adapted to
the gravitation theory, deduced in Chapter 4. In Chapter 6 a quantum cosmological model is studied which appeared
in the empty Universe approximation
with the Casimir energy dominance. In Chapter 7 the procedure of quantization of gravitons in terms of Cartan's forms is implemented and
the vacuum creation of affine gravitons is considered. In Chapter 8 the operator of creation and evolution of the quantum Universe
is constructed as a joint irreducible unitary representation of affine and
conformal groups of symmetries. In Chapter 9 the creation of matter from
vacuum is formulated in the considered model of the quantum Universe with a discussion of conformal modification of S-matrix
as a consequence of solutions of constraint equations in
the joint theory of gravitation and the Standard Model of elementary particles. In Chapter 10, within the frame of this model, we
describe the spontaneous chiral symmetry breaking in QCD via normal ordering of products of operators of the gluon and quark
fields, and also derive the quark--hadron duality and the parton model as one of the consequences of conformal modification of S--matrix.
In Chapter 11 a conformal modification of the Standard Model of elementary particles without the Higgs potential is presented. Chapter 12
is devoted to the vacuum creation of electroweak bosons; and the origins of anisotropy of temperature of CMB radiation and the
baryon asymmetry of the Universe are discussed. In Chapter 13 a cosmological modification of the Schwarzschild solution and
Newton's potential is presented.
In Chapter 14, in the framework of this cosmological modification of the Newton dynamics, the evolution of galaxies and their
superclusters is discussed.
In Chapter 15 (Postface), the list of the results is presented and the problems that arise in the model of the quantum Universe are discussed.

In conclusion, the authors consider as a pleasant duty to express deep gratitude to Profs.
A.B. Arbuzov, B.M. Barbashov, D. Blaschke, A. Borowiec, K.A. Bronnikov, V.V. Burov, M.A. Chavleishvili,
A.Yu. Cherny, A.E. Dorokhov, D. Ebert, A.B. Efremov, P.K. Flin, N.S. Han,
Yu.G. Ignatiev, E.A. Ivanov, E.A. Kuraev,
J. Lukierski, V.N. Melnikov, R.G. Nazmitdinov, V.V. Nesterenko, V.B. Priezzhev,
G. Roepke, Yu.P. Rybakov, S.I. Vinitsky, Yu.S. Vladimirov, M.K. Volkov,
A.F. Zakharov, A.A. Zheltukhin
for stimulating discussions of the problems which we tried to solve in this manuscript.
One of the authors (V.P.) is particularly thankful to Profs. Ch. Isham and Т. Kibble for discussions of the problems of the Hamiltonian
approach to the General Relativity and for hospitality at the Imperial College, Prof. S. Deser, who kindly informed about his papers on
Conformal theory of gravity, Prof. H. Kleinert for numerous discussions at the Free University of Berlin, Prof. М. McCallum for discussion of
physical contents of solutions of the Einstein equations, Profs. H. Leutwyler and W. Plessas for discussions of mechanisms of chiral symmetry
breaking in QCD, Prof. W. Thirring for discussion about predictions yielded by the General Relativity and the considered theory of gravity,
concerning motions of bodies in celestial mechanics problems.
V.P. is also thankful to his former  post-graduate students and coauthors  D. Behnke, I.А. Gogilidze,
А.А. Gusev, N. Ilieva, Yu.L. Kalinovsky, A.M. Khvedelidze, D.M. Mladenov, Yu.P. Palij, H.-P. Pavel, M. Pawlowski,
K.N. Pichugin, D.V. Proskurin, N.A. Sarikov, S. Schmidt, V.I. Shilin, S.A. Shuvalov, M.I. Smirichinski,
N. Zarkevich, V.A. Zinchuk, A.G. Zorin
for
helpful collaboration. The authors are grateful to Profs. S. Dubnichka, M.G. Itkis, W. Chmielowski, V.A. Matveev, V.V. Voronov for support of
collaboration with international scientific centers, also to B.M. Starchenko and Yu.А. Tumanov for presented photos. One of the authors
(А.P.) is grateful to the Directorate of JINR for hospitality and possibility to work on the monograph. The results
of the investigations, presented in the book, are implemented under partial support of the Russian Foundation of Basic Research
(grants 96-01-01223, 98-01-00101), аnd also grants of the Heisenberg -- Landau, the Bogoliubov -- Infeld, the
Blokhintsev -- Votruba and the Max Planck society (Germany).
\vspace{.3cm}

\begin{flushright}
V.N. Pervushin\\
A.E. Pavlov
\end{flushright}
\vspace{.1cm}
\begin{flushleft}
June, 2013\\
Dubna
\end{flushleft}

\chapter{Introduction}

\makeatletter

\renewcommand{\@evenhead}{\raisebox{0pt}[\headheight][0pt]

{\vbox{\hbox to\textwidth{\hfil \strut \thechapter .
Introduction \quad\rm\thepage}\hrule}}}
\renewcommand{\@oddhead}{\raisebox{0pt}[\headheight][0pt]
{\vbox{\hbox to\textwidth{\strut \thesection .\quad  What is this book about? \hfil \rm\thepage}\hrule}}}
\makeatother

\section{What is this book about?}


In his remarkable book\footnote{Weinberg, S.: \textit{The First Three Minutes: A Modern View of
the Origin of the Universe.} Basic Books, New York (1977).} the Nobel laureate in Physics Steven Weinberg
considers problems of Genesis according to the laws of classical cosmology. In the Epilogue he gives predictions of further
life of the Universe resulted from these laws.
``\textit{However all these problems may be resolved, and whichever cosmological model proves correct, there is not
much of comfort in any of this. It is almost irresistible for humans to believe that we have some
special relation to the universe, that human life is not just a more-or-less farcical outcome of a
chain of accidents reaching back to the first three minutes, but that we were somehow built in from the beginning.
As I write this I happen to be in an aeroplane at 30,000 feet, flying over Wyoming en route home from
San Francisco to Boston. Below, the earth looks very soft and comfortable -- fluffy clouds here and there, snow
turning pink as the sun sets, roads stretching straight across the country from one town to another.
It is very hard to realize that this all is just a tiny part of an overwhelmingly hostile universe.
It is even harder to realize that this present universe has evolved from an unspeakably unfamiliar early
condition, and faces a future extinction of endless cold or intolerable heat.
The more the universe seems comprehensible, the more it also seems pointless.
But if there is no solace in the fruits of our research, there is at least some consolation in the research itself.
Men and women are not content to comfort themselves with tales of gods and giants, or
to confine their thoughts to the daily affairs of life; they also build telescopes and satellites and accelerators,
and sit at their desks for endless hours working out the meaning of the data they gather. The effort to
understand the universe is one of the very few things that lifts human life a little above the level of farce,
and gives it some of the grace of tragedy}''.
One of these acts of the tragedy is dramatic events of last years in cosmology and physics of elementary particles:
expanding of the Universe with acceleration and the intriguingly small value of the Higgs particle mass. These events
throw discredit upon or leave without any hopefulness for success a lot of directions of modern theoretical
investigations.

\begin{figure}
\includegraphics[width=0.85\textwidth,clip]{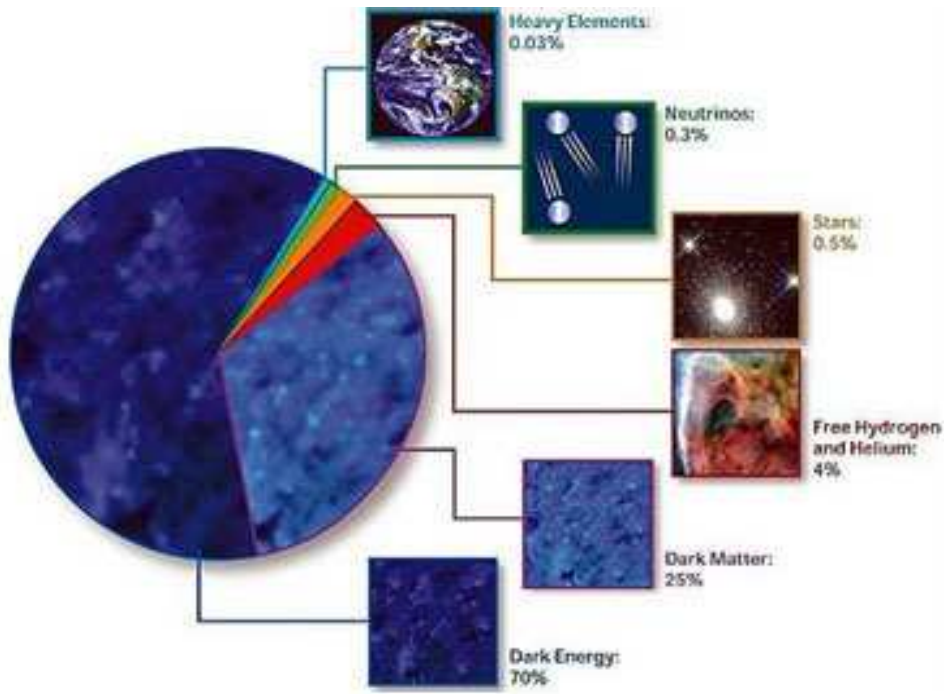}
\caption{\small According to NASA diagram, 25\% of the Universe is dark matter,
70\% of the Universe is dark energy about which practically nothing is known.}
\label{NASA-1}
\end{figure}

In recent years, two independent collaborations ``High Supernova'' and ``Supernova Cosmology Project'' obtained
new unexpected data about cosmological evolution at very large distances -- hundreds and thousands megaparsecs
expressed in redshift values $z=1\div 1.7$ \cite{Riess1998-1, Perlmutter1999-1, Riess2001-1}.
Surprisingly, it was found that the decrease of brightness with distance, on an average, happen noticeably faster than it
is expected according to the Standard cosmological model with the matter dominance.
Supernovae are situated at distances further than it was predicted.
Therefore, according to the Standard cosmological model, in the last period, the cosmological expansion proceeds with acceleration.
Dynamics, by
unknown reasons passes from the deceleration stage to an acceleration one of expansion. Observable data (see Fig.\ref{NASA-1})
testify that the Universe is filled mainly not with massive dust that can not provide accelerating expansion but with unspecified enigmatic
substance of other nature---``dark energy'' \cite{WeinbergCosmology-1}. Cosmic acceleration, at the present time, is provided
by some hypothetic substance called as quintessence. This term is borrowed from ancient Greece when philosophers constructed their
world view from five elements: earth, water, air, fire, and quintessence as a cosmic substance of celestial bodies.
In the modern cosmology, this substance means a special kind of cosmic energy.
Quintessence creates negative pressure (antigravitation) and leads to accelerating expansion. In classical cosmology
it is necessary, once again, for rescue of the situation, to put $\Lambda$-term into the Einstein's equations.
The problem is that the energy density of accelerating expansion at the beginning of the Universe evolution differs
$10^{57}$ times from the modern density. Up to now there is no such dynamical model that should be able to describe and
explain the phenomenon of such \textit{dynamical inflation} \cite{Steinhardt}.

\begin{figure}[t]
\begin{center}
\includegraphics[width=0.8\textwidth,clip]{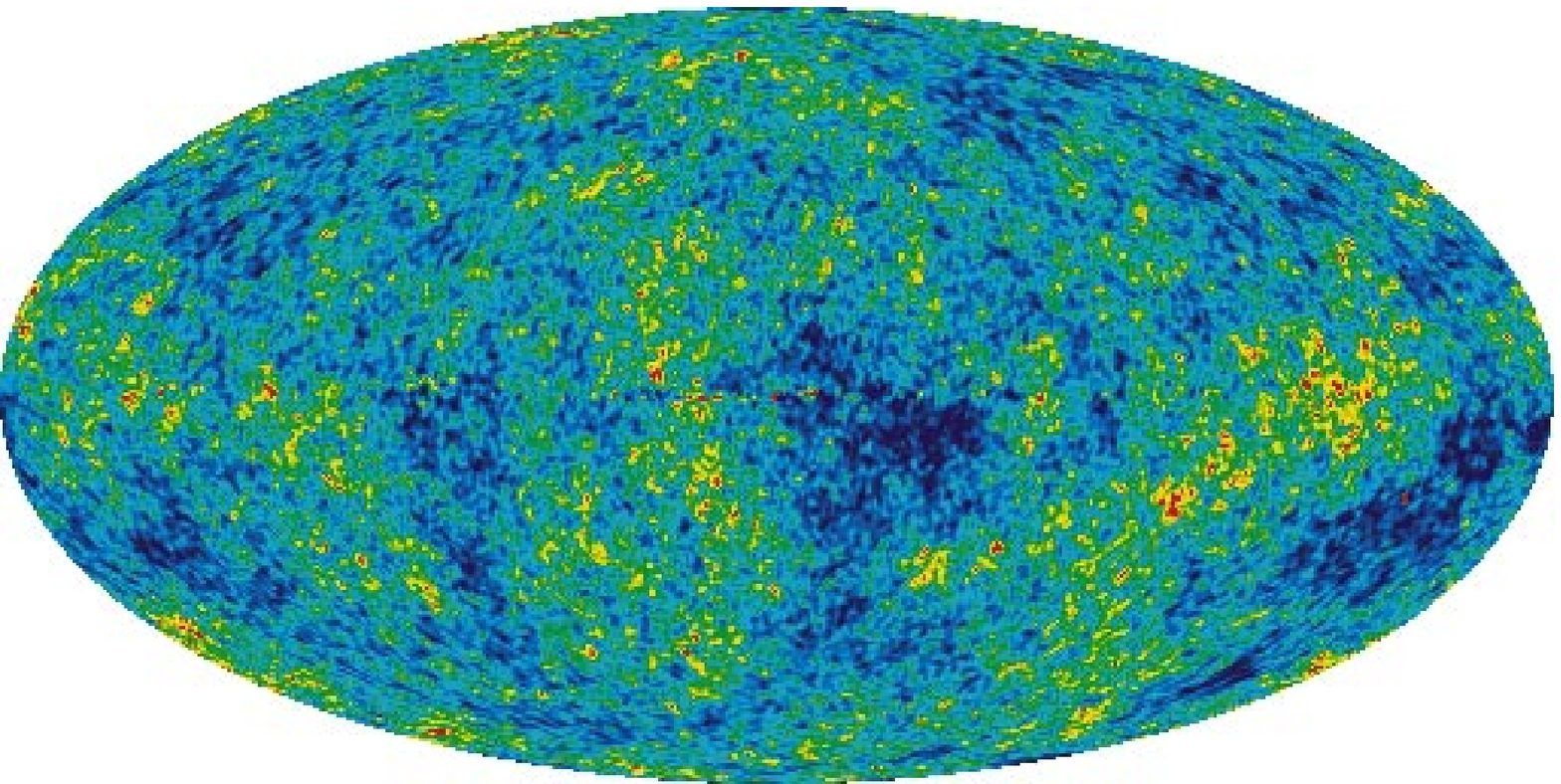}
\caption{\small Map of Cosmic Microwave Background was lined up
according to the data from the Wilkinson Microwave Anisotropy Probe (WMAP) apparatus.
Previously, the first detailed map was done by data according to the COBE apparatus, however, its resolution
is essentially (35 times) inferior to the data obtained by WMAP. The WMAP data show that the temperature
distribution of Cosmic Microwave Background by the celestial sphere has definite structure, its fluctuations
are not absolutely random. The angle anisotropy of Cosmic Microwave Background is presented, {\it id est}
dependence of photon temperature of directions of their coming.
The average photon temperature $T_0=2,725\pm 0,001$ К, and the dipole component
$\delta T_{\mbox{dipole}}=3,346$ мК are subtracted. The picture of temperature variation is shown at the level
$\delta T\sim 100\mu$К, so
$\delta T/T\sim 10^{-4}\div 10^{-5}$ (see http://map.gsfc.nasa.gov).
}
\label{WMAP-1}
\end{center}
\end{figure}

\begin{figure}[ntb]
\begin{center}
\includegraphics[width=0.5\textwidth,clip]{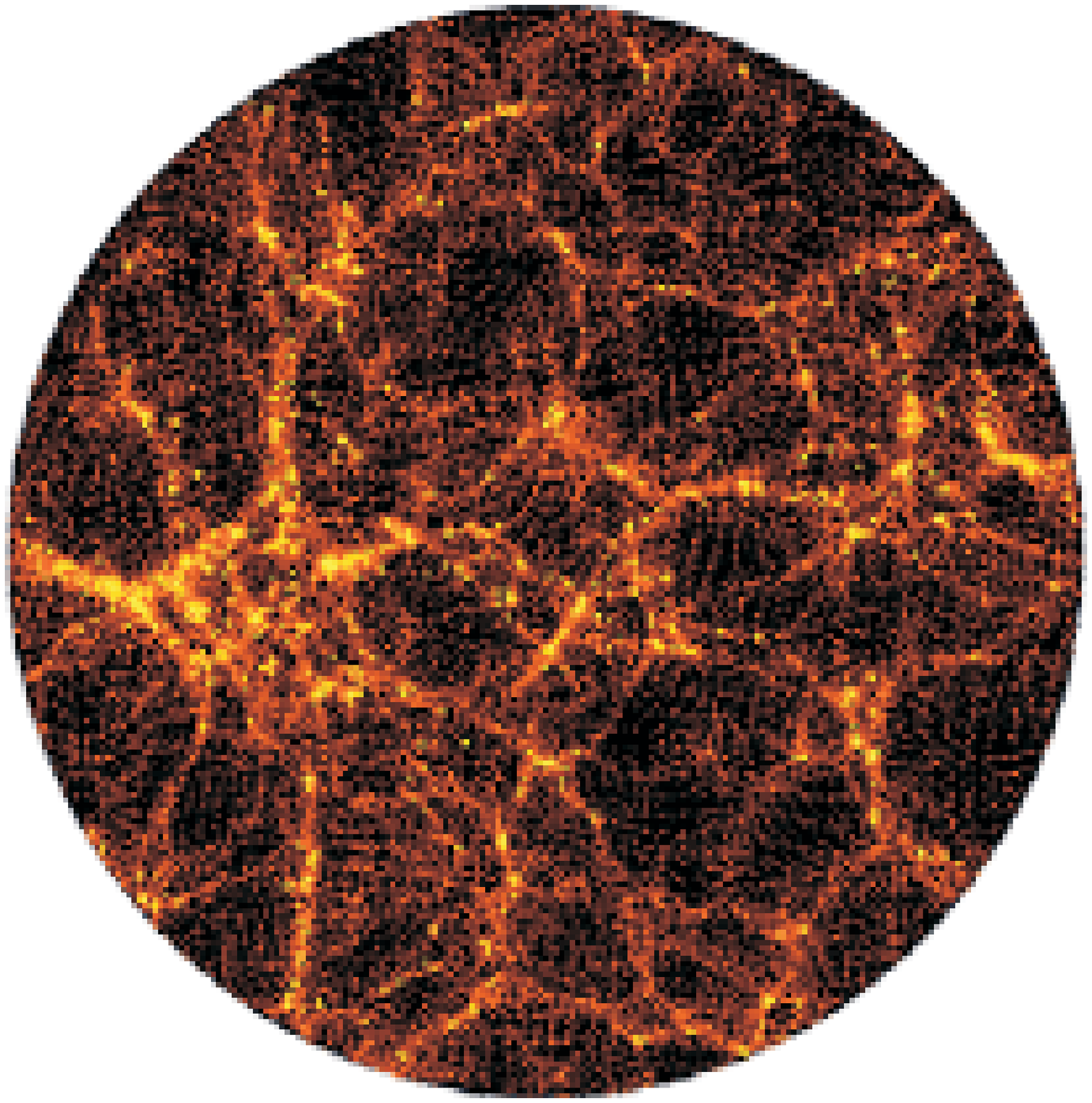}
\caption{\small
Large-scale structure of the Universe represents a complex of sufficient plane ``leaves'' divided by regions
where the luminous matter is practically absent. These regions (voids) have sizes  on the order of hundred
megaparsecs. At the scales on the order of 300 megaparsecs the Universe is practically homogeneous.
}
\label{UniverseScale}
\end{center}
\end{figure}

The crisis of the Standard cosmology enables us to reconceive its principles.
In this critical situation these new observational data
(see, for example, Figs. \ref{NASA-1}, \ref{WMAP-1}, \ref{UniverseScale})
look like a challenge for theoretical cosmology. In the present book this challenge is considered
as a possibility to
construct the cosmological model that can explain modern observational facts
at the level of well-known fundamental principles of relativity and symmetry
without whatever dynamical inflation mechanism.

Let us recall that the theory of gravitation and corresponding cosmological models of the Universe are based on the
classical papers of Einstein, Hilbert, Weyl, Dirac, Fock and other researchers, who postulated
geometrical principles, including scale and conformal symmetries. In particular, the Lagrangian of Weyl's theory
is an invariant with respect to conformal transformations \cite{Weyl1918-1}. P. Dirac in the year 1973 constructed a conformal
gravitation theory where scale transformations of a scalar dilaton compensated scale transformations of other
fields \cite{Dirac1973int-1}. In the framework of this theory of gravitation, the volume of the Universe conserves during its
evolution and the forthcoming collapse, inevitable in the Standard cosmology, does not occur. The Conformal
gravitation theory with a scalar dilaton is derived from the finite group of symmetry of initial data via the
method of Cartan's linear forms \cite{og1974-1}.

The Conformal gravitation theory in terms of Cartan's forms, keeping all achievements of the General Relativity for describing
the solar system, admits a quantum formulation by quantization of initial data immediately for these linear forms. A
remarkable possibility is given to test predictions of such a quantum theory of gravitation and its ability to describe the new
data presented by observable cosmology and solutions of its vital problems.

The goal of the present book is a consistent treatment of groups of symmetries of initial data, Cartan's method
of linear forms, derivation of the Conformal gravitation theory, its Hamiltonian formulation and quantization, and also the
description and interpretation of the new observational data in the framework of the quantum theory.

\makeatletter
\renewcommand{\@evenhead}{\raisebox{0pt}[\headheight][0pt]
{\vbox{\hbox to\textwidth{\hfil \strut \thechapter . Introduction \quad\rm\thepage}\hrule}}}
\renewcommand{\@oddhead}{\raisebox{0pt}[\headheight][0pt]
{\vbox{\hbox to\textwidth{\strut \thesection .\quad  Program \hfil \rm\thepage}\hrule}}}
\makeatother
\section{Program}

One of the main problems of theoretical physics is a classification of observational and experimental data
which form a space of all events (as an assembly of all measurements).
Measurable and observational data have always the primary status everywhere. In the process of analysis of these data
also theoretical concepts are arising, such as the Faraday -- Maxwell fields and groups
of symmetry of their equations, identified with the laws of nature. Classification of observational and experimental data,
according to Copernicus, turned out to be sufficiently simple in some definite frame of reference.
Indeed, classification of planet trajectories is appreciably simplified in the frame of reference,
connected with the Sun, called heliocentric. Copernicus' simplification considerably helped Galileo, Kepler, and
Newton in formulation of laws of the celestial mechanics\footnote{In formal, all frames of reference are
equal. By means of authors (Einstein, А., Infeld, L.: \textit{The Evolution of Physics:
From Early Concepts to Relativity and Quanta.} Touchstone, New York (1967)),
if people understood relativity, there was not such dramatic, in the history of mankind, changing of
world outlook, where the Earth was the center of the world. It is not hard to guess that the following, after
heliocentric, was the galaxy-centric frame of reference. But, in every concrete problem, there is the privileged
frame of reference, in which the contents of the problem are clarified.}.

In cosmology, there are also two privileged frames of reference: the cosmic one, where the Universe with matter is created
which is memorized by temperature of Cosmic Microwave Background, and the other frame of reference of an observer (Earth frame)
with its devices \cite{Barbashov_03-1}. Let us remember a hierarchy of motions
in which our planet is involved as it seems in our time\footnote{Chernin, А.D.: \textit{Cosmic vacuum.}
Physics--Uspekhi. {\bf 44}, 1099
(2001).}. In the galactic frame of reference\footnote{Zeroth latitude (b) in galactic system corresponds to galactic
equatorial plane, and zeroth longitude (l) corresponds to direction to the center of the galaxy, located in Archer.
Galactic latitude is measured from galactic equator to North (+) and to South (-), galactic latitude is measured in direction to
West along galactic plane from galactic center.} $l=90^0$, $b=0^0$ the Earth rotates around the Sun with the velocity 30 km/sec;
the Sun rotates with the velocity 220 km/sec around the center of our galaxy. In turn, the center of our galaxy (Milky Way)
moves with the velocity 316 $\pm$ 11 km/sec  to the center of the local group of galaxies\footnote{Local group of galaxies
includes the Milky Way, Large and Small Magellanic Clouds, Giant galaxy Andromeda Nebula (М31) and approximately 2-3
dozens of dwarf galaxies. For information:
overall size of local groups is order 1 Мpc=3,0856 $\times 10^{19}$ km.
1 parsec (pc) is a distance, with whom an object of size 1 astronomical unit
(1 a.u.=$1,5\times 10^{13}$ сm is a mean distance from the Earth to the Sun) is seen at an angle 1 second:
1 pc= $2,1\times 10^5$ a.u. = 3,3 (l. yr.). A light year (l. yr.) is a distance, traversed by photon per one Earth's year.
} in direction
$l=(93 \pm 2)^0$, $b=(-4\pm 1)^0$.
Finally, we get the velocity of the galactic center relative to the center of the local group of galaxies is 91 km/sec
in direction $l=163^0$, $b=-19^0$.

The centers of our galaxy and Andromeda Nebula (galaxy М31) under action of mutual gravitational attraction come close with the velocity
120 km/sec. Suppose, that our galaxy and Andromeda contribute substantial loading to the common mass of the local group
and the mass of our galaxy two times less than the mass of Andromeda, we get that our galaxy moves to Andromeda with the velocity
80 km/sec. A measurement of dipole anisotropy of Cosmic Microwave Background radiation (CMB radiation), implemented by the
American cosmic apparatus COBE, established that the velocity of the Sun relative to CMB radiation is order (370 $\pm$ 3) km/sec
in direction $l=(266,4 \pm 0,3)^0$, $b=(48,4\pm 0,5)^0$. This anisotropy is responsible for motion of the observer relative to
``global (absolute)'' frame of reference. Inasmuch as movement of the Sun relative to the local group and its movement relative to
the ``absolute'' frame of reference connected with the Cosmic Microwave Background radiation have, practically, opposite directions,
the velocity of the center of the local group of galaxies relative to the CMB radiation happens to be sufficiently large:
approximately (634$\pm$12) km/sec in direction $l=(269 \pm 3)^0$, $b=(48,4\pm 0,5)^0$.

In summary, the center of the local groups moves in the following directions:

a) in direction to Virgo $l=274^0$, $b=75^0$ with the velocity 139 km/sec;

b) in direction to the Great Attractor $l=291^0$, $b=17^0$, disposed at the distance  44 Мpc, with the velocity 289 km/sec;

c) in direction opposite to the local empty region, $l=228^0$, $b=-10^0$ with the velocity 200 km/sec.

Taking into account all these movements, one can affirm that the local group moves with the velocity 166 km/sec in direction
$l=281^0$, $b=43^0$. Inasmuch as errors of defining of individual velocities are in the order of 120 km/sec, the
local group is able to be, practically, at the rest relative to far galaxies.

At the Beginning, at the instant of the Universe creation from vacuum, there were neither massive bodies, nor relic radiation,
there was chosen a frame of reference, comoving to the velocity of the empty local element of volume. Such frame
was introduced by Dirac in 1958 year, as the condition of a minimal three-dimensional surface, imbedded into four-dimensional
space-time \cite{Dirac1958-1}.

In all cases in passing from the cosmic frames of reference to the frame of reference of devices of an observer,
it is necessary to have transformations of physical observables, including, the interval itself. These transformations require
the formulation of the General Relativity in the tetrad formalism.

In the present book we demonstrate that the choice of a frame of reference, co-moving to the velocity of the empty
local element of volume, simplifies the classification of observational data on redshifts of spectral lines
of far Supernovae and helps us to formulate the principles of symmetry of the unified theory of interactions and quantum
mechanisms of their violations, as well as, in olden times, the Copernicus frame helped Newton to formulate the laws of
celestial mechanics. The observable data on redshifts in the frame of reference of an empty volume testify about
conformal symmetry of laws of gravitation and Maxwell electrodynamics\footnote{Conformal invariance of the Maxwell
equations was proved for the first time in papers: Bateman, H.: \textit{The conformal transformations of a space of four
dimensions and their applications to geometric optics.} Proc. London Math. Soc. {\bf 7},  70
(1909); Cuningham, E.: \textit{The principle of relativity in electrodynamics and an extension of
the theory.} Proc. London Math. Soc. {\bf 8}, 77 (1909). P.A.M. Dirac in his paper
(Dirac, P.A.M.: \textit{Wave equations in conformal space}. Ann. of  Math. {\bf 37}, 429 (1936)) resulted
alternative, more simple proof.} and dominance of Casimir vacuum energy for empty space in the considered model of the Universe.

When we speak of the nature of dark energy and dark matter that is unknown, we mean that
these quantities are not included into the classification of fields by irreducible representations of the Lorentz group and the
Poincar$\acute{\rm e}$ group in some frame of reference. The theme of the present work is to describe observational and experimental data on
redshifts of Supernovae, including dark energy and dark matter, in the framework of the well-known classification of fields by irreducible
representations of the Lorentz, Poincar$\acute{\rm e}$ \cite{Schweber-1} and Weyl \cite{Ramond-1} groups.

Fundamental physical equations (Newton, Maxwell, Einstein, Dirac, Yang -- Mills,
Weinberg -- Salam -- Glashow, {\it et al.}) are able to treat as invariant structural relations of the corresponding
group of symmetry of initial data. The complete set of initial data includes all possible measurements in the field space of events
\cite{WDWint-1}. The question now arises of what is more fundamental: equations of motion called the laws of nature that are
independent on initial data, or finite-parametric groups of symmetry of frames of reference of initial data?

There is a point of view that is developed according to which all physical laws of nature can be obtained from the
corresponding group of symmetry of initial data. The history of frames of reference of initial data,
starting from Ptolemy and Copernicus, is considerably more ancient than the history of equations of motion.
Let us overlook for the historical sequence of using in physics
the groups of transformations of initial data with the finite number of parameters. Galileo group assigns transitions in the class
of inertial frames of reference; the six-parametric group of Lorentz describes rotations and boosts in Minkowskian space;
the Poincar$\acute{\rm e}$ group, including Lorentz group as its subgroup, is complemented by four translations in the space-time;
the affine group of all linear transformations consists of Poincar$\acute{\rm e}$ group and ten symmetrical proper affine transformations;
Weyl group includes the Poincar$\acute{\rm e}$ group complimented by a scale transformation; the fifteen-parametric
group of conformal transformations includes the eleven-parametric Weyl group and four inversion transformations.

After creation of the special theory of relativity, for a decade, Albert Einstein searched for the formulation of
the theory of relativity, extending the Poincar$\acute{\rm e}$ group of symmetry of the Special Relativity to a group of general
coordinate transformations.
The searching for covariant description led Einstein to the tensor formulation of his theory. So, he named the theory of gravitation
as the General Relativity. This name reflects the general heuristic principle of the
relativistic theory of gravity. After the theory was constructed, in the following period of re-thinking of its foundation\footnote{According to
V.A. Fock (Fock, V.A.:
\textit{The Theory of Space, Time and Gravitation.} Pergamon press, London (1964)),
principles, laid in the basis of the theory are following. The first basic idea is to unify space and time in one whole space.
The second basic idea is to reject uniqueness of Minkowskian metrics and to pass to Riemannian metrics. The metric of
space--time depends on events that take place in space--time, in first order, from the distribution and motion of masses.},
the group of general coordinate transformations gained a status of the gauge group of symmetry as in modern gauge theories. The group of
general coordinate transformations in the General Relativity is used for descriptions of interactions, while the
Poincar$\acute{\rm e}$ group serves for classification of free fields.

For definition of the variables which are invariant relative to diffeomorphisms, and thereby, elimination of gauge arbitrariness in solutions of
equations of the theory, it is necessary to separate general coordinate transformations
(which play the role of gauge ones) from the Lorentzian ones.
The solution of the problem of separation of general coordinate transformations from relativistic transformations of systems of reference
was suggested by Fock \cite{Fock1929-1} in his paper on introduction of spinor fields in the Riemannian space. In fact, instead of a metric
tensor, Fock introduced tetrads defined as ``square root'' of the metric tensor, with two indices. One index relates to Riemannian space,
being the base space, and the second -- to a tangent Minkowskian space. Tetrad components are coefficients of decomposition of Cartan's forms
via differentials of coordinate space. These differential forms, by definition, are invariants relative to general coordinate transformations,
and have a meaning as measurable geometric values of physical space, and integrable non-invariant differentials of coordinate space
considered as auxiliary mathematical values of the kind of electromagnetic potentials in electrodynamics.


According to {\it Ogievetsky's theorem} \cite{Ogievetsky-1}, the {\it invariance, under the infinite-parameter
generally covariant group, is equivalent to simultaneous invariance under the affine and the conformal group.}
The proof of the theorem is based on the note, that infinite-dimensional algebra of general transformation
of coordinates  is the closure of the finite-dimensional algebras of $SL(4,\mathbb{R})$
and conformal group\footnote{The generator of special conformal transformations in the coordinate space
$$K_\mu=-\imath (x^2\partial_\mu-2x_\mu (x_\lambda\partial^\lambda ))$$
is quadratic in the coordinates. The result of its commuting with the generator
$-\imath x_\mu\partial_\nu$ is again quadratic in $x.$ Then, commuting the resulting operators with one
another, we arrive at operators of the third degree in $x$, {\it et cetera.} In this way, step by step, we get all
generators of the group of arbitrary smooth transformations of coordinates
$\delta x_\mu=f_\mu (x)$, the parameters of which are coefficients of expanding of functions $f_\mu (x)$ in series by
powers of coordinates. The algebra of this group has infinite number of generators
$$L^{n_0 n_1 n_2 n_3}=-\imath x_0^{n_0} x_1^{n_1} x_2^{n_2} x_3^{n_3} \partial_\mu.$$}.
Thereby, there is a new approach where the formulation of the theory of gravity on the basis of finite-parametric
groups is essentially simpler than on the basis of the group of arbitrary coordinate transformations.

The novel approach can be based on some more elementary objects of the space--time. These elementary objects are
fundamental representations of conformal transformations group which Roger Penrose associated with
\textit{twistors}. A space--time is constructed as adjoint representation of conformal group
by means of twistors, just as pions are constructed as adjoint representations of the quark symmetry group in
the theory of strong interactions. In physics of strong interactions there are energies, wherein pions are disassociated
into elementary quarks. From this analogy it follows that a space--time also is able to be disassociated into elementary
twistors under sufficient energies. In the next sections, on examples of Einstein's General Relativity, we present the
derivation of physical laws from affine and conformal groups of symmetry and try to find a confirmation of the
program by the last observable data both from cosmology and physics of elementary particles.

\makeatletter
\renewcommand{\@evenhead}{\raisebox{0pt}[\headheight][0pt]
{\vbox{\hbox to\textwidth{\hfil \strut \thechapter . Introduction \quad\rm\thepage}\hrule}}}
\renewcommand{\@oddhead}{\raisebox{0pt}[\headheight][0pt]
{\vbox{\hbox to\textwidth{\strut \thesection .\quad Does the creation and
evolution of the Universe depend on an observer?
\hfil \rm\thepage}\hrule}}}
\makeatother
\section{Does the creation and evolution of the\\ Universe depend on an observer?}
Interpretation of classical and quantum theories, in particular, the dependence of an object of observation
on the observer,
at all times, up to the present day, has been a subject of very fierce disputes. Albert Einstein asked a question\footnote{Wheeler, J.A.:
{\it Albert Einstein (1879--1955). A Biographical Memoir by John Archibald Wheeler.} National Academy of Sciences.
Washington. D.C. (1980).}:
``{\it When a person such as a mouse observes the universe, does that change the state of the
universe}?'' Let us show here some fragments of dramatic history of the
Universe observers, including Copernicus, Tycho Brahe, Galileo, Kepler, Descartes, Newton,
Lagrange, Faraday, Maxwell, Einstein, Weyl, Dirac, Fock, Wigner, Blokhintsev, and Wheeler.

\begin{center}
\bf{Copernicus (1473 --- 1543)}
\end{center}
$\ldots$ Italy at the end of the 15th century, University in Bologna, tomes of ancient manuscripts put up for sale,
and books of theories by Pythagoras of Samos, Eudoxus of Cnidus, Heraclitus of Pontica, Aristarchus of Samos,
Hipparchus of Nicaea, Claudius Ptolemy and others, where breath-taking harmony of celestial spheres and
Divus plan of the Universe were being opened. May be there, in Bologna\footnote{In 1496, Copernicus took leave and travelled
to Italy, where he enrolled in a religious law program at the University of Bologna.},
the idea came to the young student
Nicolaus Copernicus to give up the traditional concept of the Earth as the center of the Universe. In order to reveal
the nature of visible motions of planets, Copernicus imaginatively placed his observer into the Sun and recalculated
in heliocentric system of reference the trajectories of all planets.
Copernicus' major theory was published in {\it ``De revolutionibus orbium coelestium''}
(``On the Revolutions of the Celestial Spheres''), in the year of his death, 1543, where he considered the Earth as one of
ordinary planets rotating around the Sun.
In the new Heliocentric frame of reference, the complicated character of planet motions described in the Geocentric frame
of reference by Ptolemy epicycles, becomes essentially simpler. Just the mathematical simplicity of Copernicus'
theory under description of motions of bodies of the Solar system opens the path to Kepler, Galileo, and Newton to
creation of celestial mechanics, whose perfectibility has been proved by all practice of investigation of
interplanetary environment and accuracy of predictions of celestial phenomena.

\begin{center}
\bf{Tycho Brahe (1546 --- 1601)}
\end{center}
The King of Denmark and Norway Frederick $II$, by his decree, an island Hven disposed near Copenhagen,
granted to Tycho Brahe in possession for life and also assigned great sums for building of an observatory and for its keeping. It was the
first building in Europe specially constructed for astronomical observations. Tycho Brahe's observers were fishermen
and sailors: his observatory was in existence of their duty. In the Universe of Tycho Brahe all planets, excluding the
Earth, rotated round the Sun, while the Sun, together with these planets, rotated round the Earth.
It is the very thing, that was and has been observed until the present days by all mariners. Tycho Brahe worked for his
taxpayers, measured every day the position of Mars on the celestial sphere with great, even for our time, precision.
Later on, Tycho Brahe leaved for Prague and served to the emperor Rudolf $II$ as the palace astronomer and astrologer.
The Geo--heliocentric system of the world had important advantage compared to Copernicus' one, especially after the
trial of Galileo: it did not provoke any objections of the Inquisition.

\begin{center}
\bf{Galileo (1564 --- 1642)}
\end{center}
It began with Galileo, the modern physics as a science of measurements. Galileo in his book about a would-be
dialogue between Ptolemy and Copernicus introduced a plethora of observers with their inertial systems
of reference. Coordinates of bodies and time in different systems of reference are connected by transformations of
Galileo's group. Galileo's principle of relativity of constant motion was demonstrated by using of an imaginary experiment
with systems of reference of two ships. Physical phenomena that happen inside the stationary ship do not differ from analogous
phenomena inside the ship of constant motion, relative to the first one. Galileo introduced the main kinematic characteristics of
a classical body moving rectilinear with constant velocity, and moving rectilinear with constant acceleration.
Observations for falling bodies in the
gravity field of the Earth led him to the conclusion that all bodies falling to the Earth had one and the same
gravitational acceleration. According to Galileo's principle of relativity, all inertial systems of reference
mathematically and physically are equivalent. Galileo's relativity means that all observers in one ship of the Universe
measure the same \textit{phenomena} (in our case, \textit{trajectories of planets}), as the observers in the other ship of the
Universe that moves with any constant velocity with regards to the first one. Observers of Ptolemy in non-inertial system of
reference connected with the Earth, observe \textit{trajectories of planets} which belong to another class of curves, in contrast to
the observers of Copernicus who connect their system of reference with the Sun.
In the ``Dialogue Concerning the Two Chief World Systems'' (1632) the Copernican system is compared with the traditional
Ptolemaic system\footnote{The {\it ``Dialogue''} was published in Florence under a formal license from the Inquisition.
In 1633, Galileo was convicted of ``grave suspicion of heresy'' based on the book, which was then placed on the
Index of Forbidden Books, from which it was not removed until 1835 (after the theories it discussed had been permitted in print in 1822.)
In an action that was not announced at the time, the publication of anything else he had written or ever might write was also banned.}.
Ptolemy's and Copernicus' systems physically are not equivalent. Formally, in mechanics, all systems of reference are
equivalent, and trajectories of bodies obtained in one system of reference is possible to be recalculated in the other system.
Just the same recalculating was the main matter of work and scientific achievement of Nicolaus Copernicus.
Copernicus singled out a system of reference where \textit{equations of planet motions} have the first integrals of
motion, called in Newton's celestial mechanics as a conserved energy and an angular momentum of system of bodies,
that was characteristic for central forces. In summary, analysing phenomena observed from different points of view,
we come to a conclusion that the formal mathematical equivalence of systems of reference does not imply their
physical equivalence.

\begin{center}
\bf{Kepler  (1571 --- 1630)}
\end{center}
A pupil of Tycho Brahe, Johannes Kepler got treasured data from his teacher; he recountered a trajectory of Mars on the
celestial sphere for Copernicus' system and obtained later three laws of rotations of planets around the Sun. Kepler
published these laws in his treatises {\it ``Astronomia Nova''} (1609) and {\it ``Harmonices Mundi''} (1619)
(``The Harmony of the World''), and so
promoted the establishment and development of Copernicus' doctrine of heliocentric system of reference.
It became apparent that planets did not move by circles, as Copernicus thought, but along ellipses, in one focus
of which the Sun was located. Galileo, in turn, decisively rejected Kepler's ellipses. In 1610 Galileo reported to Kepler
about the discovery of Jupiter satellites. Kepler met this message with mistrust and in his polemical paper
{\it ``Dissertatio cum Nuncio Sidereo''} (``Conversation with the Starry Massenger'') (1610)
disagreed with humour: Logically, by his means, then Jupiter must be inhabited by intelligent beings.
Why else would God have endowed Jupiter with the feature?
Later, Kepler got his example of the telescope and confirmed the existence of satellites and was
engaged in theory of lenses himself. The result was not only an improved telescope but a fundamental paper {\it ``Dioptrice''} (1611).
Kepler's system of the world meant not only to discovery laws of planet motions, but much more.
Analogous to Pythagoreans, in Kepler's mind, the world is the realization of some numerical harmony, simultaneously
geometrical and musical; revelation of structure of this harmony could take the answers to the very deep questions:
Kepler was convinced that\footnote{Kepler, Johannes: {\it The Harmony of the World.} American Philosophical Society (1997).}
{\it ``Great is our God, and great is His excellence and there is no count of His wisdom. Praise Him heavens; praise Him, Sun,
Moon, and Planets, with whatever sense you use to perceive, whatever tongue to speak of your Creator; praise Him, heavenly
harmonies, praise him, judges of the harmonies which have been disclosed; and you also, my soul, praise the Lord your Creator as
long as I shall live. For from Him and through Him and in Him are all things, ``both sensible and intellectual'', both those of which
we are entirely ignorant and those which we know, a very small part of them, as there is yet more beyond.
To Him be the praise, honor and glory from age to age. Amen''.}

\begin{center}
\bf{Descartes (1596 --- 1650)}
\end{center}
The observer of Descartes thought logically (according to Aristotle) in some priori concepts of space and time,
populating them with cosmic objects and leaving the concern of creation of these concepts to the Lord: {\it ``Cogito ergo sum''.}
In absolute space the \textit{coordinate system} is set named as the Descartes one.
He wrote about Galileo's condemnation to Mersenne\footnote{Jonathan Bennett: {\it Selected Correspondence of Descartes}. (2013).

www.earlymoderntexts.com/pdfbits/deslet1.pdf}:
``{\it But I have to say that I inquired in Leiden and Amsterdam whether Galileo’s World System was available,
for I thought I’d heard that it was published in Italy last year.
I was told that it had indeed been published but that all the copies had immediately been burnt at Rome, and that
Galileo had been convicted and fined.
I was so astonished at this that I almost decided to burn all my papers or at least to let no-one see them.
For I could not imagine that he -- an Italian and, as I understand, in the good graces of the
Pope -- could have been made a criminal for any reason except than that he tried, as he no doubt did, to establish that
the earth moves. I know that some Cardinals had already censured this view, but I thought I’d heard it said that it
was nevertheless being taught publicly even in Rome. I must admit that if the view is false then so are the foundations of
my philosophy, for it clearly follows from them; and it’s so closely interwoven in every part of my treatise that I
could not remove it without damaging the whole work.  But I utterly did not want to publish a discourse in which a single word
would be disapproved of by the Church; so I preferred to suppress it rather than to publish it in a mutilated form}''.
In {\it ``Principia Philosophiae''} (1644) there were formulated the main theses of Descartes:

$\bullet$
God created the world and laws of nature, then the Universe acted as an independent mechanism.

$\bullet$
There is nothing in the world, beside moving matter of various kinds. Matter consists of elementary particles,
local interactions of these execute all phenomena in nature.

$\bullet$
Mathematics is a powerful and universal method of studying nature, and an example for other sciences.

\begin{center}
\bf{Newton (1643 --- 1727)}
\end{center}
Isaak Newton, using Copernicus' reference system, for the first time, formulated laws of nature in the form of
differential equations and separated them from the initial data. Newton postulated the priority of laws of nature
and reduced all mechanics to mathematical equations that are independent of the choice of initial data
(and inertial reference systems). They predict evolution for all time of coordinates of a particle, if its initial
position and initial velocity are set. For Newton's observer, to explain the world in terms of classical
mechanics means to solve Newton's equations with initial data (Cauchy's problem). Here it is appropriate to remember
Laplace's colorful expression\footnote{Laplace, Pierre Simon:
{\it A Philosophical Essay on Probabilities}. John Wiley \& Sons. (1902).}:
{\it ``We may regard the present state of the universe as the effect of its past and the cause of its future.
An intellect which at a certain moment would know all forces that set nature in motion, and all positions of all
items of which nature is composed, if this intellect were also vast enough to submit these data to analysis,
it would embrace in a single formula the movements of the greatest bodies of the universe and those of the tiniest atom;
for such an intellect nothing would be uncertain and the future just like the past would be present before its eyes''.}
{\it``Philosophiae Naturalis Principia Mathematica''}
(``The Mathematical Principles of Natural Philosophy''\footnote{Newton, I.:
{\it The Mathematical Principles of Natural Philosophy}. Encyclopedia Britanica (1952).})
of Newton, first published in 1687 year, absorbed all
previous human experience of observations of motions of celestial and earth matter, demonstrated the same power of
clarity, accuracy and efficiency of scientific methods of natural science as Euclidean principles in geometry.
The weak place in Newton's gravitation theory was, by opinions of scientists of that time, the absence of description
of nature of the invisible force, that was able to act over vast distances.
Newton stated only mathematical formalism, and left the questions of cause of gravitational
attraction and its carrier open. On this occasion, Newton stated: ``hypotheses non fingo'', that became his famous expression.
For scientific community educated on Descartes' philosophy, the approach
was unusual and challenging, and only triumphal success of celestial mechanics in the 18th century forced physicists
temporarily come to terms with the Newtonian theory. Physical basic concepts of the theory of gravitation were cleared only
more than two centuries later, with appearance of the General Relativity. Newton's theory absolutized sharp differences
of concepts of time, space, and matter, and the universal law of conservation of energy seemed to gain perpetual
persistent status in philosophy. Newton introduced absolute space and time. They are the same for all observers.
The first physical theory was constructed by Newton, based upon the name of its book, by analogy with Euclidean
``Principles''. Theological manuscripts of Isaak Newton tell us that Newton searched justification of principles of logical
constructions of the first physical theory and concepts of absolute space and time, and, hence, absolute units of their
measurements, in arduous discussions with gnoseology officially accepted in Trinity College where he was
a professor\footnote{Gnoseology, officially accepted in ``Trinity College'' and rejected by Newton, affirmed that
a studied object must possess some realities, each described by their self consistent logics of Aristotle.
According to the theory, the existence of two complementing each other confirmations was possible about one and the same
object of cognition, under condition, that these confirmations refer to different realities of this object
(John Meyendorff: {\it Byzantyne Theology. Trends and Doctrinal Themes.} N.Y. (1979); G. G. Florovsky:
{\it Eastern Fathers of the 4th century.} Inter. Publishers Limited. (1972).}. Newton's mechanics assigned the structure
of mathematical formulation of modern fundamental physical theories, including Einstein's theory of gravitation, the
Standard Model of elementary particles, {\it et cetera}, where equations as laws of nature were put as foundation.

\begin{center}
\bf{Lagrange (1736 --- 1813)}
\end{center}
Joseph--Louis Lagrange re-wrote Newton's differential equations of motion in a covariant form, introducing generalized
coordinates. He noticed in the Preface to his treatise ``The M$\acute{\rm e}$canique Analytique''\footnote{Lagrange, Joseph-Louis:
{\it M$\acute{e}$canique Analytique.} Cambridge University Press. (2009).}:
``{\it The reader will find no figures in this work. The methods which I set forth do not require either constructions or
geometrical or mechanical reasonings: but only algebraic operations, subject to a regular and uniform rule of procedure}''.
Lagrange was one of the creators of the calculus of variations, he derived the so-called Euler -- Lagrange equations
as the conditions for the extremum of functionals.
Using the principle of the least action, he obtained equations of dynamics.
He also extended the variational principle for systems with holonomic constraints, using the so-called method of Lagrange multipliers.
Nonholonomic dynamics will be discovered later, only in the 20th century.
The mathematical formalism of the calculus of variations will be necessary for theoretical physicists to formulate
the equations of the gravitational field, at first, in the Lagrangian covariant form, then as generalized Hamiltonian dynamics.

\begin{center}
\bf{Faraday (1791 --- 1867)}
\end{center}
The first steps in creating the modern relativistic physics were taken by Michael Faraday. The great amount of
scientific discoveries belong to him, such as a laboratory model of the electric motor that, in future, changed the life of the modern civilization.
With impressive sequencing, Faraday demonstrated, by experiments, and developed his concept of field nature of matter
and unity of all physical fields of nature -- guiding ideas of physics of the 20th century where all particles are treated as
excitements of physical fields. Faraday created the field concept of the theory of electricity and magnetism\footnote{In 1938 year,
in an archive of the Royal Society there was found Faraday's letter, written in 1832 year, which he asked to open after 100 years,
where he predicated of electromagnetic nature of light (let us remember, that Maxwell was born in 1831 year).}.
Before him, the presentation of a direct and instantaneous interaction between charges and currents through empty space dominated.
Faraday experimentally proved that matter carrier of this interaction is the electromagnetic field.
The fact that Faraday was unaware of the Newton mathematical formalism in mechanics was not a barrier on the way of the experimentalist,
but helped him
to formulate new basic concept of modern physics and  predict the field nature of matter and unity of fields of nature, which
physicists discovered in the 20th century.
Remember that modern physical theories are based on the concepts of field theory, not Newton's mechanical ones.

\begin{center}
\bf{Maxwell  (1831 --- 1879)}
\end{center}
Maxwell had to ``dress'' (as Heinrich Rudolf Hertz picturesquely noted)
Faraday's theory into aristocratic clothes of mathematics.
The first paper of Maxwell on the theory of electromagnetic field is entitled:
``On Faraday's lines of force''.
Maxwell set a goal of translating the basic Faraday's treatise
``Experimental Researches in Electricity''\footnote{Faraday, Michael: {\it Experimental Researches in Electricity}.
J.M. Dent \& Sons. Ltd. London. (1914).}
into the language of mathematical formulae. The Maxwell theory turned out to be  universal
in electromagnetic phenomena as Newton's theory in celestial phenomena. Electrodynamic formulae, written down in
the language of mathematical field theory, became to live their own life, displaying their symmetric structure.
The observer of Maxwell discovered the dependence of description of results of experimental measurements of
electromagnetic phenomena from the definition of measured values in the field theory from the choice of a
standard of their measurement.
In Preface of his ``Treatise on Electricity and Magnetism''\footnote{Maxwell, James Clerk:
{\it Treatise on Electricity and Magnetism}. Clarendon Press, Oxford (1873).}
Maxwell wrote:
``\textit{The most important aspect of any phenomenon from a mathematical point of view is that of a measurable quantity.
I shall therefore consider electrical phenomena chiefly with a view to their measurement, describing the methods of measurement,
and defining the standards on which they depend}''.
In Preliminary of his book he continued:
``\textit{Every expression of a Quantity consists of two factors or components. One of these is the name
of a certain known quantity of the same kind is the quantity to be expressed, which is taken as a standard of reference.
The other component is the number of times the standard is to be taken in order to make up the enquired quantity.
The standard quantity is technically called the Unit, and the number is called the Numerical Value of the quantity}''.
The Maxwell theory, its symmetries and concepts are prototypes of all working relativistic quantum
theories of the 20th century where all elementary particles are interpreted as oscillatory excitations of corresponding fields.
The scientific works of Maxwell were not appreciated by his contemporaries. Only after Heinrich Hertz's experimental proof of the
existence of electromagnetic waves predicted by Maxwell the theory of electromagnetism got the status of {\it consensus omnium}.
It happened only ten years after Maxwell's death.

\begin{center}
\bf{Einstein  (1879 --- 1955)}
\end{center}
Geometries of Lobachevski and Riemann, field theory of Faraday and Maxwell disturbed confidence to the absolute
space and time, and the 20th century became a century of relativity and principles of symmetries of quantized fields of matter.
Einstein is a creator of two theories of relativity. The first one of these theories is the Special Relativity.
It is based on the group of relativistic transformations of Maxwell's equations obtained by Lorentz and Poincar$\acute{\rm e}$.
The Special Relativity is an adaptation of Newton's classical mechanics to relativistic transformations. The generally accepted form of
the Special Relativity is the version of Einstein and Minkowski which opened a path to creation of modern quantum field theory.
Any experimentalist of high energy physics knows that life-time of unstable particle, measured in the laboratory frame
of reference, differs from life-time of the same particle measured in a frame moving together with the particle.
If the particle is put into a train moving along the station, the Driver in the train and the Pointsman in the station
measured different life-times of the particle. These times are connected by relativistic transformations obtained by
Einstein.
\begin{figure}[hp]
{\vbox{
\psfig{figure=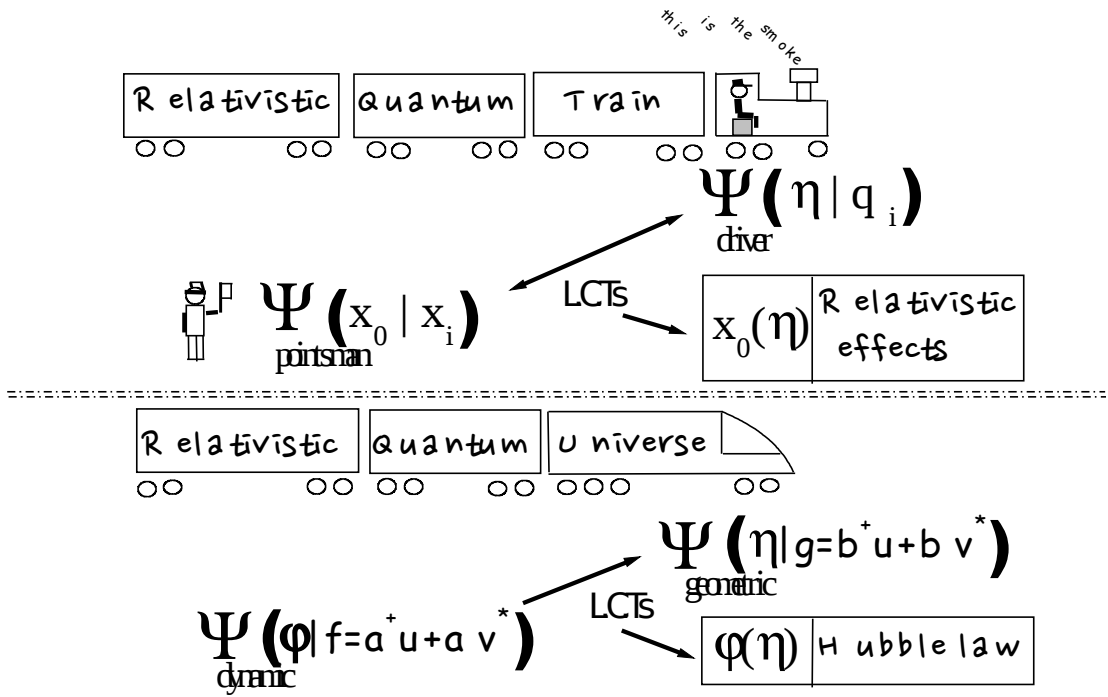, width=6in, bbllx=0pt, bblly=0pt,
bburx=430pt, bbury=270pt,
clip=}}}
\par
\caption{\small
At the top of the Figure a relativistic train is depicted with an unstable particle, moving with
velocity of 200 000 km per sec passing the Pointsman. If life-time of the particle, measured by the
Driver, is 10 sec, then life-time of the same particle, measured by the Pointsman, is equal to
$10/\sqrt{1-(2/3)^2}\simeq$ 14 sec. In the quantum field theory, that describes the process of creating
a particle, these times are complementary, not  contradictory. The Driver, being created together with
the particle, could not be a twin to the Pointsman. The first measures the length of geometric interval
(10 sec), and the second -- dynamical parameter of evolution in the space of events (14 sec).
At the bottom of the Figure a Universe is pictured, where a cosmological parameter of evolution $\varphi$
plays the role of dynamical parameter of evolution in the space of events, and the conformal time $\eta$
plays the role of the length of geometric interval. One and the same observer has two different
measurement procedures  of dynamical parameters of evolution (redshift) and the length of geometric
interval (distance to cosmic objects). These two observers (the Pointsman and the Driver) of the
relativistic object in \textit{quantum geometrodynamics} do not contradict, but complement each other.
}
\label{statics1}
\end{figure}
From the Newton mechanics point of view, these two different statements
about life-time of the same particle are in contradiction. To avoid it, according to Trinity College doctrine, one should
confirm that the particle has one reality for the Driver, and another -- for the Pointsman; then one should
construct two noncontradictory mechanics: the mechanics of the Driver and the mechanics of the Pointsman and the relation between
them as a new element of the theory. Just by the very way of existence of two realities of one and the same particle,
the development of relativistic quantum field theory went on.
Einstein laid the foundation to this development, who understood that the Lorentzian symmetry of the theory of Maxwell
meant equality of time and space coordinates of a relativistic particle. Such equality supposes that time and space
form the unified space-time named as the Minkowskian space of events. Hermann Minkowski
proclaimed\footnote{Minkowski, Hermann.: {\it Raum und Zeit}, Physikalische Zeitschrift. {\bf 10}, 75 (1908).}:
``{\it The views of space and time which I wish to lay before you have sprung from the soil of experimental physics, and therein lies their strength.
They are radical. Henceforth space by itself, and time by itself, are doomed to fade away into mere shadows,
and only a kind of union of the two will preserve an independent reality}''.
Under its motion in this space, the particle depicts a world line, where the geometric interval plays a role
of the parameter of evolution. The existence of two times of one and the same particle supposes, that
for the complete description of motion of the particle in the world space of events, one needs
as minimum, two observers to
measure its initial data (see. Fig. \ref{statics1}).
One of them is at rest, the other is co-moving with the particle. The first one measures the time with his watch as
a variable of the world space of events,
and the second one measures the time with his watch as a geometric interval on the world line of the particle
in this space of events. A new element of the theory appeared -- an equation of constraint of four components
of the vector of momentum, one of which is the energy. A solution of this equation of constraint for a particle at rest
$E=mc^2$ lies in the base of the atomic energetics. The second of Einstein's theories generalizes the field
paradigm of Faraday to gravitational interactions and it is named the General Relativity. The first attempts of
generalizing of Faraday's field paradigm to other interactions were undertaken at the beginning of the last century.
The searching of principles of symmetry was Einstein's underlying concept that differed him from other researchers.
The basic ideas of the General Relativity were prepared by all history of development of non-Euclidean geometry of the 19th
century by Gauss\footnote{Gauss, C.F.: {\it General Investigations of Curved Surfaces of 1827 and 1825}. Princeton University (1902)},
Bolyai\footnote{Bolyai, J.: {\it The Science Absolute of Space. Independent of the Truth or Falsity of Euclid's Axiom XI
(which can never be decided {\rm a priori}}. Austin, Texas (1896).},
Lobachevsky\footnote{Lobachevski, N.I.: {\it Complete Collected Works}. I--IV. Kagan, V.F. (Ed.).
Moscow--Leningrad (1946)--(1951).}, Clifford\footnote{Clifford, W.K.: {\it Mathematical Papers}. MacMillan, New York--London (1968)},
Riemann\footnote{{\it Bernhard Riemann's gesammelte mathematische Werke und
wissenschaftlicher Nachlass.} Teubner, Leipzig (1876).}.
Einstein declared that observational results of his
theory did not depend on parameters of a very wide class of coordinate transformations. That is why Einstein
named his theory the General Relativity. Einstein's dynamical equations are derived from Hilbert's variational principle
delivered in Hilbert’s report ``Foundations of Physics'' presented on November 20th, 1915 to
the G\"ottingen Mathematical Society\footnote{In addition, Hilbert first formulated the theorem that was later referred
to as the second Noether theorem.
This theorem leads to the interpretation of general coordinate
transformations as gauge ones and, therefore, to all consequences concerning both a decrease in the number
of independent degrees of freedom and the appearance of constraints imposed on initial data.
The first Noether theorem states that any differentiable symmetry of the action of a physical system has a corresponding conservation law.

Noether, E.: {\it Invariante Variationsprobleme.} Nachr. D. K\"onig. Gesellsch. D. Wiss. Zu G\"ottingen, Math-phys. Klasse. 235 (1918).
}.

\begin{center}
\bf{Weyl (1885 --- 1955)}
\end{center}
The main goal of theoretical physics is to establish several physical principles to explain all observable
effects just as a few Euclidean axioms and logical laws make it possible to prove many theorems in geometry.
In modern physics, such fundamental principles are principles of symmetry. The following statement
of Hermann Weyl is worth being reminded\footnote{
Weyl, H.: {\it Symmetry.} Princeton University Press, Princeton. (1952).}:
{\it ``What we learn from our whole discussion and what has indeed become a guiding principle in modern mathematics is this lesson:
Whenever you have to do with a structure-endowed entity $\Sigma$ try to determine its group of automorphisms, the group of those
element -- wise transformations which leave all structural relations undisturbed. You can expect to gain a deep insight into the
 constitution of $\Sigma$ in this way''.}
From this viewpoint, transformations of reference frames form an automorphism group in mechanics, while the equations
of motion derived by variation of action are invariant structure relationships.
The guiding principle of modern physical theories is to define the transformation groups of reference
frames (treated as manifolds of initial data) that preserve the equations of motion. The Galileo group in
Newtonian mechanics and the Poincar$\acute{\rm e}$ group in the Special Relativity, groups of classification of elementary particles,
gauge groups of symmetries led to equations of constraints of fields and their initial data.
Weyl proposed
a principle of scale symmetry of laws of nature: according to this proposition gravitation equations
are independent of choice of measure units and differed from the General Relativity ones. In Weyl's geometry
lengths of objects under motion over a closed contour are not integrable, and non-integrability
is connected with the presence of electromagnetic field.

\begin{center}
\bf{Fock (1898 --- 1974)}
\end{center}
Fock was the first to introduce a tangent space of Minkowski into the General Relativity. Now, all observers in the Universe are able
to measure two parameters of evolution: a proper time interval, measured in the tangent space, and a parameter of evolution in the
field space of events. The same two observers (we called them a ``Pointsman'' and a ``Driver'') could be also introduced into the General
Relativity. In the Special Relativity hardly  anybody is interested in what the fate of proper time of a particle should be, measured by a
``Driver'', after the causal quantization, which introduced the vacuum in the space of events, with the help of
changing of the operator of creation of a particle with negative energy to the operator of annihilation of a particle with
positive energy. We should see further that as a result of such change on the world line of a particle, the positive arrow and
absolute point of the beginning of geometrical interval of time measured by a ``Driver'', appear. Such quantum anomaly of
geometrical interval, in fact, means that if there is no particle there is no world line of a particle either nor
the interval on this world line. Fock, solving the problem of particles in the General Relativity, introduced a privileged,
the so-called, harmonic reference frame for solving Einstein's equations.
Fock understood that despite the formal point of view, all reference frames are equivalent, but investigating
concrete problems, one should choose  the most acceptable reference frame.

\begin{center}
\bf{Dirac (1902 --- 1979)}
\end{center}
Dirac's observer solves problems of consequent probability interpretation of the wave function that
satisfy Dirac's equations, and stability of the quantum system via the secondary quantization and
filling all states with negative energies (Dirac's sea).
A solution of the Hamiltonian constraint
both in the Special Relativity and the General Relativity has two signs. The negative sign of energy in the Special Relativity
was associated with the existence of anti-particle -- positron by Dirac. Theorists,  headed by Dirac and Fock,
solved the problem  of negative energy via two quantizations of a particle: primary one, when
generalized coordinates and conjugated momenta became operators in the equation of constraint, acting to the wave
function, which was identified with Faraday-type field; and secondary one, when the same Faraday's field, interpreted as a
sum of operators of creation of a particle with positive energy (+) and annihilation of a particle with positive
energy (--) as well. The most important element of Dirac's theory is a vacuum as a state with minimum of energy, which disappears
if the operator of annihilation acts to it.
This quantization in the modern field theory was called causal quantization, and the theory got the name: the quantum field
theory.
Dirac, following to Weyl, introduced an observer \cite{Dirac1973int-1}: ``\textit{There are reasons
for believing that the
gravitational constant varies with time. Such a variation would force one to modify Einstein's theory
of gravitation. It is proposed that the modification should consist in the revival of Weyl's geometry,
in which lengths are non-integrable when carried around closed loops, the lack of integrability being
connected with the electromagnetic field}''.
Dirac formulated a new action principle \cite{Dirac1973int-1}: ``\textit{A new action principle is set up,
much simpler that Weyl's, but requiring a scalar field function to describe the gravitational field,
in addition to the $g_{\mu\nu}$}''. This scalar field function got the name a \textit{dilaton}. According to
Weyl and Dirac, a standard of measurement of length is chosen being expanded with measured lengths together with its unit.
If the standard is also expanding, the results of measurements of all lengths do not contain the cosmological scale factor.
So, for an observer with the relative standard, the Universe is static, and all masses are proportional
to the cosmological scale factor associated with the dilaton, and become vanishingly small at the beginning of
the Universe origin. The modern cosmology, actually, uses a double standard under description of cosmic evolution
of photons emitted by massive matter from the far cosmic object: \textit{absolute} (world interval) and
\textit{relative} (conformal interval).
Friedmann was the first who used geometric intervals with relative units of measurements
(a coordinate distance to a cosmic object and a conformal time of emitting of a photon), to solve his equations.
These relative variables are used in the observable cosmology for description of motion of cosmic photons. They
left their ``footsteps'' in the form of spectral lines in photographic plates put in the telescope.
Comparing these photographic plates with those where spectral lines left photons of the Earth atoms at the present time,
astrophysicists measure redshifts of spectra of energies of cosmic atoms. Energy spectrum of atoms, as known from quantum
mechanics, is defined by masses of particles of which these atoms are combined. An acceptance of a \textit{relative} standard
leads to changing of masses, and an \textit{absolute} standard --- to changing of geometric intervals.
Dirac's observer, as, in olden times, Copernicus' observer,
could choose himself a standard of measurement, and define, according to Copernicus, what standard gives the most simplest
classification of the observational data. All standards in Einstein's theory are equivalent from the formal mathematical point of view,
in the same way, as the frames of reference of Ptolemy and Copernicus in the celestial mechanics are, formally, equivalent.
However, the phenomena are described by solutions of the equations of motion, where the initial data are demanded.
For definition of the initial data in the General Relativity, it is necessary to pass to conformal variables.
This fact was known to Andre Lichnerowicz yet in the year 1944\footnote{Lichnerowicz, Andre: {\it L'integration des equations
de la gravitation ralativiste et le probleme des $n$ corps}. J. Math. Pures Appl. {\bf 23}, 37 (1944).}.
The transition to conformal variables in cosmology means
recalculating of all observational data from an absolute unit to relative ones, multiplying these data to the cosmological
scale factor in power equal to corresponding conformal weights. Just this recalculating demonstrates, as we shall see further on,
describing the data on Supernovae, physical nonequivalence of standard variables and the conformal ones.
\textit{Standard cosmology}, recalculated in terms of conformal variables is called \textit{Conformal cosmology}.
In particular, in \textit{Conformal cosmology} measured lengths are always longer than in
\textit{Standard cosmology}. Data on Supernovae correspond to another equation of state of matter,
identified to Casimir energy in an empty Universe, as we will show in the next Chapters.

\begin{center}
\bf{Wigner (1902 --- 1995)}
\end{center}
Eugene Wigner showed that the principle of relativity of velocities offered by Copernicus for description of motion of planets,
generalized by Poincar$\acute{\rm e}$ and Einstein for motion of relativistic particles, turns out to be the principle of classification
of all observable and measurable physical objects. As a result, physicists have got the classification of states of a particle according to
its mass and spin. Such classification lays the basis of the quantum field theory. At the present time, physicists come to a
conclusion to include the Universe itself to this scheme.
Wigner explained \cite{WignerProc}:
{\it ``The world is very complicated and it is clearly impossible for the human mind to understand it
completely. It has therefore devised an artifice which permits the complicated nature of the world to
be blamed on something which is called accidental and thus permits him to abstract a domain in which
simple laws can be found. The complications are called initial conditions, the domain of regularities,
laws of nature. Unnatural as such a division of the world's structure may appear from a very detached
point of view, and probable though it is that the possibility of such a division has own limits,\footnote{The artificial nature
of the division of information into ``initial conditions'' and
``laws of nature'' is perhaps most evident in the realm of cosmology. Equations of motion which purport
to be able to predict the future of a universe from an arbitrary present state clearly cannot have an
empirical basis. It is, in fact, impossible to adduce reasons against the assumption that the laws of
nature would be different even in small domains if the universe had a radically different structure.
One cannot help agreeing to a certain degree with E.A. Milne who reminds us ({\it Kinematic relativity}
Oxford Univ. Press, 1948, page 4) that, according to Mach, the laws of nature are a consequence of the
universe. The remarkable fact is that this point of view could be so successfully disregarded and that
the distinction between initial conditions and laws of nature has proved so fruitful.} the
underlying abstraction is probably one of the most fruitful ones that human mind has made. It has made
science possible''.}
An observer of Wigner divides three levels of development of a physical theory:
\textit{phenomena, laws of dynamics, and principles of symmetry} and prioritizes to the principles of
symmetry, whereof the laws of dynamics and the description of nature can be deduced. It  is wonderful that these
principles have in their base the symmetry of the very same \textit{accidental} initial data, of which the laws of nature
are independent.

\begin{center}
\bf{Blokhintsev (1908 --- 1979)}
\end{center}
Dmitry Blokhintsev's papers make us reflect about eternity of knowledge. How do the terms and concepts arise
and how far are they absolute? How can we divide absolute and relative in learning of the world? What is
``physical reality?'' How can one define ``boundaries of applicability of concepts''.
Especially, Blokhintsev's words amaze us about impossibility of simultaneous, with arbitrary accuracy, measurement of a coordinate
and conjugated momentum of a particle: impossibility means not restriction of our cognition, but limitation of the
terms themselves (a coordinate and a corresponding momentum).
Blokhintsev said\footnote{Private communication.}:
``\textit{It is wrong to think that modern physical experiment has
insufficient accuracy for measurements of ``true'' simultaneous values of momentum and coordinate of a microparticle.
On the contrary, it is accurate enough to prove that for the microparticles the pair does not exist in nature}''.
Separation of relative values from the objective ones and
definitions of boundaries of their applicability, according to the quantum theory, gives amazing predictions of new effects
and values of physical magnitudes that describe these effects. In such a manner, Dmitry Blokhintsev as a physicist can
``explain in layman's terms'' values of physical magnitudes and predict fine effects, such as the Lamb shift discovered only 10
years later and inaugurated the birth of the quantum electrodynamics\footnote{In 1938 year Blokhintsev
delivered a lecture at the seminar on theoretical physics in Lebedev Physical Institute (Moscow), where he showed, that
taking into consideration an interaction of electrons with the field of radiation, it is able to lead to
shift of their energetic levels.}.
Blokhintsev's observer treats a transition from a classical particle to a quantum one as quantization of initial data, but
not of dynamical variables. Let us recall that Blokhintsev associated a set of all possible free initial data in quantum field
theory with a statistical ensemble. The existence of the ensemble of quantum states of initial data is a reason of
probabilistic interpretation of a wave function. For problems of cosmological particle creations,
Blokhintsev's quantization of initial data turns out more productive,
from the point of view of classification of observational data
than quantization of dynamical variables, because it forces us to search the complete set
of free initial data as constants of motion in a definite frame of reference.
In particular, in cosmology, where the particles are described as oscillators,
squeezed by a cosmological scale factor, it is possible to quantize only the ensemble of initial data, as
constants of motion, which are established via Bogoliubov's transformations. These transformations can be fulfilled
at the level of classical theory, then it is possible to obtain the complete set of free initial data. For
the theory with the quantized initial data, the Bogoliubov transformations provide a quantitative description of the
Universe creation (mysterious for the variables without initial data), identified as the Big Bang.

\begin{center}
\bf{Wheeler (1911 --- 2008)}
\end{center}
According to Wheeler's geometrodynamics, there are the same realities -- dynamical and geometrical -- in the modern
theory of gravitation, where the Hubble's law appears as a pure relativistic relation between a cosmological scale factor played role
as a time-like variable in the field space of events introduced by Wheeler in 1967 and a geometric
interval of time on the world hypersurface in this space of events. A choice of a relative standard, as we already marked,
transfers a cosmic evolution to masses, changing a fundamental parameter of Einstein's theory -- Planck's mass into dynamical
variable in Wheeler -- De Witt's space of events, which has accidental initial data, as well as any variable in Newton's
mechanics. So, the relative standard deprives Planck's mass of a fundamental status in Einstein's theory, as well as Copernicus'
relativity deprived a fundamental status of the Earth position in Ptolemy's frame of reference. It could seem, that in the General Relativity
there is no absolute value in natural units of measurement. The scale symmetry of the classical theory can be
broken via its quantization and normal ordering of operators, leading to Casimir's energy and Casimir's condensate.
A secondary causal quantization of universes with a postulate of existence of vacuum as the state of minimal energy,
in full analogy with quantum field theory, leads to a cosmological creation of universes and to a positive arrow of a geometric interval
of time. This time has the absolute origin that we, the Universe's inhabitants, conceive as the point of the Universe's creation,
with the equation of state established from the data obtained with the Hubble's telescope. In relativistic cosmology,
the positive arrow of the geometric time and its origin are consequences of quantum vacuum postulate in the field space of events,
{\it id est} they are evidences of quantum nature of our Universe.
So, answering the question of what there was until the creation of the Universe, we are
able to say, after Augustine of Hippo and Immanuel Kant, that there was not the time itself.
Only in the limit of infinitely large Universe
and infinitely large energy of motion of the Universe in the field space of events, the theory of early Universe becomes the
Einstein classical theory and modern quantum field theory of elementary particles,
accessible to our classical comprehension. The quantum theory confutes the Laplace's point of view:
at some time, having the knowledge of locations and velocities of all particles, simultaneously, in the Universe.
Wheeler asked\footnote{Our free translation from {\it ``Centenario di Einstein'' (1879--1979).
Astrofisica e Cosmologia Gravitazione Quanti e Relativit\`a.}
Giunti Barb\`era. Firenze (1979).}:
{``\it In short, whether the Universe is, in some strange sense, a sort of ``self-induced circuit?''
Giving rise in some limited stage of its existence observers--members, whether the Universe acquires, in turn,
through their observations that tangibility, which we call reality?
Is not the mechanism of its existence?
And whether from these reasonings to deduce the nature and necessity of the quantum principle?''}
Wheeler continued to discuss:
{\it ``Today textbooks on quantum mechanics tells us how to proceed in
situations where one observer is involved.
Scientific articles have dealt with the idealized
experiments in the spirit of Einstein, Podolsky and Rosen, which involved two observers.
We do not have an
idea of what to do in an extreme situation, when a very large number of
observers -- the participants and a very large number of observations play a huge role''.}


\vskip 1cm

\begin{center}
\bf{A contemporary observer (1973 --- till the present time)}
\end{center}
We marked the date of birth of a contemporary observer by 1973 \footnote{In 1973 year the famous Dirac's
paper (Dirac, P.A.M.: \textit{Long range forces and broken symmetries}. Proc.
Roy. Soc. London. {\bf A 333}, 403 (1973)) was appeared.}, to bound him with ideas, models, and theories
proposed till 1973 \footnote{It turned, by the way, exactly 500 years, --
half of the Millennium, -- if you count the time interval from the year of birth of Copernicus.} by the
\textit{giants} of physics and mathematics, \textit{on whose shoulders he climbs}, to get the simple classification
of all data on measurements and observations of physical and cosmological values obtained to the present time
(2013). We deprive the observer of having the ability of using uncommitted in time ideas of his contemporaries, and,
moreover, of having his own ideas.
We assume that our observer knows everything from Copernicus to Dirac, including three levels of classification of
physical data marked by Wigner: \textit{phenomena, laws of dynamics and principles of symmetry}, to the extent to which the
knowledge has
been outlined above. At all these levels, adequate choice of the reference frame of
initial data and standards of their measure can essentially simplify the classification of physical
data and, therefore, facilitate further development of our knowledge of the Universe and theories describing
dynamics of processes, that occur therein. To explain the dynamics of the processes in the Universe,
contemporary observer has the ability to use quantum theory of \textit{phenomena}. The quantum theory of
\textit{phenomena}, described by unitary irreducible representations of a group of
\textit{symmetry} of initial data, may be much simpler than the classical theory that is based on solutions of
classical \textit{dynamical laws}, {\it id est} equations of motion. In other words, to describe \textit{phenomena},
a classical observer has \textit{laws of dynamics} as equations of Newton, Maxwell, Einstein, equations of the
Standard Model of elementary particles and modern unified theories; at the same time, a contemporary observer has
unitary irreducible representations of finite-parametric groups of
\textit{symmetry} of initial data for description
of the same \textit{phenomena}. Just therefore, we have a unique opportunity to further build our
classification of data using unitary irreducible representations of these groups without applying to the
classical \textit{laws of dynamics} as the assumptions of the physical theory, or deduce classic
\textit{laws of dynamics} from the first principles of symmetry found up to 1973.

\newpage

\makeatletter
\renewcommand{\@evenhead}{\raisebox{0pt}[\headheight][0pt]
{\vbox{\hbox to\textwidth{\hfil \strut \thechapter . Introduction \quad\rm\thepage}\hrule}}}
\renewcommand{\@oddhead}{\raisebox{0pt}[\headheight][0pt]
{\vbox{\hbox to\textwidth{\strut \thesection .\quad Content
\hfil \rm\thepage}\hrule}}}
\makeatother

\section{Contents}

Since the time of Newton's physical theory, the dynamic processes have been described by solutions of differential equations
and some unique initial data. The initial data are set by fitting of
the results obtained via the observations or direct measurements with devices that are
identified with a certain frame of reference\footnote{Any measurement or observation suggests the existence of two selected
frames of reference -- the first is connected with the instruments of measurement, while the second (co-moving)
is associated with the object whose parameters are measured. In particular, in modern cosmology the rest frame
of instruments uniquely associated with the Earth, and the accompanying reference system of the Universe --
with the Cosmic Microwave Background radiation.
The rest frame of reference differs from the co-moving one by non-zero dipole component of temperature
fluctuations $10^{-3}$.}.
The question is what are the criteria and principles for the selection of the initial data for the Universe?
To answer these questions, in this paper we consider a cosmological model in which the initial state of the Universe
is given by vacuum of particles. To such a state of vacuum the concept of ``temperature'' can not even be applied
as it arises when describing the motion of the particles after the creation from the cosmological vacuum.

In this case, the starting point of the observable Universe is the instance of the creation of primary
particle-like irregularities, the size of which is determined by their masses in the Standard Model of
elementary particles. Since the Compton wavelength of the particle can not be larger than the horizon of the
Universe at the time of its creation, the very instance of creation of a particle can be estimated by equating its
Compton wavelength and the horizon of the Universe. Thus, the initial value of the cosmological
scale factor is given by the condition that the Compton wavelength of the primary particles is equal to the horizon
of the Universe (or equal its mass to Hubble's parameter) at the time of creation of the primary cosmological
particles from the vacuum. This condition follows from the uncertainty principle, which limits changing of
energy in the Universe by finite time of its existence.

In the Standard Model of elementary particles candidates for the role of the primary particles are massive vector
particles and minimally-interacting scalar particles. Cosmological creation of these primary particles is
described in the Conformal model of a homogeneous Universe, which differs from the Standard
cosmological model in the principles of relativity of initial data, relativity of time, and
relativity of units that should be set forth below.

Let us outline conclusions of this Chapter.

\begin{enumerate}
 \item
The initial value of the cosmological scale factor given by the condition of coincidence of
Compton wavelength of the primary particles with the horizon of the Universe (or their mass with
the Hubble parameter) at the instance of cosmological creation of the primary particles from the vacuum.
This condition follows from the uncertainty principle, which limits the change of energy in the
Universe by its finite time of existence.

\item  The principle of relativity of time is the second difference from the accepted Standard
     cosmological model. Recall that the relativity of time in Special Relativity suggests that
     time coordinate is also a degree of freedom of the particle, so that a complete set of
     degrees of freedom forms the space of events introduced by Minkowski. Instead of a single
     Newton's time there are three: the time-like variable in the space of events, time-geometric interval
     on the trajectory of the particles in the space of events, and a coordinate evolution parameter,
     reparametrization of which leads to a Hamiltonian constraint of momenta in Minkowski space.
    The resolution of this constraint regarding the time-variable momentum gives the energy of the particle in the
     Special Relativity (namely, the one used in the calculation of the energy in nuclear reactions).
     The primary and secondary quantization of the Hamiltonian constraint in the Special Relativity allows
     us to formulate a quantum field theory with the postulate of the vacuum as the state with the lowest energy.
    In this book the quantum model of the Universe is obtained in the theory of gravity as a generalization
    of the above construction of the quantum field theory from the Special Relativity
    using primary and secondary quantization of the Hamiltonian
    constraint with the postulate of a vacuum.

\item The third difference from the accepted Standard cosmological model is the Weyl's
     relativity of units (or scales) of measurements, which means that the physical devices
     measure a dimensionless ratio of the interval of time or space to the measurement unit,
     defined by the standard mass. Weyl's measured quantities (mass, density, temperature and so on)
     associated with the measured values of the Standard cosmology by multiplying the latter on a
     cosmological scale factor in power defined by conformal weight of each of these units.
{\it Standard cosmology, expressed in terms of measured values, is called the Conformal
cosmology.}
Since in Conformal cosmology a distance identified with measurable one is always greater than
the distance in the Standard cosmology, the same Supernovae data correspond to different equations of states
in Conformal cosmology and the Standard cosmology.
\end{enumerate}
Thus, the above-mentioned principles of relativity explain the origin of all matter in the
Universe as a decay product of the primary scalar and vector bosons created from
vacuum, and the arrow of time as an inevitable consequence of the primary and secondary quantization
of the Hamiltonian constraint.
To describe such creation of particles from the vacuum
we construct the creation operator of evolution of the quantum Universe
as a joint and irreducible unitary representation of the affine and conformal symmetry groups.
Expected average values of this operator between the states of matter (according to their classification
of the Poincar$\acute{\rm e}$ group in the tangent Minkowskian space) are used to describe
modern experimental and observational data.

\chapter{Initial data and frames of reference}
\renewcommand{\theequation}{2.\arabic{equation}}
\setcounter{equation}{0}
\section{Units of measurement}
\makeatletter
\renewcommand{\@evenhead}{\raisebox{0pt}[\headheight][0pt]
{\vbox{\hbox to\textwidth{\hfil \strut \thechapter . Initial data and frames of reference \quad\rm\thepage}\hrule}}}
\renewcommand{\@oddhead}{\raisebox{0pt}[\headheight][0pt]
{\vbox{\hbox to\textwidth{\strut \thesection .\quad Units of measurement
\hfil \rm\thepage}\hrule}}}
\makeatother
All units of measurement can be expressed in terms of three basic: units of length, mass and time.
Two fundamental physical theories of the last century have reduced the number of basic units from three to one.
In the Special Relativity, it was found that there is a fundamental limiting speed of propagating of physical processes
which is equal to the speed of light in the vacuum $c$.
In the quantum theory, it was appeared a new fundamental constant -- the quantum of action $\hbar$.
If we choose a system of units in which $c=1$ and $\hbar=1$ all three main units -- length, mass and time
can be expressed via any one of them.
Several examples of various physical quantities expressed in terms of mass $M$ are shown in Table
(\ref{units})
\cite{NarlikarViolent}.
\begin{table}[ht]\label{units}
\centering
\begin{tabular}{|c|c|c|}
\hline
Magnitude&Units $M, L, T$&Units of mass $M$\\[.35cm]
\hline
Length&$L$&$M^{-1}$\\[.35cm]
Velocity& $LT^{-1}$&$M^{0}$\\[.35cm]
Force&$MLT^{-2}$&$M^2$\\[.35cm]
Electric charge&$M^{1/2}L^{3/2}T^{-1}$&$M^0$\\[.35cm]
Magnetic field&$M^{1/2}L^{-1/2}T^{-1}$&$M^2$\\[.35cm]
Angular momentum&$ML^{2}T^{-1}$&$M^0$\\[.35cm]
Gravitational constant&$M^{-1}L^{3}T^{-2}$&$M^{-2}$\\[.35cm]
\hline
\end{tabular}
\caption{\label{Table:1}
Physical quantities in units of mass $(c=1, \hbar=1)$.}
\end{table}
Once we choose the unit of measurement of mass $M$, then all other units are identified.

In theoretical physics, the natural units of measurement \cite{PlanckLength} are the Planck time
$$T_{Pl}=\sqrt{\frac{\hbar G}{c^5}}\approx 5.4\times 10^{-44} \mbox{sec},$$
the Planck length
$$L_{Pl}=\sqrt{\frac{\hbar G}{c^3}}\approx 1.6\times 10^{-33} \mbox{cm},$$
and the Planck mass
$$M_{Pl}=\sqrt{\frac{\hbar c}{G}}\approx 2.2\times 10^{-5} \mbox{g}.$$
In the future we will use the natural system of measurements $c=\hbar=1$.

\section{Nonrelativistic mechanics of a particle}
\makeatletter
\renewcommand{\@evenhead}{\raisebox{0pt}[\headheight][0pt]
{\vbox{\hbox to\textwidth{\hfil \strut \thechapter . Initial data and frames of reference \quad\rm\thepage}\hrule}}}
\renewcommand{\@oddhead}{\raisebox{0pt}[\headheight][0pt]
{\vbox{\hbox to\textwidth{\strut \thesection .\quad Nonrelativistic mechanics of a particle
\hfil \rm\thepage}\hrule}}}
\makeatother
First, we shall review the initial notions by using a simple example of one-dimensional nonrelativistic mechanics,
in Lagrange formulation, defined by the action functional
 \be\label{1}
 S_L=\int dt
 L(X(t),{dX(t)/dt})
 \ee
with the Lagrangian
$$L(X(t),{dX(t)/dt})=\frac{m}{2}\left[\frac{dX(t)}{dt}\right]^2.$$
Here $X(t)$ is a variable, describing a particle trajectory, $t$ is a time coordinate, and $m$ is a mass of a particle
treated as a fundamental parameter of the theory. A condition of extremum of the action (\ref{1})
$$\delta S_L=0,$$
under fixed boundary conditions
$$\delta X(t_0)=\delta X(t_1)=0,$$
yields the differential equation of motion of the particle
 \be\label{2} m\frac{d^2X(t)}{dt^2}=0.
 \ee

The general solution
 \be\label{mex}
  X(t)=X_I+\frac{P_I}{m}(t-t_I)
  \ee
of this equation depends on the initial data of the particle: its position $X_I$ and momentum $P_I$
$$X(t_I)=X_I,~~~~~~~~~~\frac{dX(t)}{dt}\equiv \frac{P_I}{m},$$
given at the initial time $t_I$.
The initial data are measured with a set of physical devices (in this example, with a ruler and a watch in a fixed
space-time point), associated with a reference frame. The reference frames, moving with constant
relative velocities, are referred to as {\it inertial frames}. The transformation
$$X\mapsto \widetilde{X}=X+X_g+v_g(t-t_I),$$
turns a fixed reference frame with its origin at the point
$$X(t_I)=X_I,$$
into the reference frame, moving with the velocity  $v_g$ and with its origin at a point
$$X_g(t_I)=X_I+X_g.$$
This transformation group for reference frames in Newtonian mechanics is referred to as the {\it Galileo group.}
The differential equation (\ref{2}) is independent of initial data and, therefore, of a frame of reference.
The independence of the equations treated as laws of nature on initial data is referred to as the
{\it principle of relativity} \cite{E175-2}.
In the Hamiltonian approach, the action (\ref{1}) takes the form
\be\label{1.3}
 S_{H}=\int dt\left[P(t)\frac{d X(t)}{{dt}}-H \right],
\ee
where $P(t)$ is the momentum of a particle, and the canonical variables $\{P,X\}$ are coordinates of the {\it phase space}.
{\it Hamiltonian function}
\be\label{1.h}
H(P)=\frac{P^2}{2m}
\ee
is the generator of the {\it phase flow}, and its value on a phase trajectory is the energy of the particle
$$E= H(P_I).$$
The equations of motion of the particle obtained by variation of the action (\ref{1.3}) by the canonical variables
are the first-order differential equations:
\be\label{1.h1}
 P(t)=m\frac{d X(t)}{{dt}},~~~~~~~~~~~\frac{d P(t)}{{dt}}=0,
 \ee
rather than the second-order differential equation (\ref{2}).
According to Newtonian mechanics, all observers in different frames of reference use the same absolute time $t$.

\section{Foundations of Special Relativity}
\makeatletter
\renewcommand{\@evenhead}{\raisebox{0pt}[\headheight][0pt]
{\vbox{\hbox to\textwidth{\hfil \strut \thechapter . Initial data and frames of reference \quad\rm\thepage}\hrule}}}
\renewcommand{\@oddhead}{\raisebox{0pt}[\headheight][0pt]
{\vbox{\hbox to\textwidth{\strut \thesection .\quad Foundations of Special Relativity
\hfil \rm\thepage}\hrule}}}
\makeatother


\subsection{Action of a relativistic particle}

As it was shown above, the notion of the
spatial coordinates $X_{(i)},~i=1,2,3$ in Newtonian
mechanics as dynamical variables is clearly separated
from the absolute time $t$, treated as an \textit{evolution parameter}.

The relativistic mechanics was constructed after the Maxwell electrodynamics. Symmetry group of the electrodynamics was obtained by
Lorentz\footnote{Lorentz, H.A. Versl. Kon. Akad. v. Wet. Amsterdam. S. 809 (1904).}
and Poincar$\acute{\rm e}$\footnote{In 1905 (published 1906) it was noted by Henri Poincar$\acute{\rm e}$ that
the Lorentz transformations can be regarded as rotations of coordinates in a four-dimensional Euclidean space with three real space
coordinates and one imaginary coordinate representing time as $\sqrt{-1} ct$.
Poincar\'e presented the Lorentz transformations in terms of the familiar Euclidean rotations.}\cite{poi-2}.
The time $t=X_{(0)}$ and spatial coordinates
$X_{(i)},i=1,2,3$ are treated in this group as coordinates $X_{(\alpha)}, \alpha=0,1,2,3$ of unified
space of events or the Minkowskian space-time\footnote{Hermann Minkowski reformulated the Special Relativity in four dimensions.
His concept of space of events as a unified four-dimensional space-time continuum arose. He did not use the imaginary time coordinate,
but represented the four variables $(x, y, z, t)$ of space and time as coordinates of four dimensional affine space.
Points in this space correspond to events in the space-time. In this space, there is a defined light-cone associated with each point,
and events beyond the light-cone are classified as space-like or time-like.}
\cite{Minkowski-2}
with the scalar product of any pair of vectors
$$A_{(\alpha)}B_{(\alpha)}\equiv A_{(0)}B_{(0)}-A_{(i)}B_{(i)}.$$
Relativistic particles are described in the Special Relativity by the
action
\be\label{2s}
 S_{\rm \tiny SR}=-m\int
 d\tau\sqrt{\left(\frac{d X_{(\alpha)}}{d\tau}\right)^2}.
 \ee
This action is invariant with respect to transformations of the Poincar\'e group
$$\overline{X}_{(\alpha)}={X_{I(\alpha)}}+\Lambda_{{(\alpha)}(\beta)}X_{(\beta)},$$
which is the transformation group of reference frames.
Its subgroup of rotations $\Lambda_{{(\alpha)}(\beta)}X_{(\beta)}$
is referred to as the Lorentz group.

Fixing the indices ${(0)},(i)$ in this \textit{space
of events}  $[X_{(0)}|X_{(i)}]$ implies the choice of a specific
Lorentz reference frame.
It should be noted that the Special Relativity contains a
new symmetry with respect to the transformations that
do not change the initial data; namely, the action (\ref{2s}) is
invariant with respect to the reparametrization of the
coordinate evolution parameter
\be \label{tau}
  \tau
 ~~\longrightarrow~~~\widetilde{\tau}
 =\widetilde{\tau}(\tau),
 \ee
It results in originating a constraint between the variables. This transformation group is referred to as a
{\it gauge group}, while the quantities invariant with respect to gauge transformations are called {\it observables}.
{\it The geometric time interval}
\be\label{ds}
 s(\tau)=\int_0^\tau d\tilde{\tau}
 \sqrt{\left(\frac{d X_{(\alpha)}}{d\tilde{\tau}}\right)^2}
 \ee
on the world line of a particle in the space of events  $X_{(\alpha)}$ can be taken as an observable that is invariant with
respect to the time reparametrization. This interval is measured by a co-moving observer. {\it The time variable} of
the space of events $X_{(0)}$  is the time measured by an external observer.
The goal of the theory is to solve the equations describing trajectories in the space of events in terms of {\it gauge invariants.}

The non-covariant variational principle for a relativistic particle
was proposed by Max Planck \cite{Planck-2}.
He delivered a lecture ``The principle of relativity and the fundamental equations of mechanics''
to the Deutsche Physical Society in 1906.

\subsection{Dynamics of a relativistic particle}
Any reference frame in SR is defined by a unit time-like vector $l_{(\mu)}:$ with
$$l^2_{(\mu)}=l^2_{(0)}-l^2_{(i)}=1,$$
which will be referred to as a time axis. These vectors form a complete
set of the Lorentz reference frames. The time in
each frame is defined in Minkowskian space $X_{(\mu)}$ as a scalar
product of the time axis vector and the coordinate:
$$X_{(0)}=l_{(\mu)}X_{(\mu)}.$$
The spatial coordinates are defined on the
three-dimensional hypersurface
$$X^{\bot}_{(\mu)}=X_{(\mu)}- l_{(\mu)}(l_{(\nu)}X_{(\nu)}),$$
perpendicular to the time axis $l_{(\nu)}$.

Without the loss of generality, the time axis can be chosen
in the form
$$l_{(\mu)}=(1,0,0,0),$$
defining the observer
rest reference frame. After solving the equations, the arbitrary
Lorentz frame can be introduced. Taking out the
factor $d X_{(0)}/d\tau$ from the radical in Eq. (\ref{2s}), we arrive at
the action integral in the Planck's non-covariant formulation:
 \be\label{1905}
 S_{\rm\tiny SR}=-m\int
 d\tau  \frac{dX_{(0)}}{d\tau}\sqrt{1-\sum_i\left[\frac{d X_{(i)}}{d
 X_{(0)}}\right]^2}=
 \ee
$$= -m\int d X_{(0)}\sqrt{1-\sum_i\left[\frac{d X_{(i)}}{d
 X_{(0)}}\right]^2}.$$

Expressing the momentum
\be\label{05-1}
 P_{(i)}=\frac{\partial L}{\partial V_{(i)}}=\frac{mV_{(i)}}{\sqrt{1-V^2_{(k)}}},
 \ee
in terms of the velocity
$V_{(i)}={d X_{(i)}}/{d X_{(0)}}$
entering into the variation of the Lagrangian
(\ref{1905}) $$L=-m\sqrt{1-V^2_{(i)}}$$
one can obtain the Hamiltonian function
\be \label{05-2}
 H(P_{(i)})=P_{(i)}V_{(i)}-L=\sqrt{m^2+P^2_{(i)}(X_{(0)})}
 \ee
and rewrite action  (\ref{1905}) in the Hamiltonian form
\be \label{05-3}
 S_{\rm \tiny SR}=\int
 d X_{(0)}\left[P_{(i)}\frac{d X_{(i)}}{d X_{(0)}}-H(P_{(i)})
 \right].
 \ee
The energy of a particle is defined as a value of the Hamiltonian function on the trajectory:
$$E=H(P_{I(i)})=\sqrt{m^2+P^2_{I(i)}}.$$
The famous
formula  $E=mc^2$  (with $c=1$) is a consequence of the
definition of physical observables from the correspondence to the classical mechanics and follows from the low energy
expansion of the Hamiltonian function in powers of dynamical variables:
\be \label{05-4}
 H(P_{(i)})=\sqrt{m^2+P^2_{(i)}}=m+\frac{P^2_{(i)}}{2m}+\cdots .
 \ee
Variation of action (\ref{05-3}) with respect to canonical
momenta  $P_{(i)}$ and variables  $X_{(i)}$ yields, respectively, the
velocity in terms of the momenta,
\be\label{05-5}
 V_{(i)}=\frac{P_{(i)}}{\sqrt{m^2+P^2_{(i)}}},
 \ee
and the momentum conservation law:
$$\frac{dP_{(i)}}{dX_{(0)}}=0.$$
The solution of these equations determines the particle
trajectory in the space of events:
 \be\label{05-6}
 X_{(i)}(X_{(0)})=X_{(i)}(X_{I(0)})+V_{(i)}[X_{(0)}-X_{I(0)}],
 \ee
where $X_{I(0)}$ is the initial time relative to the observer rest
frame.

The transformation to any reference frame is
described by the corresponding Lorentz transformation
and equivalent to an appropriate choice of a time axis.
Every reference frame has its proper time, energy, and
momentum. The relationship between dynamical variables
and times in different reference frames is treated
as the relativity principle formulated, most clearly, by
Einstein  \cite{ein-2}. According to the Einstein relativity principle,
the Lorentz transformations contain extra information
on {\it relativistic effects}, as compared to solutions (\ref{05-6})
of the dynamical equations derived by variation of the action (\ref{05-3}).
Therefore, the appearance of relativistic effects due
to the Lorentz kinematic transformations ({\it id est}, transformations
of reference frames) means that the Einstein
theory significantly differs from the Newtonian mechanics.
In the latter, all the physical effects are to be deduced
from the equations of motion by variational method
with due regard to the initial data. In this case, the Galileo group in Newtonian mechanics contains nothing
new beyond the solutions of the equations of motion.

The following question arises: Can a relativistic particle
theory be formulated in such a way that all physical consequences,
including relativistic effects, are described by
a variational equation?
We will prove that such a relativistic particle theory
can be formulated with perfect analogy to Hilbert’s
``Foundations of Physics''  \cite{Hilbert1915-2}, {\it id est}, as {\it geometrodynamics.}
According to this theory, the description of a
physical system is based on the action functional, geometric
interval, symmetry of reference frames, gauge symmetry,
equations of motion, and constraint equations
for initial data.

\subsection{{Geometrodynamics} of a relativistic particle}

\begin{figure}[t]
 \includegraphics[width=0.70\textwidth,height=0.45\textwidth]{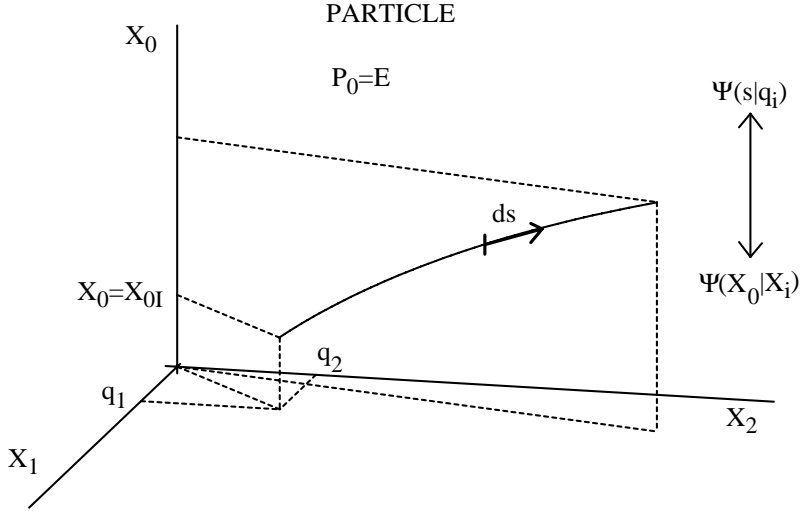}\hspace{-5mm}
\caption{\small
Motion of an unstable relativistic particle in a world line in the space of events. The motion is completely
described by the two Newtonian-like sets of observables, {\it dynamical} and {\it geometric}. Each has its proper time and
wave function $\Psi$. The two measured life-times of the particle (the time as either a {\it dynamical} variable $X^0$
or a {\it geometric} interval $s$) are interrelated by the equations of motion following
from the action of Hilbert-type {\it geometrodynamics} rather than by Lorentz transformations.
 }
\label{particle}
\end{figure}

Following to the ideas of Hilbert \cite{Hilbert1915-2}, one can construct the {\it geometrodynamics} of a
relativistic particle. The covariant approach is based on two principles: the
action  \cite{pprel-2, bpprel-2}
\be\label{1915}
 S_{\rm \tiny SR}=
 -\frac{m}{2}\int
 d\tau e(\tau) \left[\left(\frac{{d X}_{(\alpha)}}
 {e(\tau)d\tau}\right)^2+1\right]
\ee
for the variables ${X}_{(\alpha)}=[X_{(0)}|X_{(i)}]$ forming the space of
events of the moving particle and the \textit{geometric interval}
 \be\label{1915-2}
 ds=e(\tau)d\tau
  \ee
in the Riemannian one-dimensional space on the world
line of the particle in this space (see Fig. \ref{particle}). Here, $e(\tau)$
is the only metric component, the so-called lapse-function of the
\textit{coordinate evolution parameter} (ein-bein).

Variation of the action with respect to the function
$e(\tau)$ yields the equations of geometrodynamics
\be\label{1915-1-2}
  [e(\tau)d\tau]^2= {{d X}_{(\alpha)}^2}
  \equiv{d X}_{(0)}^2-{d X}_{(1)}^2-{d X}_{(2)}^2-{d X}_{(3)}^2.
 \ee
Solving these equations in  $e(\tau)$, we arrive at
 \be\label{1915-3}
  e(\tau)=\pm \sqrt{\left(\frac{d X_{(\alpha)}}{d{\tau}}\right)^2}.
 \ee
It is seen that the action  (\ref{1915}) coincides in these solutions with the initial action (\ref{2s})
of the relativistic particle up to a sign. The negative sign of  $e(\tau)$ in Eq.~(\ref{1915-3}) implies
the change of the mass sign in the action (\ref{2s}) for an antiparticle.
Equation (\ref{1915-1-2}) is referred to as a constraint equation.
For the Hamiltonian relativistic-particle theory with
constraints, the corresponding action can be derived
from (\ref{1915}) by introducing the canonical momenta
 $$P_{(\alpha)}=\partial L_{\rm \tiny SR}/\partial \dot
 X_{(\alpha)}:$$
 \be \label{b3sr}
 S_{\rm \tiny SR}=\int\limits_{\tau_1}^{\tau_2}d\tau
 \left[-P_{(\alpha)} \frac{d X_{(\alpha)}}{d{\tau}}+
 \frac{e(\tau)}{2m}\left(P_{(\alpha)}^2-m^2\right)\right].
  \ee
Here, the lapse-function $e(\tau)$  of the coordinate
evolution parameter $\tau$ defines geometric interval  (\ref{1915-2}):
\be \label{b6sr}
 ds=e(\tau)d\tau~~~\longmapsto~~~s(\tau)=
 \int\limits_{0}^{\tau}d\overline{\tau}e(\overline{\tau}).
\ee
The action  (\ref{b3sr}) and the interval  (\ref{b6sr}) are invariant with
respect to the reparametrization of the coordinate evolution parameter $\tau$:
 \be \label{b1sr}
 \tau ~~~\longrightarrow~~~\widetilde{\tau}=\widetilde{\tau}(\tau).
 \ee
Therefore, SR could be referred to as a one-dimensional
GR, with the reparametrization group of coordinate
evolution parameter  (\ref{b1sr}) serving as a group of
gauge (general coordinate) transformations. The equation
for the auxiliary lapse-function
${\delta S_{\mbox{\tiny SR}}}/{\delta e}=0$
determines
the Hamiltonian constraint imposed on the particle
momenta $P_{(0)},P_{(i)}$:
\be\label{2np}
P_{(0)}^2-P_{(i)}^2=m^2,
 \ee
the so-called a mass surface equation.

The equations
 \be \label{b5sr}
 P_{(\alpha)}=m \frac{dX_{(\alpha)}}{ed\tau}\equiv m \frac{dX_{(\alpha)}}{ds},
 ~~~~~~~~~~~~  \frac{dP_{(\alpha)}}{ds}=0,
 \ee
for the variables $P_{(\alpha)},~X_{(\alpha)}$ derived by a variation of
action  (\ref{b3sr}) are gauge-invariant.

The solution
\be \label{b7sr}
 X_{(\alpha)}(s)=X_{I (\alpha)}+\frac{P_{I (\alpha)}}{m}s,
 \ee
of these equations in terms of geometric interval (\ref{b6sr}) is
a generalization of the solution of the Newtonian equations (\ref{mex})
in the Minkowskian space. In this case, the geometric
time interval serves as an evolution parameter, while ${P_{I (\alpha)}},X_{I (\alpha)}$
 are initial data for the four variables at the
point  $s=0$:
\be \label{1915-4}
 X_{(\alpha)}(s=0)=X_{I (\alpha)}.
 \ee
These equations contain three new features, as compared
to Newtonian mechanics, namely, the momentum
constraint (\ref{2np}), the time component in solution (\ref{b7sr}) of
the equation of motion, and the initial value $X_{I (0)}$ of time
as a variable.

\subsection{Reduction of geometrodynamics to the\\ Planck's relativistic dynamics (1906)}
The action (\ref{b3sr}) and the interval (\ref{b6sr}) were above referred to as geometrodynamics of a particle.
The geometrodynamics of a particle is characterized by
two times in every reference frames namely, the
time as a \textit{geometric interval} measured by an observer
on the world line and the time as a \textit{dynamical variable}
measured by an fixed observer.
The physical interpretation of solutions (\ref{2np}) and
(\ref{b7sr}) of geometrodynamics is determined by the choice
of a specific Lorentz reference frame $P_\mu=(P_{(0)},P_{(i)})$, the so called
observer rest frame. The solution
$P_{(0)}$
\be\label{8sr}
 P_{(0)\pm}=\pm\sqrt{P_{(i)}^2+m^2}=\pm H
\ee
of constraint equation  (\ref{2np}) in the zero momentum component
$P_{(0)}$ in this reference frame is the Hamiltonian
function in the spatial dynamical variables  $[P_{(i)},X_{(i)}]$.
According to the principle of correspondence to Newtonian
mechanics, these variables belong to the so-called
{\it reduced phase space} \cite{Lanczos-2}. The variable $X_{(0)}$  is the evolution
time relative to the observer rest frame.

In a given Lorentz reference frame, the time component
of solution (\ref{b7sr}),
\be\label{0np}
X_{(0)}(s)-X_{I(0)}=\frac{{P_{(0)}}_{\pm}}{m}s.
\ee
of geometrodynamics has no analogy in Newtonian
mechanics. In this case, the formula  (\ref{0np}) is a pure kinematic
relation between the two times noted above,
namely, the dynamical variable  $X_{(0)}$  and the geometric
interval  $s$:
\be \label{9sr}
 s=[X_{(0)}-X_{I(0)}]\frac{m}{P_{(0)\pm}}.
 \ee
This equation will be referred to as a {\it geometric ratio} of the two times of a relativistic particle,
namely, the time $[X_{(0)}]$ as a variable and the time $s$ as an interval.

The substitution of geometric ratio (\ref{9sr}) into spatial part
\be\label{1np} X_{(i)}(s)=X_{I(i)}+\frac{P_{(i)}}{m}s
 \ee
of solution (\ref{b7sr}) gives the relativistic equation of motion
in the reduced phase space  $[P_{(i)},X_{(i)}]$,
\be \label{10sr}
 X_{(i)}=X_{I(i)}+\frac{P_{I(i)}}{P_{(0)+}}[X_{(0)}-X_{I(0)}].
 \ee
with the time $[X_{(0)}]$ as a variable.

Thus, \textit{geometrodynamics} in a specific reference
frame consists of constraint-free ``particle dynamics''
(\ref{10sr}) and ``geometry'' (\ref{1np}) describing purely relativistic
effects by the equations of motion in the same reference
frame  \cite{pprel-2, bpprel-2}.

The formula (\ref{05-3})
 for the action describing a moving particle
can be derived by substitution of solution  (\ref{8sr}) into
\textit{geometrodynamic} action (\ref{b3sr}). Such a substitution also
gives the action for a particle with negative energy (\ref{8sr}):
\be \label{b13sr}
 S_{\rm \tiny SR}|_{P_{(0)}=P_{(0)-}}=\int\limits_{X_{(0)}}^{X_{(0I)}}d\overline{X}_0
 \left[-P_{(i)}\frac{d{X}_{(i)}}{d\overline{X}_{(0)}}-
 \sqrt{P_{(i)}^2+m^2}\right].
  \ee
The equations corresponding to this action have the
solutions
 \be \label{b14sr}
 X_{(i)}=X_{I(i)}+\frac{P_{(i)}}{P_{(0)-}}[X_{I(0)}-X_{(0)}(s)]
 \ee
$$ =X_{(i)}+\frac{P_{(i)}}{P_{(0)+}}[X_{(0)}(s)-X_{I(0)}].$$

The problem of the negative energy was solved later, under construction of the relativistic quantum
field theory \cite{bsh-2}.

\subsection{Quantum anomaly of  geometric interval}

It is known that quantum relativistic mechanics is defined as the quantization of energy constraint (\ref{2np})
$$P_{(0)}^2-P_{(i)}^2=m^2,$$
by substituting the particle momentum
$P_{(\alpha)}=(P_{(0)},P_{(i)})$ by its operator
$\hat P_{(\alpha)}=-\imath\partial_{(\alpha)}$. The quantization
yields the Klein -- Gordon -- Fock equation for a wave function
 \be\label{q11}
 \left[\hat P_{(\alpha)}^2-m^2\right]\Psi[P_{(\alpha)}|X_{(\alpha)}]=0
 \ee
as a quantum analog of constraint equation (\ref{2np}). This equation has the normalized solution
\be\label{q1-13}
 \Psi[P_{(\alpha)}|X_{(\alpha)}]
=\frac{1}{\sqrt{2\, |P_{(0)}|}}\times
\ee
$$
\times \left[a^+\Psi_{P_{(0)+}}\theta(X_{(0)}\!-\!{X_{I(0)}})\!+\!
 a^-\Psi_{P_{(0)-}}^*\theta({X_{I(0)}}\!-\!X_{(0)})\!\right],
$$
where $\theta$ is the Heaviside step function. Two linear-independent terms with the coefficients $a^+$ and $a^-$
correspond to two classical solutions of constraint equation (\ref{2np}) with positive
and negative energies (\ref{8sr}).
Quantum field theory is known to be formulated as
the quantization of the coefficients $a^+,a^-$, {\it id est} as the
second quantization of relativistic particles  \cite{bsh-2}. In
this case, to exclude negative energies,  $-|P_{(0)}|$ and
therefore to ensure the stability of quantum systems, the coefficients $a^+$ and $a^-$  are to be treated as creation
and annihilation operators, respectively, for particles
with positive energy \footnote{Moreover, the initial data $X_{I(0)}$ is treated in the quantum theory as a
point of creation or annihilation of a particle.}. This treatment is equivalent to the postulate of the
existence of the vacuum as a lowest-energy state in the space of events. The postulate imposes a constraint on
the motion of a classical particle in the space of events,
namely, the particle with the energy  $P_{(0)+}$ ($P_{(0)-}$) which moves forward (backward):
\be \label{b12sr}
 P_{(0)+}~ \to ~X_{I{(0)}}\leq {X_{(0)}};~~~
 ~~~P_{(0)-}~ \to ~X_{I{(0)}}\geq {X_{(0)}}.
  \ee
The following question arises: How does the {\it causal quantization} (\ref{q1-13}) with the restriction
(\ref{b12sr}) influence geometric interval $s$ (\ref{b6sr})?

To answer this question we perform the Lorentz transformation
from the rest reference frame to the comoving
frame: $[\overline{X}_{(0)}|\overline{X}_{(i)}]$, where $\overline{P}_{(i)}=0$ and
$\overline{P}_{(0)\pm}=\pm m$.
It follows from (\ref{9sr}) and (\ref{b12sr}) that the time  $\overline{X}_{(0)}$ in
the comoving frame is related to the geometric interval
 $s$ by the equation
\be \label{b15sr}
 s(\overline{X}_{{(0)}}|\overline{X}_{I{(0)}})=
\ee
$$
=(\overline{X}_{{(0)}}\!-\!\overline{X}_{I{(0)}})
 \theta(\overline{X}_{{(0)}}\!-\!\overline{X}_{I{(0)}})
 \theta(\overline{P}_{{(0)}})\!+\!
 (\overline{X}_{I0}\!-\!\overline{X}_{0})\theta(\overline{X}_{I{(0)}}
 -\overline{X}_{{(0)}})\theta(-\overline{P}_{{(0)}}).
$$
This expression for the geometric interval $s$  in quantum
field theory looks like the Green’s causal function
of the comoving time:
 \be \label{b15sr1}
 \frac{d^2s(\overline{X}_{{(0)}}|\overline{X}_{I{(0)}})}{d\overline{X}^2_{{(0)}}}=
 \delta(\overline{X}_{{(0)}}-\overline{X}_{I{(0)}}).
 \ee
Therefore, the {\it positive geometric arrow of time}  $s\geq 0$ is a consequence of the postulate of existence of the
vacuum as a lowest-energy state of the system, which leads to the existence of the absolute time origin $s=0$.
The positive arrow of time implies breaking of classical symmetry
with respect to the transformation $s$  to $-s$. In contrast
to the classical symmetry, breaking of symmetry in quantum theory is referred to a {\it quantum
anomaly}\footnote{The anomaly associated with Dirac fields also follows from the
vacuum existence postulate. This fact was first pointed out by Jordan \cite{j-2}
 and then rediscovered by the authors \cite{elp-2, Ilieva-2}. The vacuum
existence postulate is verified by a number of experimental
effects, in particular, anomalous decays of pseudoscalar bound
states (neutral pion and para-positronium) into two photons.}.
Under the assumption of the existence of the vacuum
as a physical lowest-energy state, the second quantization
of an arbitrary relativistic system leads to the
absolute geometric-time origin $s=0$ in this system. The question on what was before the creation of a relativistic
particle or a universe has no physical sense for an observer measuring time because time $s=0$ is created
together with the quantum relativistic universe as a consequence of the universe stability.

We have seen that in quantum theory in each frame of reference there are two measurable times:
{\it the proper time} (\ref{b15sr}) at the world line and {\it the relative time}
$[X_{(0)}]$ (\ref{b15sr}) in the space of events $[X_{(0)}, X_{(j)}]$.
 In quantum theory, these two times are supplementary, and they cannot be identified.
Therefore, the so-called {\it twin paradox} that appears in classical relativistic mechanics
does not take place in the quantum theory.

\subsection{How does the invariant reduction \\differ from
the choice of gauge?}

We now compare the gauge invariant method of describing the field dynamics  \cite{pprel-2, bpprel-2}
to the gauge-noninvariant method, assuming that the coordinate time  $\tau$ becomes
observed\footnote{The assumption of the coordinate time $x^0$ as an observable was used in GR under description of the island
gravitational objects \cite{fadpop-2}.}. In the case of SR under consideration,
this assumption implies the use of the synchronous gauge $e(\tau)=1$ in action  (\ref{b3sr}):
 \be \label{bb3sr}
 S_{\mbox{\tiny SR}}=\int\limits_{\tau_1}^{\tau_2}d\tau
 \left[-P_{(\alpha)}\frac{d X_{(\alpha)}}{d{\tau}}+
 \frac{e(\tau)}{2m}\left(P_{(\alpha)}^2-m^2\right)\right].
  \ee
This yields a constraint-free theory:
\be \label{22b4sr}
 S_{\rm SR}|_{[e=1]}=
\ee
$$= \int\limits_{\tau_1}^{\tau_2}d\tau
 \left[P_{(i)}\frac{d X_{(i)}}{d{\tau}}-P_{(0)}\frac{d X_{(0)}}{d{\tau}}-
 \frac{1}{2m}\left(-P_{(0)}^2+P_{(i)}^2+m^2\right)\right].
$$
From the viewpoint of quantization, Eq. (\ref{22b4sr})
describes an unstable system because it contains the
variable $X_{(0)}$ making a negative contribution to the
energy
$$E=\frac{1}{2m}(-P_{(0)}^2+P_{(i)}^2+m^2),$$
 which is defined in
the interval $(-\infty<E<\infty)$. The particle action on the
three-dimensional hypersurface defined by the condition $P_{(0)}=0$
(the similar constraint in GR is referred to
as a minimal surface  \cite{fadpop-2}) coincides with the Newtonian
action
\be \label{newton}
 S_{\rm SR}|_{[e=1,P_{(0)}=0]}=S_{\mbox{\tiny Newton}}=
 \int\limits_{\tau_1}^{\tau_2}d\tau
 \left[P_{(i)}\dot X_{(i)}-
 \frac{P_{(i)}^2}{2m}\right].
  \ee
up to a constant factor, as well as with the Einstein
action (\ref{05-3}) in the nonrelativistic limit, where the time
X(0) in the rest frame coincides with the time interval $s$.
Theory (\ref{22b4sr}) under the constraint
$$P^2_{(0)}=P^2_{(i)}+m^2$$
reduces to SR because the gauge symmetry is restored.

Einstein \cite{ein-2}  found that, in contrast
to classical mechanics, just two observers are
needed to completely describe the motion of a relativistic
particle: the first is at rest, and the second moves
with the particle. For example, every Einstein's observer measures its proper lifetime of an unstable particle.
Therefore, time is a relative quantity. Einstein described this time relativity as a pure kinematic effect
by using the Lorentz transformations from a fixed reference frame to a moving one.
As was shown above, there exists a geometrodynamic generalization of the Einstein dynamics of 1905 (\ref{05-3})
to a gauge theory with the constraint (\ref{b3sr}). This generalization allows us to describe the two-time relativity as a consequence
of the equations of motion rather than the Lorentz kinematic transformations. This geometrodynamic description
defines the new two-time relativity as a ratio of the dynamical particle evolution parameter  $X_{(0)}$ to geometric
interval $s$ (\ref{b6sr}).
We now illustrate this inference with a mini-universe. In this case, purely relativistic effects can not be
described kinematically by transformations of the Lorentz type variables.

\section{Homogeneous approximation of\\ General Relativity
}
\label{PhysicalCosmology2-2a}
\makeatletter
\renewcommand{\@evenhead}{\raisebox{0pt}[\headheight][0pt]
{\vbox{\hbox to\textwidth{\hfil \strut \thechapter . Initial data and frames of reference \quad\rm\thepage}\hrule}}}
\renewcommand{\@oddhead}{\raisebox{0pt}[\headheight][0pt]
{\vbox{\hbox to\textwidth{\strut \thesection . Homogeneous Approximation of General Relativity
\hfil \rm\thepage}\hrule}}}
\makeatother
\subsection{Radiation-dominated cosmological model}
As was shown in the preceding Section, relativistic effects in variational equations can be dynamically
described within the framework of SR formulated by analogy to the Hilbert’s variational description of GR
\cite{Hilbert1915-2}. According to Hilbert, GR geometrodynamics is based on two basic notions: the action
\be \label{02-01e}
W_{\rm H}=-\int d^4x\sqrt{-g}\dfrac{R^{(4)}(g)}{6}
\ee
in the units
$$\sqrt{\frac{3}{8\pi}}M_{\rm Pl}=c=\hbar=1$$
and the geometric interval of the Riemannian coordinate manifold
\be \label{02-02e}
ds^2=g_{\mu\nu}dx^\mu dx^\nu.
\ee
Both the action (\ref{02-01e}) and the interval (\ref{02-02e}) are invariant with
respect to the general coordinate transformations
\be \label{02-03e}
x^\mu \to \widetilde{x}^\mu= \widetilde{x}^\mu(x^0, x^1, x^2, x^3).
\ee
They serve as a generalization of the action considered
above and as an interval for a relativistic particle,
invariant with respect to the reparametrization group of
coordinate time.

In the case of the homogeneous approximation,
\be \label{02-04e}
ds^2=g_{\mu\nu}dx^\mu dx^\nu\simeq
\left\{\underbrace{g_{00}(x^0)[dx^0]^2}_{(dt)^2}-\underbrace{|g^{(3)}(x^0)|^{1/3}}_{a^2(t)}
[dx^j]^2\right\}
\ee
one can keep only two metric components:
$$g_{00}=|g^{(3)}|^{1/3}N_0^2$$
and the spatial metric determinant $|g^{(3)}(x^0)|$.
 In Friedmann's notations this metric takes the following form
\be \label{dse3}
{ds^2}=d t^2 - a^2(t)(d {\bf r})^2.
\ee
Here $t$ is the \textit{world time}, $a(t)$ is the \textit{cosmological scale factor}, and
\be \label{02-06e}
r\equiv\sqrt{x_1^2+x_2^2+x_3^2}
\ee
is the \textit{coordinate distance} to a considered cosmic object.
They are invariant with respect to reparametrization of the
coordinate evolution parameter
\be \label{02-05e}
x^0 \to \widetilde{x}^0= \widetilde{x}^0(x^0).
\ee
 From the light cone interval  (\ref{dse3})
$$d t =  a(t)d r$$
one gets a relation between the \textit{coordinate distance} and the {\it conformal time} $\eta$:
\be\label{dse4}
r(\eta)=\int\limits_{t_I}^{t_0}\frac{dt}{a(t)}\equiv\eta_0-\eta.
\ee
Here $\eta_0$ is the present day value of the conformal time, for which
the cosmological scale factor is equal to unit $a(\eta_0)=1$, and $\eta$ is the time
of emission of a photon by an atom at a cosmic object, that is at the coordinate distance
$r$ to the Earth. In other words, this coordinate distance $r$ is equal to a difference between
$\eta_0$ and $\eta$
\be\label{1dse4}
r=\eta_0-\eta,~~~{\rm or}~~~~\eta=\eta_0-r.
\ee
In the case of the homogeneous
approximation, (\ref{dse3})
the GR action (\ref{02-01e}) reduces to the
 cosmology action \cite{WheelerVision-2, Narlikar-2} 
\be\label{H-3}
W_{\rm H} = -V_0\int dx^0 N_0\left[ \left(\frac{da}{N_0dx^0}\right)^2+\rho_{\rm rad}\right]
=\int dx^0 L,
\ee
that can contain  an additional matter  term, in particular, the energy density of radiation
$\rho_{\rm rad}=$ constant. Here  $V_0$ is the volume,
 $L$ is the Lagrangian, and
$$N(x^0)=a^{-1}\sqrt{g_{00}}$$
is the {\it lapse function}.
This action keeps the reparametrization time invariance. As shown above, the reparametrization
group of the coordinate parameter means that one of the variables (here, the only variable
is $a$) is identified with time as a variable, while its canonical momentum
\be\label{energyP}
P_a=\frac{\partial L}{\partial (da/dx^0)}=-2V_0\frac{da}{N_0 dx^0}
\equiv-2V_0\frac{da}{ d\eta},
\ee
is taken as the corresponding Hamiltonian
function, with its value on the equations of motion
becoming the energy of events.
The  action (\ref{H-3}) reduces to the Hamiltonian cosmology
action
\be\label{energyF}
W_H=\int dx^0\left( P_a\frac{da}{dx^0}-N_0\left[-\frac{P_a^2}{4V_0}+V_0\rho_{\rm rad}\right]
\right).
\ee
Variation of the action (\ref{energyF}) with respect to the lapse
function $N_0$:
$$\dfrac{\delta W_H}{\delta N_0}=0,$$
yields the energy constraint equation
\be\label{energyUniverserad}
\frac{P_a^2}{4V_0}=V_0\rho_{\rm rad}.
\ee
Solutions of this constraint take the form
\be\label{energyU}
P_a=\pm E;~~~~~~~~~~~~~~~~~E =2V_0\sqrt{\rho_{\rm rad}}.
\ee
The Hubble law
\be \label{Hubble}
\left(\frac{da}{d\eta}\right)^2=\rho_{\rm rad}
\ee
follows from Eqs. (\ref{energyP}) and (\ref{energyUniverserad}),
and it yields the relation
between the two times in the form of the Friedmann differential
equation:
\be\label{interval}
\eta_0-\eta_I=\int\limits_{a_I}^{a_0}\, \frac{da}{\sqrt{\rho_{\rm rad}}}=
\frac{(a_0-a_I)}{\sqrt{\rho_{\rm rad}}}.
\ee
This relation  describes classical cosmology, {\it id est} the Hubble law, and is an auxiliary relation to the
Wheeler -- De Witt quantum cosmology, provided that this cosmology is defined
as a quantization of the constraint equation (\ref{energyUniverserad}).
The quantization is defined by the substitution of variables by operators,
$\hat P =\imath d/da$ acting on the Wheeler -- De Witt wave function $\Psi$:
\be\label{q1-13UDV}
 \left[\frac{d^2}{da^2}+4V_0^2\rho_{\rm rad}\right]\Psi_U(a)=0.
 \ee
As we have shown above, both the classical and the quantum cosmologies followed from the Hilbert geometrodynamics.
This allows us to combine them to settle their troubles, namely, the quantization of classical
cosmology and the description of the Hubble law in quantum cosmology.
The wave function can be presented in the form of the sum of two terms
\bea\label{q1-13U}
\Psi_U(a)=\frac{1}{\sqrt{2\, |E|}}\times
\eea
$$\times
 \left[ A_U^+\exp(\imath E a)\theta(a_{(0)}-{a_{I(0)}})+
 A_U^-\exp(-\imath E a)\theta({a_{I(0)}}-a_{(0)})\right],
$$
where $A_U^+$ and $A_U^-$ are treated as the creation and annihilation operators
of the Universe with the positive energy in accordance with the vacuum postulate $A_U^-|0\rangle=0$,
by analogy with the wave function of a relativistic particle (\ref{q1-13}).

In the Standard cosmology this wave function describes the Universe in the epoch of
the radiation dominance, where the energy density corresponds to the number of the CMB photons $\sim 10^{87}$
and their mean energy and wave length $\sim 1$ mm.
The question is: How can we derive these quantities from the first principles?

\subsection{Arrow of conformal time as quantum anomaly }
The vacuum existence postulate $A_U^-|0\rangle=0$ restricts the motion of the
universe in the field space of events, and it means that the
universe moves forward ($a > a_I$) or backward ($a < a_I$)
if the energy of events is positive ($P_a\geq 0$) or negative
($P_a \leq 0$), respectively, where $a_I$ is the initial value. In quantum
theory, the quantity $a_I$ is considered as a creation
point of the universe with positive energy $P_a\geq 0$ or as
an annihilation point of the anti-universe with negative
energy $P_a \leq 0$. We can assume that the singular point
$a = 0$ belongs to the anti-universe: $P_a \leq 0$. A universe
with positive energy of events has no cosmological
singularity $a = 0$.
According to the vacuum existence postulate, the conformal time (\ref{interval}) as a solution of
 the Hubble law  is  positive for both the universe and anti-universe
\be\label{interval-An}
\eta_0-\eta_I 
\geq 0.
\ee

\section{Standard cosmological models}
\makeatletter
\renewcommand{\@evenhead}{\raisebox{0pt}[\headheight][0pt]
{\vbox{\hbox to\textwidth{\hfil \strut \thechapter . Initial data and frames of reference \quad\rm\thepage}\hrule}}}
\renewcommand{\@oddhead}{\raisebox{0pt}[\headheight][0pt]
{\vbox{\hbox to\textwidth{\strut \thesection .\quad Standard cosmological models
\hfil \rm\thepage}\hrule}}}
\makeatother
We now consider the equation (\ref{Hubble}) of the universe evolution
\be \label{eq:plan}
\left[\frac{da}{d\eta}\right]^2 = \rho_c(a).
\ee
The universe is
filled with homogeneous matter with the density $\rho_c(a)$.
In the Standard cosmology, the conformal density is the sum of those depending on the scale $a$
\be\label{uncol}
\rho_{c}(a)=\rho_{\rm rigid}a^{-2} +
\rho_{\rm rad}+\rho_{M}a+\rho_{\Lambda}a^4.
\ee
Here, $\rho_{\rm rigid}$ is the contribution of the equation of rigid
state for which the density is equal to pressure
\be\label{rigidstate}
\rho_{\rm rigid}=p_{\rm rigid},
\ee
 The densities $\rho_{\rm rad}, \rho_{M}$, and $\rho_{\Lambda}$
describe the contributions of radiation, baryon matter, and Lambda-term, respectively.
For each of these states, the Eq.(\ref{eq:plan}) can be solved, under the initial conditions
$$a(\eta_0)=1,~~~~~~~~~a'(\eta_0)=H_0 :$$
in terms of the conformal time $\eta$:
\be\label{CC1}
a_{\rm rigid}(\eta)=\sqrt{1 - 2H_0 r},~~~~~~~~
a_{\rm rad}(\eta)=1 -  H_0r,
\ee
\be\label{CC2}
a_{M}(\eta)={\left[1 - \frac{1}{2} H_0r\right]}^{2},~~~~~~~~
a_{\Lambda}(\eta)={\displaystyle \frac{1}{1+H_0r}}.
\ee
The conformal time  $\eta$ is defined in
observational cosmology as the instant of time at which
the photon is radiated by an atom at a cosmic object
moving with the velocity c = 1 in a geodetic line on the
world cone
$$ds^2=a^2[(\eta)^2-dr^2]=0.$$
This allows us to find the relation of the distance
\be \label{distance}
r=\eta_0-\eta.
\ee
In other words,
$\eta_0$ is the present-day conformal time of the photon measured
by a terrestrial observer with $a(\eta_0) = 1$, and $\eta$ is
the time instant at which the photon is radiated by an
atom at the distance $r$ from the Earth. Therefore, $\eta$ is
the difference between the present-day conformal time
$\eta_0$ and the time of a photon flight to the Earth, coinciding
with the distance (\ref{distance}). Equation (\ref{distance}) gives
\be \label{distance-1}
\eta=\eta_0-r.
\ee
In an observational cosmology, the density  (\ref{uncol}) can be
expressed in terms of the present-day critical density $\rho_{\rm cr}$:
\bea\label{1uncol}
\rho_{c}(a)&=&\rho_{\rm cr}\Omega (a),\\
\Omega (a) &=& \Omega_{\rm rigid}a^{-2} +\Omega_{\rm
rad}+\Omega_{M}a+\Omega_{\Lambda}a^4 \label{def:Omega}
\eea
and the relative ones
$\Omega_{\rm rigid}$, $\Omega_{\rm rad}$, $\Omega_{M}$, $\Omega_{\Lambda}$
satisfying the condition \cite{Bahcall-2}
$$\Omega_{\rm rigid} +\Omega_{\rm rad}+\Omega_{M}+\Omega_{\Lambda}=1.$$
The Classical cosmology describes the redshift of the
radiation spectrum $E(\eta)$ at a cosmic object as relative to
the spectrum $E(\eta_0)$ at the Earth. The redshift is defined as the scale factor
versus the coordinate distance (given by conformal time (\ref{distance-1})) to the object.

Taking these relations into account and substituting $a=1/(1+z)$
 and $\eta=\eta_0-r$, we can write down the scale evolution
equation (\ref{uncol}) at the light ray geodetic line ${dr}/{d\eta} = -1$
 in the form
$$ \frac{1}{H_0}\frac{dz}{dr}=(1\!+\!z)^2\sqrt{\rho_{\rm
cr}\left[\Omega_{\rm rigid}(1\!+\!z)^{2}\! + \!\Omega_{\rm
rad}\!+\!\Omega_{M}(1\!+\!z)^{-1}\!
+\!\Omega_{\Lambda}(1\!+\!z)^{-4}\right]},
$$
where $H_0=\sqrt{\rho_{\rm cr}}$.

The solution
\be\label{1rs}
H_0 r(z)=\int\limits_{1 }^{1+z }
\frac{dx}{\sqrt{\Omega_{\rm
 rigid}x^6 +\Omega_{\rm rad}x^4+\Omega_{M}x^3+\Omega_{\Lambda}}},
 \ee
coinciding with solution (\ref{distance}) of this equation, determines the coordinate distance as a function of redshift
$z$ and gives formulae (\ref{CC1}), (\ref{CC2}) for every state. The formula (\ref{1rs}), being a basis of the observational cosmology
(for example, see \cite{Narlikar-2}), is used to determine
the equation of state of matter in the Universe according to the astrophysical measurements of the redshift in
assumption of flat space.
The formula is universal for all standards of measurement.
The Friedmannian distance $R(z)$ in the Standard cosmology is connected with conformal distance $r(z)$
in Conformal cosmology by the relation
\be\label{scale}
  R(z) = a(z) (\eta_0-\eta) = a(z) r(z), ~~~~~ a  = \frac{1}{1+z},
\ee
following from the definition of the metric
$${ds^2}=a^2(\eta)\left[d\eta^2 -(d {\bf r})^2\right]$$
and the relation (\ref{1dse4}).
Thus, different standards for the same data on the dependence of the redshift of
distances correspond to different equations of state of matter in the Universe.
%

\section{Summary}
\makeatletter
\renewcommand{\@evenhead}{\raisebox{0pt}[\headheight][0pt]
{\vbox{\hbox to\textwidth{\hfil \strut \thechapter . Initial data and frames of reference
\quad\rm\thepage}\hrule}}}
\renewcommand{\@oddhead}{\raisebox{0pt}[\headheight][0pt]
{\vbox{\hbox to\textwidth{\strut \thesection .\quad Summary and literature
\hfil \rm\thepage}\hrule}}}
\makeatother

Thus, the unified geometrodynamic formulation of
both theories (SR and GR), which is based on the
Hilbert variational principle \cite{Hilbert1915-2}, makes it possible to
quantize the cosmological models similarly to the first
and second quantization of a relativistic particle. The
latter is a basis of the modern quantum field theory \cite{bsh-2},
which is verified by a great number of high-energy
experiments. A similar approach to the quantization
within the framework of GR was first formulated by
Wheeler \cite{WheelerVision-2} and De Witt \cite{DeWitt-2}. They assumed that cosmological
time, treated as a variable, is identical with the
cosmological scale factor. Moreover, they introduced
into GR the concept of a field space of events, in which
the relativistic universe moves, by analogy to the motion of a relativistic particle
of the Minkowskian space. However, the Wheeler -- De Witt
formulation \cite{WheelerVision-2, DeWitt-2}
does not contain time as a geometric
interval and, therefore, its scale-factor dependence
(interpreted in the Friedmann cosmology as the Hubble
law). Thus, as noted above, classical cosmology fails to
quantize \cite{Narlikar-2}, while quantum cosmology fails to
describe the Hubble law  \cite{WheelerVision-2, DeWitt-2}.
In this Section, we use the invariant reduction of the
Wheeler -- De Witt cosmology, considered as a relativistic
universe geometrodynamics, to restore the
relation of observational cosmology ({\it id est} the Hubble
law) to the first and second quantization of the universe
and calculate the distribution of created universes. This
reduction allows us to solve a series of problems,
namely, Hubble evolution, universe creation from the
vacuum, arrow of time, initial data, and elimination of
the cosmological singularity, under the assumption that
the Hamiltonian is diagonal and the universe is stable.
In what follows, we consider a similar invariant reduction
in the General Relativity in order to define physical observables,
quantize gravity, and formulate a low-energy perturbation theory.
\chapter{Principles of symmetry of physical theories \label{sect_3}}
\renewcommand{\theequation}{3.\arabic{equation}}
\setcounter{equation}{0}
\section{Irreducible representations of\\ Lorentz group}
\makeatletter
\renewcommand{\@evenhead}{\raisebox{0pt}[\headheight][0pt]
{\vbox{\hbox to\textwidth{\hfil \strut \thechapter . Principles of symmetry of physical theories \quad\rm\thepage}\hrule}}}
\renewcommand{\@oddhead}{\raisebox{0pt}[\headheight][0pt]
{\vbox{\hbox to\textwidth{\strut \thesection .\quad
Irreducible representations of Lorentz group
\hfil \rm\thepage}\hrule}}}
\makeatother

The Lorentz group is determined by the requirement of invariance of the speed of light in all inertial reference systems.
It is a generalization of the Galilean transformations, and includes those that are mixed up spatial
and time coordinates of a particle.
The set of linear transformations, preserving the invariant form of the interval
$$ds^2=c^2 dt^2 - dx^2 - dy^2 -dz^2\equiv ({dx_0})^2- ({dx_1})^2- ({dx_2})^2- ({dx_3})^2,$$
is called the Lorentz group.
Transformations of the group are defined as
\be\label{LorenzL}
x_\mu^{'}=\Lambda_{\mu\nu}x_\nu,
\ee
where $\Lambda\in O(3,1).$
We introduce the Hermitian generators of the Lorentz transformations
$$L_{\mu\nu}=\imath(x_\mu\partial_\nu-x_\nu\partial_\mu).$$
The generators $L_{\mu\nu}$ form a Lie algebra $so(3,1)$:
\be\label{Lorenz} [L_{\mu\nu},
L_{\rho\tau}]=\imath(g_{\mu\rho}L_{\nu\tau}-g_{\mu\tau}L_{\nu\rho}-
g_{\nu\rho}L_{\mu\tau}+g_{\nu\tau}L_{\mu\rho}).
\ee
The most common representation of operators,
satisfying the commutation relations (\ref{Lorenz}), has the form
$$M_{\mu\nu}\equiv \imath(x_\mu\partial_\nu-x_\nu\partial_\mu)+\Sigma_{\mu\nu},$$
where the spin operators $\Sigma_{\mu\nu}$ form the same Lie algebra (\ref{Lorenz}) and commute with
operators $L_{\mu\nu}$.
The Hermitian generators $M_{ij}$ form an algebra of rotations $su(2):$
\be\label{LorenzMij}
[M_{ij}, M_{kl}]=-\imath\delta_{jk}M_{il}+\imath\delta_{ik}M_{jl}+
\imath\delta_{jl}M_{ik}-\imath\delta_{il}M_{jk}.
\ee
We introduce the operators of space rotations
$$J_i\equiv\frac{1}{2}\varepsilon_{ijk}L_{ik},$$
where $\varepsilon_{ijk}$ is the Levi--Civita symbol, antisymmetric in all indices,
and boost operators
$$K_i\equiv L_{0i}.$$
From the algebra (\ref{Lorenz}) we get
\be\label{Lorenz1} [J_i, J_j]=\imath\varepsilon_{ijk}J_k,\qquad
[K_i, K_j]=-\imath\varepsilon_{ijk}J_k,\qquad [J_i, K_j]=\imath\varepsilon_{ijk}K_k.
\ee
The commutation relations (\ref{Lorenz1}) is possible to dissociate by introducing the linear combinations
$$N_i\equiv\frac{1}{2}(J_i+\imath K_i),\qquad N_i^+\equiv\frac{1}{2}(J_i-\imath K_i)$$
with the algebra
\be\label{Lorenz2}
[N_i, N_j^+]=0,\qquad
[N_i, N_j]=\imath\varepsilon_{ijk}N_k,\qquad
[N_i^+, N_j^+]=\imath\varepsilon_{ijk}N_k^+.
\ee
Therefore, in the new generators, the Lie algebra (\ref{Lorenz}) is represented as the
direct sum of complex--conjugated spin algebras:
$$su(2)\oplus su(2).$$

There are two Casimir operators
$N_i N_i$, $N_i^+ N_i^+$,
belonging to a universal
enveloping algebra \cite{Wigner1939-3, Barut-3} with eigenvalues
$n(n+1)$, $m(m+1)$.
States within a considered representation differ by eigenvalues of the operators $N_3$ и $N_3^+$
of the corresponding algebras.
According to the Schur's lemma, the operators, commuting with all the generators of an algebra, are
proportional to the unit.
Therefore, the obtained representations can be numbered by pairs of numbers $(n, m)$
that take integer and half-integer values:
$n, m=0,\, 1/2,\, 1,\, 3/2,\, 2,\ldots$.

For example, let us consider the following representations combined by a pair of integer and half-integer numbers:

1. $(0,0)$: spin is equal to zero---scalar or pseudo-scalar particle;

2. $(1/2, 0)$: spin is equal to 1/2, left Weyl's spinor;

3. $(0, 1/2)$: spin is equal to 1/2, right Weyl's spinor;

4. $(0, 1/2)\oplus (1/2, 0)$: Dirac's spinor;

5. $(1/2, 0)\oplus (0, 1/2) = (0, 0)\oplus (1, 0)$:
In this case, the inner product is given by the antisymmetric product.
A new representation (1,0) is described by anti-symmetric self-dual
second-rank tensor. The representation (0,1) corresponds to the anti-selfdual tensor;

6. $(0, 1)\oplus (1, 0)$: Maxwell's tensor of the electromagnetic field.
%

\section{Irreducible representations of\\Poincar\'e group}
\makeatletter
\renewcommand{\@evenhead}{\raisebox{0pt}[\headheight][0pt]
{\vbox{\hbox to\textwidth{\hfil \strut \thechapter . Principles of symmetry of physical theories\quad\rm\thepage}\hrule}}}
\renewcommand{\@oddhead}{\raisebox{0pt}[\headheight][0pt]
{\vbox{\hbox to\textwidth{\strut \thesection .\quad
Irreducible representations of Poincar\`e group
\hfil \rm\thepage}\hrule}}}
\makeatother

An additional requirement of invariance of an isolated physical system with respect to uniform
translations in space and time leads to a generalization of the six-parameter Lorentz group (\ref{LorenzL}) to
a ten-parameter Poincar$\acute{\rm e}$ group \cite{Wigner1939-3}
\be\label{LorenzP}
x_\mu^{'}=\Lambda_{\mu\nu}x_\nu+a_\mu,
\ee
where $\Lambda_{\mu\nu}\in SO(3,1), a_\mu\in \mathbb{R}.$

Hermitian generators of translations $P_\mu =-\imath\partial_\mu$ commute with each other:
\be\label{PoincareP}
[P_\mu, P_\nu]=0,
\ee
but do not commute with the generators of the Lorentz group:
\be\label{PoincarePL}
[M_{\mu\nu}, P_\rho]=-\imath g_{\mu\rho}P_\nu + \imath g_{\nu\rho}P_\mu.
\ee
Algebra of Poincar$\acute{\rm e}$ is the semidirect sum of an ideal\footnote{In the theory of Lie algebras,
the ideal is a maximal commutative subalgebra.} (\ref{PoincareP})
and the Lorentz algebra $so(3,1)$.
As mentioned above, all irreducible representations are characterized by the eigenvalues of the
Casimir operators that commute with all the generators of the algebra of the group.
\begin{center}
{\vspace{0.31cm}}
\parbox{0.5\textwidth}{
\includegraphics[height=8.truecm,
angle=-0]{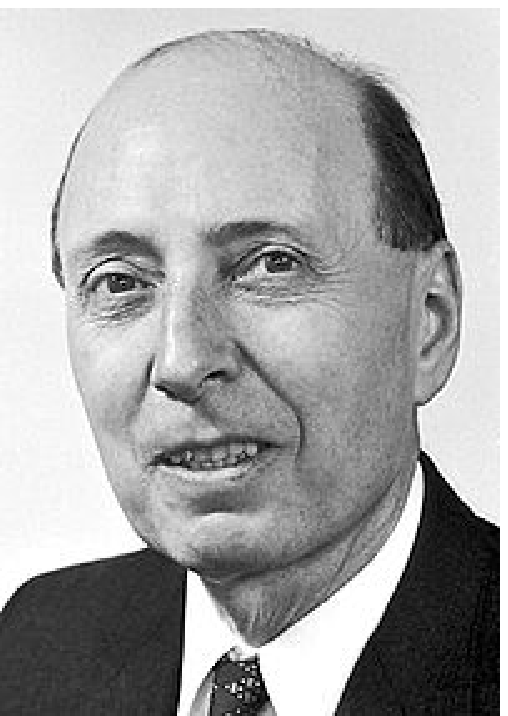}}
\parbox{0.48\textwidth}
{\small
Eugene Wigner (November 17, 1902, Budapest — January 1, 1995, Princeton, USA) was a Hungarian American
theoretical physicist and mathematician.
He received a share of the Nobel Prize in Physics in 1963 ``for his contributions to the theory of the
atomic nucleus and the elementary particles, particularly through the discovery and application of fundamental
symmetry principles''; the other half of the award was shared between Maria Goeppert--Mayer and J. Hans D. Jensen.
Wigner is important for having laid the foundation for the theory of symmetries in quantum mechanics
as well as for his research into the structure of the atomic nucleus.
Wigner developed the theory of irreducible representations of the Poincar$\acute{\rm e}$ group
as the theory of the classification of elementary particles.
}
{\vspace{0.31cm}}
\end{center}

The Casimir operator is the square of the four-momentum operator $P_\mu P^\mu$, because of its invariance with respect to the
Lorentz transformations. The second Casimir operator is constructed from the vector Pauli -- Luba$\acute{\rm n}$ski
$W^\mu$:
\be\label{Lyubanski}
W^\mu\equiv\frac{1}{2}\varepsilon^{\mu\nu\rho\sigma}P_\nu M_{\rho\sigma},
\ee
where
$\varepsilon^{\mu\nu\rho\sigma}$ is the antisymmetric tensor of Levi--Civita.
Considering
(\ref{PoincareP}) and (\ref{PoincarePL}), we obtain the commutation relations for the vector:
\be
[W_\mu, P_\nu]=0,\qquad [M_{\mu\nu}, W_\rho]=-\imath g_{\mu\rho}W_\nu+\imath g_{\nu\rho}W_\mu.
\ee
From here we see the square of the length of the vector $W^\mu W_\mu$ is the Casimir operator.
Wigner's representations are infinite-dimensional that corresponds to an unlimited momenta.
From physical point of view, the following representations of the group are of special interest.

1. An eigenvalue of the operator $P_\mu P^\mu\equiv m^2$ is a real positive number.
An eigenvalue of the operator $W^\mu W_\mu$ is $-m^2 s (s+1),$ where $s$ is a spin,
with values $s=0,\, 1/2,\, 1,\ldots$.
States within the representation differ by the third
component of the spin
$s_3= -s,\, -s+1,\ldots, s-1,\, s$
and continuous eigenvalues $p_i.$
The state corresponds to a particle with a mass $m$, a spin $s$, three-dimensional momentum $p_i$ and the projection
of the spin $s_3$.
Massive particles with the spin $s$ possess $2s+1$ degrees of freedom.

2. An eigenvalue of the operator $P^\mu P_\mu$ is equal to zero, that corresponds to a particle with zero rest mass.
An eigenvalue of the operator $W^\mu W_\mu$ is equal to zero. The scalar product of operators
$P^\mu$ and $W^\mu$ is equal to zero: $P^\mu W_\mu=0$. The coefficient of proportionality is called
helicity and equals $\pm s$, where $s=0,\, 1/2,\, 1,\ldots$ is a spin of the representation. Examples of particles:
a photon with spin 1 and two states with helicity $\pm 1$, a neutrino with helicity $\pm 1/2$
and a metric graviton with two states of helicity $\pm 2$.

3. An eigenvalue of the operator $P_\mu P^\mu\equiv m^2$ is a real negative number.
Hypothetical particles with imaginary mass are called \textit{tachyons} \cite{KirzhnitsPapers-3}.
They are widely met in the physical world, appearing as quasi-particles in complex systems, having
lost stability at phase transitions. In the theory of elementary particles tachyons
make the vacuum state of the system unstable that leads to its restructuring, providing the appearance of
mass of the elementary particles. In the Standard cosmology, the tachyon unstable vacuum state of
the scalar field is used in an inflationary scenario of the expansion of the Universe.
Further, as we shall see in Chapter 7, in cosmological models of the Universe, the
gravitons acquire tachyon mass, equal to a Hubble parameter.
Classification of fields by the Poincar$\acute{\rm e}$ group differs from the classification of fields by the Lorentz group,
primarily because that involves the selection of the reference frame, where the time coordinate is separated from the space
coordinates.

In quantum electrodynamics, under the classification by irreducible representations of the Poincar$\acute{\rm e}$ group,
the time and spatial components of the field are not equal, and
satisfy different equations and describe different physical phenomena.
In particular, the time component of the field is treated as a Coulomb
potential of charges, forming simultaneously quantum states.
And only the transverse field spatial components are treated as independent electromagnetic waves (photons), which
give the radiative corrections to the spectrum of bound states.
In the case of free massless photons, one can select such a frame of reference where conditions are imposed:
the velocity of the longitudinal component is zero and the longitudinal component is zero itself.
This last condition is called Coulomb gauge, or the choice of radiation variables.

In a certain frame of reference a massive field is divided into a time component and three spatial components.
Time component is non-dynamic and plays a role of the Yukawa potential, comoving to
an appropriate charge. Three spatial components of the massive vector field are
divided into two perpendicular to the direction of the wave vector, and one is a longitudinal.
All three components are independent dynamical variables describing the degree of freedom
with some initial data.

In Einstein's theory of gravity a separation of the time coordinate is
called 4 = 3 +1 splitting of the space--time \cite{Dirac1959-3}.
At the same time, at every point of a pseudo--Riemannian space it is possible to construct an appropriate
tangent Minkowskian space where there are transformations of the Poincar$\acute{\rm e}$ group.
In the General Relativity, from ten components of the metric only two spatial components of the metric describe
independent degrees of freedom of gravitons, while all the others give Newton's potentials and their
generalizations in the General Relativity.


\section{Weyl group}
\makeatletter
\renewcommand{\@evenhead}{\raisebox{0pt}[\headheight][0pt]
{\vbox{\hbox to\textwidth{\hfil \strut \thechapter . Principles of symmetry of physical theories \quad\rm\thepage}\hrule}}}
\renewcommand{\@oddhead}{\raisebox{0pt}[\headheight][0pt]
{\vbox{\hbox to\textwidth{\strut \thesection .\quad
Weyl group
\hfil \rm\thepage}\hrule}}}
\makeatother

Weyl group \cite{Barut-3} includes, along with the Poincar$\acute{\rm e}$ group, an Abelian group of scale
transformations\footnote{ Scale transformations were experienced by Alice (Carroll, Lewis: \textit{Alice's
Adventures in Wonderland}. Macmillan and Co., London (1865)) in order to penetrate through a small door. Scientists have thought about
this, as seen that every country adopted its measure lengths, weights. Thus, Galileo,
reflecting on the invariance of the laws of nature for a change of scale, wrote in the book
``{\it Dialogue Concerning Two New Sciences}'' (1638) the following considerations. If you increase the size of the animal two
times, its weight will increase eight times, proportional to the volume. The same cross-sectional size of its
bones grow four times the square of the resolution. Consequently, they can
only withstand four times the load.}.
The theory is scale -- invariant if its classical action does not contain dimensional constants.
If coordinates of a space transform under a scale transformation
\be
x_\mu\rightarrow x_\mu'=e^\lambda x_\mu,\qquad \lambda>0,
\ee
then a scalar field transforms as:
\be \varphi (x)\rightarrow\varphi'(x')=e^{\lambda n}\varphi
(x),
\ee
where $n$ is a conformal weight of the field.

\begin{center}
{\vspace{0.31cm}}
\parbox{0.5\textwidth}{
\includegraphics[height=8.truecm,
angle=-0]{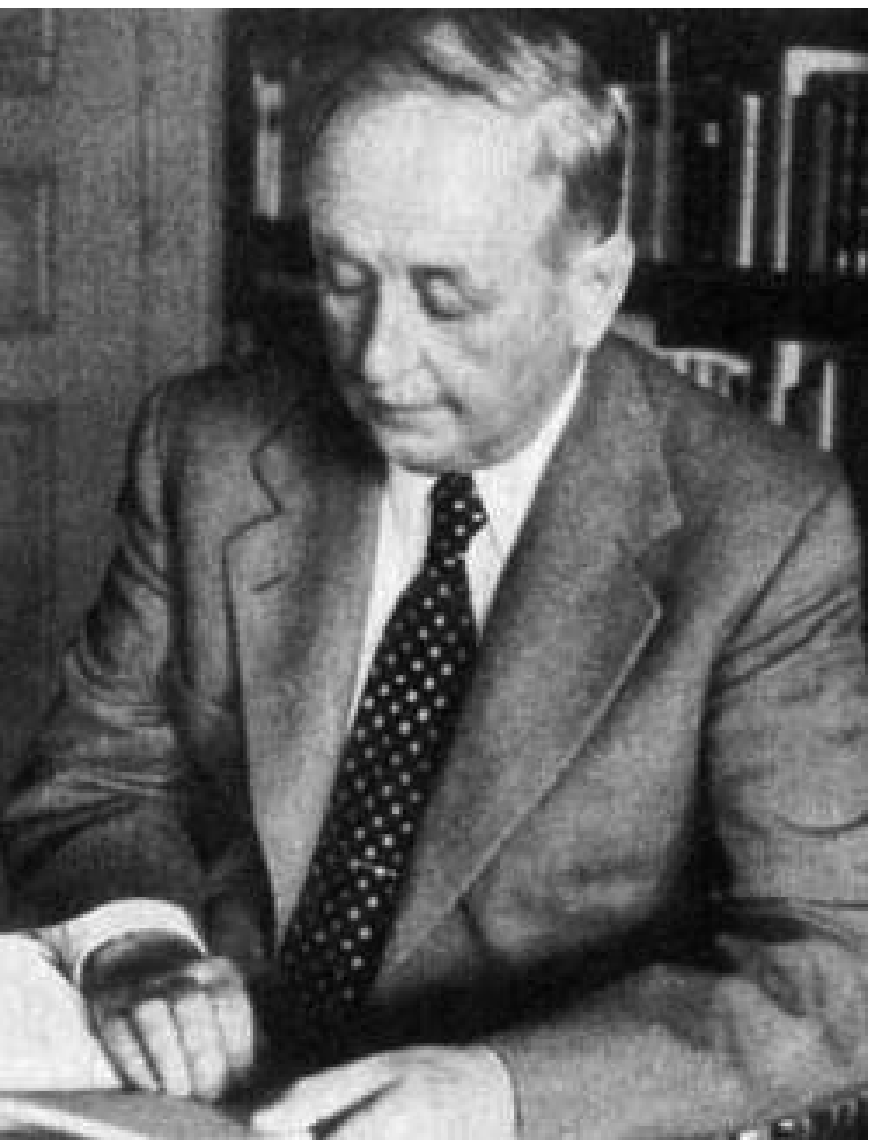}}
\parbox{0.48\textwidth}
{\small
Hermann Klaus Hugo Weyl, (9 November 1885 -- 8 December 1955) was a German mathematician,
theoretical physicist and philosopher. Although much of his working life was spent in Z\"urich, Switzerland and then Princeton, the USA
he is associated with the University of G\"ottingen tradition of mathematics, represented by David Hilbert and Hermann Minkowski.
After the creation of the General Theory of Relativity by Einstein, he turned to the unified field theory.
Although the unified theory of gravity and electromagnetism failed, his theory of gauge invariance assumed great significance.
Weyl is also well known for the use of the group theory to quantum mechanics.
}
{\vspace{0.31cm}}
\end{center}
In the infinitesimal form, under $\exp (\lambda)\approx 1+\lambda$
the law of transformation takes a form
$$
{\varphi}'(e^\lambda x)\approx{\varphi}'(x+\lambda x)\approx {\varphi}'(x)+
\lambda x^\mu\frac{\partial \varphi'}{\partial x^\mu}(x),
$$
consequently,
\be\label{varphi}
\delta\varphi\equiv\varphi'(x)-\varphi(x)=\lambda x^\mu \frac{\partial \varphi'}{\partial x^\mu}(x).
\ee
From here we get a generator of scale transformation (\textit{generator of dilatation}) $D$:
$$D\equiv - \imath x^\mu\frac{\partial }{\partial x^\mu}.$$

Commutation relations of the generator of dilation $D$ with the generators of the Poincar$\acute{\rm e}$ algebra:
$$[D, P_\mu]=-\imath P_\mu,\qquad [D, M_{\mu\nu}]=0.$$
Representations of the Weyl group are characterized by integer and half-integer
numbers -- conformal weights. Conformal weights $n$ of various fields (scalar $n=-1$, spinor $n=-3/2$,
vector $n=0$, tensor $n=2$) are possible to be calculated, assuming the sum of conformal
weights of all the factors in the action for these fields to be zero.

Free actions of these fields and their interaction with each other may include dimensional parameters,
such as the mass. Then one talks about the \textit{hard} violation of the scale symmetry of the theory.
Breaking of a scale symmetry of the theory is called \textit{soft}
if such violation has occurred as a result of quantization of the original scale-invariant classical theory.
Then one talks about the quantum anomalies. An example of such anomalies is the quantum vacuum fluctuations due to
creation and annihilation of particles of quantum fields. The source of the anomaly can be taken for the
separation of the field on the positive and negative frequency parts and the subsequent interpretation
of the coefficient of the wave function particles with negative energy as the annihilation operator
of the particles with positive energy. This treatment of the particles with
negative energy, nowadays, in quantum field theory is the only unique way to build a vacuum state with
the lowest energy \cite{Rivista-3}.

In this way, the separation of a field to the positive and negative frequency parts, led to anomal
meson decays, the above-mentioned Casimir energy, and dimensional condensates of fields in quantum field theory (QFT),
classical versions of which did not contain all of these effects, experimentally validated. One can say
that the very construction of a stable vacuum in quantum field theory, as the state of the lowest energy, and even a
hypothesis about its existence, suggests the possibility of quantum anomalies.

\section{Conformal group C}
\makeatletter
\renewcommand{\@evenhead}{\raisebox{0pt}[\headheight][0pt]
{\vbox{\hbox to\textwidth{\hfil \strut \thechapter . Principles of symmetry of physical theories \quad\rm\thepage}\hrule}}}
\renewcommand{\@oddhead}{\raisebox{0pt}[\headheight][0pt]
{\vbox{\hbox to\textwidth{\strut \thesection .\quad
Conformal group C
\hfil \rm\thepage}\hrule}}}
\makeatother

A conformal transformation is an invertible mapping of the space-time coordinates $x^\mu \to x'^\mu (x)$ such that it leaves the
metric tensor invariant up to a local scale factor
$$g_{\mu\nu} (x) \to g'_{\mu\nu} (x') =\Omega^2 (x) g_{\mu\nu} (x).$$
A classical theorem of Liouville \cite{Dubrovin-3}
states that any conformal, or angle-preserving map, between open subsets of $\mathbb{R}^n$,
for $n\geq 3$, is a composition of an inversion, dilatation, and isometries.
So, the aforementioned Weyl's group, supplemented by special conformal transformations
\be
{x_\mu}\rightarrow \frac{{x_\mu}+\beta_\mu {x^2}}{1+2(\beta_\lambda x^\lambda)+\beta^2 x^2},
\quad x^2\equiv x_\mu x^\mu,\quad \beta_\mu\in\mathbb{R},
\ee
defines the group of conformal transformations. The special conformal transformation consists of an inversion
on a unit hyperboloid
$$x^\mu\to\frac{x^\mu}{x^2},$$
followed by a translation
$$x^\mu\to x^\mu+\beta^\mu,$$
followed by a second
inversion on a unit hyperboloid
$$x^\mu\to\frac{x^\mu}{x^2}.$$

An infinitesimal special conformal transformation
$$x^\mu \to x'^\mu= x^\mu-2(\beta_\lambda x^\lambda) x^\mu +\beta^\mu x^2,\qquad |\beta^\mu|\ll  |x^\mu|,$$
such that
$$\frac{\partial x'^\mu}{\partial x^\nu}=\delta^\mu_\nu-2(\beta_\lambda x^\lambda)\delta^\mu_\nu+2
(x_\nu\beta^\mu-x^\mu\beta_\nu).$$

A generator of dilatation $D$, generators of special conformal transformations $K_\mu$
$$K_\mu=-\imath (x^2\partial_\mu-2 x_\mu (x_\lambda\partial^\lambda)),$$
and the above presented generators of the Poincar$\acute{\rm e}$ group $P_\mu$, $M_{\mu\nu}$,
besides the algebra of Poincar$\acute{\rm e}$, satisfy the following commutation
relations \cite{MackSalam-3}:
$$[D, P_\mu]=-\imath P_\mu,\qquad [D, M_{\mu\nu}]=0,$$
$$[D, K_\mu]=\imath K_\mu,\qquad [K_\mu, K_\nu]=0,$$
$$[K_\mu, P_\nu]=-2\imath(g_{\mu\nu} D+M_{\mu\nu}),\quad
[K_\rho, M_{\mu\nu}]=\imath(g_{\rho\mu} K_\nu-g_{\rho\nu}K_\mu).$$
Let a $\varphi$ is a scalar field with conformal dimension $d.$ Then
$$[P_\mu,\varphi(x)]=-\imath\partial_\mu\varphi(x),~~~~~~~
[M_{\mu\nu},\varphi(x)]=-\imath (x_\mu\partial_\nu-x_\nu\partial_\mu)\varphi(x),$$
$$[D,\varphi(x)]=-\imath (x_\mu\partial_\mu+d)\varphi(x),$$
$$[K_\mu,\varphi(x)]=-\imath (-x^2\partial_\mu+x_\mu (x_\nu\partial_\nu+d))\varphi(x).$$

In order to identify the structure of algebra we introduce the following notations for the generators
$$J_{\mu\nu}=M_{\mu\nu},\quad J_{65}=D,\quad J_{5\mu}=\frac{1}{2}(P_\mu-K_\mu),\quad
\quad J_{6\mu}=\frac{1}{2}(P_\mu+K_\mu).$$
Then we obtain the commutation relations
$$[J_{\rm KL}, J_{\rm MN}]=
\imath(g_{\rm KN}J_{\rm LM}+g_{\rm LM}J_{\rm KN}-g_{\rm KM}J_{\rm LN}-g_{\rm LN}J_{\rm KM})$$
with the diagonal 6-dimensional metric tensor
$$g_{\rm AA}=(+ - - -, - +), \quad {\rm A}= 0,1,2,3,5,6.$$
This shows that the commutation relations define the algebra $so(4,2)$ of the group of orthogonal
rotations in the pseudo-Euclidean space, which is isomorphic to the algebra $su(2,2)$ of the fundamental
representation of a twistor space $\mathbb{C}^4$:
 $$so(4,2)\approx su(2,2).$$
The generators of the six-dimensional self-representation are given by \cite{Strathdee-3}
$$(J_{\rm AB})_{\rm CD}=\imath (g_{\rm AC}g_{\rm BD}-g_{\rm AD}g_{\rm BC}).$$

\section{Conformal invariant theories\\ of gravitation}
\makeatletter
\renewcommand{\@evenhead}{\raisebox{0pt}[\headheight][0pt]
{\vbox{\hbox to\textwidth{\hfil \strut \thechapter . Principles of symmetry of physical theories \quad\rm\thepage}\hrule}}}
\renewcommand{\@oddhead}{\raisebox{0pt}[\headheight][0pt]
{\vbox{\hbox to\textwidth{\strut \thesection .\quad
Conformal invariant theories of gravitation
\hfil \rm\thepage}\hrule}}}
\makeatother
The equations of free massless fields --
Maxwell, Klein -- Gordon, Dirac, Yang -- Mills are conformal invariant.
Let us consider some attempts of generalizations the General Relativity.

In Weyl geometry there is no absolute way to compare elements of length at points spaced of each
other, but it preserves the angles between the vectors during the conformal mapping. A comparison can be
held for infinitely close points \cite{Weyl1918chapter2-3}.
Let us consider a vector of length $s$ in a point
with coordinates $x^\mu.$ We transfer it parallel to itself to a point with coordinates $x^\mu+\delta x^\mu$.
The change of its length is proportional to $s$ and $\delta x^\mu$:
\be \label{Amu} \delta s=s \kappa_\mu\delta x^\mu,
\ee
where $\kappa_\mu$ is some vector. Suppose that the standard of length is changed so that the length
is multiplied by $\lambda (x),$ depending of coordinates. Then $s$ becomes equal to
$s'=\lambda (x)s,$ and $s+\delta s$ changes as
$$s'+\delta s'=(s+\delta s)\lambda (x+\delta x)=(s+\delta s)\lambda (x)+
s\frac{\partial\lambda}{\partial x^\mu}\delta x^\mu,$$
where we neglect the values of the second order. One obtains
$$\delta s'=\lambda\delta s+s\frac{\partial\lambda}{\partial x^\mu}\delta x^\mu=
\lambda s\left(\kappa_\mu+\frac{\partial\phi}{\partial x^\mu}\right)\delta x^\mu,$$ where
\be\label{phigradient}
\phi\equiv\ln\lambda.
\ee
Thus, we obtain
$$\delta s'=s'{\kappa_\mu}'\delta x^\mu,$$
where
\be\label{gradient}
{\kappa_\mu}'=\kappa_\mu+\frac{\partial\phi}{\partial x^\mu}.
\ee
If the vector is parallelly carried around the closed loop,
the change in its length will be expressed by the following formula:
$$\delta s= s F_{\mu\nu}\delta S^{\mu\nu},$$
where
\be\label{FMUNU}
F_{\mu\nu}\equiv \frac{\partial \kappa_\mu}{\partial x^\nu}-\frac{\partial \kappa_\nu}{\partial x^\mu},
\ee
and $\delta S^{\mu\nu}$ is an element of area enclosed by the loop. The antisymmetric tensor (\ref{FMUNU})
is invariant under gauge transformations of the form (\ref{gradient}).
The vector, carried along the contour, changes its length, so the geometry
underlying in the base of the theory, is non-Riemannian.

From the point of view of the analytical description of the geometry, quadratic and linear differential forms
$$ds^2=g_{\mu\nu} dx^\mu dx^\nu,~~~~~~~~~~~~ \omega^1= \kappa_\mu dx^\mu$$
equivalent to the corresponding forms
$$\lambda g_{\mu\nu} dx^\mu dx^\nu,~~~~~~~~~~~~ \kappa_\mu dx^\mu + d \ln \lambda.$$

In the Weyl theory, field quantities $\kappa_\mu$, appeared in (\ref{Amu})
taken as the electromagnetic potentials. They are subject to gauge transformations
(\ref{gradient}), not associated with changes in geometry, but only a change in length standards.
These quantities (\ref{FMUNU}) have a geometrical meaning, independent of length standards,
and comply with the electromagnetic field tensor.
Thus, the Weyl geometry, in the author's opinion, describes the electromagnetic field by the geometric language.

The dynamical equations are constructed from the variational principle of minimal action. The Lagrangian density
of the gravitational field should be a magnitude with a conformal weight $-2.$
Weyl chose it as the square Riemannian curvature similar to the electromagnetic field
$${\EuScript L}=R^\mu_{\nu\alpha\beta}R_\mu^{\nu\alpha\beta}.$$

Einstein criticized the Weyl's theory.
Despite the remarkable properties of the theory \cite{Bergmann-3}, it was not accepted by physicists
because it contradicts the quantum theory -- quantum phenomena give us an absolute standard of length.
The terminology, however, is caught in physics: gauge, gauge transformations, gauge invariants.

\begin{center}
{\vspace{0.31cm}}
\parbox{0.5\textwidth}{
\includegraphics[height=8.truecm,
angle=-0]{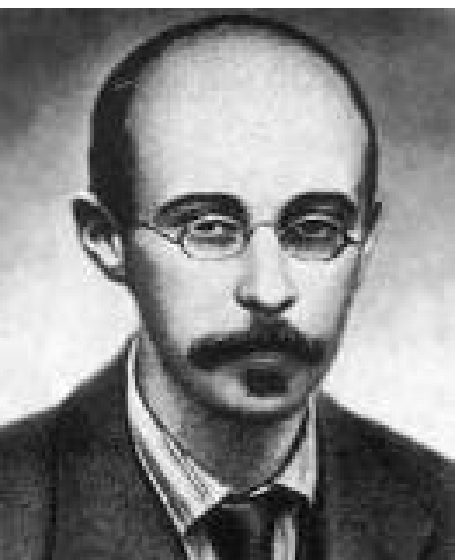}}
\parbox{0.48\textwidth}
{\small
Alexander A. Friedmann (16 June 1888 -- 16 September 1925), an outstanding Russian theoretical physicist.
In 1923 his book ``The World as Space and Time'' [in Russian]
was issued; it informed the public about the new physics.
Friedmann predicted the expansion of the Universe. The first non-stationary solutions of Einstein's equations
received by him in 1922--1924
in the study of relativistic models of the Universe began the development of the theory of non-stationary
Universe.
The scientist studied the non-stationary homogeneous isotropic models with space, of first, positive
and then negative curvature filled with dust matter (zero pressure).}
{\vspace{0.31cm}}
\end{center}

In the case of cosmology, there are only two ways to select the units of lengths of geometric
intervals -- absolute units when the length of interval
$$dl^2={g}_{ij}^{(3)}dx^idx^j$$
measured by the energy scale, and the relative, where such similar units are measured by similar, {\it id est},
intervals
$$\widetilde{dl}^2=\widetilde{g}_{ij}^{(3)}dx^idx^j$$
-- by intervals, and energies -- by energies. In case of choosing the absolute units, the space expands,
and the sizes of space objects remain unchanged. In case of choosing the relative units, the space
remains unchanged, and the sizes of space objects are reduced. Both of these features are discussed in the
book of A. A. Friedmann \cite{Fried65-3}, dedicated to the cosmology of the Universe,
that binds a second chance with the principle of scale invariance of the laws of nature.
A. A. Friedmann finds the following remarkable words about the principle of
scale invariance\footnote{Our free translation from Russian.}:
``{\it
...moving from country to country, we have to change the scale, {\it id est}, measured in Russia --- by arshins,
Germany --- meters, England --- feet. Imagine that such a change of scale we had
to do from point to point, and then we got the above operation of changing of scale.
Scale changing in the geometric world corresponds, in the physical world, to different ways of measuring of the length.
\ldots Properties of the world are divided into two classes:
some are independent of the above said change of scale, better to say, do not change their shape under any
changes of scale, while others under changing of the scale, will change their shape.
Let us agree on their own properties of the world, belonging to the first class, and call \textit{scale invariant}.
Weyl expands the invariance postulate, adding to it the requirements that
\textit{all physical laws were scale-invariant properties of the physical world}.
Consistent with such an extension of the postulate of invariance, we have to demand that the world equations
would be expressed in a form satisfactory to not only coordinate, but the scale invariance}''.

Fundamental physical constants allow us to set the system of absolute units of
distance, time, mass, {\it et cetera}. These constants are more than it is necessary for that purpose, and one can
build a dimensionless combination. The ratio of the electrical and gravitational forces, acting between the electron
and proton
$$\frac{e^2}{G m_e m_p}\,,$$
is order $\sim 10^{39}$, the ratio of the mass of the Universe to the mass of the proton is order $\sim 10^{78}$.
If we express the age of the Universe $\sim 10^9$ years in atomic units
$$\frac{e^2}{m_e c^3}\,,$$
then we get a numeric value close to $\sim 10^{39}$.
This leads to the idea that large numbers should not be viewed as constants, but the functions of
time expressed in atomic units, {\it id est}, up to prime factors $t, t^2$ and so on,
where $t$ is time in the modern era in atomic units.
P.A.M. Dirac expressed the new principle named by him as {\it the Hypothesis of Large Numbers}:

``{\it Any two of the very large dimensionless numbers occurring in Nature are connected by a simple
mathematical relation, in which the coefficients are of the order of magnitude unity}''
\cite{Dirac1938-3}.

The gravitational constant changes\footnote{Current observations on change of
fundamental constants are presented in paper: Melnikov, V.N.:
{\it Fields and constants in the theory of gravitation}. CBPF--MO--002/02.
Rio de Janeiro (2002).} simultaneously with time $t$ of aging epoch inversely to $t.$

Conformal invariant scalar - tensor theory of gravitation was built by S. Deser in 1970
\cite{Deser:1970chapter2-3}.
Let us present his further arguments. Under the conformal transformation
$$g_{\mu\nu}=\bar{g}_{\mu\nu}\phi^{-2},~~~~~~~~~~\sqrt{-g}=\sqrt{-\bar{g}}\phi^{-4}$$
with some conformal factor $\phi$ we have
$$\frac{1}{6}\sqrt{-g}R(g)\phi^2=\frac{1}{6}\sqrt{-\bar{g}}R(\bar{g})-
\sqrt{-\bar{g}}\bar{g}^{\mu\nu}\phi_{;\mu}\phi_{;\nu}\phi^{-2}$$
and
$$\sqrt{-{g}}{g}^{\mu\nu}\phi_{;\mu}\phi_{;\nu}=\phi^{-2}\sqrt{-\bar{g}}\bar{g}^{\mu\nu}\phi_{;\mu}\phi_{;\nu}.$$
Consequently, in the expression
$$\frac{1}{2}\int\, d^4x \sqrt{-g}\left[\phi_{;\mu}\phi_{;\nu} g^{\mu\nu}+\frac{1}{6}R\phi^2\right]=
\frac{1}{12}\int\, d^4x \sqrt{-\bar{g}}R(\bar{g}),$$
the scalar field has been removed from the degrees of freedom.

A scalar field, added to the theory, is coupled non-minimally with the metric field of gravity
$$W(\phi)=-\frac{1}{2}\int\, d^4x\sqrt{-g}
\left(g^{\mu\nu}\phi_{;\mu}\phi_{:\nu}+\frac{1}{6}R\phi^2\right).$$

The theory of Brans -- Dicke modifies the Einstein's theory of gravity by introducing a scalar field $\phi$
\cite{Brans-Dicke-3},
which is related to the density of mass in the Universe.
The authors of the new theory proceeded from Mach's principle, which states that the phenomenon of inertia
is a consequence of accelerations of bodies relative to the total mass distribution in the Universe.
A variation of the action
$$
\delta\int d^4 x\sqrt{-g}\left(\phi R-\xi\frac{\phi_{;\alpha}\phi^{;\alpha}}{\phi}\right)=0,
$$
where $\xi$ is some dimensionless constant, leads to the following fields equations
$$
R_{\mu\nu}-\frac{1}{2}g_{\mu\nu}R=\frac{\xi}{\phi^2}
\left(\phi_{;\mu}\phi_{;\nu}-\frac{1}{2}g_{\mu\nu}\phi_{;\alpha}\phi^{;\alpha}\right)+
\frac{1}{\phi}(\phi_{;\mu;\nu}-g_{\mu\nu}\square\phi).
$$
The model of Deser is obtained from Brans -- Dicke model at $\xi=-3/2.$

\begin{center}
{\vspace{0.31cm}}
\parbox{0.5\textwidth}{
\includegraphics[height=8.truecm,
angle=-0]{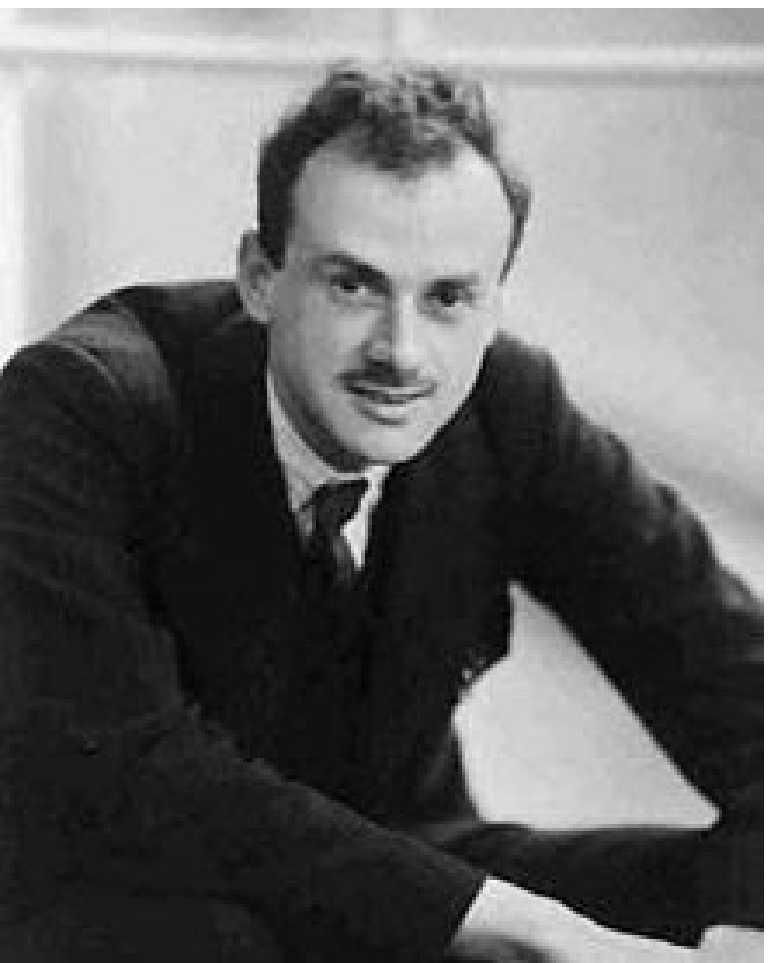}}
\parbox{0.48\textwidth}
{\small
Paul Adrien Maurice Dirac (8 August 1902 – 20 October 1984) was an English theoretical physicist who made fundamental
contributions to the early development of both quantum mechanics and quantum electrodynamics.
Among other discoveries, he formulated the Dirac equations which describe the behavior of fermions and predicted
the existence of antimatter. Dirac shared the Nobel Prize in Physics for 1933 with Erwin Schr\"odinger,
``for the discovery of new productive forms of atomic theory''.
He also did work that forms the basis of modern attempts to reconcile General Relativity with quantum mechanics.
}
{\vspace{0.31cm}}
\end{center}

The scale-invariant theory of gravity, which retains all the achievements of Einstein's theory, was formulated
by Dirac in the famous paper \cite{Dirac1973chapter2-3}.
For this purpose, he developed the analysis in the conformal geometry.
Under any change of scale a length $ds$ is multiplied by a factor $\lambda (x)$:
$ds' = \lambda ds.$
If the local value $\varphi$ is transformed by the law
$\varphi' = \lambda^n \varphi$,
one says, that its conformal weight is $n.$
From the expression for the interval
$ds^2 = g_{\mu\nu} dx^\mu dx^\nu$
it follows that the metric tensor $g_{\mu\nu}$ has a conformal weight $2$, because $dx^\mu$ are not affected by a scale
transformation. Contravariant tensor
$g^{\mu\nu}$ has a conformal weight $-2$, and $\sqrt{-g}$ has a conformal weight $4$.
Following Dirac, let us obtain generalized covariant derivatives. Take first a scalar $S$ of power $n.$
Under scale changing its covariant derivative (which is the usual derivative) $S_{\mu}$ is transformed by the formula
$$S_{\mu}^{'}=(\lambda^n S)_{,\mu}=\lambda^n S_{\mu}+n\lambda^{n-1}\lambda_{\mu}S=
\lambda^n\left[S_{\mu}+n(\kappa_\mu^{'}-\kappa_\mu)S\right],$$
where we used (\ref{phigradient}), (\ref{gradient}). From here we get
\be\label{scalarWeyl}
\left(S_{\mu}-n \kappa_\mu S\right)^{'}=\lambda^n \left(S_{\mu}-n \kappa_\mu S\right),
\ee
and the definition of the covariant derivative of a scalar:
\be\label{covS}
S_{*\mu} = S_{\mu}-n \kappa_\mu S.
\ee
Note that it, according to (\ref{scalarWeyl}), has a conformal weight $n$.

For getting covariant derivatives of vectors and tensors, we introduce modified symbols of
Christoffel ${}^{*}\Gamma^\alpha_{\mu\nu}$,
which are determined through the usual symbols
$\Gamma^\alpha_{\mu\nu}$ by the following way:
\be\label{modKrist}
{}^{*}\Gamma^\alpha_{\mu\nu}=\Gamma^\alpha_{\mu\nu}-g_\mu^\alpha \kappa_\nu-
g_\nu^\alpha\kappa_\mu+g_{\mu\nu}\kappa^\alpha.
\ee
Symbols ${}^{*}\Gamma^\alpha_{\mu\nu}$ are invariant with respect to gauge transformations.
Let $A_\mu$ be a vector with conformal weight $n.$ An expression
$$A_{\mu,\nu}-{}^{*}\Gamma^\alpha_{\mu\nu}A_\alpha$$
is a tensor.
Under the gauge transformations it transforms as follows:
$$\left(A_{\mu,\nu}-{}^{*}\Gamma^\alpha_{\mu\nu}A_\alpha\right)'=
\lambda^n A_{\mu,\nu}+n\lambda^{n-1}\lambda_\nu A_\mu-
{}^{*}\Gamma^\alpha_{\mu\nu}\lambda^n A_\alpha=$$
$$=\lambda^n\left(A_{\mu,\nu}+n(\kappa_\nu^{'}-\kappa_\nu)A_\mu-
{}^{*}\Gamma^\alpha_{\mu\nu}\lambda^n A_\alpha\right).$$
Consequently, the covariant derivative of the vector has the form:
$$A_{\mu *\nu}=A_{\mu,\nu}-n\kappa_\nu A_\mu-{}^{*}\Gamma^\alpha_{\mu\nu}A_\alpha ,$$
or, using the definition (\ref{modKrist}), we rewrite it as
\be
A_{\mu *\nu}=A_{\mu ;\nu}-(n-1)\kappa_\nu A_\mu + \kappa_\mu A_\nu -
g_{\mu\nu}\kappa^\alpha A_\alpha.
\ee
In a similar way to a contravariant vector $B^\mu$ of power $n$ we get
\be
B^\mu_{*\nu}=B^\mu_{;\nu}-(n+1)\kappa_\nu B^\mu + \kappa^\mu B_\nu -
g_{\nu}^\mu\kappa_\alpha B^\alpha.
\ee
Then you can form a covariant derivative for tensors with different upper and
lower indices by the same rules. The covariant derivative has the same degree as the
initial value. The Leibnitz rule for the product of two tensors is also performed
$$(TU)_{*\alpha}=T_{*\alpha} U+T U_{*\alpha},$$
as well as the consistency condition:
$$g_{\mu\nu*\alpha}=0,~~~~~~~~~~g^{\mu\nu}_{*\alpha}=0.$$
Now we find the second covariant derivative of a scalar $S$ of a power $n$
$$S_{*\mu*\nu}=S_{*\mu;\nu}-(n-1)\kappa_\nu S_{*\mu}+\kappa_\mu S_{*\nu}-
g_{\mu\nu}\kappa^\sigma S_{*\sigma}.$$
Substituting here the formula for the first covariant derivative (\ref{covS}), we get
the following expression
$$S_{*\mu*\nu}=S_{\mu;\nu}-n\kappa_{\mu;\nu}S-n\kappa_\mu S_\nu-n\kappa_\nu
(S_\mu-n\kappa_\mu S)+\kappa_\nu S_{*\mu}+\kappa_\mu S_{*\nu}-
g_{\mu\nu}\kappa^\sigma S_{*\sigma}.$$
So far as $S_{\mu;\nu}=S_{\nu;\mu},$ then
$$S_{*\mu*\nu}-S_{*\nu*\mu}=-n(\kappa_{\mu;\nu}-\kappa_{\nu;\mu})S=-n F_{\mu\nu} S.$$
For a vector $A_\mu$ of power $n$ we have
$$A_{\mu*\nu*\sigma}=$$
$$=A_{\mu*\nu;\sigma}-n\kappa_\sigma A_{\mu*\nu}+
(g_\mu^\rho\kappa_\sigma+g_\sigma^\rho\kappa_\mu-g_{\mu\sigma}\kappa^\rho)A_{\rho*\nu}+
(g_\nu^\rho\kappa_\sigma+g_\sigma^\rho\kappa_\nu-g_{\sigma\nu}\kappa^\rho)A_{\mu*\rho}.$$

To get a curvature tensor, we calculate the difference between the derivatives of the vector $A_\mu$
$$A_{\mu*\nu*\sigma} - A_{\mu*\sigma*\nu} =$$
$$=\!\!\left(
{}^{*}B_{\mu\nu\sigma\rho}\!+\!\frac{1}{2}(g_{\rho\nu}F_{\mu\sigma}\!+\!g_{\mu\sigma}F_{\rho\nu}\!-\!
g_{\rho\sigma}F_{\mu\nu}\!-\!g_{\mu\nu}F_{\rho\sigma})
\right)A^\rho\!-\!(n-1)F_{\nu\sigma}A_\mu.$$
The tensor ${}^{*}B_{\mu\nu\sigma\rho}$ has the conformal weight $2$ and symmetries
under permutations of the indices
$${}^{*}B_{\mu\nu\sigma\rho}=-{}^{*}B_{\mu\sigma\nu\rho}=-{}^{*}B_{\rho\nu\sigma\mu}=
{}^{*}B_{\nu\mu\rho\sigma},$$
and also
$${}^{*}B_{\mu\nu\sigma\rho}+{}^{*}B_{\mu\sigma\rho\nu}+{}^{*}B_{\mu\rho\nu\sigma}=0.$$
It can be called the Riemann tensor of the Weyl space.
The Ricci tensor is obtained by contraction of the Riemann tensor by indices
$${}^{*}B_{\mu\nu}={}^{*}B^\sigma_{\mu\sigma\nu}=R_{\mu\nu}+
\kappa_{\mu;\nu}+\kappa_{\nu;\mu}+g_{\mu\nu}\kappa^\sigma_{;\sigma}+2\kappa_\mu\kappa_\nu-
2g_{\mu\nu}\kappa^\sigma\kappa_\sigma.$$
It has conformal weight equal to zero. Contracting once more, we obtain a curvature
$${}^{*}R={}^{*}R^\sigma_{\sigma}=R+6\kappa^\sigma_{;\sigma}-6\kappa^\sigma\kappa_\sigma,$$
which is the scalar of power $-2.$

The action of scalar - tensor theory of gravitation is proposed to be taken as a
$$
W=\int\, d^4x\sqrt{-g}\left(\frac{1}{4}F_{\mu\nu}F^{\mu\nu}-\beta^2 R+
6\beta^{;\mu}\beta_{;\mu}+c\beta^4
\right),
$$
where $\beta$ is a scalar field, $c$ is a constant, and the first term is the contribution from
the electromagnetic field.
Introduced by Dirac, the scalar field has been called a {\it dilaton} \cite{1998-3}
that means \textit{expansion} because of the dilaton $D$ plays the role of a very cosmological
scale factor as a parameter of evolution in the space of field degrees of freedom,
where the motion of the Universe is given. Unlike the standard General Relativity, the Dirac's dilaton
does not \textit{expand} lengths, but increases masses.

\section{Affine group $A(4)$}
\makeatletter
\renewcommand{\@evenhead}{\raisebox{0pt}[\headheight][0pt]
{\vbox{\hbox to\textwidth{\hfil \strut \thechapter . Principles of symmetry of physical theories \quad\rm\thepage}\hrule}}}
\renewcommand{\@oddhead}{\raisebox{0pt}[\headheight][0pt]
{\vbox{\hbox to\textwidth{\strut \thesection .\quad
Affine group $A(4)$
\hfil \rm\thepage}\hrule}}}
\makeatother

Affine group $A(4)$ consists of all linear transformations of the space - time:
$$x_\mu^{'}=a_{\mu\nu}x_\nu+c_\mu.$$

Affine group is a semidirect product of the group $L(4,\mathbb{R})$
and the translation group and contains the Poincar$\acute{\rm e}$ group as a subgroup.
Algebra of generators of the affine group consists of four translations $P_\mu$,
six generators of the Lorentz group $M_{\mu\nu}$ and ten generators of properly
affine transformations $R_{\mu\nu}$
$$R_{\mu\nu}=-\imath (x_\mu\partial_\nu+x_\nu\partial_\mu),$$
together with dilatations, it has a form:
$$
[M_{\mu\nu}, M_{\rho\tau}]=\imath
(g_{\mu\rho}M_{\nu\tau}-g_{\mu\tau}M_{\nu\rho}- g_{\nu\rho}M_{\mu\tau}+g_{\nu\tau}M_{\mu\rho}),
$$
$$[M_{\mu\nu}, R_{\rho\tau}]=\imath (g_{\mu\rho}R_{\nu\tau}+g_{\mu\tau}R_{\nu\rho}-
g_{\nu\rho}R_{\mu\tau}-g_{\nu\tau}R_{\mu\rho}), $$
\bea \label{A-1}
[R_{\mu\nu}, R_{\rho\tau}]=\imath (g_{\mu\rho}M_{\nu\tau}+g_{\mu\tau}M_{\nu\rho}+
g_{\nu\rho}M_{\mu\tau}+g_{\nu\tau}M_{\mu\rho}),
\eea
$$[M_{\mu\nu}, P_\rho]=\imath (g_{\mu\rho}P_\nu-g_{\nu\rho}P_\mu), $$
$$[R_{\mu\nu}, P_\rho]=\imath (g_{\mu\rho}P_\nu+g_{\nu\rho}P_\mu).$$
In the vector representation, the generators $M_{\mu\nu}$ and $R_{\mu\nu}$ are
defined as
$$(M_{\mu\nu})_{\alpha\beta}=-\imath (g_{\mu\alpha} g_{\nu\beta}- g_{\mu\beta} g_{\nu\alpha}),\qquad
(R_{\mu\nu})_{\alpha\beta}=-\imath ( g_{\mu\alpha} g_{\nu\beta}+ g_{\mu\beta} g_{\nu\alpha}).$$

The self-linear and self-conformal transformations do not correspond to the main conservation laws.
Therefore, these symmetries must be dynamic ones, spontaneously broken.

\section{Fundamental elements of \\ base Minkowskian space ${\mathbb{M}}$}

\makeatletter
\renewcommand{\@evenhead}{\raisebox{0pt}[\headheight][0pt]
{\vbox{\hbox to\textwidth{\hfil \strut \thechapter . Principles of symmetry of physical theories \quad\rm\thepage}\hrule}}}
\renewcommand{\@oddhead}{\raisebox{0pt}[\headheight][0pt]
{\vbox{\hbox to\textwidth{\strut \thesection .\,
Fundamental elements of base Minkowskian space
\hfil \rm\thepage}\hrule}}}
\makeatother

We associate a vector $x=(x^0, x^1, x^2, x^3)$ of the space ${\mathbb{M}}$ with the Hermitian matrix
$(2\times 2)$ using quaternions:
$$X=\left(
\begin{array}{cc}
x^0+x^3& x^1-\imath x^2\\
x^1+\imath x^2& x^0-x^3
\end{array}
\right) = x^0 I_2 +\sum\limits_{i=1,2,3} x^i \sigma_i,$$
where $I_2$ is a unit matrix $(2\times 2)$, and
$\sigma_i$ are matrices of Pauli. On the light cone, where $\det X=0$,
this matrix can be presented as a direct product of a two-dimensional column
$Q=\left(
\begin{array}{c}
\xi \\
\eta
\end{array}
\right) $ to a complex - conjugated line $Q^+=(\bar\xi,\bar\eta)$
$$\frac{X}{\sqrt{2}}=Q \otimes Q^+=\left(\!\!\!
\begin{array}{cc}
\xi\bar\xi& \xi\bar\eta\\
\eta\bar\xi &\eta\bar\eta
\end{array}
\!\!\!\right),$$ where $\xi,\eta$ are two complex numerics.
Thus, the Lorentz group can be described by the spinor language.

Analogously, fundamental elements of the base Minkowskian space-time of $\mathbb{M}$,
on which relativistic fields were built, were introduced by Roger Penrose and named by him {\it twistors}
\cite{PenroseProgramme-3}. Points of the space-time are represented by two-dimensional linear
subspaces of the four-dimensional complex vector (twistor) space, on which Hermitian form of signature $(+ + - -)$
is defined.

Then the matrix $X$ can be associated with the matrix $(4\times 2)$:
$\left(
\begin{array}{c}
\imath X\\
I_2
\end{array}
\right), $
where $I_2$ is a unit matrix. Now we consider a two-dimensional plane in a
complex space $\mathbb{C}^4$, spanned on two four-dimensional
column - vectors of the matrix. The obtained two-dimensional complex plane is the image of the point
$x\in {\mathbb{M}}$
in the complexified space--Grassmannian $\mathbb{CM}$.
Twistors themselves are the elements of the fundamental representation of the group $SU(2,2).$
A twistor $Z^\alpha$ with components
$(Z^0,Z^1,Z^2,Z^3)$ belongs to $\mathbb{C}^4:$
$$(Z^0,Z^1,Z^2,Z^3)\in\mathbb{C}^4.$$

\section{Summary}
\makeatletter
\renewcommand{\@evenhead}{\raisebox{0pt}[\headheight][0pt]
{\vbox{\hbox to\textwidth{\hfil \strut \thechapter . Principles of symmetry of physical theories \quad\rm\thepage}\hrule}}}
\renewcommand{\@oddhead}{\raisebox{0pt}[\headheight][0pt]
{\vbox{\hbox to\textwidth{\strut \thesection .\,
Summary and literature
\hfil \rm\thepage}\hrule}}}
\makeatother


Is it possible for the classification of modern observational data (within the concept of quantum relativistic Universe)
to identify the wave function of the Universe with some unitary irreducible representation of any finite-parametric
symmetry group?
To answer this question in this Chapter candidates for the role of such symmetry were considered:
15-parametric group
of conformal transformations and 16-parametric group of affine transformations as natural extensions of the
Poincar$\acute{\rm e}$ group. We would like to
recall that the 16-parametric group of affine transformations of the coordinates of the Minkowskian space
includes 4 shifts, 6 Lorentz (antisymmetric) transformations and 10 properly affine (symmetric) transformations.
The fundamental representation of the conformal group, called twistors, allows us to suggest
that the space-time on the light cone,
as the adjoint representation of the conformal group, consists of more elementary elements -- twistors,
just as in the theory of strong interactions mesons consist of quarks.
In the following chapters we will assume that the analogy of the theory of gravity to the theory of strong interactions
has deeper roots, and construct a theory of gravitation as nonlinear realization of affine and conformal symmetries
in the image and likeness of building of chiral
phenomenological Lagrangians, that were successfully operated for description of the experimental low-energy data of meson physics.

\chapter{Nonlinear realizations of symmetry groups}\label{sect_4}
\renewcommand{\theequation}{4.\arabic{equation}}
\setcounter{equation}{0}
\section{Differential forms of Cartan}
\makeatletter
\renewcommand{\@evenhead}{\raisebox{0pt}[\headheight][0pt]
{\vbox{\hbox to\textwidth{\hfil \strut \thechapter . Nonlinear realizations of symmetry groups
\quad\rm\thepage}\hrule}}}
\renewcommand{\@oddhead}{\raisebox{0pt}[\headheight][0pt]
{\vbox{\hbox to\textwidth{\strut \thesection .\quad
Differential forms of Cartan
\hfil \rm\thepage}\hrule}}}
\makeatother

A space of affine connection is built in the following way \cite{Cartan-4}.
Let us consider $n$--dimensional manifold. In every point
$$M(x^1, x^2,\ldots, x^n)$$
we define an affine frame by $n$ linear independent vectors
${\bf I}_i (M),$ $i=1,\ldots, n$
and consider it to be imbedded to
$n$--dimensional affine space $A_n$. The space has with our manifold a mutual point $M$ and mutual vectors in the point $M$.
Any vector $\xi$ in point $M$ can be expanded by vectors of the frame
${\bf\xi} = \xi^k {\bf I}_k (M)$.
The manifold  is said to be {\it a space of affine connection} if the affine correspondence between local affine spaces
$A_n$ and $A_n',$ attached to infinitely close points
$$M(x^1, x^2,\ldots, x^n), \qquad M'(x^1+dx^1, x^2+dx^2,\ldots, x^n+dx^n)$$
of our manifold is set.
\begin{center}
{\vspace{0.31cm}}
\parbox{0.5\textwidth}{
\includegraphics[height=8.truecm,
angle=-0]{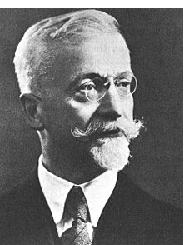}}
\parbox{0.48\textwidth}
{\small
$\acute{\rm E}$lie Joseph Cartan (9 April 1869 – 6 May 1951), a famous French mathematician.
The main theme of his works was the theory of Lie groups.
He worked over the foundational material on the complex simple Lie algebras.
This proved definitive, as far as the classification went,
with the identification of the four main families and the five exceptional cases.
He defined the general notion of anti-symmetric differential form, in the style used now.
Cartan added the exterior derivative, as an entirely geometric and coordinate-independent operation.
With these basics — Lie groups and differential forms — he went on to produce a very large body of work,
and also some general techniques, such as moving frames that were gradually incorporated into the mathematical mainstream.
}
{\vspace{0.31cm}}
\end{center}
This correspondence can be set by pointing out the place for the frame of space
$A_n'$ to be placed after mapping of $A_n'$ to $A_n.$ The vector of displacement of point $\overrightarrow{MM'}=d{\bf r}$
we expand by basis
\be
\label{Cartan1}
d{\bf r}=\omega^k {\bf I}_k (M).
\ee
The mapped basis vectors  ${\bf I}_i (M')$ differ infinitely from ${\bf I}_i (M)$, so
$${\bf I}_i (M')={\bf I}_i (M)+d{\bf I}_i,$$
and also $d{\bf I}_i$ we expand by basis vectors:
\be\label{Cartan2}
d{\bf I}_i=\omega_i^k{\bf I}_k (M).
\ee
Coefficients of expanding $\omega^k$ and $\omega_i^k$ define the affine connection. They depend on the choice of points
$M, M'$, so they are expressed as linear differential forms $\omega^k (d)$ and $\omega_i^k (d)$ of coordinates.
If in the space of affine connection a curve $x^i=x^i (t)$ is set, the coefficients $\omega^k$, $\omega_i^k$
are represented as functions of the parameter $t$, multiplied by $dt.$ Let us integrate the system of
differential equations (\ref{Cartan1}) for unknown vector functions ${\bf r}$ and ${\bf I}_i$. Initial conditions in the
initial point of the path under $t=0$: ${\bf r}=0, {\bf I}_i={\bf I}_i^0$.
As a result of integration, ${\bf r}, {\bf I}_i$ will be vector functions of $t$ in the space $A_n^0$ of the point $M_0$
and give the affine mapping of the space $A_n$ at the arbitrary point of the path $M(t)$ to $A_n^0.$

An especially interesting case occurs when a path is closed and we return to the initial point $M_0.$
Then we get a mapping of the space $A_n^0$ to itself. The group derived at the $A_n^0$ by these affine mappings is called a
\textit{holonomy group} of the space of affine connection. For recognition of characteristics of the considered geometry,
let us consider an infinitely small  transformation of holonomy group, corresponding to infinitely small closed path
of going around. Let the cycle be as a small parallelogram. There are two differential Cartan forms
$$\omega (d)=a_i dx^i,\qquad \theta (\delta)=b_i \delta x^i.$$
An external product of these forms  $\omega$ and $\theta$ is called the anti-symmetric product
$$\omega (d)\wedge \theta (\delta)\equiv\omega (d) \theta (\delta) - \theta (d)\omega (\delta)=
(a_i b_k - b_i a_k) dx_i \delta x_k.$$
An external differential of the form $\omega$ is the following expression
$$\omega'\equiv d\omega (\delta)-\delta\omega (d)\equiv da_i\delta x^i-\delta a_i dx^i=
\left(\frac{\partial a_i}{\partial x^k}-\frac{\partial a_k}{\partial x^i}\right) dx^k\delta x^i.$$
An ordinary differentiation corresponds to the shift along one of coordinate axes, whereas the external
differentiation corresponds to passing along the closed infinitely small cycle.
In case the Cartan form is full differential, the external differentiation of this form is zero identically
(Poncar$\acute{\rm e}$'s lemma).

Cartan's equations of structure are obtained by external differentiation of equations (\ref{Cartan1}) and (\ref{Cartan2}).
For the Euclidean space
$$0 = d{\bf I}_i\wedge\omega^i(\delta) + {\bf I}_i(\omega^i)',$$
$$0 = d{\bf I}_j\wedge\omega_i^j (\delta)+{\bf I}_j(\omega_i^j)'.$$
Substituting (\ref{Cartan2}) into the obtained equations, we find
$$0 = {\bf I}_i\left((\omega^i)' + \omega_j^i\wedge\omega^j\right),$$
$$0 = {\bf I}_j\left((\omega_i^j)' + \omega_k^j\wedge\omega_i^k\right).$$
From the linear independence of ${\bf I}_i$ the structural equations of Euclidean space are the following:
$$(\omega^i)' + \omega_j^i\wedge\omega^j=0,$$
$$(\omega_i^j)' + \omega_k^j\wedge\omega_i^k=0.$$
In the Euclidean space, the frame connected with point $M$ is not changed in the contour path.

In the general case of the Riemannian geometry, a frame under the infinitely small contour undergoes displacement
$$(\omega^i)' + \omega_j^i\wedge\omega^j=\Omega^i,$$
$$(\omega_i^j)' + \omega_k^j\wedge\omega_i^k=\Omega_i^j.$$
Additional movement to return the frame into the initial place defines a torsion and a curvature of the
Riemannian space. The torsion is set by the shift
$$\Omega^i=-\frac{1}{2}T^i_{jk}\omega^j\wedge\omega^k,$$
in order to return the origin of the frame into the initial placement, and the curvature of the
Riemannian space -- by additional rotation of the frame into the initial placement by the value
$$\Omega_i^j=-\frac{1}{2}R^j_{ikl}\omega^k\wedge\omega^l.$$
Here $T^i_{jk}$---the tensor of torsion, and $R^j_{ikl}$---the tensor of curvature.

Let us consider the Riemannian space with zero torsion $T=0, R\neq 0.$ Any point $M$ in sufficiently
small area of the point $O$ lies in the definite geodesic line with origin $O.$ Let $a^i$ be the
direction cosines of its tangent line in $O$ and $t$ -- a length of geodesic line $OM.$ Then
normal coordinates of point $M$ are called $n$ values, defined by equations $x^i=a^i t.$ In point $O$
the orthogonal frame is set. In every point $M$ in the area of point $O$ the orthogonal frame is set by
parallel transfer along the geodesic curve $OM.$ We find the forms $\omega^i$, $\omega_i^j$, setting
infinitesimal displacement and rotation under transition from the frame in point $M$ to the frame
connected to infinitesimally close point $M'.$ We use variables $a^i, t$,
assuming finally $t=1, a^i=x^i.$ If we put $a^i=const$ and change $t,$ the frame transfers in a parallel
way
$$
\omega^i = a^i dt;~~\omega^{j}_i \equiv 0.
$$
We define $\bar{\omega}^i$  $\bar{\omega}^{j}_i$ as values of the form  $\omega^i$ and $\omega^{j}_i$
under $dt=0$ and changing $a^i.$ Then
\be\label{1.6a}
\omega^i (t, a^i; dt, d a^i) = a^i dt + \bar{\omega}^i (t, a^i; d a^i);
\ee
\be \label{1.6}
\omega^j_i (t, a; dt, d a) = \bar{\omega}^j_i (t, a^i; d a^i).
\ee
Now let us define the form $\bar{\omega}^i$  $\bar{\omega}^{j}_i$ as functions of $t,$
consider $a^i, da^i$ as parameters. The initial point of the following reasonings
are equations of structure of space with zero
torsion:
\be\label{1.7a}
(\omega^i)^{\prime} = \omega^k\wedge\omega^{i}_k;
\ee
\be \label{1.7}
(\omega^{j}_i)^{\prime} = \omega^{k}_i\wedge\omega^{j}_k
-\frac{1}{2} R^{j}_{ikl}\omega^{k}\wedge\omega^{l}.
\ee
Substituting here ${\omega}^i$,  ${\omega}^{j}_i$ with expressions (\ref{1.6a}), (\ref{1.6}) and  separating terms
containing $dt,$ one gets
$$
[da^i, dt] +[dt, \bar{\partial \omega}^{i} / \partial t] + d_{a} \bar{\omega}^{i} =
[a^{k} dt + \bar{\omega}^{k},  \bar{\omega}^{i}_{k}]
$$
$$
[dt, \bar{\partial \omega}^{j}_{i} / \partial t] + d_{a} \bar{\omega}^{j}_{i} =
[\bar{\omega}^{k}_{i}, \bar{\omega}^{j}_{k}] - R^{j}_{ikh}[a^{k}dt +
\bar{\omega}^k], a^{k}dt + \bar{\omega}^h]/2,
$$
where $d_a$ denotes differentiation by all $a^i$ under condition $t=const.$
Comparing terms with multiplier $dt,$ we get equations which Cartan named as fundamental:
\be\label{1.8-4a}
\frac{\partial \bar{\omega}^{i}}{\partial t}= d a^i + a^k \bar{\omega}^{i}_{k}
\ee
\be \label{1.8-4}
\frac{\partial \bar{\omega}^{j}_{i}}{\partial t} =
- \frac{1}{2} R^{j}_{ikh}(a^{k}\bar{\omega}^h - a^{h} \bar{\omega}^k)
\ee
The solutions of these equations under $t=1$, $a^i = x^i$ have the following form:
\be\label{1.9a}
\omega^{i}(x, dx)|_{t=1} = \sum^{\infty}_{1} (m^{n})^{i}_{k}\frac{(-1)^n}{(2n+1)!}dx^k;
\ee
\be \label{1.9}
\omega^{i}_{j}(x, dx)|_{t=1} =
- \sum^{\infty}_{0} (m^{n})^{l}_{p}\frac{(-1)^n}{(2n+2)!}dx^p R^{i}_{jkl}x^k,
\ee
where
$$m^{i}_{k}\equiv R^{i}_{nlk}x^{n}x^{l};~~~~~~~~~(m^{2})^{i}_{j} \equiv m^{i}_{k_1}m^{k_1}_j.$$

Symbolically, the expressions (\ref{1.9a}), (\ref{1.9}) are able to be written down shorter:
\be\label{1.10a}
\omega^{i}(x, dx) = (\sin\sqrt{m}/\sqrt{m})^{i}_{k}dx^k;
\ee
\be \label{1.10}
\omega^{i}_{j}(x, dx) = R^{i}_{jkl}x^{k}[(1 - \cos\sqrt{m}) / m]^{l}_{p}d x^p.
\ee
For the Euclidean space $R^{i}_{jkl} \equiv 0$ and the Cartan forms in the normal coordinates:
$$\omega^{\prime}(x, dx) = dx^i;~~~~~~~~~~ \omega^{i}_{j} \equiv 0.$$
The square interval of length between infinitely close points is defined by the expression \cite{Helgason-4}
\be \label{1.11}
ds^2 = \omega^{i}(x, dx)\omega_{i}(x, dx) \equiv g_{ab}(x) dx^{a} dx^{b}
\ee
in accordance with geometric sense of the form $\omega^i$. The group of transformations of space
that keeps the quadratic form (\ref{1.11}) invariant, is called the \textit{movement group of the
Riemannian space.}

Let us state connection with standard concepts of differential geometry---metric tensor and symbols of
Christoffel. For this, we should pass to natural frames
${\bf N}_a:$
$$({\bf N}_{a}, {\bf N}_{b})=
g_{ab};~~~~{\bf I}_i = e^{a}_{i}(x) {\bf N}_a;~~~~e^{a}_{i}e^{j}_{a} = \delta_{ij};~~~~e^{a i}=e^{a}_{i},$$
where
\be \label{1.12}
d{\bf r} = {\bf N}_{a}(M)d x^a,
\ee
$e^{a}_{i}(x)$ are coefficients of decompositions of Cartan's forms $\omega^{i} (x, dx)$ by differentials
$d x^a$:
$$\omega^{i} (x, dx) = e^{i}_{a}(x) d x^a.$$
Laws of changing of an arbitrary vector ${\bf A}$ on the natural basis have form
$$
d(A^{i}{\bf I}_i) = (d A^i + A^j \omega^{i}_{j}) \equiv d (A^{a} {\bf N}_a) =
d A^{a} N_{a} + A^{a} d(e^{i}_{a} I_{i}) = $$
$$=[d A^{b} + A^{a}(d e^{i}_{a} e^{b}_{i} + e^{i}_{a}
\omega^{i}_{j} e^{b}_{i})]N^{j} =
[d A^{b} + A^{a} \Gamma^{b}_{ac} dx^c]N^{j},
$$
where
$$\Gamma^{b}_{ac} dx^c =
(e^{b}_{i} d e^{i}_{a} + e^{b}_{i} \omega^{i}_{j} e^{j}_{a}  ).$$

%

Let us consider some finite continuous group $G,$ dependent of $n+r$ parameters
$a_1, a_2,\ldots, a_n;$ $\eta_1, \eta_2,\ldots, \eta_r.$
One can consider parameters $a, \eta$
as coordinates of point $A$ in $n+r$--dimensional space called \textit{group space}.
Latin indices are used for notifications of values connected with generators $X_i$, and Greek ones---with
generators $Y_{\alpha}$.
Here $a^k$, $\eta^{\alpha}$ are parameters of the group;  $Y_{\alpha}$ are generators of transformations that
belong to subgroup $H;$ $X_k$ are generators that complement $H$ to the full group $G$, {\it id est} generators
of factor-space $G/H$. These generators obey the algebraic commutation conditions
$$
[Y_{\alpha}, Y_{\beta}] = \imath C^{\gamma}_{\alpha \beta} Y_{\gamma};~~~~~
[X_{k}, Y_{\alpha}] = \imath C^{i}_{k \alpha } X_{i};~~~~~
[X_{i}, X_{k}] = \imath C^{\alpha}_{i k} Y_{\alpha}
$$
Generators of the group can be considered as analog of basis vectors of Cartesian frame placed to the
origin of coordinates. Definition of equality of vectors in the group space allows one to introduce a transformation
corresponding to infinitesimal vector with origin in the arbitrary point of the space $(a, \eta).$

Every point of this space $A(a, \eta)$ assigns a transformation of group $G_{a,\eta}=G_A$ and vise versa.
Let us agree to name a point corresponding to identical transformation as the initial point of space.
A pair of points set a vector. A base point for investigation of geometry of group space is the definition
of equality of vectors. We shall say that two vectors $A_1 A_2$ and $B_1 B_2$ are equal, if an element
$G_{A_1}$ maps to $G_{A_2}$ and an element $G_{B_1}$ maps to $G_{B_2}$ by one and the same transformation $G(g)$,
acting by the rule:
$$G(g) G_{A_1} = G_{A_2};~~~~~~~~~~G(g) G_{B_1} = G_{B_2}.$$ From here we get
$$G_{A_2} G^{-1}_{A_1} = G_{B_2} G^{-1}_{B_1}.$$

Any point infinitely close to the initial point is analytically defined by infinitesimal transformation of the group.
Any infinitesimal transformation is expressed linearly by $n + r$ generators $X_k,~Y_{\alpha}$:
$$
G_{(d a^k, d \eta^{\alpha})} = I + d G_{(a^{k}, \eta^{\alpha})}
$$
\be \label{2.1}
G_{(d a^k,  \eta^{\alpha})} = \imath[da^k X_k + d \eta^{\alpha} Y_{\alpha}];~~~~
k = 1, ..., n;~~\alpha = 1, ..., r.
\ee

A vector $(0, 0; d a^{\prime}, d \eta^{\prime})$ is
equal to a vector
$(a, \eta; a + da, \eta + d\eta),$
if a point $(0,0)$ maps to a point $d a^{\prime},~d \eta^{\prime}$, a point $(a, \eta)$
maps to a point $(a + da, \eta + d \eta)$ by one and the same transformation
$$
G(g) \equiv G(a + d a, \eta + d \eta)G^{-1} (a, \eta) =
G (d \eta^{\prime}, d a^{\prime}) G^{-1}(0, 0) = G (d \eta^{\prime}, d a^{\prime}).
$$
It means that with every point of the space $(a, \eta)$ it is possible to connect such Cartesian frame, equal
(in the group sense) to the frame connected with the point of origin.
A vector $(a, \eta; a + da, \eta + d\eta)$ has the same analytic expression as (\ref{2.1}). Denoting
$$d \eta^{\prime i} = \omega^{\prime}(a, \eta; da, d\eta),~~~~~~~~~
d\eta^{\prime\alpha} = \theta^{\alpha}(a, \eta; da, d\eta),$$
we get
\be \label{2.2}
dG_{a, \eta}  G^{-1}_{a, \eta} =
\imath(\omega^{i} X_{i} + \theta^{\alpha} Y_{\alpha}).
\ee

\section{Structural equations}
\makeatletter
\renewcommand{\@evenhead}{\raisebox{0pt}[\headheight][0pt]
{\vbox{\hbox to\textwidth{\hfil \strut \thechapter . Nonlinear realizations of symmetry groups
\quad\rm\thepage}\hrule}}}
\renewcommand{\@oddhead}{\raisebox{0pt}[\headheight][0pt]
{\vbox{\hbox to\textwidth{\strut \thesection .\quad
Structural equations
\hfil \rm\thepage}\hrule}}}
\makeatother

Let $f$ be a function of variables in the space of irreducible representations of some group.
An infinitesimal action of elements of the group on the function $f$ has a form
\be \label{2.3}
df = \imath[\omega^{i}(d)X_i + \theta^{\alpha}(d)Y_{\alpha}]f~.
\ee
Let us build a bilinear differential
$$
\delta df = \imath[\delta \omega^{i}(d)X_i + \delta \theta^{\alpha}(d)Y_{\alpha}]f +
\imath[\omega^{i}(d)\delta(X_i f) + \theta^{\alpha}(d)\delta(Y_{\alpha} f)],
$$
where the differential of functions $X_i f$ and $Y_{\alpha} f$ is defined according to (\ref{2.3}).
An action of external differential to the left side of the equality (\ref{2.3}) results to zero:
$$ (df)^{\prime} = \delta df - d \delta f = 0.$$

Equating the coefficients of similar linear independent generators in (\ref{2.3}) leads to the system
of the structural equations:
$$
(\omega^i)^{\prime} = C^{i}_{k \beta}\omega^{k}\wedge\theta^{\beta};
$$
$$
(\theta^\gamma)^\prime=\frac{1}{2}C^\gamma_{\alpha\beta}\theta^\alpha\wedge\theta^\beta+
\frac{1}{2}C^\gamma_{ki}\omega^k\wedge\omega^i;$$
where
$$
(\omega^k)^{\prime} = - \delta \omega^{k}(d) + d \omega^{k}(\delta);
$$
$$
\omega^{k}\wedge\theta^{\beta} = \omega^{k}(d)\theta^{\beta}(\delta) -
\omega^{k}(\delta)\theta^{\beta}(d).
$$
Having obtained the equations, one can pass to such structural equations where there is dependence only
on Latin indices connected with the space of parameters $a^i$ of coset $G/H$. For this, we
define new differential forms
\be \label{2.4}
\omega^i_{k} = C^{i}_{k \beta}\theta^{\beta}
\ee
and use the Jacobi identities leading to
$$
C^{k}_{\alpha \beta} C^{l}_{j k} =
C^{l}_{\beta k} C^{k}_{\alpha j} - C^{l}_{\alpha k} C^{k}_{\beta j}.
$$
Finally we get for the differential forms
$\omega^i$, $\omega^i_{k}$ the following equations
\be \label{2.5}
(\omega^i)^{\prime} = \omega^{k}\wedge\omega^i_{k},
\ee
\be\label{2.5a}
(\omega^{i}_{j})^{\prime} =-\frac{1}{2}
 R^{l}_{j k i}\omega^{k}\wedge\omega^i + \omega^{k}_{j}\wedge\omega^{l}_{k},
\ee
where
$$- R^{l}_{j k i} = C^{l}_{j \gamma} C^{\gamma}_{k i};$$
and dependence is left only on the Latin indices of the coset space.

The equations (\ref{2.5}), (\ref{2.5a}) coincide in form with structural equations of Cartan for the Riemannian
$n$--dimensional space with nonzero curvature. Further, we consider only a group space of parameters
$a^i$, setting the parameters equal to zero: $\eta_\alpha = 0$. One can treat the differential forms
$\omega^i$ as components of infinitesimal shift of the frame origin relatively from the frame in the
point $a$, and $\omega^i_{j}$ as changing of the frame components. According to such geometric interpretation
of the forms $\omega^i$ and $\theta^{\gamma}$ as shift and rotation, it is natural to consider that transformation
of group $G$ is a rotation if it belongs to subgroup $H$, and a shift if it is originated by infinitesimal
transformation $\omega^i X_i$. The subgroup $H$ of transformations leaves fixed the origin of the frame of group
space and is named as \textit{stationary subgroup of the space.} One can get realizations of the transformations
representing a common group transformation $G$ in the form of multiplication
\be \label{2.6}
G = K(a)H(\eta),
\ee
where $K(a)$ is a transformation that belongs to the left coset $G/H$ of group $G$ by subgroup $H.$
Acting from the left on the element of group $G$  by arbitrary $G(g)$ and factorizing the obtained element
according to (\ref{2.6}):
\be \label{2.7}
G(g)K(a)H(\eta) = K(a^{\prime}(a, g))H(\eta^{\prime}(\eta, a, g)),
\ee
one can define in which manner the parameters $a, \eta$ are being transformed. A parametrization of $K(a)$,
or, in other words, the explicit form of finite group transformations, can be quite arbitrary. It corresponds to
arbitrary  motions of frames in the differential geometry of Cartan. Every parametrization $K$ is equivalent
to definite choice of coordinates in the coset $G/H$.

\section{Exponential parametrization}
\makeatletter
\renewcommand{\@evenhead}{\raisebox{0pt}[\headheight][0pt]
{\vbox{\hbox to\textwidth{\hfil \strut \thechapter . Nonlinear realizations of symmetry groups
\quad\rm\thepage}\hrule}}}
\renewcommand{\@oddhead}{\raisebox{0pt}[\headheight][0pt]
{\vbox{\hbox to\textwidth{\strut \thesection .\quad
Exponential parametrization
\hfil \rm\thepage}\hrule}}}
\makeatother

Let us consider the exponential parametrization of groups
\be \label{3.1}
K(a) = \exp(\imath a^{j}X_{j}).
\ee
We define explicitly the Cartan forms and infinitesimal transformations in this case.
The equations for the differential forms are the following
$$
\exp (- \imath X_{k}a^{k}) d \left[\exp (\imath X_{k}a^{k})\right] =
\imath [\omega^{i}(a, da)X_{i} + \theta^{\alpha}(a, da) Y_{\alpha}].
$$
Let us introduce a parameter $t$ to (\ref{3.1}) using a substitution $a^{k} \rightarrow a^{k}t$:
$$
\exp (- \imath X_{k}a^{k}t) d\left[\exp(\imath X_{k}a^{k}t)\right] =
\imath [\omega^{i}(ta, tda)X_{i} + \theta^{\alpha}(ta, tda) Y_{\alpha}].
$$

Differentiating by $t$, the left and the right sides of the obtained equality, we get the system of equations:
$$
\frac{\partial \omega^i}{\partial t} = d a^i + a^{k} \theta^{\beta}C^i_{k \beta}
$$
$$
\frac{\partial \theta^{\alpha}}{\partial t} = a^i\omega^{l} C^{\alpha}_{i l}.
$$
After substitution (\ref{2.4}) in terms of the forms $\omega^i, \omega^i_{j}$ these equations coincide
with the fundamental Cartan's equations, which describe the motion of the frame along the geodesic lines
and define Cartan's forms in the normal coordinates. Consequently, the exponential parametrization of
the finite group transformations are equivalent to the choice of normal coordinates in coset $G/H$:
$$
\omega^i (a, da) = (\sin \sqrt{m} / \sqrt{m})^{i}_{k} da^{k};
$$
$$
\theta^{\alpha}(a, da) =[(1 - \cos \sqrt{m} / m)]^{i}_{k} da^{k} C^{\alpha}_{i l} a^{l};
$$
$$
m^{i}_{l} = - C^{i}_{j \alpha} C^{\alpha}_{k l} a^{j} a^{k}.
$$

\section{Algebraic and dynamical\\ principles of symmetry}
\makeatletter
\renewcommand{\@evenhead}{\raisebox{0pt}[\headheight][0pt]
{\vbox{\hbox to\textwidth{\hfil \strut \thechapter . Nonlinear realizations of symmetry groups
\quad\rm\thepage}\hrule}}}
\renewcommand{\@oddhead}{\raisebox{0pt}[\headheight][0pt]
{\vbox{\hbox to\textwidth{\strut \thesection .\quad
Algebraic and dynamical principles of symmetry
\hfil \rm\thepage}\hrule}}}
\makeatother

According to Wigner \cite{Wigner-4}, all groups of symmetry are divided into two classes:
\textit{algebraic symmetries} that reflect laws of conservations and are used for classification of free physical
objects -- particles and fields, universes and their quantum analogous,
and \textit{dynamical symmetries}\footnote{By vivid expression of E. Wigner,
(Wigner, E.: {\it Symmetries and Reflections}. Indiana University press, Bloomington -- London (1970)),
the algebraic symmetries belong to the area of \textit{terra cognita},
and dynamical symmetries---to the area of \textit{terra incognita}.},
which allow one to define interactions between these objects, and also constrains of initial data and their quantization.
Progress in understanding of the role and the essence of dynamical symmetries is connected to studying of spontaneous
symmetry breaking phenomenon of vacuum. Firstly, the effects of spontaneous
symmetry breaking phenomena were considered in the theory of many particles by N.N. Bogoliubov \cite{NN-1-4},
in the relativistic theory  --- by Nambu \cite{Nambu-60-4} and Goldstone \cite{Godstone-61-4}.

Symmetry under a group is said spontaneously broken if the vacuum of the system with an invariant Lagrangian
as a state with minimal energy is stable only under transformations of subgroup $H$ of the full group $G$.
In such a case the subgroup $H$ is an \textit{algebraic} group of classifications of fields and particles of the theory.
Spontaneous breaking of symmetry of the vacuum is accompanied by creation of separate fields with zero mass,
called as \textit{Goldstone} fields (Bogoliubov's theorem in statistical physics and Goldstone's theorem in field
theory).

In particular, in the theory of strong interactions in the capacity of \textit{dynamical symmetry} the
\textit{chiral symmetry}\footnote{Chira -- in Greek is a hand which is traditionally used for illustration
of right and left particle helicity.} takes place. According to this symmetry, the strong interactions are invariant
under actions of transformation groups, including those with isotopic transformations with algebra of
generators
\be\label{I-1} [I_i,I_j]=\imath \varepsilon_{ijk}I_k
\ee
also generators $K_j$ with algebra
\be\label{I-1a} [I_i,K_j]=\imath
\varepsilon_{ijk}K_k,~~~~[K_i,K_j]=\imath \varepsilon_{ijk}I_k,
\ee
changing states with different parity. An example of linear representation of \textit{chiral symmetry} is right
and left neutrino. There is nonlinear realization of \textit{chiral} symmetry -- chiral phenomenological
Lagrangians, with which the low-energy results in QCD during 1967--72 were obtained, before formulation of
the theory of QCD in 1973--74. In the method of nonlinear realization of
\textit{chiral} symmetry with six parameters, three isotopic parameters belong to the subgroup
$H$ of vacuum stability, and three rest are proper chiral transformations, changing states with different parity.
The latter three chiral parameters are identified with three \textit{Goldstone fields}. These fields set coordinates of coset $K=G/H$
and their linear forms by rules
$$\exp(-\imath K_i\pi^i)\partial_\mu\exp(\imath K_i\pi^i)=\imath \left[
\omega^i(\partial_\mu)K_i+\theta^j(\partial_\mu)I_j\right]$$
according to commutation relations between generators of infinitesimal small transformations of the group $G$.

Shifts $\omega^i(\partial_\mu)$ and rotations $\theta^i(\partial_\mu)$ describe various movements of orthogonal frames
in the coset space. Chiral phenomenological Lagrangians of fields interactions are built univalently
in the coset $K=G/H$ out of these linear forms. These Lagrangians allow one to describe numerous processes in low-energy physics
of hadrons in satisfactory agreement with experimental data \cite{Weinberg-4, Coleman-4, Callan-4, Volkov-4, 1978-4}.

We consider nonlinear realizations of group $A(4)$, which become linear on its subgroup -- Poincar$\acute{\rm e}$ group.
Let us have a look more closely at realization in coset space $A(4)/L$, where $L$ is the Lorentz group.
We take a symmetric tensor field $h_{\mu\nu}$ and define an action of an element of the group $g:$
$$g\exp(\imath x_\mu P_\mu)\exp\left(\frac{\imath}{2}h_{\mu\nu}R_{\mu\nu}\right)=$$
$$\exp\left(\imath x'_\mu P_\mu\right)\exp\left( \frac{\imath }{2}h'_{\mu\nu}(x')R_{\mu\nu}\right)
\exp\left(\frac{\imath}{2} U_{\mu\nu}(x')L_{\mu\nu}\right),$$
where
$x_\mu', h'_{\mu\nu}(x')$ and $U_{\mu\nu}(x')$ depend of parameters of transformation $g$ and field $h_{\mu\nu}$.
Let $\Psi$ be an arbitrary field which is a linear representation of the Lorentz group.
Then, an action of the group $A(4)$ to the field $\Psi$ is defined as
$$g \Psi =\Psi'(x)=\exp\left(\frac{\imath}{2}U_{\mu\nu}(h(x), g)L_{\mu\nu}^\Psi\right)\Psi,$$
where $L_{\mu\nu}^\Psi$ is a matrix generator in linear representation of the Lorentz group.
Then arbitrary frame movements (shifts and rotations) in coset space $A(4)/L$ are described by Cartan's forms
$\omega$ as coefficients of expansion of infinitesimal transformations of generators of algebra $A(4)$ (\ref{A-1}):
$$\left[\exp\left(-\frac{\imath}{2}h_{\alpha\beta} R_{\alpha\beta}\right)
\exp(-\imath x_\mu P_\mu)\right] \,d \,
\left[\exp(\imath x_\mu P_\mu)\exp\left(\frac{\imath}{2}h_{\alpha\beta}R_{\alpha\beta}\right)\right]=
$$
 \bea \label{1-0}\nonumber
 =GdG^{-1}&=&\imath[\underbrace{P_{(\alpha)}\cdot
 \omega^P_{(\alpha)}+R_{(\alpha)(\beta)}\cdot
 \omega^R_{(\alpha)(\beta)}}_{{ {\rm shifts}}~K=A(4)/L
 }
 +\underbrace{L_{(\alpha)(\beta)}\cdot
 \omega^L_{(\alpha)(\beta)}}_{{\rm rotations }~K=A(4)/L
 }],
 \eea
Forms
 \bea 
\label{pR}
 \omega^P_{(\alpha)}(d)&=&e_{(\alpha)\mu}dx^\mu,\\
  \label{tR}
 \omega^R_{(\alpha)(\beta)}(d) &=& \frac{1}{2}
 \left(e^{\mu}_{ (\alpha)}de_{(\beta)\mu }
 +e^{\mu}_{(\beta)}de_{(\alpha)\mu }\right),
 \\ \label{tL}
 \omega^L_{(\alpha)(\beta)} (d) &=& \frac{1}{2}
 \left(e^{\mu}_{ (\alpha)}de_{(\beta)\mu }
 -e^{\mu}_{ (\beta)}de_{(\alpha)\mu }\right).
\eea
define covariant differentials of coordinates and Goldstone fields and are used for definition of
covariant differential of fields $\Psi$. Here $e_{(\alpha)\mu}$ are components of tetrads with two indexes.
One index belongs to Riemann space $\mu$, and the second $(\alpha)$ -- to tangent Minkowski space.
Components of tetrads are coefficients of expansion of Cartan forms by differentials of coordinate space.

For descriptions of fermions in Riemann space the Fock frame in tetrad formalism is used \cite{Fock:1929-4}.
The action of fermion field is set as
 \be\label{m-f1}
 W_{\rm matter}[g,\Psi]=\int d^4x \sqrt{-g}\left[-{\overline{\Psi}}\imath\gamma_{(\beta)}
 D_{(\beta)}\Psi-m_0\overline{\Psi} \Psi\right],
 \ee
where
$$\gamma_{(\beta)}=\gamma^\mu e_{(\beta)\mu}$$
-- Dirac $\gamma$--matrices, summarized with tetrads $e_{(\beta)\nu}$, and $m_0$ is a fermion mass at the present time.
Covariant differentials of set of fields $\Psi$ are defined by the formula
 \bea\label{fock}
D_{(\gamma)}\Psi&=&\frac{D\Psi}{\omega^P_{(\gamma)}}=
 \left[\partial_{(\gamma)}
 +\frac{\imath}{2}
v_{(\alpha) (\beta) ,(\gamma)} L^{\Psi}_{(\alpha) (\beta) }\right]\Psi,
 \eea
where
$$\partial_{(\gamma)}=(e^{-1})_{ \mu
(\gamma)}\partial_{\mu},$$
and
$$L^{\Psi}_{(\alpha)(\beta)}=[\gamma_{(\alpha)},\gamma_{(\beta)}]$$
-- are generators of the Lorentz group,
a linear form
$v_{(\alpha)  (\beta),(\gamma)}$
is built by Cartan forms (\ref{tR}) и (\ref{tL}):
 \bea
\label{carta1}
v_{(\alpha) (\beta),(\gamma)}= \left[\omega^L_{(\alpha)(
\beta)}(\partial_{(\gamma)})+ \omega^R_{(\alpha)(\gamma)}(\partial_{(\beta)})
-\omega^R_{(\beta )(\gamma)}(\partial_{(\alpha)})\right].
 \eea

\section{Theory of gravitation as nonlinear\\ realization of $A(4)\otimes C$}
\makeatletter
\renewcommand{\@evenhead}{\raisebox{0pt}[\headheight][0pt]
{\vbox{\hbox to\textwidth{\hfil \strut \thechapter . Nonlinear realizations of symmetry groups
\quad\rm\thepage}\hrule}}}
\renewcommand{\@oddhead}{\raisebox{0pt}[\headheight][0pt]
{\vbox{\hbox to\textwidth{\strut \thesection .\quad
Theory of gravitation as nonlinear realization of $A(4)\otimes C$
\hfil \rm\thepage}\hrule}}}
\makeatother

\subsection{Derivation of action of General Relativity}

\begin{center}
{\vspace{0.31cm}}
\parbox{0.5\textwidth}{
\includegraphics[height=9.truecm,
angle=-0]{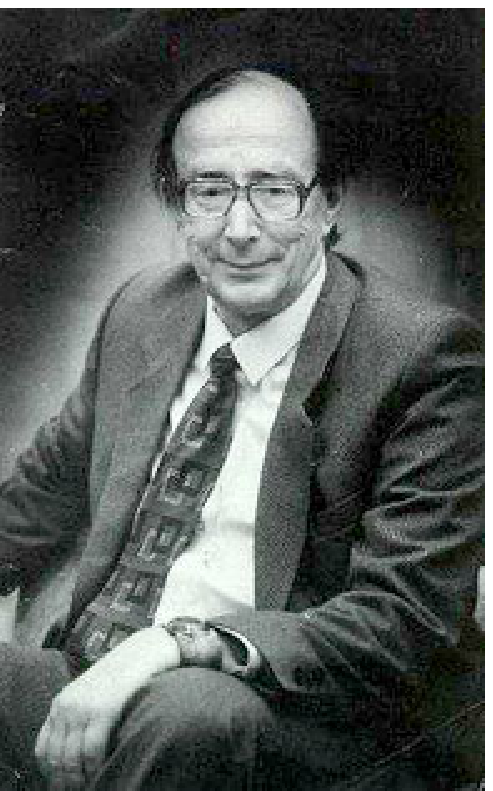}}
\parbox{0.48\textwidth}
{\small At the end of the 1950s - beginning of the 1960s of the last century, in close collaboration with I.V. Polubarinov,
V.I. Ogievetsky obtained a number of pioneer results in the range of theory of fields treatment of gauge
theories and gravitation.
The very bright achievement was the new comprehension of the theory of gravitation as
nonlinear realization of two spontaneously broken space-time symmetries -- conformal and affine ones, and a
graviton as corresponding to the Goldstone particle. Until the last years of his life he was a chief of the
``Supersymmetry'' sector in the Laboratory of Theoretical Physics named after of N.N. Bogoliubov, JINR.}
{\vspace{0.31cm}}
\end{center}

Using an analogy with phenomenological chiral Lagrangians \cite{1978-4},  it is possible to obtain phenomenological
affine Lagrangian as nonlinear joint realization of affine and conformal symmetry groups.
Such nonlinear joint realization was constructed in \cite{bor-og-chapter4}. The authors of the paper affirm that
the theory coincides with the Einstein General Relativity with Hilbert action,
if one chooses Lorentz subgroups as capacity of subgroup of vacuum
stability, and ten gravitons identify with ten parameters of coset space of proper
affine transformations
$$G=e^{\imath P\cdot x}e^{\imath R\cdot h}.$$

Covariant expression for action of Goldstone fields can be obtained
with the aid of commutator of covariant differentiation of a field $\Psi$ (\ref{fock})
 \be \left[
 D_{(\delta)}D_{(\gamma)}-D_{(\gamma)}D_{(\delta)}\right]\Psi
 =\imath R^{(4)}_{(\alpha)(\beta),(\delta)(\gamma)} L^{\Psi}_{(\alpha) (\beta)}
 \frac{\Psi}{2},
 \ee
where
$$R^{(4)}_{(\alpha)(\beta),(\gamma)(\delta)} =
\partial_{(\gamma)} \; v_{(\alpha) (\beta),(\delta)}+
v_{(\alpha) (\beta) ,(\zeta)} \; v_{(\delta) (\zeta),(\gamma)}
 + v_{(\alpha)(\zeta),(\delta)} \; v_{(\beta) (\zeta),(\gamma)}-$$
\bea
\label{curv}
-((\gamma)~\leftrightarrow~(\delta))
 \eea
is a tensor of curvature. Then from the Cartan forms one can get the Hilbert action for the General Relativity
\footnote{Recall that we use here and hereafter natural units
\be   \nonumber
 \label{units1}
  M^*_{\rm Pl}\equiv M_{\rm Pl}\sqrt{3/(8\pi)}=c=\hbar=1.
 \ee
}
 \be\label{m-1}
 W_{\rm H}(g)=-\int d^4x \left[\sqrt{-g}\frac{R^{(4)}(g)}{6}\right]
\ee
with the interval
\be\label{ginterval}
ds^2=g_{\mu\nu}dx^\mu dx^\nu.
\ee
However, the action of the General Relativity (\ref{m-1}) is not invariant of conformal group of symmetry.
The conformal invariant version of the General Relativity is obtained from the action (\ref{m-1}), if we implement
substitution of variables
$$g_{\mu\nu}=e^{-2D}\widetilde{g}_{\mu\nu}$$
and choose another definition of measured interval.
In this case the curvature takes a form
$$R^{(4)}(g=e^{-2D}\widetilde{g})=e^{-D}\left(R^{(4)}(\widetilde{g})-
6\widetilde{{\square}}\right)e^{-D},$$
where
$$\widetilde{{\square}}\equiv
\frac{1}{\sqrt{-\widetilde{g}}}
\frac{\partial}{\partial x^\mu}\left(
\sqrt{-\widetilde{g}}\widetilde{g}^{\mu\nu}\frac{\partial}{\partial x^\nu}
\right)$$
is the operator of D'Alembert in metric $\widetilde{g}$ with interval
\be\label{cginterval}
\widetilde{ds}^2=\widetilde{g}_{\mu\nu}dx^\mu dx^\nu.
\ee
After this substitution the action (\ref{m-1}) takes a form
\bea
 \label{1-1}
W_{\rm C}(\widetilde{g},D)=
\eea
$$
 =-\int\limits_{ }^{ }\! d^4x\,
 \biggl[
  \dfrac{\sqrt{-\widetilde{g}}}{6}\,{R}^{(4)}(\widetilde{g})
 \,e^{-2D}\,-e^{-D}\,\frac{\partial}{\partial x^\mu}
\, \left(\sqrt{-\widetilde{g}}\,\widetilde{g}^{\mu\nu}\,\frac{\partial}{\partial x^\nu} e^{-D}\,
\right) \biggr],
$$
where $D$ is a scalar dilaton field, scale transformation of which compensates transformations
of another fields.

Let us prove the conformal invariance of the action (\ref{1-1}) explicitly. For this, we take once more
conformal transformation
$$\widetilde g_{\mu\nu}=e^{-2\lambda}\widehat{g}_{\mu\nu}.$$
Since the Ricci scalar transforms under the conformal transformations as \cite{Eisenhart-4}
$$\frac{1}{6}\sqrt{-\widetilde{g}}\widetilde{R}^{(4)}=\frac{1}{6}e^{-2\lambda}\sqrt{-\widehat{g}}\widehat{R}^{(4)}
-e^{-\lambda}\frac{\partial}{\partial x^\mu}
\left(\sqrt{-\widehat{g}}\,\widehat{g}^{\mu\nu}\,\frac{\partial}{\partial x^\nu} e^{-\lambda}\,\right),$$
the action (\ref{1-1}) takes the form:
\bea
W_{\rm C}(\widehat{g},D, \lambda)=
\eea
$$=-\int\limits_{ }^{ }\! d^4x
 \biggl[
  \dfrac{\sqrt{-\widehat{g}}}{6}{R}^{(4)}(\widehat{g})
 e^{-2(D+\lambda)}-e^{-(2D+\lambda)}\frac{\partial}{\partial x^\mu}
\left(\sqrt{-\widehat{g}}\,\widehat{g}^{\mu\nu}\frac{\partial}{\partial x^\nu} e^{-\lambda}
\right) \biggr]-
$$
$$-\int\limits_{ }^{ }\! d^4x\, e^{-D}
\frac{\partial}{\partial x^\mu}
\left(e^{-2\lambda}\sqrt{-\widehat{g}}\,\widehat{g}^{\mu\nu}\frac{\partial}{\partial x^\nu} e^{-D}
\right).
$$
We transform the last two terms in the resulting expression, after selecting a common factor
$\exp{(-(D+\lambda))}$:
$$e^{-D}\frac{\partial}{\partial x^\mu}
\left(\sqrt{-\widehat{g}}\,\widehat{g}^{\mu\nu}\frac{\partial}{\partial x^\nu} e^{-\lambda}\right)+
\frac{\partial}{\partial x^\mu}
\left(\sqrt{-\widehat{g}}\,\widehat{g}^{\mu\nu}\, e^{-\lambda}\,\frac{\partial}{\partial x^\nu} e^{-D}\right)+
$$
$$
+\sqrt{-\widehat{g}}\,\widehat{g}^{\mu\nu}\,\frac{\partial}{\partial x^\mu}e^{-\lambda}
\frac{\partial}{\partial x^\nu}e^{-D}=
\frac{\partial}{\partial x^\mu}
\left(\sqrt{-\widehat{g}}\,\widehat{g}^{\mu\nu}\frac{\partial}{\partial x^\nu} e^{-(D+\lambda)}
\right).
$$
Now, we finally obtain
\bea
W_{\rm C}(\widehat{g},D,\lambda)=
\eea
$$
 =-\int\limits_{ }^{ }\! d^4x
 \biggl[
  \dfrac{\sqrt{-\widehat{g}}}{6}{R}^{(4)}(\widehat{g})
 e^{-2(D+\lambda)}-e^{-(D+\lambda)}\frac{\partial}{\partial x^\mu}
 \left(\sqrt{-\widehat{g}}\widehat{g}^{\mu\nu}\frac{\partial}{\partial x^\nu} e^{-(D+\lambda)}
\right) \biggr].
$$
The requirement of invariance of the action (\ref{1-1}) determines the dilaton field transformation:
$$D+\lambda= \widehat{D}.$$
Thus we proved the conformal invariance of the action (\ref{1-1})
\bea
W_{\rm C}(\widehat{g},D,\lambda)=W_{\rm C}(\widehat{g},\widehat{D}).
\eea
In the action of conformal invariant theory (\ref{1-1}) a number of variables is
the same as in the theory of Einstein (\ref{m-1}). Moreover, in the infinite volume all solutions
of classical equations of the theory (\ref{1-1}) correspond to the solutions of classical equations of
the theory (\ref{m-1}). However, the observed data irrefutably testify the finite space volume and
finite time interval of life of the Universe, finite energy and finite energy density. All these
finite values can be defined in a concrete system of reference. In conformal theory there is a
system of reference with unit determinant of space metric with conformal interval (\ref{cginterval}). It is just
the system we use for the classification of observed data.

\subsection{Differences between the standard General Relativity and the nonlinear realization of $A(4)\otimes C$}
The Lagrangian of joint nonlinear realization of product of groups is an analog of phenomenological Lagrangians.
The corresponding theory conserves all observed predictions of the General Relativity in the Solar system scales.
Nevertheless, the obtained theory differs from the metric formulation of the standard General Relativity.
Let us enumerate these differences.

\begin{enumerate}

\item
All measured fields and observables of the conformal theory
$\widetilde{F}^{(n)}\equiv F_c^{(n)}$,
including the metrics, are connected with the corresponding fields and observables of the standard General Relativity
${F}^{(n)}\equiv F_s^{(n)}$
by the scale transformation
 \bea \label{D-0}
F_c^{(n)}=e^{nD}F_s^{(n)},
 \eea
where $D$ is a scalar dilaton field,
and $(n)$ is a conformal weight. In particular, the metric tensor of the General Relativity $g_{\mu\nu}$ differs from
the metric tensor of the Conformal theory $\widetilde{g}_{\mu\nu}$:
   \be\label{D-g}
    g_{\mu\nu}=e^{-2D}\widetilde{g}_{\mu\nu}.
    \ee

\item
In distinction from an ordinary scalar field, the dilaton $D$ has indefinite metrics in the Hilbert space,
{\it id est} negative probability. The dilaton field $D$ can be decomposed by harmonics
 \bea \label{D-1g}
 D(x^0,x^1,x^2,x^3)=\langle D\rangle(x^0)+ \overline{D}(x^0,x^1,x^2,x^3),
 \eea
where $\langle D\rangle(x^0)$ is zeroth dilaton harmonics and $\overline{D}(x^0,x^1,x^2,x^3)$ is a sum of other nonlinear
harmonics with condition
$$\int_{V_0} d^3x \overline{D}=0.$$

\item
The zeroth dilaton harmonics $\langle D\rangle(x^0)$ is defined as average of dilaton by the finite volume
$V_0=\int_{V_0} {d^3x}$ 
 \bea
  \label{1-avd}
 &&\langle D \rangle(x^0)=V_0^{-1}\int_{V_0} {d^3x}
 D(x^0,x^1,x^2,x^3).
\eea
The zeroth dilaton harmonics describes \textit{luminosity} defined in observational cosmology and astrophysics as
(with minus sign) logarithm of cosmological scale factor
\bea \label{tau0}
  \langle D \rangle= - \ln a = \ln(1+z),
\eea
where $z=(1-a)/a$ is a redshift. The zeroth dilaton harmonics plays a role of time in the field space of events.
The corresponding canonical momentum of the zeroth dilaton harmonics becomes the energy of the Universe in this
field space of events and whereby the problem of nonzero energy in the General Relativity is solved.

\item
Non-zeroth dilaton harmonics $D$, in force of orthogonality with zeroth harmonics, have zeroth momenta and
become Newton-type potentials, just as the ones which twice increase an angle of deviation of light beam
by Solar field of attraction, in comparison with the Newton theory.

\item
Relations between variables of standard ($s$) and conformal models ($c$) can be illustrated with the example of
massive part of fermion action
\be\label{mass-1}
 W_{\rm m}[g_s,\Psi_s]=-\int d^4x \sqrt{-g_s}\;\overline{\Psi}_s \Psi_s \, m_0,
\ee
and their transformations into conformal values:
\bea
&& g_{s\mu\nu}=e^{-2D}{g}_{c\mu\nu}=e^{-2D}\widetilde{g}_{\mu\nu},\qquad \Psi_s = e^{3D/2}\Psi_c.
\eea
As a result, we get
 \be \label{m-1m1}
 W_{\rm m}[g_c,\Psi_c,D]=-\int d^4x \sqrt{-{g}_c}\;
 \overline{{\Psi}}_c {\Psi}_ce^{-D} \, m_0.
 \ee
This correspondence between the General Relativity and its conform--affine version had been already established by Dirac
\cite{Dirac1973chapter4}.
All classical tests in the General Relativity, including precession of Mercury perihelion, deviation of light beam
by the Sun, gravitational redshift and gravitational lensing are implemented in full.

\item
The result (\ref{m-1m1}) means that in the conform-affine version (\ref{1-1}) cosmological scale factor
(\ref{tau0}) changes not the interval but particle masses. Instead of expansion of space with constant sizes
of cosmic objects in the Standard cosmology, the
Conformal cosmology leads to the constant space with decreasing cosmic objects' sizes. So, the transition
to conformal variables as observables has the same radical paradigm as the transition to
heliocentrical system of reference in the Middle Ages. The consequence of this passage consists in
more simple classification of observable data in the system of reference of an observer where the
evolution of himself, together with objects of his observation, occurs. In the heliocentrical system
the observer himself rotates with the Earth around the Sun. In the class of system of references
and observable variables of the Conformal theory the observer himself experiences the cosmic evolution of his
mass instead of external space.

\item
In addition, in every point of the Riemannian space the tangent Minkowskian space in terms of Fock's frame
is introduced.
The interval takes a form of sum of products of Fock's tetrad components in the tangent Minkowskian space
with the metrics
$$\eta^{(\alpha)(\beta)}=sign:(1,-1,-1,-1):$$
 $$\widetilde{g}_{\mu\nu}dx^\mu dx^\nu=\omega_{(\alpha)} \otimes \omega_{(\beta)}
\eta^{(\alpha)(\beta)}.$$

\end{enumerate}

The components of Fock's frame $\omega_{(\alpha)}$ are invariants  respective to general coordinate
transformations. That is why, as we demonstrate later, graviton in the theory (\ref{1-1}) has only single
component, contrary to the standard General Relativity.

\newpage
\section{Summary}
\makeatletter
\renewcommand{\@evenhead}{\raisebox{0pt}[\headheight][0pt]
{\vbox{\hbox to\textwidth{\hfil \strut \thechapter .
Nonlinear realizations of symmetry groups \quad\rm\thepage}\hrule}}}
\renewcommand{\@oddhead}{\raisebox{0pt}[\headheight][0pt]
{\vbox{\hbox to\textwidth{\strut \thesection .\quad
Summary and literature
\hfil \rm\thepage}\hrule}}} \makeatother

For the purpose of construction of the quantum operator of creation and evolution of the Universe
as unitary irreducible representation of conformal and affine groups of symmetry, one represents
general elements of the theory of nonlinear realizations of groups of symmetry developed by $\acute{\rm E}$lie Cartan
\cite{Cartan-4}. Then the  derivation of classical theory of gravitation as nonlinear joint realization of
conformal and affine symmetries \cite{bor-og-chapter4} is presented by analogy with the derivation of the chiral
phenomenological Lagrangian for pions \cite{1978-4}.

The derived theory of gravitation contains, besides known physical effects of the General Relativity for the Solar system,
all elements of further development of Einstein's ideas suggested by his contemporaries and followers,
including Hilbert's variational principle of action (1915), Fock's frames \cite{Fock:1929-4} in the tangent space of Minkowski,
Dirac's conformal interval, where the determinant of metrics is identified with the scalar dilaton.


\chapter{Hamiltonian formulation of the theory of gravity\label{S-5}}

\section{Foliation 4=3+1\label{S-5a}}
\renewcommand{\theequation}{5.\arabic{equation}}
\setcounter{equation}{0}
\makeatletter
\renewcommand{\@evenhead}{\raisebox{0pt}[\headheight][0pt]
{\vbox{\hbox to\textwidth{\hfil \strut \thechapter . Hamiltonian formulation of the theory of gravity
\quad\rm\thepage}\hrule}}}
\renewcommand{\@oddhead}{\raisebox{0pt}[\headheight][0pt]
{\vbox{\hbox to\textwidth{\strut \thesection .\quad
Foliation 4=3+1
\hfil \rm\thepage}\hrule}}}
\makeatother

There is a one-to-one correspondence between solutions of the Conformal dilaton Dirac's theory,
(\ref{1-1}) and classical solutions of Einstein's equations in the General Relativity
$$\dfrac{\delta W_{\rm H}}{\delta g_{\mu\nu}}=0$$
in terms of components of a metrics $g_{\mu\nu}$.
Metric components are objects of arbitrary general coordinate transformations.
In particular, the group of general coordinate transformations (diffeomorphisms) of the Hamiltonian approach
contains the following coordinate transformations
\bea \label{gt}
 x^0 \rightarrow \tilde x^0&=&\tilde x^0(x^0);\\
 \label{kine}
 x_{i} \rightarrow  \tilde x_{i}&=&\tilde x_{i}(x^0,x_{1},x_{2},x_{3}).
 \eea

This group of transformations preserves the family (congruence) of
hypersurfaces $x^0=\rm{const}$,
and is called a {\it kinemetric subgroup} \cite{zelmanov-5} of the
group of general coordinate transformations
$$x_{\mu} \rightarrow  \tilde x_{\mu}=\tilde x_{\mu}(x^0,x_{1},x_{2},x_{3}).$$

\begin{figure}[hpb!]
\begin{centering}
\includegraphics[width=3.5in]{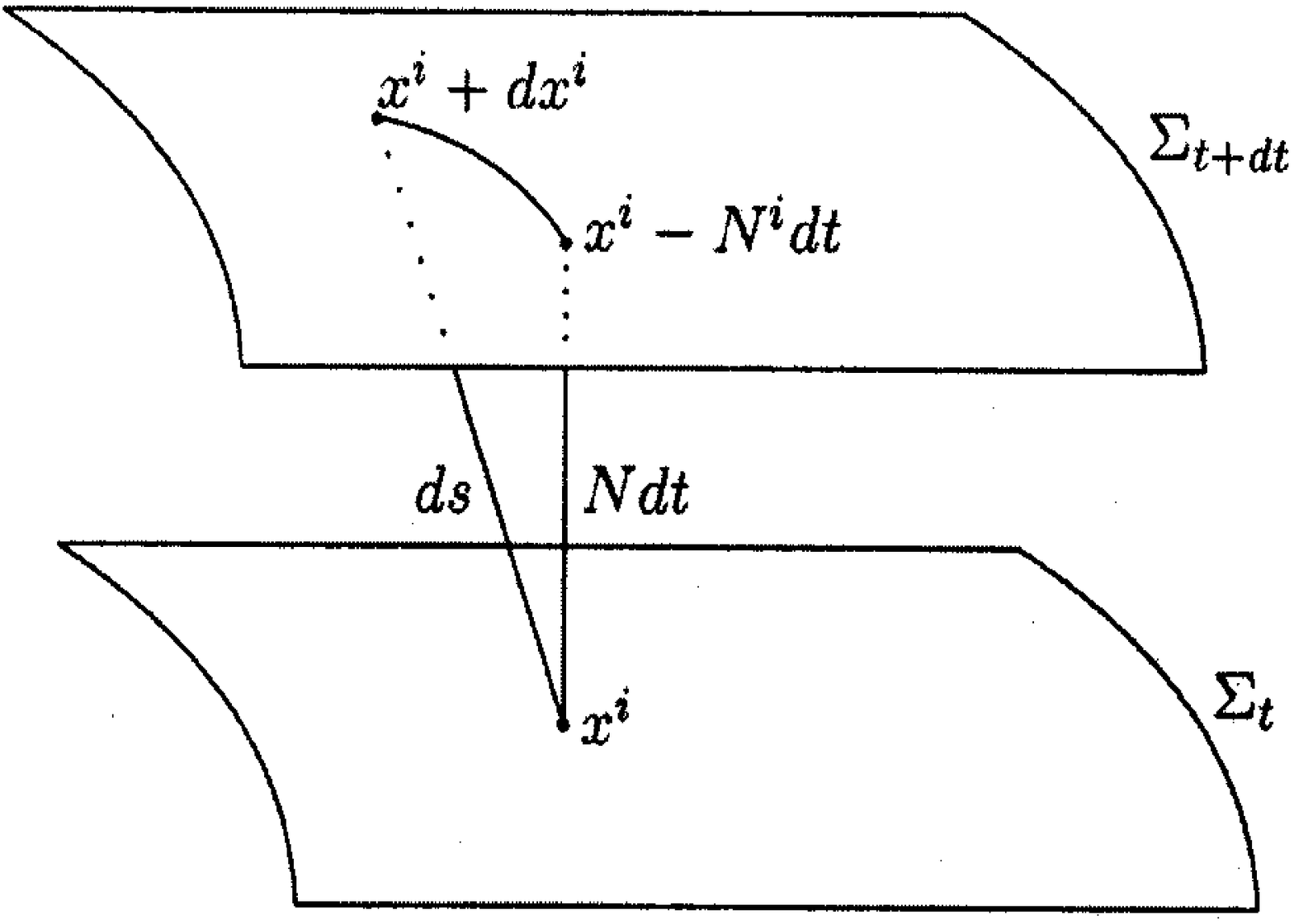}
 \label{Fig.5.1}
\vskip 0.5cm
\end{centering}
{\small
Figure \ref{Fig.5.1} shows a line of time and two space-like three-dimensional hypersurfaces
through which this line of time is spent in the General Relativity.
A transition from a $\sum_t$ to a hypersurface $\sum_{t+dt}$ is described by a lapse function $N$ and
a shift-vector $N_i$.
The family of all the space-like three-dimensional hypersurfaces is called a congruence,
and the appropriate parametrization of the metric component is called 4 = 3 + 1 bundle of the space-time.
}
\end{figure}

The group of kinemetric transformations contains reparametrizations of a {\it coordinate time} (\ref{gt})
in class of functions, depending only on a coordinate time, which we call as global.
Whereas, the transformations (\ref{kine}) we call as local. Thus, the subgroup of
diffeomorphisms of the Hamiltonian formulation of the General Relativity (\ref{gt}) and (\ref{kine}) is
composed of one global and three local transformations, {\it id est}, the structure of the
kinemetric subgroup is of the form
$1G\oplus3L$.

Identification and extraction of physical degrees of freedom is one of the most important problems of the
theory of gravity, which stimulated Dirac to create a generalized Hamiltonian formulation of systems
with constraints \cite{Dirac1979-5}, and, later, to develop this formulation by many authors
\cite{Sundermeyer1982-5, Faddeev1984-5, Henneaux1992-5}.
The solution to this problem consists in extraction of true evolution of the observed dynamical and geometrical
magnitudes from general coordinate (gauge) transformation (\ref{gt}) and (\ref{kine}).

Introduced above, the formalism of Cartan allows us to formulate the theory of gravity in terms of invariants
relative to general coordinate transformations
through the transition to invariant components of Fock's frames.
A foliation of the space-time 4=3+1 (see Fig.\ref{Fig.5.1}) involves the introduction of components of Fock
frames $\omega_{(\alpha)}$ in the following form
 \bea \label{dg-2aa-chapter5}
&&{{\omega}}_{(0)}=e^{-2{D}}N dx^0,
\\
 \label{dg-3a-chapter5} &&{{\omega}}_{(b)}={\bf e}_{(b)i}dx^i+{N}_{(b)}dx^0.
\eea
Here $N$ is the lapse function in the theory (\ref{1-1}),
$${N}_{(b)}=N^j{\bf e}_{(b)j}$$
are components of the shift-vector;
${\bf e}_{(b)i}$  are orthonormal triad components with unit determinant:
$${\bf e}_{(b)i}{\bf e}^j_{(b)}=\delta^j_i;~~~~~~~{\bf e}_{(a)j}{\bf e}^j_{(b)}=\delta_{(a)(b)}.$$

\begin{center}
{\vspace{0.31cm}}
\parbox{0.5\textwidth}{
\includegraphics[height=8.truecm,
angle=-0]{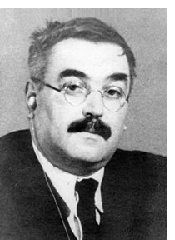}}
\parbox{0.48\textwidth}
{\small V.A. Fock was born in Saint - Petersburg.
After graduating from high school in Petrograd (1916), he entered the faculty of physics and mathematics
of the University of Petrograd.
His primary scientific contribution lies in the development of quantum physics, although he also contributed
significantly to the fields of mechanics, theoretical optics, theory of gravitation, physics of continuous media.
In 1926 he derived the Klein –- Gordon equation. He gave his name to Fock space,
the Fock representation and the Fock state, and developed the Hartree -- Fock method in 1930.
He wrote the first textbook on quantum mechanics ``Foundations of Quantum Mechanics'' (1931)
and a very influential monograph ``The Theory of Space, Time and Gravitation'' (1955).
}
{\vspace{0.31cm}}
\end{center}

In the action of affine - conformal theory of gravity (\ref{1-1}), expressed through the Maurer -- Cartan
forms, differentials of coordinates of the Riemannian space are not directly measurable quantities
$dx^0$ and $dx^i$,
but invariant under general coordinate transformations of
orthogonal components of the frame in the tangent space (\ref{dg-2aa-chapter5}) and (\ref{dg-3a-chapter5}).
These components are, in general, non-integrable linear forms.
The dependence of the linear forms from the coordinates of the tangent space
$$X_{(b)}=x^i {\bf e}_{(b) i}$$
can be found by using the Leibnitz rule
$$AdB=d [AB]-[AB]d\ln A$$
and the condition of orthogonality of triads
$${\bf e}_{(a)\,i}{\bf e}^{j}_{(a)}=\delta^j_i.$$
Substituting these expressions to the linear form
$${\overline{\omega}}_{(b)}(d) = {\bf e}_{(b)i}dx^i,$$
we get
$$d[x^i]{\bf e}_{(b)i}=d[x^i{\bf e}_{(b)i}]
-x^id[{\bf e}_{(b)i}]=d[x^i{\bf e}_{(b)i}]
-[x^i{\bf e}_{(a)\,i}][{\bf e}^{j}_{(a)}]d[{\bf e}_{(b)j}].$$
Then, using the definition of the observables
$X_{(b)} = x^i {\bf e}_{(b)i}$,
it is possible to find the sought-for dependence:
 \bea
 \label{P-1e}
 &{\overline{\omega}}_{(b)}(d)={\bf e}_{(b)i}dx^i
 = dX_{(b)}-X_{(c)}{\bf e}^{i}_{(c)}d{\bf e}_{(b)i} \nonumber\\
  &=dX_{(b)}-X_{(c)}[\omega^R_{(c)(b)}(d)+\omega^L_{(c)(b)}(d)],
 \eea
where $$\omega^R_{(c)(b)}(d)=\frac{1}{2}({\bf e}^{i}_{(c)}d{\bf e}_{(b)i}+
{\bf e}^{i}_{(b)}d{\bf e}_{(c)i}),$$
$$\omega^L_{(c)(b)}(d)=\frac{1}{2}({\bf e}^{i}_{(c)}d{\bf e}_{(b)i}-
{\bf e}^{i}_{(b)}d{\bf e}_{(c)i})$$
are the Cartan's forms
(coefficients of the spin connection), describing strong gravitational waves.
The factor $X_{(c)}$ in Eq.(\ref{P-1e}) means that a hypersurface, perpendicular to the wave vector
of a gravitational wave experiences an expansion or a contraction of the Hubble type \cite{grg12chapter5},
known in the Standard cosmology.

\section{Hamiltonian formulation of GR\\ in terms of Cartan forms}
\renewcommand{\theequation}{5.\arabic{equation}}
\makeatletter
\renewcommand{\@evenhead}{\raisebox{0pt}[\headheight][0pt]
{\vbox{\hbox to\textwidth{\hfil \strut \thechapter . Hamiltonian formulation of the theory of gravity
\quad\rm\thepage}\hrule}}}
\renewcommand{\@oddhead}{\raisebox{0pt}[\headheight][0pt]
{\vbox{\hbox to\textwidth{\strut \thesection .\quad
Hamiltonian formulation of GR in terms of Cartan forms
\hfil \rm\thepage}\hrule}}}
\makeatother

Let us reformulate the standard description of the General Relativity in terms of Cartan's forms.
The Hilbert action with the electromagnetic field
$F_{\mu\nu}=\partial_\mu A_\nu -\partial_\nu A_\mu$
and the scalar field $Q$ has a form\footnote{Remember that we use the natural system of units
$$\hbar=c=M_{\rm Planck}\sqrt{{3}/({8\pi}})=1.$$}:
\bea \label{1-1dc}
 {W}[g,A,Q]=
\eea
$$= -\int d^4x {\sqrt{-g}}\;
 \Bigg{( }\dfrac{1}{6} R^{(4)}(g)
- \frac{1}{4}F_{\mu\alpha}F_{\nu\beta}g^{\mu\nu}g^{\alpha\beta}
 +\partial_\mu Q\partial_\nu Q g^{\mu\nu}\Bigg{)}.
$$
Passing to conformal variables (\ref{D-g})
$g_{\mu\nu}=e^{-2D}\widetilde{g}_{\mu\nu}$,
we get the action
\bea \label{1-1dcc}
 {W}[\widetilde{g},A,{Q}]= -\int d^4x {\sqrt{-\widetilde{g}}}\;
 \Bigg{[ }e^{-D}\left(\dfrac{1}{6} R^{(4)}(\widetilde{g})-\Box\right)e^{-D}\\
  - \frac{1}{4}F_{\mu\alpha}F_{\nu\beta}\widetilde{g}^{\mu\nu}\widetilde{g}^{\alpha\beta}
 +\partial_\mu {Q}\, \partial_\nu {Q}\, \widetilde{g}^{\mu\nu}\Bigg{]},
 \eea
where \be\label{1-1ddc}
\Box \equiv\frac{1}{\sqrt{-\widetilde{g}}}\partial_\mu\left(\sqrt{-\widetilde{g}}
\widetilde{g}^{\mu\nu}\partial_\nu \right)\ee
is the operator of D'Alembert.
By determining the tetrad components (\ref{dg-2aa-chapter5}) and (\ref{dg-3a-chapter5}),
the action (\ref{1-1dc}) is rewritten in the following form
  \bea
  \label{1-3dgl}
 {W}&=&\int\limits_{}^{}
 d^4x{N}\left[{\cal L}_D+{\cal L}_g+{\cal L}_A+{\cal L}_Q\right].
 \eea
Here
 \bea\nonumber
{\cal L}_D&=&-{v^2_{{D}}}-\frac{4}{3}e^{-7D/2}\triangle e^{-D/2},
 \\\nonumber
{\cal L}_g&=&\frac{1}{6}\left[v_{(a)(b)}v_{(a)(b)}-e^{-4D}R^{(3)}({\bf
e})\right],
 \\\nonumber
{\cal L}_A&=&\frac{1}{2}\left[v^2_{(b)(\rm A)}-F_{ij}F^{ij}\right],
 \\\nonumber
{\cal L}_Q&=&e^{-2D}{(v_{{Q}})^2}
 -e^{-2D}\left(\partial_{(b)}{{Q}}
\right)^2;
 \eea
Lagrangian densities and
 \bea
 \nonumber
 v_{Q}&=&\frac{1}{N}\left[(\partial_0-N^l\partial_l){{Q}}
+\partial_lN^l/3\right],\\\label{1-3-d}
 v_{{D}}&=&\frac{1}{N}\left[(\partial_0-N^l\partial_l){D}
+\partial_lN^l/3\right],\\\label{1-3-dD}
 \label{1-3-dvvc}
 v_{(a)(b)}&=&\frac{1}{N}\biggl[\omega^R_{(a)(b)}(\partial_0-N^l\partial_l)+
 \partial_{(a)}N^{\bot}_{(b)}
 +\partial_{(b)}N^{\bot}_{(b)} \biggr],
 \\\nonumber
 \label{1-28}
 v_{(b)(\rm A)}&=&\frac{1}{N}{\bf e}^i_{(a)}
 \left[\partial_0A_i-\partial_iA_0+F_{ij}N^j\right]
 \eea
are velocities of the metric components and fields, and
$R^{(3)}({\bf{e}})$ is a three-dimensional spatial curvature, expressed in terms of triads ${\bf e}_{(a)i}$
\bea \label{1-17a}
 R^{(3)}&=&R^{(3)}({\bf e})-
 \frac{4}{3}{e^{7D/2}}\triangle e^{-D/2},
\eea
\bea \label{1-17}
R^{(3)}({\bf e})&=
\eea
$$=-2\partial^{\phantom{f}}_{i}
 [{\bf e}_{(b)}^{i}\sigma_{{(c)|(b)(c)}}]-
 \sigma_{(c)|(b)(c)}\sigma_{(a)|(b)(a)}
 + \sigma_{(c)|(d)(f)}^{\phantom{(f)}}
 \sigma^{\phantom{(f)}}_{(f)|(d)(c)},
$$
where
$$
 \sigma_{{(c)}|(a)(b)} = [\omega^L_{(a)(b)}(\partial_{(c)})+
 \omega^R_{(a)(c)}(\partial_{(b)} )-
 \omega^R_{(b)(c)}(\partial_{(a)})],
$$
\bea\label{1-19}
 \omega^R_{(a)(b)}(\partial_{(c)})&=&\frac{1}{2}\left[
 {\bf e}^j_{(a)}\partial_{(c)}{\bf e}^j_{(b)}
 +{\bf e}^i_{(b)}\partial_{(c)}{\bf e}^i_{(a)}\right],
\eea
\bea\label{1-20}
 \omega^L_{(a)(b)}(\partial_{(c)})&=&\frac{1}{2}\left[
 {\bf e}^j_{(a)}\partial_{(c)}{\bf e}^j_{(b)}
 -{\bf e}^i_{(b)}\partial_{(c)}{\bf e}^i_{(a)}\right],
 \eea
and
$$\triangle \equiv\partial_i[{\bf e}^i_{(a)}{\bf e}^j_{(a)}\partial_j]$$
is the operator of Beltrami -- Laplace.

With the help of Legendre transformation $v^2/N=pv-Np^2/4$ we define momenta
 \bea \label{1-17g}
 p_{(a)(b)}&=&\frac{v_{(a)(b)}}{3},
 \\
 \label{1-17D}
 p_{D}&=& 2{v_{{D}}},
 \\
 \label{1-17Q}
 p_{Q}&=&2{v_{{Q}}},
 \\
 \label{1-17A}
 p_{\rm A(b)}&=&{v_{\rm A(b)}}.
 \eea
As a result, the action (\ref{1-3dgl}) takes the Hamiltonian form
\bea\label{1-3dgc}
  {W}&=&\\\nonumber
&=& \int\limits_{ }^{ }\!\! d^4x \biggl[
 p_Q\partial_0{{{Q}}}
 + p_{(a)(b)}\omega_{(a)(b)}^R(\partial_0)
 +p_{A{(b)}}\partial_0A_{(b)} -p_D\partial_0{{D}}
-{\cal C}\biggr],\\\label{1-2c}
 {\cal C} &=& \\\nonumber
&=& N{\cal H}+N_{(b)}T_{(b)}
 +A_{(0)}\partial_{(b)}p_{A{(b)}}
 + \lambda_{(0)}p_D + \lambda_{(b)}\partial_k {\bf e}^k_{(b)}
 + \lambda_{A}\partial_{(b)} A_{(b)},
\eea
where $N$, $N_{(b)}$ and $A_{(0)}$
are Lagrangian multipliers, variation in which yields
the first kind constraints on the classification of Dirac \cite{Dirac1979-5}, and
$\lambda_{(0)}$, $\lambda_{(b)}$ and $\lambda_{A}$
are Lagrangian multipliers for the second kind of constraints
\bea
 \label{h-2g}
 && \partial_k {\bf e}^k_{(b)} = 0,\\
 \label{D-1}
 && p_{{D}} =  0.
 \eea
First three constraints (\ref{h-2g}) fix space coordinates, and the constraint (\ref{D-1}) is
known as the minimum condition of three-dimensional hypersurface embedded in a four-dimensional
pseudo-Riemannian space.
In the Lagrangian formulation the constraint (\ref{D-1}) looks like an equation
on the divergence of the shift-vector
\bea\label{h-2g1-shift}
   \partial_0 (e^{-3D})+\partial_l (N^le^{-3D})=0.
 \eea
Magnitudes
 \bea \label{CC-3}
 {\cal H}& =&-\frac{\delta {W}}{\delta N}={\cal H}_D+{\cal H}_g
 +{\cal H}_A+{\cal H}_Q,
 \eea
where
 \bea
 \label{CC-4a}
 {\cal H}_D&=&-\frac{p^2_{D}}{4}-\frac{4}{3}e^{-7D/2}\triangle e^{-D/2},\\\label{3g}
 {\cal H}_g&=&\left[6{p^2_{(a)(b)}}+
 \frac{e^{-4D}}{6}R^{(3)}({\bf e})\right],\\\label{3A}
 {\cal H}_A&=&
 \frac{e^{-2 D}}{2}\left[{p_{i(\rm A)}p^i_{(\rm A)}}+
F_{ij}F^{ij}\right],\\\label{3Q}
 {\cal H}_Q&=&e^{-2D}\!\left[\!e^{2D}\frac{p_{{Q}}^2}{4}
 + e^{-2D}\left(\!\partial_{(b)}{Q}\!
 \right)^2\right], \eea
and
 \bea\label{3T}
T_{(0)(a)}& = &-{\bf e}_{(a)}^i\frac{\delta W} {\delta
{N}_{i}}=
  -\partial_{(b)}p_{(b)(a)}+\widetilde{T}_{(0)(a)},
 \eea
where
 \bea  \label{CC-4a1}\widetilde{T}_{(0)(a)} =
\sum\limits_{F=\overline{\phi},\overline{Q},\widetilde{F}}^{}p_F\partial_{(a)}F
\eea
are components of the tensor of energy -- momentum.

Conditions of the components of the energy -- momentum are equal to zero
  \bea \label{CC-3h}
 {\cal H}&=&0,\\\label{CC-3t}
 T_{(0)(a)}&=&0
 \eea
and were called by Dirac as primary constraints of the first kind.
In accordance with it, the first of these conditions (\ref{CC-3h}) is called a {\it Hamiltonian constraint},
by analogy with the corresponding condition for a relativistic particle.
Recall that the Hamiltonian of a relativistic particle is a solution of the Hamiltonian constraint,
relative to the momentum canonically conjugated to the evolution parameter in the space of events.
In this way of explicit solutions of the primary constraints of the first kind
there is one of the central problems of relativistic theories of gravity -- a choice of
a parameter of evolution in the field space of events.

As for the explicit solution of the second constraint (\ref{CC-3t}), it is convenient to use the expansion
 \bea
 \label{dv-7}
 N_{(b)}&=&N^{|\!|}_{(b)}+N^{\bot}_{(b)},\\
 \label{dv-8}
 \partial_{(b)}N^{|\!|}_{(b)}&=& \partial_jN^j,\\
 \label{dv-9}
 \partial_{(b)}N^{\bot}_{(b)}&=&0,
 \\ \label{c-6}
 p_{(b)(a)} &=& p^{\bot}_{(b)(a)}+\partial_{(a)}
 f^{\bot}_{(b)}+\partial_{(b)} f^{\bot}_{(a)}.
 \eea
Square of the momentum in the equation (\ref{3g}) is possible to be represented as
 \be \label{c-7}
 p^2_{(b)(a)}=(p^{\bot}_{(a)(b)})^2+[\partial_{(a)}
 f^{\bot}_{(b)}+\partial_{(b)} f^{\bot}_{(a)}]^2,
 \ee
where  $f^{\bot}_{(a)}$ satisfies the equation
 \be \label{c-8}
 \left[\triangle f^{\bot}_{(a)}+\partial_{(a)}\partial_{(b)} f^{\bot}_{(a)}\right]=\widetilde{T}_{(0)(a)},
 \ee
which follows from (\ref{CC-3t}) after a substitution of (\ref{c-6}).

The constraint of the second class (\ref{D-1}) leads to one more secondary constraint
$$\frac{\delta W}{\delta D}=-T_D=0,$$
namely,
 \bea\label{e2t}
 && (\partial_\tau-{\cal N}_{(b)}\partial_{(b)})p_{{D}} = T_D,
\eea
where
$$
T_D=\frac{4}{3}\left[7{\cal N}e^{-7D/2}\triangle e^{-D/2}
 +  e^{-D/2}\triangle[{\cal N}e^{-7D/2}]\right]-$$
$$ - {\cal N}\partial_D[{\cal H}_g+{\cal H}_A+{\cal H}_Q].
$$
We here just adapted the standard Hamiltonian formulation \cite{ADM-5} of the theory of gravitation
to Cartan's forms. In terms of these forms, the curvature takes the bilocal form.
The action of such a theory describes a physical system as a squeezed oscillator
\cite{Arbuzov:2010fz5-5}.
It gives hope to construct a quantum theory of such a system, if we can solve the problems of the
standard Hamiltonian formulation at the level of Cartan's forms.
\newpage
\section{Problems of Hamiltonian formulation}
\renewcommand{\theequation}{5.\arabic{equation}}
\makeatletter
\renewcommand{\@evenhead}{\raisebox{0pt}[\headheight][0pt]
{\vbox{\hbox to\textwidth{\hfil \strut \thechapter . Hamiltonian formulation of the theory of gravity
\quad\rm\thepage}\hrule}}}
\renewcommand{\@oddhead}{\raisebox{0pt}[\headheight][0pt]
{\vbox{\hbox to\textwidth{\strut \thesection .\quad
Problems of Hamiltonian formulation
\hfil \rm\thepage}\hrule}}}
\makeatother

Let us list these problems.
\begin{enumerate}
\item
The first of them is a problem of unique determination of non-zero Hamiltonian as the generator of evolution.
The fact that the General Relativity is a singular theory with primary and secondary first-class constraints.
The Hamiltonian as a constraint is equal to zero. As a result, under the constraints ${\cal C}=0$
the action takes a form
\bea\label{1-3dgc1}
  {W}_{C=0}=
\eea
$$
 =\int\limits_{ }^{ }\!\! d^4x \biggl[p_{(a)(b)}\omega_{(a)(b)}^R(\partial_0)
 +p_Q\,\partial_0{{{Q}}}
 +p_{A{(b)}}\partial_0A_{(b)} -p_D\partial_0{{D}}
\biggr],
$$
where all of the canonical momenta and velocities satisfy the conditions of constraints \cite{Isham-5}.
This fact complicates the unambiguous definition of the generator of the evolution for a quantum state
in a corresponding quantum theory.

\item
The second problem is the self-consistency of the perturbation theory.
As it was noticed as early as by K. Kucha$\check{\rm r}$ \cite{Ku1970-5}, the lapse function $N$
generally is not included in the linearized constraint equations.
That is non-self-consistency that in turn greatly impedes formulation of a perturbative quantum theory.
Indeed, the metric representation of the functional of the state is based
on the assumption that the components of the metric tensor are
taken as independent variables.
In the classical theory this hypothesis was formulated as a ``thin sandwich theorem'',
according to the initial values of the metric tensor together with its derivatives that
uniquely (under suitable boundary conditions) determine the space-time metric.
It is supposed that by setting the initial hypersurface the metric tensor together with its derivatives
and using four constraint equations, one can determine the four unknown ones -- the lapse function and
the shift-vector,
that is to define a fully 4-metric of space-time.
In the linear approximation, this theorem is violated and it is necessary somehow to fix the lapse function
and the shift-vector.
From here one can conclude that in the linear approximation there is not enough information to
determine, for example, the lapse function by a given metrics and its time derivative.

\item
One more problem is the problem of reduction. It means a separation of the dynamical variables of the theory on
the constraints surfaces of the excess parameters of gauge transformations.
Of course, this problem is related to the previous two.
There are two ways to solve this problem.
The first consists of imposition of additional gauge conditions to exclude extra variables.
The second way is solving of the constraints.
To the advantages of the first method one should notice its convenience and simplicity, because, as usual,
such conditions are chosen that substantially enable calculations,
but its drawback is in quite narrow applicability
of such gauge and lack of confidence that the particular
gauge does not spoil the ``true'' dynamics.
The method of solving of the constraints, if it could be carried out,
should be ideal for researchers \cite{PS-1-5, S-2-5, S-1-5}.

\item
Structure of the four local constraints of the first kind (\ref{h-2g}) and (\ref{D-1})
does not reflect the structure of diffeomorphisms of the
Hamiltonian formulation of the theory of gravity.
Recall that in the Hamiltonian formulation
we have one global (\ref{gt}) and three local diffeomorphisms (\ref{kine}).

\item
Of course, on the way of constructing the quantum gravity there are
a number of other problems of fundamental as well  as technical
character. Among them we mention nonrenormalizability of the theory
associated with the dimension of Newton's constant, and the questions of
interpretation of the vector of quantum state \cite{KucharTime-5} describing the quantum Universe. For
the latter, there are no outside classic instruments.

\item
The local condition of the minimal surface $v_D=0$ (\ref{D-1})
leads to the absence of any dynamics including cosmological\footnote{In the General Relativity, this statement
can be summarized as follows:
in non static space of the General Relativity with a closed family of hypersurfaces $t = const$ and non-zero
energy -- momentum tensor of matter, there is no global time-like congruence
({\it id est}, a continuous family of time-like lines)
that a field of unit tangent vectors to this congruence would satisfy to the following properties:
1) tensor of the angular velocity is equal to zero,
2) trace of the tensor of velocities of deformations is also equal to zero \cite{PS-1-5, S-2-5}.}.

\item
The class of functions of standard perturbation theory \cite{Faddeev:1973-5}
$$g_{\mu\nu}(x^0,{\bf x})=\eta_{\mu\nu}+O(1/|\textbf{x}|),$$
where
$\eta_{\mu\nu}={\mathrm{Diag}:}(1,-1,-1,-1)$,
eliminates the cosmological evolution.
Recall that since the pioneering results of Friedmann
and, continuing their modern development \cite{lifhal-5, bard-5, Giovannini-5, Barbashov:2005hu5-5},
the cosmological evolution is introduced into the theory of gravity by
nonperturbative infrared dynamics of the metric tensor
$g_{\mu\nu}(x^0)\not =\eta_{\mu\nu}$
with a finite time interval, and a finite volume of space.
\end{enumerate}
In other words, the above list of problems is
proposed to be solved with the introduction of the zero harmonic  of the dilaton  (\ref{D-1g}) to the action
(\ref{1-1}).
In the future, as we said above, we call the equation (\ref{CC-3h})
not the zero Hamiltonian, but the \textit{Hamiltonian constraint},
and solve the Hamiltonian constraint relative to one of the canonical momenta, in complete analogy with
solution of the equation of mass surface in the Special Theory of Relativity.
Then, the canonical momentum must be associated with the Hamiltonian of the reduced system
(which on the solutions of the classical equations will be identified with the energy of the system).
Canonically conjugated magnitude to the Hamiltonian
must be a scalar (or scalar density) with respect to kinemetric transformations.
Zero harmonic of the dilaton  (\ref{D-1g}) as a conformal multiplier, extracted from the metric,
is precisely such a quantity \cite{Barbashov:2005hu5-5}.

Here we list the solutions to these problems that have been given at the level of a model miniuniverse:

\begin{enumerate}
\item
The information capacity of a relativistic theory with constraints is much more than
the information contained in a non-relativistic theory.
It is sufficient to say that the relativistic theory of gravitation
(as well as the General Relativity and other theories) has \textit{three spaces}:

1) \textit{pseudo--Riemannian}, introduced
by Einstein,

2) \textit{tangent}, introduced by Fock, and

3) \textit{space of events}, introduced by Bryce De Witt.

In all of these spaces there is their own evolution parameter:

1) a coordinate time as an object of the general coordinate transformations,

2) a geometric interval (or components of the Fock's frame), and

3) dynamic evolution parameter in the space of events, respectively.

A non-zero Hamiltonian as the generator of evolution in the space of events is uniquely determined,
if we specify the dynamic evolution parameter in this space,
solve the equation of the Hamiltonian constraint and implement the primary and secondary quantization
to establish a stable vacuum.

\item
A lapse function $N$ is included into the number of the observed ones only in the form of a factor to
the differentials of the coordinate time.
(In other words, only the invariants as components of the Fock's frame are measured).

\item
A way of resolving constraints, if it is fully implemented,
is ideal for the identification of the true dynamics of relativistic
systems with constraints \cite{PS-1-5, S-2-5}. This is the method we use hereinafter.

\item
Next, we will distinguish the Hamiltonian constraint from a non-zero Hamiltonian,
which is a solution to that constraint.

\item
Questions of interpretation of the vector of quantum state, describing the quantum Universe, are
solved taking into account the fact that the role of external classical instruments was played by Casimir vacuum.
\end{enumerate}
A solving of the constraints is called a reduction of the extended phase space on the space of physical variables.
The problem of solving of the constraint equations in terms of linear forms will be discussed in the next
Section of this Chapter.

\section{Exact solution of the Hamiltonian\\ constraint \label{S-10}}

\renewcommand{\theequation}{5.\arabic{equation}}
\makeatletter
\renewcommand{\@evenhead}{\raisebox{0pt}[\headheight][0pt]
{\vbox{\hbox to\textwidth{\hfil \strut \thechapter . Hamiltonian formulation of the theory of gravity
\quad\rm\thepage}\hrule}}}
\renewcommand{\@oddhead}{\raisebox{0pt}[\headheight][0pt]
{\vbox{\hbox to\textwidth{\strut \thesection .\quad
Exact solution of the Hamiltonian constraint
\hfil \rm\thepage}\hrule}}}
\makeatother
\subsection{Statement of the problem}

From the theory of nonlinear finite-parametric representations of the symmetry groups there
was derived the action of the Conformal Theory of Gravity that contains all consequences of the General Relativity
for the Solar system.
However, the Conformal Theory of Gravity is significantly different from the General Relativity under
descriptions of cosmological data.

$\bullet$
The action functional of the theory is defined on three spaces --- Riemannian $x^\mu$,
tangent $\omega_{(\alpha)}$ and field $[D|F]$, each of them has its own evolution parameter:
$x^0$, $\omega_{(0)}$, and $\langle D\rangle$.

$\bullet$
The second difference is the identification of the observed distances with the conformal
geometric interval. In contrast to the standard interval of the General Relativity, the
geometric interval can describe all the observed data at different periods of the evolution of the Universe
by the dominant Casimir vacuum energy of the empty Universe.

$\bullet$
The action of the Conformal Theory of Gravity becomes bilocal
in terms of the Cartan forms and allows the quantization of gravitons directly
in terms of Cartan.

$\bullet$
The fourth difference is that the observed values of the theory are the components of the Fock's frame
in the tangent space of Minkowski $\omega_{(\alpha)}$, linear Cartan's forms and the field variables
of the space of events $[D|F]$,
so the solutions of the equations of the theory, including the constraints,
can be expressed only in terms of these linear forms.

\subsection{Lagrangian formalism \label{S-10-1}}

Recall that in the Hamiltonian formulation we have one global  (\ref{gt})  and three local diffeomorphisms (\ref{kine}).
In this Chapter we show that the Hamiltonian formulation of the theory of gravity
also contains three local constraints in full accordance with the structure of diffeomorphisms (\ref{kine}) and
one global constraint, as a consequence of the invariance of the theory relative to
reparametrization of the coordinate time (\ref{gt}).

Invariance of the theory with respect to the reparametrization of the coordinate time (\ref{gt}) means that a
time-like evolution parameter in the field space of events is identified with the zero harmonic of the dilaton field
$\langle D\rangle$ \cite{grg12chapter5}.
Recall that the zero harmonic is determined by ``averaging'' over the final volume $V_0=\int_{V_0} {d^3x}$
 \bea
  \label{1-avd1}
 &&\langle D \rangle(x^0)\equiv \frac{1}{V_0}\int_{V_0} {d^3x}
 D(x^0,x^1,x^2,x^3).
\eea
In astrophysics and cosmology, the zero harmonic of the dilaton (\ref{1-avd1})
describes \textit{luminosity}, defining it (with sign minus)
as the logarithm of the cosmological scale factor
\bea \label{tau0-5}
  \langle D \rangle= - \ln a = \ln(1+z),
\eea
where $z=(1-a)/a$ is a redshift. In theories of gravity, indeed,
the zero harmonic of the dilaton plays the role of a parameter of evolution in the field space of events.

The non-zero harmonics of the dilaton, which we have denoted above as $\overline{D}$, satisfy
an orthogonality condition with zero harmonic:
$$\int_{V_0} d^3x \overline{D}=0.$$
In force of the condition of orthogonality of harmonics, nonzero harmonics
from this zero harmonics are independent. This means that the non-zero harmonics $\overline{D}$ have zero velocities
(\ref{h-2g1-shift})
 \bea\label{h-2g1}
  v_{ \overline{D}} =\frac{1}{N}[\partial_0 (e^{-3D})
 +\partial_l (N^le^{-3D})]=0
 \eea
and momenta \cite{grg12chapter5}
$$p_{ \overline{D}}=2v_{ \overline{D}}=0$$
(see Eqs.~(\ref{1-3-d}), (\ref{1-17D}) and (\ref{D-1})).
A condition of zero velocities  (\ref{h-2g1})) in the Lagrangian formalism looks like an equation of
the divergence of the shift-vector.

We can choose the shift-vector divergence so that non-zero harmonics of the dilaton, as we already mentioned above,
will be Newtonian-like potentials just those that increase
twice the angle of deflection of light by the gravitational field of the Sun in comparison with the theory of Newton.
Thus, the action (\ref{1-1}) becomes the sum of two terms
\bea\label{1-3n}
 W=W_{\rm G}+\overline{W},
\eea
where
\bea\label{1-3n-1}
W_{\rm G}=- \int dx^0 \left[\frac{d\langle D \rangle(x^0)}{dx^0}\right]^2\int d^3x \frac{1}{N}
\equiv\int dx^0 L_{\rm G}
\eea
is the kinetic part of the action for the zero harmonics of the dilaton and
an expression $\overline{W}$ coincides with the action (\ref{1-3dgl}), where the
speed of a local volume element (\ref{1-3-d}) is equal to zero.
Thus, the equation of the theory of gravity, obtained by variation of the action (\ref{1-3n})
on lapse function
$$N \dfrac{\delta W_{\rm G}}{\delta N}=- N \dfrac{\delta \overline{W}}{\delta N}\equiv {N}\widetilde{{\cal H}},$$
takes a form
\bea \label{1-fbb}
 \frac{1}{N}\left[\frac{d\langle D \rangle(x^0)}{dx^0}\right]^2= N\widetilde{{\cal H}};
 \eea
here
\bea\label{7-L}
\widetilde{{\cal H}}&=&-\frac{4}{3}e^{-7D/2}\triangle e^{-D/2}+\overline{{\cal H}},
 \\\label{7-L1}
\overline{{\cal H}}&=&{\cal H}_g+{\cal H}_A+{\cal H}_Q,
\\\label{7-L2}
{\cal H}_g&=&\frac{1}{6}\left[v_{(a)(b)}v_{(a)(b)}+e^{-4D}R^{(3)}({\bf
e})\right],
 \\\label{7-L3}
{\cal H}_A&=&\frac{1}{2}\left[v^2_{(b)(\rm A)}+F_{ij}F^{ij}\right],
 \\\label{7-L4}
{\cal H}_Q&=&e^{-2D}{(v_{{Q}})^2} +e^{-2D}\left(\partial_{(b)}{{Q}}\right)^2
 \eea
-- Hamiltonian densities at zero speed of the local volume element in the expression (\ref{CC-4a}).
Averaging the equation (\ref{1-fbb}) on three-dimensional volume (see (\ref{1-avd1}))
and using the definitions
\be
 \label{CC-22}
 \left\langle \frac{1}{N} \right\rangle\equiv\frac{1}{N_0};~~~~~~~~~~
 N\equiv N_0{\cal N}~~~~~\Longrightarrow ~~~~~
 \left\langle \frac{1}{{\cal N}} \right\rangle={1}
 \ee
and $N_0dx^0=d\tau$,
we get the diffeoinvariant global constraint equation
\bea \label{1-fbb1}
 \left[\frac{d\langle D \rangle(x^0)}{N_0dx^0}\right]^2
 \equiv\left[\frac{d\langle D \rangle(\tau)}{d\tau}\right]^2=
 \left\langle {\cal N}\widetilde{{\cal H}}\right\rangle.
 \eea
Substituting it into equation (\ref{1-fbb}), we get
surprisingly simple equation for diffeoinvariant local lapse function
\bea \label{1-fbb2}
\left[\frac{d\langle D \rangle(\tau)}{d\tau}\right]^2= {\cal N}^2\widetilde{{\cal H}}.
\eea
Since the left side of the equality does not depend on the spatial coordinates,
a normalization condition
$\langle{\cal N}^{-1} \rangle=1$
allows us to express the diffeoinvariant local lapse function (\ref{CC-22}) in the explicit form
 \be
 \label{CC-2}
 {\cal N}=\frac{\left\langle \sqrt{\widetilde{{\cal H}}}\right\rangle}
 {\sqrt{\widetilde{{\cal H}}}}.
 \ee
Substituting (\ref{CC-2}) into Eq.(\ref{1-fbb2}), we obtain the required global constraint equation
\bea \label{1-fbb3}
\left[\frac{d\langle D \rangle(\tau)}{d\tau}\right]^2=
\left\langle \sqrt{\widetilde{{\cal H}}}\right\rangle^2.
\eea
The solution of the equation (\ref{1-fbb1}) gives the cosmological
dependence of the zero harmonic of the dilaton on the time interval of luminosity:
 \bea\label{Dirac-c2}
 \tau &=&\int\limits_{\langle D \rangle_I}^{\langle D \rangle_0}
 d\langle D \rangle \left\langle \sqrt{\widetilde{{\cal H}}}\right\rangle^{-1},
 \eea
where
$\langle D \rangle_I,~~\langle D \rangle_0$
are the initial and final data, correspondingly.
The dependence of the zero harmonic of the dilaton from the time interval of
luminosity, in the exact theory of gravitation, is an analogue of the Hubble law in cosmology.

The equation of motion for the zero harmonic of the dilaton
$$\frac{{\delta W}}{{\delta \langle D\rangle}}\equiv-T_{\langle D\rangle}=0$$
coincides with the equation obtained by differentiating on $\tau$ of the global
constraint equation (\ref{1-fbb1}):
$$\frac{d^2{\langle{D}\rangle}}{(d\tau)^2}=
\frac{d\langle \sqrt{\widetilde{{\cal H}}}\rangle}{d\tau}=\frac{1}{\sqrt{\widetilde{{\cal H}}}}
\frac{d\langle {\widetilde{{\cal H}}}\rangle}{d\tau}=
\frac{d\langle {\widetilde{{\cal H}}}\rangle}{d\langle{D}\rangle}.
$$
In the case of dominance of the vacuum energy, the right hand side is equal to zero, and we get an empty model of
the Universe, which is considered in detail in Chapter 6.
For non-zero harmonics an equation of motion
$$\frac{\delta W}{\delta \overline{D}}=-T_{\overline{D}}=0$$
takes a form
\be\label{e2t2}
  T_{\overline{D}}=T_D-\langle T_D\rangle=0 ,
\ee
\be\label{e2tc2}
 T_D=\frac{4}{3}\left[7{\cal N}e^{-7D/2}\triangle e^{-D/2}
 +  e^{-D/2}\triangle[{\cal N}e^{-7D/2}]\right]
 - {\cal N}\frac{\partial \overline{{\cal H}}}{\partial D}\,,
\ee
where $\overline{{\cal H}}$ is given by the equations (\ref{7-L1}) -- (\ref{7-L4}).
Thus, solving the constraints, we expressed all components of the metric through the components of the energy-momentum
tensor and linear forms of Cartan (\ref{P-1e})
\bea\label{ds-12}
 \widetilde{ds}^2= e^{-4D}\!\frac{\langle\sqrt{\widetilde{{\cal H}}}\rangle^2}
 {\widetilde{\cal H}}\!d\tau^2-
\eea
$$
 -\left(
 dX_{(b)}-X_{(c)}[\omega^R_{(c)(b)}(d)+\omega^L_{(c)(b)}(d)] -
 {\cal N}_{(b)} d\tau\!\right)^2\!\!.
$$
Square interval in diffeoinvariant form on the surface of constraint (\ref{CC-22})
depends only on the indices of the tangent space.

\subsection{Hamiltonian formalism \label{S-10-H}}
For Hamiltonian formulation of the theory, we introduce momenta of the fields according to the definitions
(\ref{1-17g}) -- (\ref{1-17A}).
The momentum of the global component of the dilaton
 \bea \label{1-17g7}
 P_{\langle D\rangle}&=&\frac{\partial L_G}{\partial (d\langle D\rangle/dx^0)}=
 -{2V_0}\frac{d\langle D\rangle}{{N_0}dx^0}\equiv V_0 p_{\langle D\rangle},
 \eea
the momentum of the scalar field
\bea
 \label{1-17D7}
 p_{Q}&=&2{v_{{Q}}}=\frac{2}{N}\left[(\partial_0-N^l\partial_l){{Q}}
+\frac{1}{3}\partial_lN^l\right],
\eea
momenta of the photon field
 \bea\label{1-17D7f}
 p_{\rm A(b)}&=&{v_{\rm A(b)}}=\frac{1}{N}{\bf e}^i_{(a)}
 \left[\partial_0A_i-\partial_iA_0+F_{ij}N^j\right],
\eea
and momenta of the gravitational field
 \bea\label{1-17g-new}
 p_{(a)(b)}&=&\frac{v_{(a)(b)}}{3}\equiv p^{\bot}_{(b)(a)}+\partial_{(a)}
 f^{\bot}_{(b)}+\partial_{(b)} f^{\bot}_{(a)},
 \\\label{1-3-dD7}
 \label{1-3-dvvcp}
 v_{(a)(b)}&=&\frac{1}{N}\biggl[\omega^R_{(a)(b)}(\partial_0-N^l\partial_l)+
 \partial_{(a)}N^{\bot}_{(b)}
 +\partial_{(b)}N^{\bot}_{(b)} \biggr].
 \eea
In the Hamiltonian formalism, the equation for the shift-vector has the form (\ref{3T})
\bea\label{3T7}
{\bf e}_{(b)}^i\frac{\delta W} {\delta{N}_{i}}
\!\!=\!\!-T_{(0)(a)}\!& = &\!
 \partial_{(b)}p_{(b)(a)}\! -\widetilde{T}_{(0)(a)}=0,
 \eea
where
 \bea  \label{CC-4a17}\widetilde{T}_{(0)(a)} =
\sum\limits_{F=A^{T}_{(a)},\overline{Q}}^{}p_F\partial_{(a)}F
\eea
-- components of the tensor of energy - momentum of the photon and the scalar field.
The condition of the transverseness of the graviton
$$\partial_{(a)}\omega^R_{(a)(b)}=0$$
allows us to express the transverse part of the shift-vector through the components of the tensor of
energy--momentum of the photon and the scalar field,
while the divergence of the shift-vector ({\it id est}, its longitudinal part) is given by the condition
of zero momentum of the local dilaton (\ref{h-2g1})
\bea\label{h-2g17}
 p_{ \overline{D}}= 2v_{ \overline{D}} =\frac{2}{N}[\partial_0 (e^{-3D})
 +\partial_l (N^le^{-3D})]=0.
 \eea
Thus, we determine all these components of the photon and the gravitational field,
except for the longitudinal component of the photon and the anti-symmetric linear form of the
gravitational field $\omega^L_{(c)(b)}(d)$.
However, these components in the kinetic terms of the action are absent,
and they are determined by the distribution of the external currents and the matter, respectively.

Thus, on the level of constraints ${\cal C}=0$ the action is
\bea\label{1-3dgc127}
  {W}_{C=0}=
\eea\label{1-PP}
$$=
  \int\limits_{ }^{ }\!\! d^3x
  \biggl[\int [p_{(a)(b)}\omega_{(a)(b)}^R(d)
 +p_Q\,d{{{Q}}}
 +p_{A{(b)}}dA_{(b)}]\biggr] - \int P_{\langle D\rangle}
 d{\langle{D}\rangle}
,
$$
where the canonical momentum of the dilaton $P_{\langle D\rangle}$
satisfies the Hamiltonian constraint
\be \label{5-C}P^2_{\langle D\rangle}=
\left[2\int d^3x \frac{d \langle D \rangle(\tau)}{d\tau}\right]^2=
\left[2\int d^3x\sqrt{\widetilde{{\cal H}}}\right]^2
\ee
and plays a role of the generator of evolution. The value of the momentum of the zero harmonics
on the solutions
of the equations of motion is the energy of the Universe in this space of events. This is one of
ways to solve the problem of non-zero energy as well in General Relativity \cite{grg12chapter5}.

Thus, if we leave the zero harmonic of the dilaton $\langle D(x^0)\rangle$,
the homogeneous lapse function $N_0(x^0)$ and the vacuum energy of quantum
oscillators, we obtain a simple dynamical system, known in literature
\cite{Misner1969-5}
as a \textit{miniuniverse} (see Appendix E).

In the next two chapters \textit{miniuniverse} will be considered as an example,
to demonstrate the ability of solving of most of the problems listed above.

\section{Summary}
\renewcommand{\theequation}{5.\arabic{equation}}
\setcounter{equation}{0}
\makeatletter
\renewcommand{\@evenhead}{\raisebox{0pt}[\headheight][0pt]
{\vbox{\hbox to\textwidth{\hfil \strut \thechapter . Hamiltonian formulation of the theory of gravity
\quad\rm\thepage}\hrule}}}
\renewcommand{\@oddhead}{\raisebox{0pt}[\headheight][0pt]
{\vbox{\hbox to\textwidth{\strut \thesection .\quad
Summary and literature
\hfil \rm\thepage}\hrule}}}
\makeatother

The quantization of any dynamical system involves a Hamiltonian description of the system.
This chapter is focused on adapting  Dirac's Hamiltonian approach to the General Relativity for the affine
Conformal Theory of Gravity, presented in Chapter 4.
All the problems of a unique definition of the energy and  time,
inherent in the Hamiltonian description of the General Relativity, are inherited by the Conformal Theory of Gravity.
However, as it has been shown, these problems have a unique solution by introducing the concept of
zero harmonic of the dilaton and postulating of the existence of vacuum as the state with the lowest energy,
in full accordance with the dimension of diffeomorphisms of the Hamiltonian evolution.
The Hamilton equation of constraint in presence of the zero harmonic of the dilaton
becomes algebraic and exactly resolved relative to the canonical momentum of the
zero harmonic. This canonical momentum becomes a generator of evolution of the
Universe in a field space of events and defines the energy of the Universe
on the solutions of the classical equations.
The zero harmonic of the dilaton and the energy of the vacuum set the empty Universe model, which
 Chapter 6 discusses.

\chapter{A model of an empty Universe\label{S-6}}
\renewcommand{\theequation}{6.\arabic{equation}}
\setcounter{equation}{0}
\section{An empty Universe}
\makeatletter
\renewcommand{\@evenhead}{\raisebox{0pt}[\headheight][0pt]
{\vbox{\hbox to\textwidth{\hfil \strut \thechapter . A model of an empty Universe
\quad\rm\thepage}\hrule}}}
\renewcommand{\@oddhead}{\raisebox{0pt}[\headheight][0pt]
{\vbox{\hbox to\textwidth{\strut \thesection .\quad
Empty Universe
\hfil \rm\thepage}\hrule}}}
\makeatother

In the first five chapters, we set out the elements of an alternative physics program which appeared
in the period
from 1915 till 1974  to describe and classify the experimental data. This program is
based on
the principles of symmetry of
the initial data.
The essence of this program is as follows.

1. There are elementary objects (such as quarks or twistors) as
fundamental representations of the group $G$: $(SU(2)\otimes SU(2))$ or $(A(4)\otimes C)$.

2.   Mesons or space - time are formed from these elementary objects as adjoint representations of the
group $G$, that allows the determination of the vacuum stability subgroup $H$ and the corresponding
coset $K=G/H.$

3. Linear Cartan forms, describing an arbitrary motion (translation and rotation) of the frame in this coset,
are derived on the algebra of $G$.

4. Lagrangians of the chiral theory and the theory of gravity are constructed with the help of these forms.

5. Cartan forms are invariant under gauge transformations.

Next, we demonstrate the ability of feasibility of this program to describe
 the observational data on the example
of \textit{miniuniverse}.
Here we will understand  the \textit{miniuniverse} as the theory of gravitation developed above, in which
we keep only the zero harmonics of the dilaton $\langle D(x^0)\rangle$, the homogeneous lapse function $N_0(x^0)$,
and the vacuum energy of the quantum oscillators
\be\label{vvsd}
{\rho^\tau_{\rm Cas}}(a)=
\frac{1}{V_0}\sum_f{\textsf{H}^\tau_{(f)\rm Cas}}=\frac{1}{V_0}\sum\limits_{q,f}\frac{\omega_{q,f}}{2},
\ee
where the vacuum energy of a finite universe in this model appears as a sum of vacuum energies of all
fields $f$.
In quantum field theory, this amount is called the Casimir energy \cite{Bordagempty-6}.
Recall that the vacuum energy arises under the normal ordering of the field operators
after separation of them to positive and negative frequency parts. In particular, the energy of the
sum of the oscillators
has the form
$$
\frac{1}{2}\sum_n \left({p_n^2 + \omega^2_n q^2_n}\right)  =
\frac{1}{2}\sum_n  \omega_n \left({a_n^+ a^-_n \!+\! a^-_n a^+_n}\right) =
$$
$$=
\sum_n  \omega_n a^+_n a^-_n +\sum_n
 \frac{\omega_n}{2}.
$$
The latter term is called the vacuum energy, defined as the state of a set of oscillators
with the lowest energy.

Let us leave the vacuum energy in the action defined by the formulae (\ref{1-3n}) и (\ref{1-3n-1}),
instead of matter fields. Then we get the cosmological model of the homogeneous empty Universe, described by
the action (up to a total derivative)
 \bea
 \label{1-3nt}
 && W_{\rm Universe}=
 -V_0\int\limits_{\tau_I}^{\tau_0} \underbrace{dx^0 N_0}_{=d\tau}
 \left[\left(\frac{d\, \langle D \rangle}{N_0dx^0}\right)^2
 +{\rho^\tau_{\rm Cas}}(\langle D \rangle)\right].
\eea
The value of the Casimir energy of  field oscillators  ${\rho^\tau_{\rm Cas}}(\langle D \rangle)$
is inversely proportional to the size of the spatial volume.
Therefore, in the classical limit of an infinite volume, the action (\ref{1-3nt}) is equal to zero
$$\lim\limits_{V_0\to \infty} W_{\rm Universe}=0.$$

Varying the action  (\ref{1-3nt}) by variables $\langle D \rangle$ and $N_0$, we get two equations
\bea \label{tau3}
&&  \frac{\delta W_{\rm Universe}}{\delta\langle D \rangle}=0~~\Rightarrow~~
 2\frac{d }{d\tau}\left[\frac{d \langle D \rangle}{d\tau}\right]
 =\frac{d{\rho^\tau_{\rm Cas}}}{d\langle D \rangle},
\\ \label{tau4}
&& \frac{\delta W_{\rm Universe}}{\delta N_0}=0 ~~\Rightarrow~~
 \left[\frac{d \langle D \rangle}{d\tau}\right]^2={\rho^\tau_{\rm Cas}}.
\eea
The second equation
is the integral of the first one and is treated as a constraint equation of the initial data (initial momentum)
 of the
dilaton. According to the second Noether's theorem, the second equation is a consequence of the invariance
of the
action (\ref{1-3nt}), relatively to reparametrization of the coordinate evolution parameter:
$$x^0 \to \widetilde{x}^0=\widetilde{x}^0(x^0).$$
The second equation, rewritten in terms of the cosmological scale factor
$a = \exp (-\langle D \rangle)$ and conformal density
\be\label{vsd}
{\rho^\eta_{\rm Cas}}(a)=
\frac{{\rho^\tau_{\rm Cas}}}{a^{2}}
\equiv \frac{H_0}{d_{\rm Cas}(a)},
\ee
coincides with the Friedmann equation
\bea \label{tau5}
\left[\frac{da}{d\eta}\right]^2=\rho^\eta_{\rm Cas}(a),
 \eea
where ${d_{\rm Cas}(a)}$ in (\ref{vsd})
is the conformal size of the Universe and $H_0$ is the Hubble parameter, and
$a=(1+z)^{-1}$ is a scale cosmological factor and $z$ is a redshift.

The solution of the Friedmann equation (\ref{tau5}) gives a conformal horizon
 \bea \label{tau51}
 d_{\rm horison}(a)=2 r_{\rm horison}(a)=2\int\limits_{0}^{{a}}
 d \overline{a}\;\;[\rho^\eta_{\rm Cas}(\overline{a})]^{-1/2}.
 \eea
The horizon is defined as the distance that the photon runs on the light cone $d\eta^2-dr^2=0$  during the life - time
of the Universe.
In this case, a conformal horizon coincides with apparent size of the Universe
 $d_{\rm Cas}({a})$ in (\ref{vsd}):
\be\label{evs}
d_{\rm Cas}(a)=d_{\rm horison}(a).
\ee
Solutions of the equations (\ref{vsd}), (\ref{tau5}), (\ref{tau51}), and (\ref{evs})
\be
 \label{rho-cr}
 d_{\rm horison}({a}) =\frac{{a^2}}{H_0} ~~~\Rightarrow~~\rho^\tau_{\rm Cas}=H_0^2\equiv\rho_{\rm cr}
\ee
give the Hubble diagram of the description of Supernovae
\cite{Riess2001empty-6, Zhu:2003sq-6}
in Conformal Cosmology
 \cite{Behnke:2001nwempty-6, Barbashov:2005huempty-6, Behnke_04empty-6, Blaschke:2004byempty-6,
 Zakharov:2010nfempty-6, Arbuzov:2010fzempty-6},
obtained as a consequence of the Dirac definition of measured intervals in the approximation of the empty
space.
In terms of conformal variables, the solution corresponds to the equation of rigid state  of an
\textit{empty space} (\ref{evs})
 \be
 \label{eta-2}
 \left[\frac{da}{d\eta}\right]^2=\frac{\rho_{\rm cr}}{{a^2}}\;.
 \ee

In terms of the values of luminosity, where  $\rho^\tau_{\rm Cas}$ is a constant,
$$\frac{d{\rho^\tau_{\rm Cas}}}{d\langle D \rangle}=0,$$
we get the inertial motion of the dilaton with an acceleration equal to zero. For an empty space,
the obtained solution (\ref{eta-2}) describes the Supernova data in the Conformal Cosmology,
where the measured distances
are \textit{longer} than those, used in the Standard Cosmology. Thus,
according to the principles of conformal and affine symmetries, namely, this \textit{remoteness} of Supernovae were
discovered by observers \cite{Riess2001empty-6, Zhu:2003sq-6}.

\section{The Supernovae data in\\ Conformal Cosmology}
\makeatletter
\renewcommand{\@evenhead}{\raisebox{0pt}[\headheight][0pt]
{\vbox{\hbox to\textwidth{\hfil \strut \thechapter . A model of an empty Universe
\quad\rm\thepage}\hrule}}}
\renewcommand{\@oddhead}{\raisebox{0pt}[\headheight][0pt]
{\vbox{\hbox to\textwidth{\strut \thesection .\quad
The Supernovae data in Conformal Cosmology
\hfil \rm\thepage}\hrule}}}
\makeatother

The Nobel Prize in Physics of 2011 was awarded to S. Perlmutter, A. Riess, and B. Schmidt for their
work \cite{Riess2001empty-6, Riess1998empty-6, Perlmutter1999empty-6, Riess2004empty-6} related to
studying of Supernovae Type Ia to determine the parameters of cosmological models. Thus
it was assumed that the maximum luminosity of Supernovae does not depend on the distance to them, but
depends on the rate of change of the luminosity according to the so-called law of the Pskovski ---
Phillips \cite{Riess2001empty-6}, that is, they are the so-called ``standard candles''.
Studying distant Supernovae from Earth, observers found that these stars are at least
a quarter fainter than predicted by the theory which means that the stars are too far away.
Thus, calculating parameters of expansion in cosmological models of Friedmann -- Robertson -- Walker
\cite{MisnerThorneWheelerempty-6} with an arbitrary equation of state of matter, the researchers found
that in frames of the Standard Cosmology, this process is proceeded with acceleration.
\begin{center}
\includegraphics[width=0.7\textwidth]{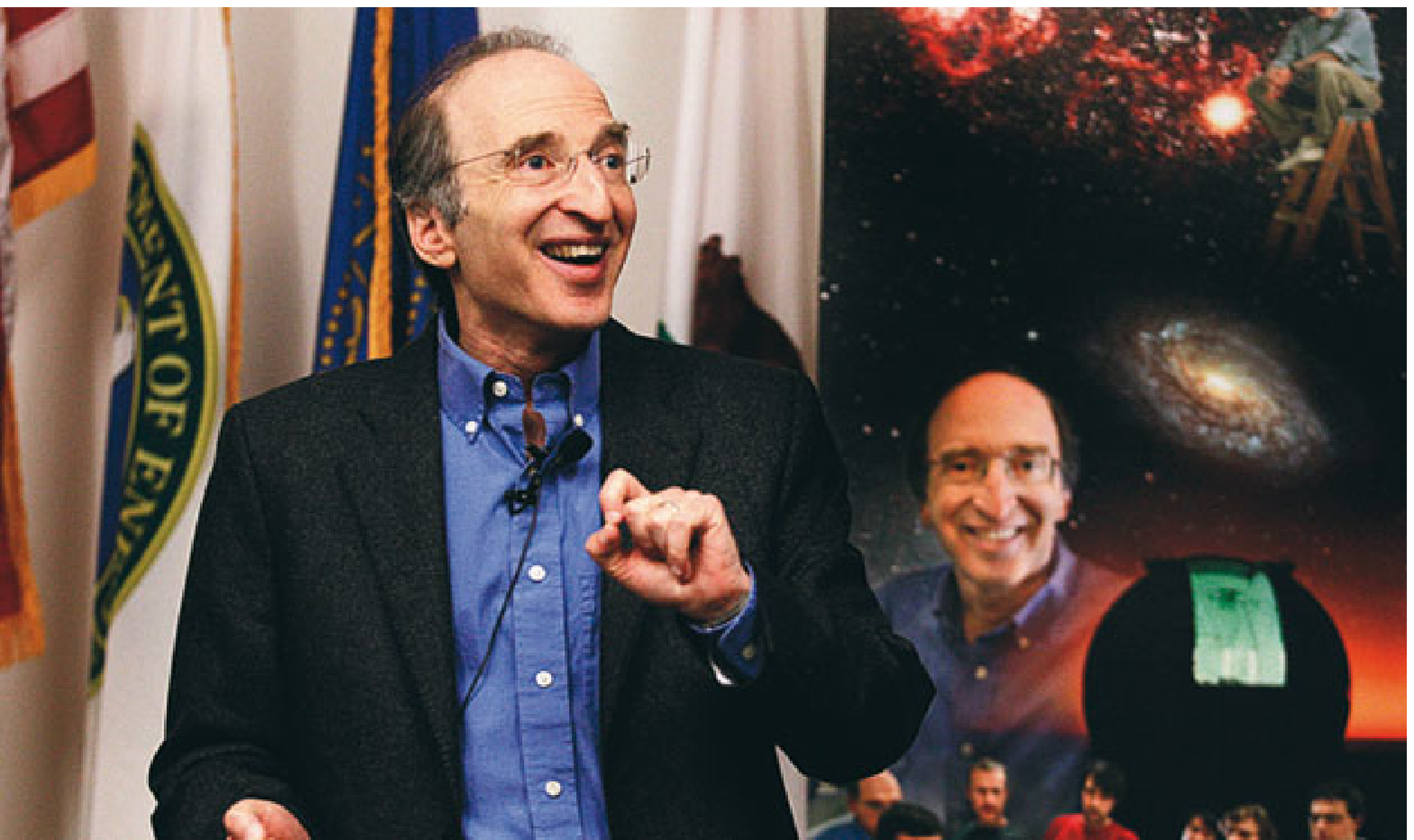}
\end{center}
{\small Saul Perlmutter (born September 22, 1959) is an American astrophysicist and Nobel Prize in Physics 2011
(together with Brian Schmidt and Adam Riess)
``for the discovery of the accelerated expansion of the Universe through observations of distant Supernovae''.
Perlmutter grew up in the Philadelphia area Mount Airy,
where he studied in an elementary school Greene Street Friends School and Germantown Friends School.
In 1981  he graduated with honors from Harvard University.
In 1986  in the University of California at Berkeley, Perlmutter received his PhD.
His thesis was devoted to the problem of detection of objects -- candidates for the role of Nemesis.
Currently, Perlmutter  heads  the project ``Supernova Cosmology Project'' at the Lawrence Berkeley National Laboratory.
His team together with a group of Brian Schmidt proved the existence of accelerated expansion
of the Universe.
Perlmutter is also a lead investigator in the Supernova/Acceleration Probe project,
which aims to build a satellite dedicated to finding and studying more Supernovae in the distant Universe.
This will help to better constrain the rate at which the Universe is accelerating.
He is also a participant in the Berkeley Earth Surface Temperature project, which aims to increase our understanding
of recent global warming through improved analyses of climate data.}

It corresponds to the presence of a non-zero Lambda term\footnote{Albert Einstein in paper (A. Einstein:
\textit{Kosmologiche Betrachtungen zur
allgemeinen Relativit\"atstheorie}. Sitzungsber. d. Berl. Akad. {\bf{1}}, 142 (1917))
was forced to introduce a universal
$\lambda$-term in the equations of the theory
from  static requirements of the cosmological solutions,
defensively only in that supplement where its covariance equations are not violated.
Later, Einstein, closely acquainted with the work of Alexander Friedmann (Friedmann, A:
\textit{\"Uber die kr\"ummung des raumes.} Zs. f\"ur Phys. {\bf 10}, 377 (1922))
admitted the erroneousness of the introduction of
$\lambda$-term, thereby, opening the way for the study of non-stationary models.
After exploring the world with positive curvature, Alexander Friedmann in paper
(Friedmann, A: \textit{\"Uber die m\"oglichkeit einen welt mit konstanter negativer
kr\"ummung des raumes.} Zs. f\"ur Phys. {\bf 21}, 326 (1924))
obtained a cosmological solution with negative curvature.}.
In this case, one speaks about the so-called dark energy.  There is still an unresolved within the Standard
cosmology,  problem of the origin of matter with similar properties.
This form of matter is not predicted even by representations of the Poincar\'e  group.

On the other hand, there is a Conformal cosmological model \cite{Behnke:2001nwempty-6},
which allows us to describe the Supernova data without lambda-term,
because in this model the observed distances are identified with conformal longer intervals.
The authors of the discovery in paper \cite{Riess2001empty-6} recognize the fact of existence of both
alternative explanations, and compare interpretations of the results of observations  with the Conformal
cosmological model \cite{Behnke:2001nwempty-6}.

As we saw above, in the Conformal cosmological model, to explain
far distances to Supernovae there are enough assumptions about the dominance of the zero energy of the vacuum.
According to quantum mechanics, in microworld, each particle has a zero energy of
fluctuations of vacuum, which is called the Casimir energy \cite{Bordagempty-6}.
\begin{center}
\includegraphics[width=0.7\textwidth]{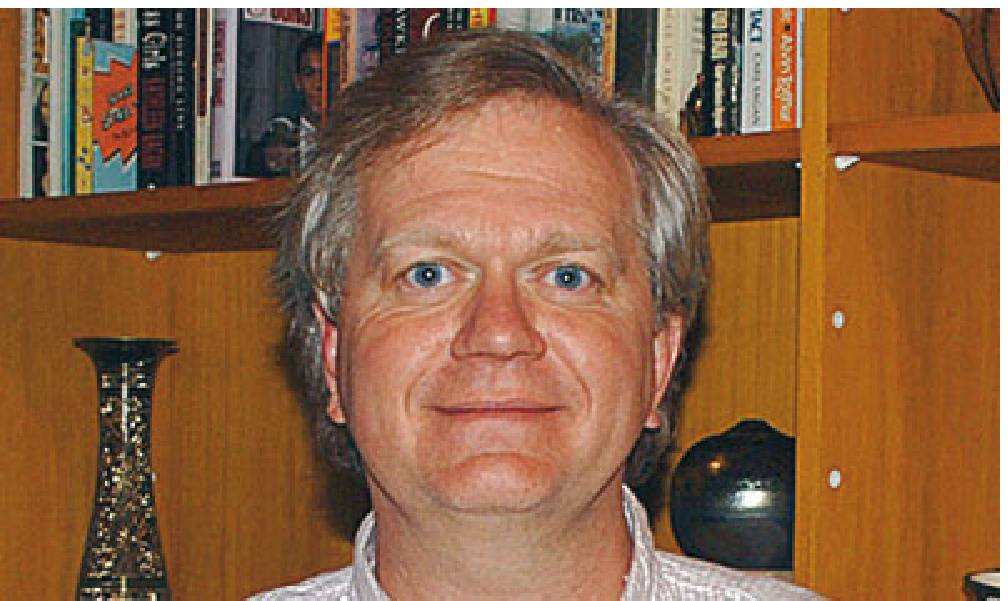}
\end{center}
{\small
Brian Paul Schmidt (born February 24, 1967) is a Distinguished Professor, Australian Research Council Laureate Fellow and astrophysicist at
The Australian National University Mount Stromlo Observatory and Research School of Astronomy and Astrophysics.
Schmidt, along with Riess and Perlmutter, jointly won the 2011 Nobel Prize in Physics for their observations
which led to the discovery of the accelerating Universe.
Schmidt attended Bartlett High School in Anchorage, Alaska, and graduated form it in 1985.
He earned his BS (Physics) and BS (Astronomy) from the University of Arizona in 1989.
He received his MA (Astronomy)
in 1992 and then PhD (Astronomy) in 1993 from Harvard University.
Schmidt was a Postdoctoral Fellow at the Harvard -- Smithsonian Center for Astrophysics (1993–1994)
before moving on to Mount Stromlo Observatory in 1995.
Schmidt led the High-Z Supernova Search Team  from Australia, and in 1998 with
Adam Riess the first evidence was presented that the Universe's expansion rate is accelerating. The
observations were contrary to the current theory
that the expansion of the Universe should be slowing down;
on the contrary, by monitoring the brightness and measuring the redshift of the Supernovae,
they discovered that these billion-year old exploding stars and their galaxies were accelerating away from our reference frame.
Schmidt is currently leading the SkyMapper telescope Project and the associated Southern Sky Survey.
}
\begin{center}
\includegraphics[width=0.7\textwidth]{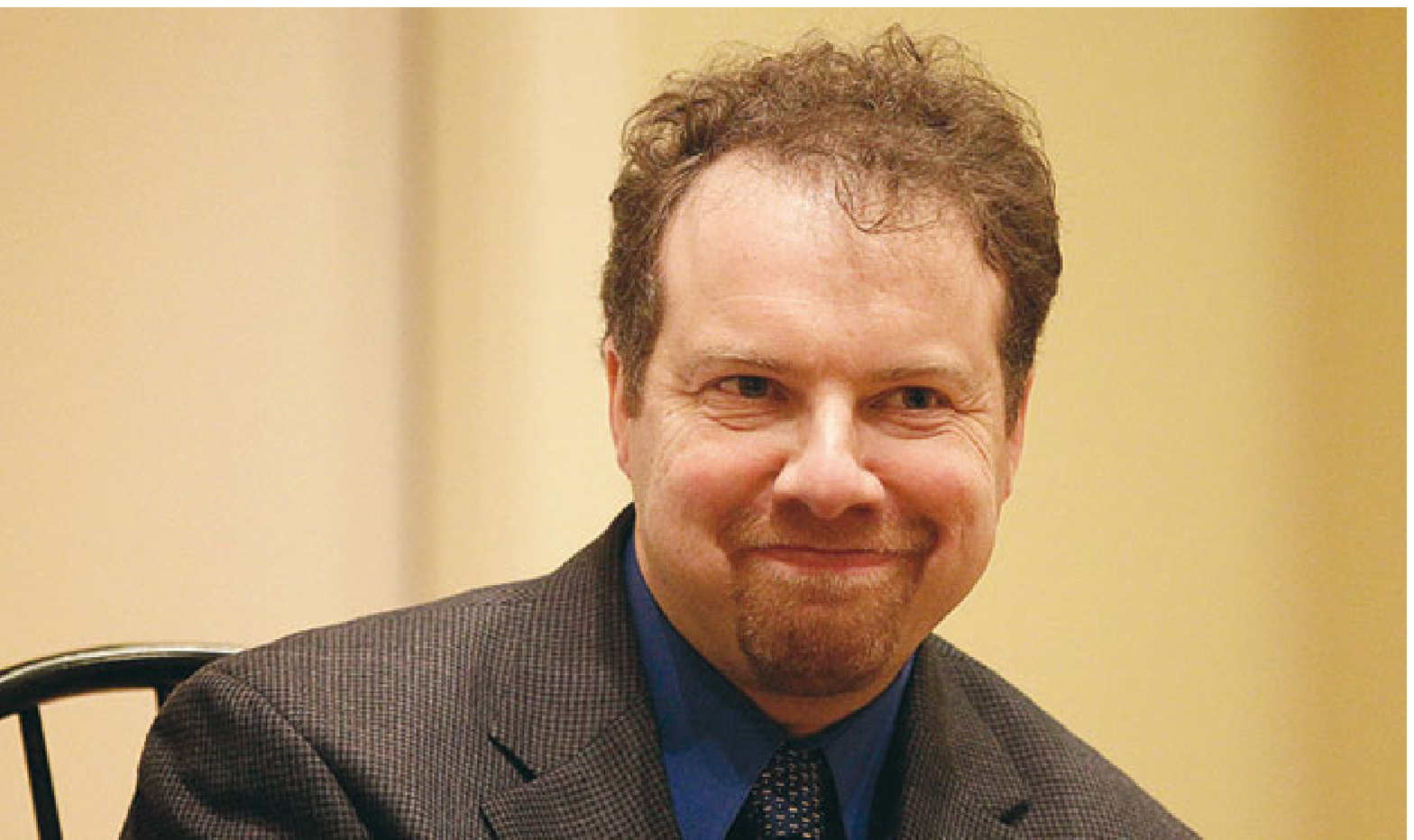}
\end{center}
{\small
Adam Guy Riess (born 16 December 1969) is an American astrophysicist at the Johns Hopkins University
and the Space Telescope Science Institute and is known for his research in using Supernovae as cosmological probes.
Riess shared the 2011 Nobel Prize in Physics with Saul Perlmutter and Brian P. Schmidt for providing evidence
that the expansion of the Universe is accelerating.
Riess was born in Washington, D.C., grew up in Warren, New Jersey.
Riess graduated from the Massachusetts Institute of Technology in 1992.
He received his PhD from Harvard University in 1996; it resulted in measurements of over twenty new type Ia Supernovae
and a method to make Type Ia Supernovae into accurate distance indicators by correcting for intervening dust and intrinsic inhomogeneities.
Riess jointly led the study with Brian Schmidt in 1998 for the High-z Supernova Search Team.
The team's observations were contrary to the current theory that the expansion of the universe was slowing down;
instead, by monitoring the color shifts in the light from supernovas from Earth,
they discovered that these billion-year old novae were still accelerating.
This result was also found nearly simultaneously by the Supernova Cosmology Project, led by Saul Perlmutter.
The corroborating evidence between the two competing studies led to the acceptance of the accelerating Universe theory,
and initiated new research to understand the nature of the Universe, such as the existence of dark matter.
}

\begin{figure}[htb!]
\begin{center}
\includegraphics[width=0.9\textwidth]{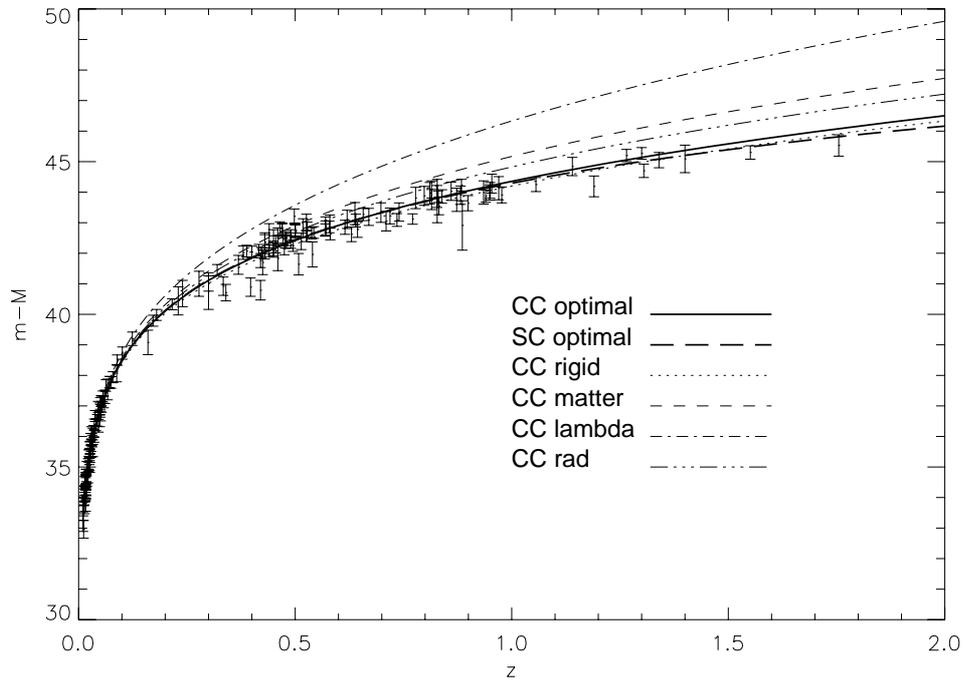}
\end{center}
\caption
{\small
Hubble diagram, based on 73 data obtained by the collaboration SNLS.
For the theoretical analysis, the model of a flat Universe in the Standard cosmology (SC)
and the Conformal cosmology (CC) is used.
The better agreement with the data demands the value of cosmological constant:
$\Omega_{\Lambda} =0,755$ and cold matter: $\Omega_m=0,245$ in case of the Standard cosmology,
while in the case of the Conformal cosmology, these data are consistent with the regime of nucleosynthesis
and domination of the rigid state condition $\Omega_{\rm rigid} =0,755$.}
\label{fig2-6-Hubble}
\end{figure}

In papers \cite{Zakharov:2010nfempty-6, PervushinZakharov2012-6}
it was shown that with account of the great number of data of Supernovae, the interpretation of
observational data
with the Conformal cosmological model
(solid curve in Fig. \ref{fig2-6-Hubble}) is almost as good as the interpretation in the framework of models of
Friedmann -- Robertson -- Walker with a non-zero Lambda member (the dashed line in Fig. \ref{fig2-6-Hubble}).
According to the Conformal model, S. Perlmutter, A. Riess and B. Schmidt discovered the physical vacuum
which is the Universe.
For comparative analysis, data of the collaboration ``Supernova Legacy Survey'' (SNLS) are used
\cite{Astier-6}. Detection of Supernova explosions was carried out with the telescope CFHT (Canada -- France -- Hawaii),
whereupon with modern telescopes, photometric and spectroscopic study of Supernovae were conducted.
It is appropriate to remind the correct statement of the
Nobel laureate in Physics Steven Weinberg\footnote{Weinberg, S.: \textit{The First Three Minutes: A Modern View of
the Origin of the Universe.} Basic Books, New York (1977).}
about interpretation of experimental data on redshifts.
\textit{``I do not want
to give the impression that everyone agrees with this interpretation of the red shift. We do not actually observe
galaxies rushing away from us; all we are sure of is that the lines in their spectra are shifted to the red, i.e.
towards longer wavelengths. There are eminent astronomers who doubt that the red shifts have anything to do with
Doppler shifts or with expansion of the universe''.}

\begin{figure}[ht]
\begin{center}
 \vspace{5mm}
\epsfig{figure=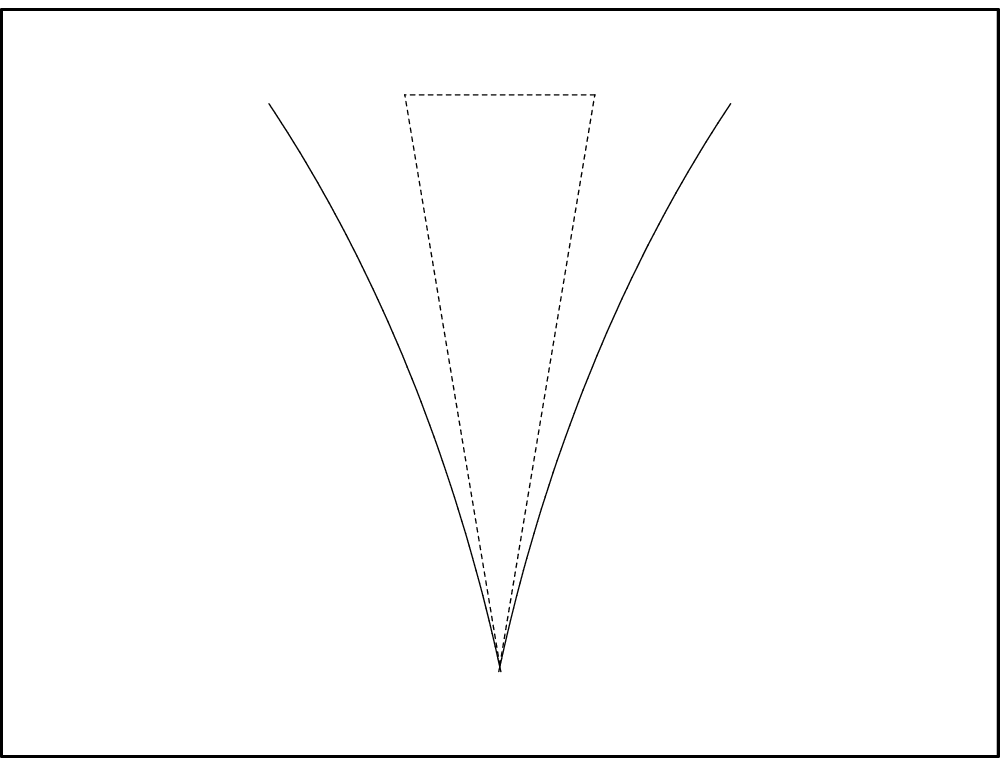,width=2 cm}
\hspace{2cm}\epsfig{figure=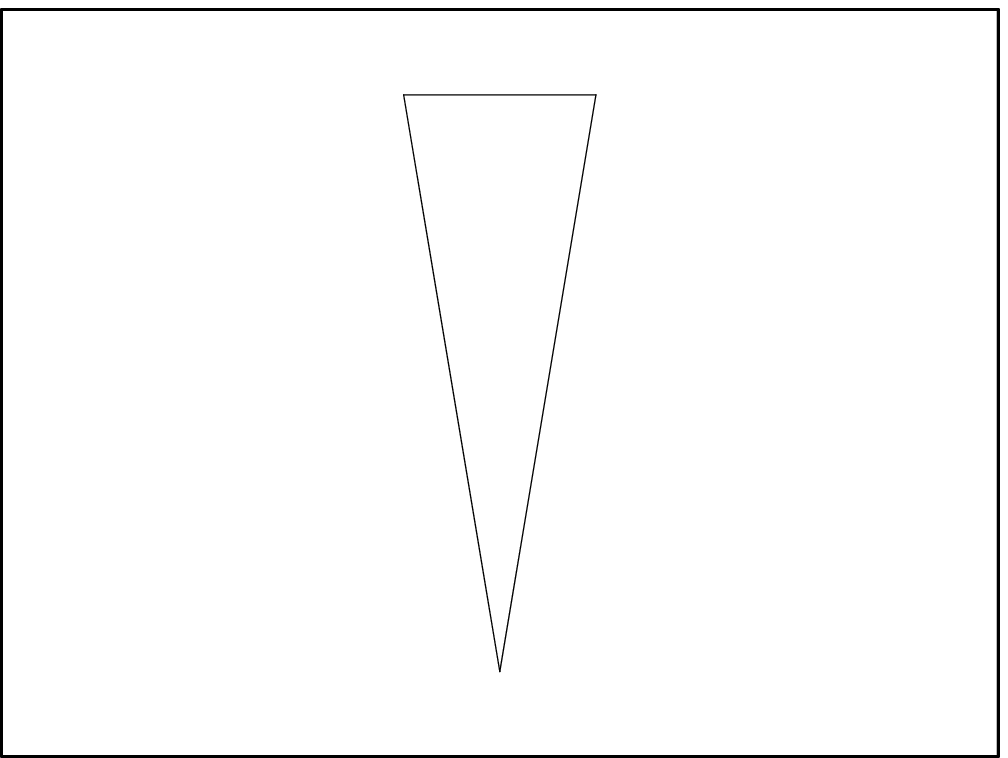,width=2 cm}

 \vspace{1cm}

\epsfig{figure=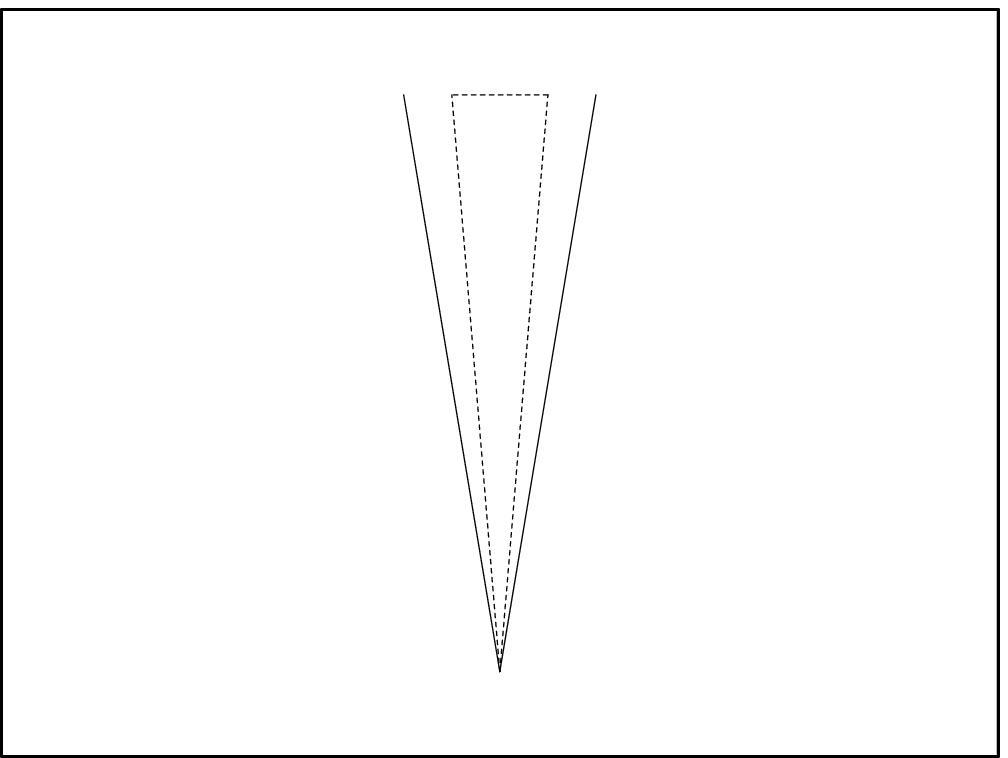,width=2 cm}
\hspace{2cm}\epsfig{figure=pr2.eps,width=2 cm}
\end{center}
 \caption{\small
Comparison of case of a stationary Universe (right panel) to the case of evolving Universe with an absolute standard
(top left panel), and with the case of the evolving Universe with the relative standard (bottom left panel).
\label{fig1-Panel}}
\end{figure}

If we identify the observed values with conformal variables (conformal time, conformal density,
conformal temperature and Planck running mass), the evolution of the lengths in cosmology
is replaced by the evolution of the masses.
This identification means choosing the equations of General Relativity (GR) and the Standard Model (SM) in
conformally invariant form, where the space scale factor scales all masses including the Planck mass.
The initial values of the masses are much less than modern ones. In this case, the Planck epoch in the early Universe
loses its absolute predestination.
It was shown \cite{Blaschke:2004byempty-6, PervushinProskurin-6} that in the case of regime of rigid state
equation the early Universe is a factory of the cosmological creation of massive vector bosons
from the vacuum, when the Compton wavelength of these bosons coincides with the event horizon of the early Universe,
so the conformal-invariant versions of the Standard Model and the General Relativity can, in principle,
explain the origin of the observed matter as the final product of decay of primary bosons.
In an evolving Universe as opposed to a stationary Universe, a part of photons is lost during their flight to the Earth.
This is due to an increase in the angular size of the cone of light of emitted
photons (absolute standard) or because of reduced angular size of the cone of light of absorbed
photons (the relative standard), as shown in Fig. \ref{fig1-Panel} for both cases.

\newcommand{\Wvalue}{50 pt}
\newcommand{\Hvalue}{25 pt}


%
%
%

To restore full luminosity for both standards (both absolute and relative)
we have to multiply by a factor of coordinate distance $(1 + z)^2$ \cite{Behnke:2001nwempty-6}, therefore the
observational cosmology uses the luminosity distance $\ell$, which is defined as the measured
distance ($r$ (\ref{1rs}) or $R$ (\ref{scale})), multiplied by a factor $(1 + z)^2$ for both standards
\bea
\label{ell:CC}
 \ell_{{\rm abs}} (z) &=& (1 + z)^2 R(z) = (1 + z)r(z),
\\
 \ell_{{\rm rel}} (z) &=& (1 + z)^2 r(z)
 \label{ell:SC}.
\eea
In the literature, the first case corresponds to the Standard cosmology (SC), the second -- to the Conformal Cosmology (CC).
So, for the relative standard we have an additional factor $(1+z)$,
and relations (\ref{ell:CC}) and (\ref{ell:SC}) mean, that observational data are described by
different regimes for different standards of measurement.
In  Fig. \ref{fig2-6-Hubble} \cite{Behnke:2001nwempty-6} the results of the Standard
and Conformal cosmologies for the relation between the effective magnitude and redshift are compared:
$$m(z) = 5 \log [H_0 \ell(z)] + {\cal M},$$
where ${\cal M}$ is a constant, according to latest data for Supernovae
\cite{Riess2001empty-6, Riess1998empty-6}.
As we see from Fig. \ref{fig2-6-Hubble}, in the region $0\leq z\ \leq 2 $ observational data,
including the last point
(SN 1997ff) with $z=1,7$ \cite{Riess2001empty-6},
are interpolated in the Standard cosmology with the absolute standard and parameters
\be
\label{mode:SC}
\Omega_{\rm rigid}= 0,~~~~~\Omega_M \geq 0,245,
~~~~~ \Omega_{\Lambda} \leq 0,755
\ee
as well as in the Conformal Cosmology with the relative standard and parameters
\be
\label{mode:CC}
\Omega_{\rm
rigid} \geq 0,755,~~~~~\Omega_{M} \leq 0,245,~~~~~ \Omega_{\Lambda} =0.
\ee
Further, in case of the relative standard of measurements, the evolution of Supernovae does not contradict
to the equation of state of primordial nucleosynthesis with the dependence of the cosmological factor from the
observed conformal time
\be
\label{data3}
\tilde a(\eta) = [z+1]^{-1}(\eta)=\sqrt{1+2H_0(\eta-\eta_0)},~~~~~~~~
\Omega_{\rm rigid}=1.
\ee

From this relation it is easy to find the coordinate distance
$r=\eta_0-\eta$ as function of $z$:
\be\label{cldr2} H_0r(z)=\frac
12\left[1-\frac{1}{(1+z)^2}\right]=\frac{1}{\left(1+z\right)^2}\left(z+\frac{z^2}{2}\right),
\ee
which implies the dependence of the luminosity distance
(\ref{ell:SC}) from the redshift $z$:
 \be\label{cldrr}
 \ell_{{\rm rel}}(z) = (1+z)^{2}r(z)=\frac{1}{H_0}
 \left[z+\frac{z^2}{2}\right].
 \ee
In Fig. \ref{fig2-6-Hubble} the function (\ref{cldrr}), arisen as a result of the solution of the equation
of rigid state, is shown by the solid line and we see that the astrophysical data on
Supernovae and primordial nucleosynthesis, recalculated in terms of the relative standard,
testify that all evolution of the Universe takes place in the regime of dominance of rigid state
(\ref{data3}) with relative density
\be\label{rigid}
\rho_{\rm rigid}(a)=\frac{\rho_{\rm cr }}{a^2}
=\frac{ H^2_0 }{a^2}.
\ee
It is singular at the zero value of the scale factor.
If this density is dominant in the modern era,
it also dominated in the primary era of the early Universe, for which the solution
(\ref{data3})
 \be\label{rigid0}
 a(\eta)=a_I\sqrt{1+2H_I(\eta-\eta_I)}
 \ee
is expressed in terms of the initial data
 \be\label{rigid01}
 a_I=a(\eta_I), ~~~~~~~~~~~~H_I=\frac{a'(\eta_I)}{a(\eta_I)},
 \ee
which are associated with modern values
$$a_0=a(\eta_0),~~~~~~~~~~H_0=\frac{a'(\eta_0)}{a(\eta_0)}$$
by the following relations
\be\label{rigid02}
a(\eta)=a_I\sqrt{1+2H_I(\eta-\eta_I)}=
a_0\sqrt{1+2H_0(\eta-\eta_0)}.
\ee
\section{The hierarchy of cosmological scales
\label{hierarchy}}
\makeatletter
\renewcommand{\@evenhead}{\raisebox{0pt}[\headheight][0pt]
{\vbox{\hbox to\textwidth{\hfil \strut \thechapter . A model of an empty Universe
\quad\rm\thepage}\hrule}}}
\renewcommand{\@oddhead}{\raisebox{0pt}[\headheight][0pt]
{\vbox{\hbox to\textwidth{\strut \thesection .\quad
The hierarchy of cosmological scales
\hfil \rm\thepage}\hrule}}}
\makeatother

Let us consider the beginning of the Universe, assuming dominance of the Casimir vacuum energy.
If in the Beginning, the Universe was quantum, we can apply the postulate of Planck's least action to
determine the initial value of the cosmological factor and consider the hierarchy (classification) of
cosmological scales according to their conformal weights.

A hypothetical observer measures the conformal horizon (\ref{tau51})
\bea
 \label{1-3hh}
 d_{\rm horizon}({a})=2r_{\rm horizon}(z)=
 2 \int\limits_{a_I\to 0}^{{a}}
 d \overline{a}\;
 \dfrac{\overline{a}}{\sqrt{\rho_{\rm cr}}}=\dfrac{a^2}{H_0}\;.
\eea
as the distance that the photon runs on its light cone
$d\eta^2-dr^2=0$
during the lifetime of the Universe. In accordance with the formula (\ref{1-3hh}), four-dimensional volume
of the early Universe, limited by the horizon
$$\eta_{\rm horizon}=r_{\rm horizon}(z)=\frac{1}{2H_0(1+z)^{2}},$$
is equal to
 \be\label{Pl-1}
 V_{\rm horizon}^{(4)}(z)=\frac{4\pi}{3}r^3_{\rm horizon}(z)\cdot \eta_{\rm horizon}(z)=
 \frac{4\pi}{3\cdot16 H_0^4(1+z)^8}.
 \ee
It is natural to assume that at the time of its creation, the Universe was quantum.
In this case, values of the action of the Universe are \textit{quantized}.
A minimal \textit{quant} of the action of the Universe under initial scale factor
$a_{\rm Pl}=(1+z_{\rm  Pl})^{-1}$
is given by the Planck postulate
 \bea
 \label{vac-a}
 W_{\rm Universe}=\rho_{\rm cr}V_{\rm horizon}^{(4)}(z_{\rm Pl})=
 \dfrac{M_{\rm  Pl}^2}{H_0^2}\dfrac{1}{32 (1+z_{\rm  Pl})^{8}}=2\pi.
\eea
Using the current data for the Planck mass and the Hubble parameter\footnote{$h=0,71\pm 0,02$ (stat) $\pm 0,06$ (syst)
is the Hubble parameter in units
100 (km/sec)/Mps \cite{Weinberg-6}.}
at $(\tau=\tau_0)$
and  $h\simeq 0.7$
\bea\label{ri-12ah}
 M_{\rm C} e^{\langle D \rangle(\tau_0)}= M_{\rm Pl}=1.2211\times 10^{19} \mbox{\rm GeV},
 \qquad \langle D \rangle(\tau_0) = 0,
\eea
\bea
\!\!\!\!\frac{d}{d\tau}\langle D \rangle(\tau_0)\!=\!H_0 \!=\! 2.1332\times 10^{-42}
\mbox{\rm GeV}\times h \!=\! 1.4332 \times 10^{-42}\mbox{\rm GeV},
 \eea
we obtain from (\ref{vac-a}) the primordial redshift value
 \bea\label{pl-2}
 a^{-1}_{\rm  Pl}=(1+z_{\rm  Pl}) \approx
 \left[\frac{M_{\rm Pl}}{H_0}\right]^{1/4}\times\left[\frac{4}{\pi}\right]^{1/8}\times\frac{1}{2}
 \simeq 0.85\times 10^{15}.
 \eea
In other words, the Planck mass and the present value of the Hubble parameter are related to each other as
the age of the Universe, expressed in terms of the primary redshift in the fourth order
$$\frac{M_{\rm Pl}}{H_0}=(1+z_{\rm  Pl})^4\simeq z_{\rm  Pl}^4.$$
We can say that the conformal weight of the Planck mass is four in the class frames of reference
associated with the time interval of luminosity $d\tau$, where the relativistic energy of the particle is
$$\omega_\tau =a^2\sqrt{{\bf k}^2+a^2M_0^2}.$$
From the expansion of the energy in powers of the cosmological scale factor
there is a classification of energies by
irreducible representations of the Weyl group \cite{Ramondempty-6}.
According to these representations, conformal weights $n=0, 2, 3, 4$ correspond to
velocity of dilaton  $v_D=H_0$, massless energy $a^2\sqrt{{\bf k}^2}$,
massive energy $M_0a^3$, Newtonian coupling constant $M_{\rm Pl}a^4$
(\ref{vac-a}), respectively.
In this classification it is possible also to include a non-relativistic particle
$$H_0\times a^{-1}_{\rm  Pl} =10^{-13}~\mbox{cm}^{-1}$$
with  conformal weight $n=1$ of its energy
$$\omega^{\rm nonr}_\tau=\frac{a^{1}{\bf k}^2}{M_0}.$$
The cosmological evolution of all these energies is given by the Hubble parameter and
can be written by a unified formula of the kind
 \be
 \label{ce-1}
 \langle\omega\rangle^{(n)}(a)=\left(\frac{a}{a_{\rm  Pl}}\right)^{(n)}H_0\;,
 \ee
According to this formula in the Beginning of the Universe, values of all these energies coincide
with the Hubble parameter. In modern time the values of all these energies are determined by
the product of the Hubble parameter to the primary (Planck) value of the redshift (\ref{pl-2})
in power equal to the corresponding conformal weight:
$$\langle\omega\rangle^{(0)}_0=H_0,\quad
\langle\omega\rangle^{(1)}_0=R_{\bigodot}^{-1},
$$
$$\langle\omega\rangle^{(2)}_0=k_{CMB},\quad
\langle\omega\rangle^{(3)}_0=M_{EW},\quad
\langle\omega\rangle^{(4)}_0=M_{0\rm Pl}.$$
As a result, Planck's postulate of minimal action
leads to a hierarchy of cosmological scales to the present days ($a=1$)
 \be
 \label{ce-2}
 \omega_0^{(n)}\equiv\langle\omega\rangle^{(n)}(a)\Big|_{(a=1)}=(1/a_{\rm
Pl})^{(n)}H_0,
 \ee
shown in  Table \ref{Table:1}.

\begin{table}
\begin{tabular}{|c|c|c|c|c|c|}\hline
 n&{n}=0&{n}=1&{n}= 2&{n}=3&{n}=4
 \\ \hline
$\omega_0^{(n)}$ &$ H_0\simeq 1,4\cdot\!10^{-42}$&
$\simeq1,2\cdot \!10^{-27}\!$&
$
\simeq 10^{-12}$&$\simeq3\cdot 10^{2}$&$\simeq4\cdot 10^{18}$\\ \hline
\end{tabular}
\caption{\label{Table:1}
The hierarchy of cosmological scales in GeV.
}
\end{table}
 Table \ref{Table:1} contains scales corresponding to the inverse size of the
Solar System for the conformal weight $(n=1)$, the average momentum of the CMB $(n=2)$,
electroweak scale of SM $(n=3)$ and Planck's mass $(n=4)$.
We conclude that the observational data suggest that the cosmic evolution (\ref{ce-1}) of
all of these energy scales with conformal weights $(n=0,1,2,3,4)$
has a common origin, which can be the Casimir energy of an empty space.

Thus, the use of minimum action postulate leads to
the primary value of the cosmological scale factor
$a_{\rm{Pl}}$, given by the equation (\ref{pl-2}) in the considered Conformal model of the Universe.
The classification of the different states of matter,
in accordance with their conformal weights,
reveals a hierarchy of energy scales, according to the observations.

Why is an empty Universe filled with particles? And why are these particles just enough
as we observe in the Universe? The answers to these questions will be discussed in the next chapters.

\section{Special Relativity -- General Relativity correspondence}
\makeatletter
\renewcommand{\@evenhead}{\raisebox{0pt}[\headheight][0pt]
{\vbox{\hbox to\textwidth{\hfil \strut \thechapter . A model of an empty Universe
\quad\rm\thepage}\hrule}}}
\renewcommand{\@oddhead}{\raisebox{0pt}[\headheight][0pt]
{\vbox{\hbox to\textwidth{\strut \thesection .\quad
Special Relativity -- General Relativity correspondence
\hfil \rm\thepage}\hrule}}}
\makeatother

The problem of \textit{understanding  creation and evolution of the Universe} is probably not in the exact solution
of the equations of the General Relativity, but in an ontology, {\it id est}, in the adequate application of the concepts
of modern relativistic and quantum physics to observational cosmology.
To demonstrate this assertion, let us consider according to Wheeler and De Witt, a
quantum theory of the Universe for a model of an empty Universe (\ref{1-3nt}), that can be solved exactly in
classical and quantum cases.

Above, there were presented arguments and evidence that
the classical exact solution provided a description of the data on the dependence of the redshift
from the distance to cosmic objects in all ages of the evolution of the Universe, including the latest data on the Supernovae.
As we show below, the quantum solution gives a positive arrow of time interval
and allows to describe the creation of the Universe from a vacuum, defined as the state with the lowest
energy, according to the postulates of quantum field theory.
We make in the action (\ref{1-3nt}) the change of variables (neglecting the total derivative)
\bea \label{qu-1}
\sqrt{2V_0}\,\langle D \rangle&=&X_U,
\\\label{qu-2}
\sqrt{2V_0}\,\rho_{\rm cr}&=&M_U.
\eea
Then the expression (\ref{1-3nt}) takes the form of the action for a relativistic particle at rest in the Special Relativity
\bea \label{qu-1Um}
\!\!\!\! W_{\rm Universe}&\!=\!&
 -\frac{1}{2}\int\limits_{\tau_I}^{\tau_0} dx^0 N_0
 \left[\left(\frac{d\, X_U}{N_0dx^0}\right)^2
 +M_U^2\right]=\int\limits_{\tau_I}^{\tau_0} dx^0 L, \\\label{qu-2U} 
\!\!\!\!dx^0 N_0&\!=\!&d\tau,
\eea
where the speed of light is set equal to unity, $c=1$.
Here, $X_{U}$ plays a role of an evolution parameter in a space of measurements,
$x^0$ is a coordinate evolution parameter in the one-dimensional Riemannian manifold as the object of  coordinate
transformations $
x^0\to \widetilde{x}^0=\widetilde{x}^0(x^0)$
with
\textit{unmeasured parameters},
$N_0(x^0)$ is a lapse function that has sense as a metrics for a geometric time interval
$N_0(x^0)dx^0={d\tau}$
for classical equations of motion.

By introducing the canonical momentum variable $X_{U}$:
  $$P_{U}=\frac{\partial L}{\partial ({\partial_0X_{U}})},$$
the action (\ref{qu-1Um}) we rewrite in the canonical form
\be\label{1-qu}
   W=\int d x^0 \left[P_{U}\frac{dX_{U}}{d x^0}+\frac{N_0}{2}\left(P^2_{U}-M_U^2\right)\right].
   \ee
The equations of motion take a form
        $$\dfrac{dP_{U}}{d\tau}=0, ~~~~~~~~~~~~~~\dfrac{dX_{U}}{d\tau}={P_{U}}.$$
The solutions of these equations
  \be\label{2qu}
        X_{U}(\tau)=X_{IU}+
       {P_{IU}}\,  (\tau-\tau_I)
   \ee
depend on initial data
$$X_{U}(\tau=\tau_I)=X_{IU},~~~~~~~{P_{IU}}=E_U.$$
A variation of the action by the metric $N_0(x^0)$ yields a formula for Hamiltonian constraint of
an initial momentum $E_U$
         \be\label{2-qu}
         E^2_{U}-M_U^2=0.
         \ee
In the General Relativity, this formula is traditionally identified with the zero-point energy of the system.
In our monograph, the \textit{energy of relativistic Universe} is the solution of the constraint
(\ref{2-qu}) in respect to $E_U$
   \be\label{3-qu}
          E_U = \pm M_U.
   \ee
A formula for the Hamiltonian constraint (\ref{2-qu}) for the \textit{energy of relativistic Universe}
is an analog of the Hamiltonian constraint for the \textit{energy of relativistic particle} at rest
$E^2-m^2=0$.
For the \textit{relativistic particle} a solution of the Hamiltonian constraint gives two values of energy:
positive and negative $E=\pm m$.
A negative value of energy means that the classical relativistic particle is unstable.
To get rid of negative energies in the relativistic theory and enter a stable vacuum as a state of
the lowest energy, one makes two quantizations of the Hamiltonian constraint (\ref{2-qu}):
the primary, when the constraint is converted into the equation for the wave function,
and the secondary, where the wave function itself becomes the operator in Fock's \textit{space of numbers of occupation}.
Let us consider these quantizations in our model of the Universe.

\section{The arrow of time as a consequence of the postulate of vacuum}
\makeatletter
\renewcommand{\@evenhead}{\raisebox{0pt}[\headheight][0pt]
{\vbox{\hbox to\textwidth{\hfil \strut \thechapter . A model of an empty Universe
\quad\rm\thepage}\hrule}}}
\renewcommand{\@oddhead}{\raisebox{0pt}[\headheight][0pt]
{\vbox{\hbox to\textwidth{\strut \thesection .\quad
The arrow of time as a consequence of the postulate of vacuum
\hfil \rm\thepage}\hrule}}}
\makeatother

For the first time such an idea on the level of quantization of the General Relativity by analogy with the quantization
of the Special Relativity was formulated by Bryce De Witt \cite{DeWitt-6}, where he identified a
\textit{parameter of evolution} in cosmology with the cosmological scale factor and introduced in the General Relativity
the concept of {\it field space of events}, where the {\it relativistic Universe} moved, by analogy with the concept of
Minkowskian space of events, where the {\it relativistic particle} moved.

The primary quantization of the constraint
$P_{U}^2=M_{U}^2,$
by replacement of a particle momentum $P_{(0)}$ to an operator
$$\hat P_{U}=-\imath\frac{d}{d X_{U}}$$
leads to an equation of the Klein -- Gordon for the wave function \cite{Vilenkin-4}
 \be\label{q11-6}
 (\hat P^2_{U}-M_{U}^2)\Psi_U=0,
 \ee
which in cosmology is called the equation of Wheeler -- De Witt (WDW).
Its solution is a sum of two terms
\bea\label{q1-U}
\!\!\!\!&& \Psi_U=\frac{1}{\sqrt{2E_U}}\times\\\nonumber
\!\!\!\!&&\times
 \left[A_I^+e^{\imath E(X_{U}-{X_{IU}})}\theta(X_{U}-{X_{IU}})+
 A_I^-e^{-\imath E(X_{U}-{X_{IU}})}\theta({X_{IU}}-X_{U})\right]
\eea
with coefficients $A_I^+,A_I^-$, according to two classical solutions of the
constraint equation with positive and negative energy.
\begin{figure}[t]
\includegraphics[width=0.90\textwidth,height=0.45\textwidth]{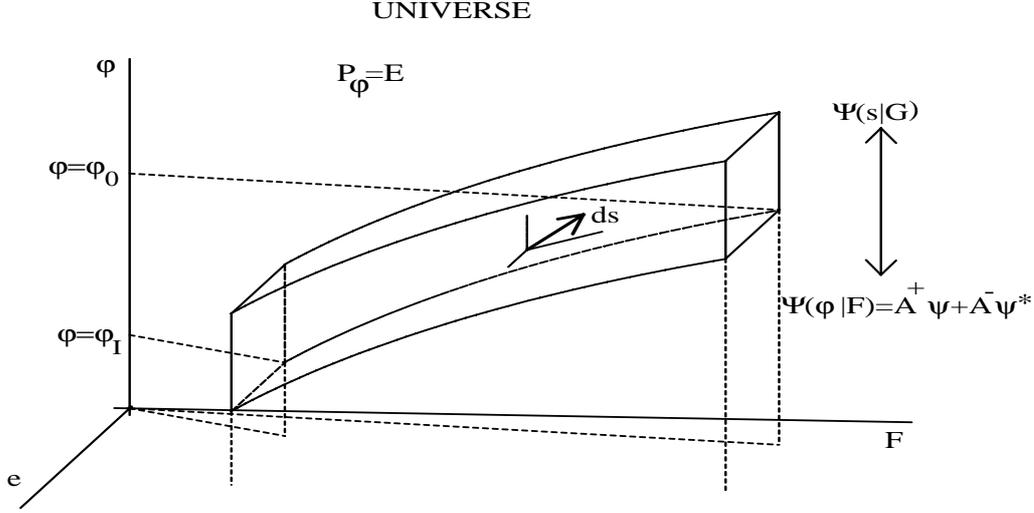}
\caption{\small
The figure shows the motion of a relativistic Universe in its field space of events.
Full description of the motion is given by the two sets of observables:
{\it dynamical} in the space of events and {\it geometrical} in the tangent Minkowskian
space. Each of these sets has its parameter of evolution and its wave function $\Psi$.
Two measured parameters of evolution ({\it dynamical} parameter
$\varphi=M_{\rm Pl}e^{-D}$ and time as
{\it geometrical} interval $s$) are connected by the Hubble law.
}
\end{figure}
The secondary quantization of the initial data
$$[\hat A_I^-,\hat A_I^+]=1$$
leads to the vacuum $A_I^-|0>=0$
as state with the lowest energy if the coefficient $A_I^+$ is
interpreted as an operator of creation of a particle with positive energy,
which flies ahead of the initial data $X_{IU}< X_{U}$,
and the coefficient $A_I^-$ as  an operator of annihilation
of a particle also with positive energy, which flies to the initial data $X_{U}< X_{IU}$.
Substituting these solutions into the expression for the trajectory of the Universe (\ref{2qu}),
we obtain, that the geometrical interval $\tau-\tau_I>0$ is always greater than zero.
This is the arrow of time.  Thus, the existence of a stable vacuum leads
to the arrow of time.

The WDW equation (\ref{q11-6}) is obtained by variation of the action of
the corresponding classical theory of field type Klein -- Gordon \cite{DeWitt-6}
\footnote{Unlike the original relativistic system (\ref{1-3nt}) with three spaces,
the formulation of the Wheeler and De Witt (\ref{uf-6}) loses {\it time as geometrical interval} and,
consequently, its dependence on the scale factor, which is interpreted
in the classical Friedmann cosmology as Hubble's law. As a result, in classical cosmology one
does not know how to quantize, and in quantum cosmology \cite{DeWitt-6} -- how to describe the Hubble law.}:
\be\label{uf-6}
W_{\rm U}=\frac{1}{2}\int dX_U
\left[\left(\frac{d\Psi_U}{dX_U}\right)^2-E^2_U\Psi_U^2\right]\equiv \int  dX_U
L_{\rm U}.
\ee
Such an approach could be called a field theory for the universes
\cite{PavlovFriedmann-6,PavlovOpen-6}.

The negative energy solutions in (\ref{3-qu}) means that the relativistic system has not a
minimum of its energy and arbitrarily small interaction makes the system unstable.
The system can be made stable in a quantum field theory, arising from the second quantization of
WDW field $\Psi_U$, if further the existence of a vacuum state with the lowest energy is postulated.
Introducing the canonical momenta
$$P_\Psi=\frac{\partial L_{\rm U}}{\partial(\partial_{X_U}\Psi_U)},$$
one can yield the Hamiltonian formulation of the theory with an action (\ref{uf-6})
\be\label{ufh}
W_{\rm U}=\int dX_U \left(P_\Psi\frac{d\Psi_U}{d X_U}-H_{U}\right),
\ee
where
\be\label{ufh1}
H_{U}=\frac{1}{2}\left[P_\Psi^2+E_U^2\Psi_U^2\right]
\ee
is the Hamiltonian. Determination of the energy $E_U$ for one particular universe
gives us the opportunity to present the Hamiltonian $H_{U}$ in standard form of
product of the energy $E_U$ and the number of occupation of excitations of Wheeler -- De Witt field,
which can be identified with the number of created \textit{universes}
 \be\label{AA}
 \hat N_U=A^+A^-,
 \ee
 \be\label{ufh1z1}
 H_{U}=\frac{1}{2}E_U\left[A^+A^-+A^-A^+\right]=E_U\left[N_U+\frac{1}{2}\right]
 \ee
by passing to holomorphic variables \cite{ps1chapter6}
 \be\label{g-2-6}
 \Psi_U=\frac{1}{\sqrt{2E_U}}(A^{+}+A^{-}),~~~~~~~~
 P_\Psi=\imath\sqrt{\frac{E_U}{2}}(A^{+}-A^{-}),
 \ee
where $A^+,~A^-$ are operators of creation and annihilation of {\it universes}, correspondingly.

To eliminate the negative energy one would have to postulate that the $A^-$ is the operator of
annihilation of  {\it universe} with positive energy, it assumes the existence of a vacuum
state as the state with the lowest energy:
 \be\label{g-1-6}
 A^{-}|0>_{\rm A}=0.
 \ee
A number of {\it universes} $N_U=A^+A^-$ (\ref{AA})
can not be saved if the energy $E_U$ depends on $X_U$.
In this case, the vacuum state (\ref{g-1-6}) becomes unstable, since a
dependence of the energy $E_U$ of the dynamic evolution parameter
$X_U$ leads to an additional term in the action (\ref{uf-6}),
if it is rewritten in terms of holomorphic variables in the functional space
\bea\label{und3}
P_\Psi \frac{d\Psi_U}{d X_U}&=&\\\nonumber
&=&  \left[\frac{\imath}{2}\left(A_q^+\frac{d A^-}{d X_U}-A^+
 \frac{d A^-}{d X_U}\right)-
 \frac{\imath}{2}\left(A^+A^+- A^-A^-\right)\triangle(X_U)\right],
\eea
where
\be\label{tri}
 \triangle(X_U)=\frac{1}{2E_U}\frac{dE_U}{dX_U}.
\ee
The last term in the expression (\ref{und3}) describes a cosmological creation of
{\it universes} from the \textit{vacuum}.
The method of describing of such cosmological creation is the Bogoliubov transformations
 \cite{ps1chapter6, b-6}.

\section{Creation of the Universe}
\makeatletter
\renewcommand{\@evenhead}{\raisebox{0pt}[\headheight][0pt]
{\vbox{\hbox to\textwidth{\hfil \strut \thechapter . A model of an empty Universe
\quad\rm\thepage}\hrule}}}
\renewcommand{\@oddhead}{\raisebox{0pt}[\headheight][0pt]
{\vbox{\hbox to\textwidth{\strut \thesection .\quad
Creation of the Universe
\hfil \rm\thepage}\hrule}}}
\makeatother

To determine a \textit{vacuum} and a set of conserving numbers, called integrals of motion,
we can use (as in the case of cosmological creation of particles \cite{ps1chapter6})
the Bogoliubov transformations \cite{b-6} of variables $(A^+,A^-)$
\be \label{u17}
A^+=\alpha
 B^+\!+\!\beta^*B^-,~~~\;\;A^-=\alpha^*
 B^-\!+\!\beta B^+~~~(|\alpha|^2-|\beta|^2=1),
\ee
so corresponding equations, expressed in terms of {\it universes} $(A^+,A^-)$:
 \be \label{1un}
\left(\imath\frac{d}{d X_U}+E_U\right)A^+=\imath A^-\triangle,~~~~~
\left(\imath\frac{d}{d X_U}-E_U\right)A^-=\imath A^+\triangle,
  \ee
take a diagonal form in terms of {\it quasiuniverses}
  $B^+,B^-$:
\be\label{2un}
\left(\imath\frac{d}{d X_U}+E_B\right)B^+=0,~~~~~
\left(\imath\frac{d}{d X_U}-E_B\right)B^-=0.
  \ee

This means that coefficients of the Bogoliubov transformations satisfy to equations
\be \label{3un}
 \left(\imath\frac{d}{d X_U}+E_U\right)\alpha=\imath\beta\triangle,~~~~~~~~
\left(\imath\frac{d}{d X_U}-E_U\right)\beta^*=\imath\alpha^*\triangle.
\ee
If we express the coefficients of the Bogoliubov transformations in the form of
\be \label{4un}
\alpha=e^{\imath\theta}\cosh r,~~~~~~~~
\beta^*=e^{\imath\theta}\sinh r,
\ee
where magnitudes $r,\theta$ are called parameters of shear and rotation, respectively,
these equations take the following form
\be \label{5un}
\left(\frac{d \theta}{d X_U}-E_U\right)\sinh 2r=-\triangle\cosh 2r\sin 2\theta,~~~~
\frac{d r}{d X_U}=\triangle\cos 2\theta,
\ee
while the energy of the {\it quasiuniverses} in equations (\ref{2un}) is given by
\be \label{6un}
 E_B=\frac{E_U-\partial_{X_U}\theta}{\cosh 2r}.
 \ee
By these equations (\ref{2un}), a number of {\it quasiuniverses}
${\cal N}_B=(B^+B^-)$
is conserved
  \be \frac{d{\cal N}_B}{dX_U}\equiv
  \frac{d(B^+B^-)}{dX_U}=0.
  \ee
Hence, we get the definition of \textit{vacuum} as state without {\it quasiuniverses}:
 \be \label{sv}
 B^-|0>_{\rm U}=0.
 \ee

\begin{center}
{\vspace{0.31cm}}
\parbox{0.5\textwidth}{
\includegraphics[height=9.truecm,
angle=-0]{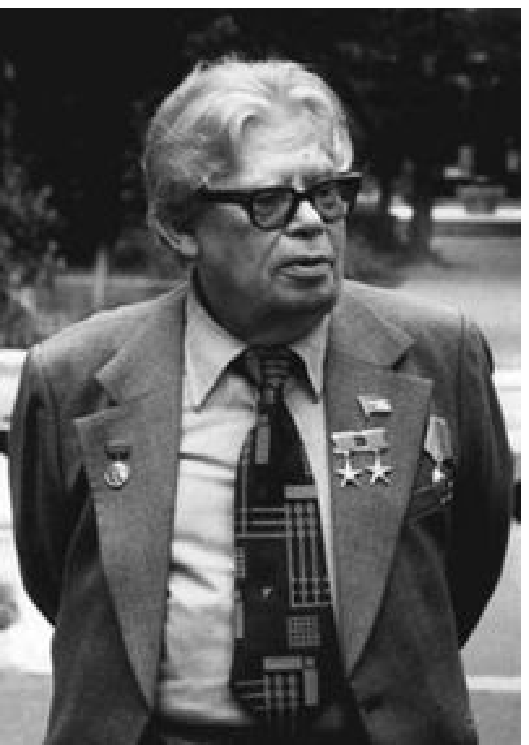}}
\parbox{0.48\textwidth}
{\small
N.N. Bogoliubov (8 (21) August 1909, Nizhni Novgorod — 13 February 1992, Moscow).
The outstanding Russian mathematician and physicist, Academician of the Russian Academy of Sciences,
founder of scientific schools in nonlinear mechanics and theoretical physics.
Since 1956 - Director of the Laboratory of Theoretical Physics, JINR, Dubna,
since 1965 till 1988  — Director of JINR.
Head of quantum field theory and statistical physics department of Moscow State University from 1966 to 1992.
The main works are devoted to asymptotic methods of nonlinear mechanics, quantum field theory,
statistical mechanics, calculus of variations, the approximate methods of mathematical analysis,
differential equations and mathematical physics, the theory of stability,
dynamical systems, and other areas of theoretical physics.
}
{\vspace{0.31cm}}
\end{center}

\begin{table}[ht]\label{tab1}
\centering
\begin{tabular}{|c|c|c|}
\hline $\mathrm{N}^{\underline{\mathrm{o}}}$&UNIVERSE&PARTICLE\\[.35cm]
\hline 1.&$x^0 \to \widetilde{x}^0=\widetilde{x}^0(x^0)$&$\tau\to\widetilde{\tau }=
\widetilde{\tau }(\tau )$\\[.35cm]
\hline 2.&$N(x^0)\,dx^0=d\tau=\dfrac{d\eta}{a^2}=\dfrac{dt}{a^3}$&$ds=e (\tau)\,\, d\tau$\\[.35cm]
\hline 3.&$[\langle D\rangle~|~\widetilde{F}]$&$[X_0~|~X_k]$\\[.35cm]
\hline 4.&$P^2_{\langle D\rangle}-E_U^2=0$&$P^2_0-E^2_0=0$\\[.35cm]
\hline 5.&$\tau_{(\pm)}=\pm\int^{\langle D\rangle_0}_{\langle D\rangle_I}
{d\langle D\rangle}~{\langle{
(\widetilde{H}})^{-1/2}\rangle}\geq 0$&$s_{\pm}=\pm\dfrac{m}{E}[X_0^0-X^0_I]\geq 0$
\\[.35cm]
\hline 6.&$E_U=\pm 2 \int d^3x(\widetilde{H})^{1/2}$&$E_p=\pm\sqrt{m^2+{\bf p}^2}$\\[.35cm]
\hline 7.&$[\hat P^2_{\langle D\rangle}-E_U^2]\Psi_{\rm WDW}=0$&$
[\hat P^2_0-E^2_0]\Psi_{\rm KG}=0$\\[.35cm]
\hline 8.&$\Psi_{\rm U}=\dfrac{A^++A^-}{\sqrt{2E_U}}$&$\Psi_{\rm KG}=\dfrac{a^++a^-}
{\sqrt{2E_0}}$\\[.35cm]
\hline 9.&$A^+=\alpha B^+\!+\!\beta^*B^-$&$a^+=\alpha b^+\!+\!\beta^*b^-$\\[.35cm]
\hline 10.&$B^-|0>_{\rm B}=0$&$b^-|0>_{b}=0$\\[.35cm]
\hline 11.&${}_{\rm B}\!\!<0|A^+A^-|0>_{\rm B}\not=0$&${}_{b}\!\!<0|a^+a^-|0>_{b}\not=0$\\[.35cm]
\hline
\end{tabular}
\end{table}

\begin{table}[ht]\label{tab2}
{\textbf{Correspondence Universe -- particle}.\\
{
$\mathrm{N}^{\underline{\mathrm{o}}}$1 -- group of diffeomorphisms, \\
$\mathrm{N}^{\underline{\mathrm{o}}}$2 -- diffeoinvariant integrals, \\
$\mathrm{N}^{\underline{\mathrm{o}}}$3 -- space of events with a parameter of evolution,\\
$\mathrm{N}^{\underline{\mathrm{o}}}$4 -- Hamiltonian constraint in the space  of events,\\
$\mathrm{N}^{\underline{\mathrm{o}}}$5 -- Hubble law,\\
$\mathrm{N}^{\underline{\mathrm{o}}}$6 -- energy in the space of events,\\
$\mathrm{N}^{\underline{\mathrm{o}}}$7 -- primary quantization, \\
$\mathrm{N}^{\underline{\mathrm{o}}}$8 -- secondary quantization,\\
$\mathrm{N}^{\underline{\mathrm{o}}}$9 -- Bogoliubov transformations,\\
$\mathrm{N}^{\underline{\mathrm{o}}}$10 -- vacuum of quasiparticles,\\
$\mathrm{N}^{\underline{\mathrm{o}}}$11 -- occupation numbers of \textit{universes} and particles.\\
}} 
\end{table}

A number of created {\it universes} from this Bogoliubov vacuum can be found by calculating the average
of the operator of {\it number of universes} (\ref{AA}) by Bogoliubov vacuum.
It can be seen that this number is proportional to square of the coefficient, given in equation (\ref{u17})
 \be\label{usv1}
 N_{\rm U}(X_U)={}_{\rm U}\!\!\!<0|A^+A^-
 |0 >_{\rm U}\equiv |\beta|^2.
 \ee
This value can be called as \textit{number of universes}
  $N_{\rm U}(X_U)$,
while the value of
\be\label{usv2}
 R_{\rm U}(X_U)=\left(\frac{\imath}{2}\right)~{}_{\rm U}\!\!<0|[A^+A^+-A^-A^-]|0 >_{\rm U}=
 \ee
$$= \imath(\alpha^*\beta^*-\alpha\beta)
 =-\sinh 2r\sin 2\theta
$$
as \textit{Bogoliubov condensate}, correspondingly.
Bogoliubov equations, expressed in terms of quantities of \textit{number of universes}
$N_{\rm U}(X_U)$ and
\textit{Bogoliubov condensate} $R_{\rm U}(X_U)$ take the form
\be\label{usv3}
 \left\{\begin{aligned}
 \frac{dN_{\rm U}}{dX_U}&=\triangle(X_U)
 \sqrt{4N_{\rm U}(N_{\rm U}+1)-R_{\rm U}^2},  \\
 \frac{dR_{\rm U}}{dX_U}&=-{2E_U(X_U)}
 \sqrt{4N_{\rm U}(N_{\rm U}+1)-R_{\rm U}^2}
 \end{aligned}\right.
 \ee
with initial data
 $$N_{\rm U}(X_{U\,I})=R_{\rm U}(X_{U\,I})=0.$$

We see that the vacuum postulate leads to a positive value of conformal time
as for the Universe
$$E_U>0, \qquad X_{U}>X_{U\,I},$$
and for an anti-Universe
$$E_U<0, \qquad X_{U\,I}<X_{U},$$
that is leading to an arrow of conformal time. Time has a beginning and the
Quantum Universe is created with his time.
\section{Summary}
\makeatletter
\renewcommand{\@evenhead}{\raisebox{0pt}[\headheight][0pt]
{\vbox{\hbox to\textwidth{\hfil \strut \thechapter . A model of an empty Universe
\quad\rm\thepage}\hrule}}}
\renewcommand{\@oddhead}{\raisebox{0pt}[\headheight][0pt]
{\vbox{\hbox to\textwidth{\strut \thesection .\quad
Summary and literature
\hfil \rm\thepage}\hrule}}}
\makeatother

We have considered in this Chapter an empty Universe model, leaving only a
zero harmonic of the dilaton and experimentally measured Casimir vacuum energy.
Casimir vacuum energy is given by the size of the space. If the size of space is equal to the horizon,
which in turn is determined by Casimir energy in the \textit{empty} Universe, there is a self-consistent
equation of state of the \textit{empty} space. The solution of this equation
gives the dependence of the density of the cosmological  scale factor.
In the Conformal cosmology, with longer intervals than in the Standard one,
the resulting dependence of the density of empty space of the cosmological scale factor
describes the recent observational data on the Supernovae.
Thus, data on the Supernovae show that we still live in a nearly empty Universe.

The Planck principle of minimal action of gravity ({\it id est} the quantum of action) for the space limited
by the size of the horizon, leads to the primary value of the cosmological
scale factor (and the corresponding redshift) $\sim 0.8 \times 10^{15}$.
Representations of the Weyl group for massive and massless particles give (together with modern values of
Hubble's parameter and the Planck mass) a hierarchy of cosmological scales in a surprising
agreement with the values of the CMB temperature (for massless particles)
and the scale of the electroweak interaction (for massive particles).

From a mathematical point of view, the Conformal theory of gravity in the approximation of the empty Universe
is a model of squeezed oscillator of the dilaton. Primary and secondary quantization of
this model with the postulate of the existence of the vacuum leads to the arrow of the geometric interval of
time and to a certain wave function of the Universe, such as the irreducible unitary
representations of the Poincar\'e group for a relativistic particle.
Now we are ready to consider the irreducible unitary representations of affine and conformal symmetry groups
in the exact theory, that will be done in Chapter 8.

\chapter{Quantization of gravitons in terms of Cartan forms
\label{S-7}}
\renewcommand{\theequation}{7.\arabic{equation}}
\setcounter{equation}{0}
\section{Affine gravitons}
\makeatletter
\renewcommand{\@evenhead}{\raisebox{0pt}[\headheight][0pt]
{\vbox{\hbox to\textwidth{\hfil \strut \thechapter . Quantization of gravitons in terms of Cartan forms
\quad\rm\thepage}\hrule}}}
\renewcommand{\@oddhead}{\raisebox{0pt}[\headheight][0pt]
{\vbox{\hbox to\textwidth{\strut \thesection .\quad
Affine gravitons
\hfil \rm\thepage}\hrule}}}
\makeatother
It is well known that the General Relativity in terms of the metric components is nonrenormalizable theory.
A renormalizable quantum gravitation theory does not exist \cite{Faddeev:1973zb-7}.
Here we show that GR  in terms of Cartan forms becomes not only renormalizable theory, but
 describes free gravitons, far from the matter sources, where
 the Newton-type potentials
can be  neglected:
 $$\bar D=0,~~~~~~~ {\cal N}_i=0,~~~~~~~ {\cal N}=1$$
(like as, in QED, photons become free far from
the charges and currents) \cite{grg12-7}.

Let us consider the graviton action~ (\ref{1-3dgc1}), where one keeps
 only the simplex components
$$\overline{\omega}_{(a)}(d)=\widetilde{{\bf e }}_{i(a)}dx^i.$$
They obey the condition of the
diffeo-invariance. It is one of the main differences of the diffeo-invariant Conformal GR from the
metric GR.
The choice of the diffeo-invariant
symmetry condition in the GR leads to the result that follows from the
theorem~\cite{Tod-7}: \textit{any arbitrary two-dimensional space metric
$$dl^2=h_{AB}dx^A dx^B,~~~~(A,B=1,2)$$ can be represented by diffeomorphisms
$$x^A\to \widetilde{x}^A=\widetilde{x}^A(x^1,x^2)$$ in a diagonal form.} The
result is in the fact that a kinemetric-invariant nonlinear plane wave
moving in the direction ${\bf k}$ with the unit determinant $det\;h\;=1$
contains only a single metric component.

In particular, in the  frame of reference  ${\bf k}=(0,0, k_3)$ one has
$$\widetilde{{\bf e}}^1_{(1)}=e^{g(x_{(3)},\tau)},~~~~~~~~
\widetilde{{\bf e}}^2_{(2)}=e^{-g(x_{(3)},\tau)},~~~~~~~~
\widetilde{{\bf e}}^3_{(3)}=1;$$
all other (non-diagonal) components $\widetilde{{\bf e}}^i_{(a)}$ are equal to zero.

Thus, we obtain
 \bea \label{dg-2a2}
 \overline{{\omega}}_{(1)}&=& 
 dX_{(1)}- [X_{(1)}]d g,
 \\\label{dg-2a3}
 \overline{{\omega}}_{(2)}&=&
 dX_{(2)} + [X_{(2)}]d g,
 \\
 \label{dg-2ax}
 \overline{{\omega}}_{(3)}&=&dx_{3}=dX_{(3)},
 \eea
where a single-component affine graviton $g=g(X_{(3)},\tau)$  is a function depending
on the time and a single spatial coordinate $X_{(3)}$
in the tangent space $X_{(b)}$.
The solutions of the equation
$$\frac{\delta W}{\delta g}=0~~~~\to ~~~~g=g(\eta, X)$$
can be expressed via the
tangent coordinates:
   \bea
   \label{X-1}X_{(1)}= e^{g(x_{(3)},\tau)}  \overline{x}^1
   \\
   \label{X-2} X_{(2)}= e^{-g(x_{(3)},\tau)}  \overline{x}^2. \eea
Eqs.~(\ref{dg-2a2}) and (\ref{dg-2a3}) mean
 an expansion (or contraction) of the hypersurface $X_{(A)}\,( A=1,2)$ perpendicular
to the direction of the  gravitational wave propagation $X_{(3)}$.
A gravitation wave changes the particle velocity   via the Hubble like law:
the more is the base, the more is the additional velocity induced by the graviton.
The exact local Hamiltonian density for the affine graviton is given by (\ref{3g})
\bea\label{3g-1}
 {\cal H}_{\rm g}&=&\left[6{p^2_{(a)(b)}}+
 \frac{1}{6}R^{(3)}(\widetilde{{\bf e}})\right]\;,
 \eea
where $R^{(3)}({{\bf e}})$ and $p^2_{(a)(b)}$  are are defined from Eqs.~(\ref{1-17}) и (\ref{1-17g}),
respectively. For the frame of reference ${\bf k}=(0,0, k_3)$,
 we have~\cite{ll-7}:
 \bea \label{dg-11}
 R^{(3)}(\widetilde{{\bf e}})&=&(\partial_{(3)}g)^2,\qquad
 {p^2_{(a)(b)}}=\frac{1}{9}\left[\partial_\tau{g}
 \right]^2.
 \eea
There is a difference of the diffeo-invariant affine graviton from the metric one
$$g^{TT}_{ij}=g^{TT}_{ji}$$
in GR \cite{ll-7}.
While the affine graviton has  a single degree of freedom,
the metric graviton has two  traceless  and
transverse components that satisfy four constraints
 \bea\label{TT-1}
 g^{TT}_{ii}&=&0,\\\label{TT-2}
 g^{TT}_{i3}&=&g^{TT}_{3i}=0.
 \eea
In the general case of the Conformal GR $\widetilde{{\bf e}}_{(b)i}={\bf e}^{T}_{(b)i}$,
both the transverse constraint
 \be\label{tr-1}
 \partial_i{\bf e}^{T}_{(b)i}=0
 \ee
and the unit determinant one
 \be\label{tr-2}
 |{\bf e}^{T}_{(b)i}|=1
 \ee
(as the analog of the Lichnerowicz gauge in the metric formalism~\cite{lich-7})
admit to generalize  Eqs.~(\ref{dg-2a2}), (\ref{dg-2a3}), and (\ref{dg-2ax}) for the
linear forms
 \bea
 \label{P-1}
 {\overline{\omega}}_{(b)}(d)&=&{\bf e}^{T}_{(b)i}dx^i
 \\\nonumber
 &=&d [{\bf e}^{T}_{(b)i}x^i]-x^jd{\bf e}^{T}_{(b)j}\\\nonumber
 &=& dX_{(b)}-X_{(c)}{\bf e}^{Ti}_{c}d{\bf e}^{T}_{(b)i}\\\nonumber
 &=&dX_{(b)}-X_{(c)}\left[\omega^R_{(b)(c)}+
 \omega^L_{(b)(c)}\right]
 \eea
in the tangent coordinate space.  Here $X_{(b)}$  can be obtained by the formal
generalization
of Eqs.~(\ref{dg-2a2}), (\ref{dg-2a3}), and (\ref{dg-2ax})
by means of the Leibnitz rule
$${\bf e}^{T}_{(b)\,i}d[x^i]=
d[{\bf e}^{T}_{(b)\,i}x^i] -x^id{\bf e}^{T}_{(b)i}.$$
 The diffeomorphism-invariance
admits the choice of the gauge in Eq.~(\ref{P-1})
 \be\label{gauge-1}
 \omega_{(b)\,(c)}^L=0.
 \ee
Similar result is valid for a general case of arbitrary wave vector
$$\textbf{k}=\dfrac{2\pi}{V^{1/3}_0}\textbf{l},$$ where
$X_{(3)}$
is replaced by
$$X_{(k)}=\frac{(\textbf{k}\cdot\textbf{X})}{\sqrt{{\textbf{k}}^2}}.$$
The single-component graviton  $g(\tau,\textbf{X})$ considered
as the tensor massless representation of the Wigner classification
of the Poincar$\acute{\rm e}$ group
 can be decomposed
into a series of strong waves (in natural units)
 \bea \label{R-1a}
 \omega_{(a)(b)}^R(\partial_{(c)})
=\imath\sum_{\textbf{k}^2\not=0}
 \dfrac{e^{\imath\textbf{\textbf{k}\textbf{X}}}}{\sqrt{2\omega_{\textbf{k}}}}
 \textbf{k}_{c}&\times&
\\\nonumber
&\times& [\varepsilon^R_{(a)(b)}(k)
 g_{\textbf{k}}^+(\eta)+\varepsilon^R_{(a)(b)}(-k)
 g_{-\textbf{k}}^-(\eta)].
\eea
 Here ${\varepsilon}^R_{(a)\,(b)}(\textbf{k})$  satisfies the constraints
 \bea\label{TT-3}
 \varepsilon^R_{(a)(a)}(k)&=&0,\\\label{TT-4}
 {\bf k}_{(a)}\varepsilon^R_{(a)(b)}(k)&=&0\;,
 \eea
similar to (\ref{TT-1}), (\ref{TT-2}). The variable
$\omega_{\textbf{k}}=\sqrt{{\textbf{k}}^2}$ is the graviton energy and the affine
graviton
 \be \label{3-1pp}
 \overline{g}_{\textbf{k}}=\dfrac{\sqrt{8\pi}}{M_{\rm Planck}V^{1/2}_0}g_{\textbf{k}}
 \ee
is normalized  to the units of volume and time (like a photon in QED~\cite{ll-7}).

In the mean field approximation
 \bea \label{mfa-11}
 {\cal N}(x^0,x^j)&=& 1, ~~~N^j=0, ~~~~\overline{D}=0,\\
 \label{mfa-6a}
 \widetilde{ds}^2&=&[d\eta]^2-[\omega_{(b)}\otimes\omega_{(b)}],
 \eea
 when one neglects all Newtonian--type
interactions, the action of an affine graviton
reduces to the form of the exact action for the strong gravitational wave~\cite{ll-7}
$$
 {W}^g_{\rm lin}=\int\limits_{}^{}d\tau\textsf{L}^g_{\tau},
$$
$$
  \textsf{L}^g_{\tau}=\dfrac{v^2_{(a)(b)}
 -\,e^{-4D}\!R^{(3)}}{6}
 = \sum\limits_{\textbf{k}^2\not =0}\dfrac{
 v^g_{\textbf{k}}v^g_{-\textbf{k}}-\,e^{-4D} \textbf{k}^2
 \overline{g}_{\textbf{k}}\overline{g}_{-\textbf{k}}}{2}=
$$
\bea\label{2-3ngw}
 =\left[\sum\limits_{\textbf{k}^2\not =0}
 p_{-\textbf{k}}^gv^g_{\textbf{k}}\right]-\textsf{H}^g_{\tau},
\eea
where $v^g_{\textbf{k}}=\partial_\tau \overline{g}_{\textbf{k}}$
is the derivative
with respect to the luminosity time interval 
 and
 \bea\label{2-10gh}
 \textsf{H}^g_{\tau}=\sum\limits_{\textbf{k}^2\not =0}
 \dfrac{p_{\textbf{k}}^gp_{-\textbf{k}}^g+ e^{-4\langle D\rangle}
 \textbf{k}^2 \overline{g}_{\textbf{k}}\overline{g}_{-\textbf{k}}}{2}
 \eea
is the corresponding Hamiltonian.

Thus, in the mean field approximation~(\ref{mfa-11}) the diffeo-invariant sector
of the strong gravitational plane waves coincides with a bilinear theory given
by Eqs.~(\ref{2-3ngw})--(\ref{2-10gh}). In
this approximation our model is reduced to a rather simple theory which is
bilinear with respect to the single-component graviton field
as discussed also in Ref.~\cite{grg12-7}.
Note that we consider here the tangential space, and the chosen variables allow us
to obtain the simple solutions. The main  postulated condition here was
the requirement of the diffeo-invariance of the graviton equation of motion.
While in the standard GR the symmetry properties are required only for the interval,
we impose  the symmetry with respect to diffeomorphisms also on the Maurer -- Cartan forms.

\section{Comparison with metric gravitons}
\makeatletter
\renewcommand{\@evenhead}{\raisebox{0pt}[\headheight][0pt]
{\vbox{\hbox to\textwidth{\hfil \strut \thechapter . Quantization of gravitons in terms of Cartan forms
\quad\rm\thepage}\hrule}}}
\renewcommand{\@oddhead}{\raisebox{0pt}[\headheight][0pt]
{\vbox{\hbox to\textwidth{\strut \thesection .\quad
Comparison with metric gravitons
\hfil \rm\thepage}\hrule}}}
\makeatother

It is instructive to compare the properties  of the affine and metric
gravitons, which was done first in Ref.~\cite{Arbuzov:2010fz-7}).

The action of metric gravitons in the accepted GR~\cite{Babak:1999dc-7}
coincides with the affine one (\ref{2-3ngw}) in the lowest order of the
decomposition over  ${\bf k}^2/M^2_{\rm Pl}$
 \bea
 \label{2-3nh}
 W^{\rm GR}_{\rm non-lin}&=&{W}^g_{\rm lin}+{W}_{\rm non-lin},
 \eea
if we keep
only diagonal graviton components. It is well-known \cite{Faddeev:1973zb-7}
that the accepted action (\ref{2-3nh})  is
highly nonlinear even in the approximation~(\ref{mfa-11}).

In the approximation  (\ref{mfa-11}),
we keep
only the dynamical part $\omega^R_{(cb)}$ (which enters into the action~(\ref{2-3ngw}))
and the present day value of the cosmological scale factor $a=e^{-\langle D\rangle}=1$.
Let us compare the affine gravitons (\ref{P-1})
 with the commonly accepted  metric gravitons, given by the decomposition
\cite{Babak:1999dc-7}
 \bea
 \label{fs-1s-7}
 \widetilde{ds}_h^2&=&(d\eta)^2  - dx^idx^j
\left(\delta_{ij}+2h^{TT}_{ij}+\cdots \right).
 \eea
In the accepted case, the graviton moves in the direction of vector
$\textbf{k}$, its wave amplitude $\cos (\omega_{\textbf{k}}x_{(k)})$ depends
on the scalar product
$$x_{(k)}=\frac{(\textbf{k}\cdot \textbf{x})}{\omega_{\textbf{k}}}.$$

The graviton changes the squared test
particle velocity
$$\left(\dfrac{ds}{d\eta}\right)^2\sim \dfrac{dx^idx^j}{d\eta\,d\eta}\varepsilon^\alpha_{ij}$$
in the plane, orthogonal to the direction of
motion. Here $\varepsilon^\alpha_{ij}$ is the traceless transverse tensor:
$$\varepsilon^\alpha_{ii}=0, ~~~~~~~~~~~~k_i\varepsilon^\alpha_{ij}=0.$$
All these effects are produced by the first order of series (\ref{fs-1s-7})
 \bea
 \label{pl-1s}
dl_{h}^2=2dx^idx^j h^{TT}_{ij}(t,\textbf{x})=
\eea
$$=dx^idx^j\varepsilon^\alpha_{ij}
 \sqrt{6}\cos\left(\omega_{\textbf{k}}x_{(k)}\right)({H_0}/{\omega_{\textbf{k}}})
 {\Omega_{\textbf{k}\rm h}^{1/2}}
 +O(h^2),
$$
where  $H_0$ is the Hubble parameter,
$$\Omega_{\textbf{k}\rm h}=\frac{\omega_{\textbf{k}}N_{\textbf{k}\rm h}}{ V_0\rho_{\rm cr}}$$
is the energy
density of the gravitons in units of the cosmological critical energy density.
One observes that in the accepted perturbation theory the  contribution of a
single gravitational wave to the geometrical intervals, Eq.~(\ref{fs-1s-7}), is
suppressed by the factor $H_0/\omega_{\textbf{k}}$.

In our version the linear term of the spacial part of Eq.~(\ref{P-1})
takes
the form
 \bea \label{2-1gg1} \nonumber
 dl_{g}^{2} = 2dX_{(b)}X_{(c)} \omega^R_{(c)(b)}
 = dX_{(b)}X_{(c)}
 \varepsilon^\alpha_{(c)(b)}
 \sqrt{6}\cos\{\omega_{\textbf{k}}X_{(k)}\}{H_0} {\Omega_{\textbf{k}\rm h}^{1/2}}.
 \eea
Evidently, two models (the GR and the CGR) differ by
an additional factor which can be
deduced from the ratio
 \be\label{ratio}
 \bigg|\frac{dl_h^2 }{dl_g^2}\bigg| =
 \bigg|\frac{ dx^idx^j\left(h^{TT}_{ij}\right)}
 {(dX_{(b)}X_{(c)}\omega^R_{(c)(b)})}\bigg|
 \simeq \frac{1}{r_{\!\!\bot}\,\omega_{\textbf{k}}}
\sim \frac{\lambda_g}{r_{\!\!\bot}}.
 \ee
Here
$$r_{\!\!\bot}\,=\sqrt{|\vec X_{\!\!\bot}\,|^2}$$
is the coordinate
distance between two test particles  in the plane perpendicular to the wave
motion direction and $\lambda_g$ is the graviton wave length. Therefore, in the CGR
there is the effect of the  \textit{expansion} of the plane perpendicular to the affine wave
motion direction.

As a result, in the CGR the total velocity of a test
classical particle in the central gravitational field of a mass $M$ and of a strong
gravitational wave is the sum of three velocities
at the cosmic evolution $a\neq 1$. The first term is the
standard Newtonian (N) velocity, the second is the velocity of the {\em graviton
expansion} (g) in the field of a gravitational wave, and the third one is the
velocity of the Hubble  evolution (H):
 \be \label{sum-1}
|\vec v|^2=\big|\dfrac{{dl}_g}{d\eta}\big|^2=
\ee
$$
\left[\underbrace{{\vec n}_{\rm N}\sqrt{\dfrac{r_g}{2R_{\bot}}}}_{\rm
Newtonian~ velocity} + \underbrace{{\vec n}_{\rm g}
\sqrt{R_{\bot}H_0}\sqrt{\Omega_g}}_{\rm graviton~ expansion} +
\underbrace{{\vec n}_{\rm H}\gamma H_0R_{\bot} }_{\rm Hubble~ evolution}
\right]^2 .
$$
Here,
$$R_{\bot}=r_{\bot}a(\eta)$$
is the Friedman distance from the central
mass, $H_0$ is the Hubble parameter,
$$r_g(R_{\bot})=\frac{M}{M_{\rm Pl}^2}$$
is a constant
gravitational radius,
and
 \be \label{vectors1} \left\{
 \begin{array} {l}
 {\vec n}_{\rm N}=(0,-1,0),\\
 {\vec n}_{\rm g}=(+1/\sqrt{2},-1/\sqrt{2},0),\\
 {\vec n}_{\rm H}=(1,0,0)\\
 \end{array}\right.
 \ee
are the unit velocity vectors.
Their scalar products are
$$({\vec n}_{\rm N}\cdot {\vec n}_{\rm g})\not=0,~~~ ({\vec n}_{\rm N}\cdot {\vec n}_{\rm H})=0,~~~
({\vec n}_{\rm N}\cdot {\vec n}_{\rm g})\not=0,~~~ ({\vec n}_{\rm N}\cdot {\vec n}_{\rm H})=0.$$
The graviton energy density $\Omega_g$ is given in units
of the cosmological critical energy density  $\rho_{\rm cr.}$

The last two terms provide possible sources of a modified Newtonian dynamics.
One observes that the interference of the Newtonian and the graviton-induced
velocities in (\ref{sum-1})
$$v_{\rm n-g~interf}\simeq \sqrt[4]{\Omega_g r_g H_0}$$
does not depend on the radius $R_{\bot}$. This term can play a role of ``dark'' matter. In this case it is
enough $\Omega_g\simeq 0.1,$ in order to get the velocity of the Sun in our Galaxy $v_{\rm n-g ~interf}\simeq 200$ km/sec.

The third term could imitate the Dark Matter effect in  COMA-type clusters
with $|R|\sim 10^{25}$cm, in accordance with the validity limit of the
Newtonian dynamics,
$$\frac{r_g}{R_{\rm limit}}<2(R_{\rm limit}H_0)^2,$$
discussed
in~\cite{Einstein:1945id-7,Gusev:2003sv-7}. The factor $\gamma=\sqrt{2}$ is defined by the
cosmological density~\cite{z-06-7}.

Thus, in our model strong gravitational waves  possess peculiar properties
which can be tested by observations and experiments.

\section{Vacuum creation of affine gravitons}
\makeatletter
\renewcommand{\@evenhead}{\raisebox{0pt}[\headheight][0pt]
{\vbox{\hbox to\textwidth{\hfil \strut \thechapter . Quantization of gravitons in terms of Cartan forms
\quad\rm\thepage}\hrule}}}
\renewcommand{\@oddhead}{\raisebox{0pt}[\headheight][0pt]
{\vbox{\hbox to\textwidth{\strut \thesection .\quad
Vacuum creation of affine gravitons
\hfil \rm\thepage}\hrule}}}
\makeatother

Here we are going to study the effect of intensive creation of affine gravitons.
We will briefly recapitulate the derivation given in Ref.~\cite{Arbuzov:2010fz-7} and further,
using the initial data of the hierarchy of cosmological scales
obtained in Sect. 6.3, 
estimate the number of created
particles.

The approximation defined by Eqs.~(\ref{2-3ngw})--(\ref{2-10gh})
can be rewritten by means of the conformal variables and coordinates,
where the action
\bea\label{W-1c}
 {W}^g_{\rm lin}=\int\limits^{\eta_0}_{\eta_I} d\eta \left[-V_0{(\partial_\eta
 \langle D\rangle)^2} e^{-2\langle D\rangle}+\textsf{L}^g_{\rm \eta}\right]
 \eea
is given in the interval $\eta_I \leq \eta \leq \eta_0$ and spatial
volume $V_0$. Here the Lagrangian and Hamiltonian
 \bea \label{2-3ng}
 \textsf{L}^g_{\rm \eta}&=& \sum\limits_{\textbf{k}^2\not =0}\,e^{-2 \langle
 D\rangle}\dfrac{ v^g_{\textbf{k}}v^g_{-\textbf{k}}- \textbf{k}^2
 \overline{g}_{\textbf{k}}\overline{g}_{-\textbf{k}}}{2}
 =\left[\sum\limits_{\textbf{k}^2\not =0}
 p_{-\textbf{k}}^gv^g_{\textbf{k}}\right]- \textsf{H}^g_{\rm \eta},
 \\ \label{2-3ngh1}
 \textsf{H}^g_{\rm \eta}&=&\sum\limits_{\textbf{k}^2\not =0}\,
 \dfrac{ e^{2 \langle D\rangle}p^g_{\textbf{k}}p^g_{-\textbf{k}}
 + e^{-2 \langle D\rangle}{\omega^2_{0\textbf{k}}}
 \overline{g}_{\textbf{k}}\overline{g}_{-\textbf{k}}}{2}
 \eea
are defined in terms of the variables $\overline{g}_{\textbf{k}}$,
their momenta, and one-particle conformal energy
 \be \label{hh-1}
 p_{\textbf{k}}^g=e^{-2\langle D\rangle}v^g_{\textbf{k}}
 = e^{-2 \langle D\rangle}\partial_\eta
 \overline{g}_{\textbf{k}},\quad
 {\omega}^g_{0{\bf k}}=\sqrt{{\bf k}^2},
 \ee
 respectively. The transformation (squeezing)
 \be \label{2-10gh1}
 p_{\textbf{k}}^g=\widetilde{p}_{\textbf{k}}^g e^{-\langle D\rangle}
 [\omega^g_{0{\bf k}}]^{-1/2}, \qquad \overline{g}_{\textbf{k}}
 = \widetilde{g}_{\textbf{k}} e^{\langle D\rangle}[\omega^g_{0{\bf k}}]^{1/2}
 \ee
leads to the canonical form
 \bea \label{2-newh-7}
 \textsf{H}^g_{\rm \eta}&=&\sum\limits_{\textbf{k}^2\not =0}
 \omega^g_{0{\bf k}}\dfrac{\widetilde{p}^g_{\textbf{k}}\widetilde{p}^g_{-\textbf{k}}
+
 {\widetilde{g}}_{\textbf{k}}{\widetilde{g}}_{-\textbf{k}}}{2}=
 \sum\limits_{\textbf{k}}^{}\underline{{\cal H}}_{\textbf{k}}^g,
 \\ \label{3-8ac-7}
 \underline{{\cal H}}_{\textbf{k}}^g&=&
 \frac{\omega^g_{0\textbf{k}}}{2}
 [\widetilde{g}_{\textbf{k}}^+\widetilde{g}_{-\textbf{k}}^-+
 \!\widetilde{g}_{\textbf{k}}^-\widetilde{g}_{-\textbf{k}}^+]\;,
 \eea
where
 \be \label{hv-1}
 \widetilde{g}^{\pm}_{\textbf{k}}=\left[\widetilde{g}_{\textbf{k}}
 \mp i\widetilde{p}_{\textbf{k}} \right]/\sqrt{2}
 \ee
are the conformal-invariant
classical variables in the holomorphic representation \cite{z-06-7}.

In virtue of Eqs.~(\ref{hh-1})--(\ref{hv-1}), the action (\ref{W-1c}) takes
the form
 \bea\label{W-3c}
 {W}^g_{\rm lin}
 &=&\int\limits^{\eta_0}_{\eta_I} d\eta \left[ -{V_0(\partial_\eta \langle
 D\rangle)^2} e^{-2\langle D\rangle}- \textsf{H}^g_{\rm
 \eta}\right]\\\nonumber
 &+&\int\limits^{\eta_0}_{\eta_I}d\eta
 \sum\limits_{\textbf{k}^2\not =0} \widetilde{p}_{-\textbf{k}}
 \left[\partial_\eta \widetilde{g}_{\textbf{k}}+\partial_\eta\langle D\rangle
 \widetilde{g}_{\textbf{k}} \right].
 \eea
The evolution equations for this action are
 \be\label{cc-5-7-7}
 \partial_\eta \widetilde{g}^{\pm}_{\textbf{k}}=\pm i \omega^{g}_{0\textbf{k}}
 \widetilde{g}^{\pm}_{\textbf{k}}+H_\eta\,\widetilde{g}^{\mp}_{\textbf{k}},
 \ee
where
 $H_\eta=\partial_\eta(\ln a)=-\partial_\eta\langle D\rangle$
is the conformal Hubble parameter (in our model $H_\eta=H_0/a^2$).

It is generally accepted to solve these equations by means of the Bogoliubov
transformations
 \bea\label{cc-6-7}
 \widetilde{g}^{+}_{\textbf{k}}&=&\alpha_{\textbf{k}} b_{\textbf{k}}^++
 \beta^*_{\textbf{-k}} b_{\textbf{-k}}^-,
 \\\label{cc-6c-7}
 \widetilde{g}^{-}_{\textbf{k}}&=&\alpha^*_{\textbf{k}} b_{\textbf{k}}^-+
 \beta_{\textbf{-k}} b_{\textbf{-k}}^+,
 \\\label{cc-6cc-7}
 \alpha_{\textbf{k}}&=&\cosh r^g_{\textbf{k}}e^{i\theta^g_{\textbf{k}}},~~~~~
 \beta^*_{\textbf{k}}=\sinh r^g_{\textbf{k}}e^{i\theta^g_{\textbf{k}}},
 \eea
where $r^g_{\textbf{k}}$ and $\theta^g_{\textbf{k}}$ are the squeezing
parameter and the rotation one, respectively 
These transformations preserve the Heisenberg algebra
$O(2|1)$ \cite{jor-63-7} and diagonalize Eqs.~(\ref{cc-5-7-7}) in the form of:
 \bea\label{cc-7-7}
 \partial_{\eta}b^{\pm}_{\textbf{k}}=
 \pm i \overline{\omega}^g_{B\textbf{k}} b^{\pm}_{\textbf{k}},
 \eea
if the parameters of squeezing $r^g_{\textbf{k}}$ and rotation
$\theta^g_{\textbf{k}}$ satisfy the following equations~\cite{z-06-7}:
 \bea \label{cc-8-7}
  \partial_\eta{r^g}_{\textbf{k}}&=&H_\eta\cos 2\theta^g_{\textbf{k}},
 \\ \label{cc-9-7}
 {\omega}^g_{0\textbf{k}}-\partial_\eta{\theta^g}_{\textbf{k}}
 &=& H_\eta\coth 2r^g_{\textbf{k}}\sin 2\theta^g_{\textbf{k}},\\
 \label{cc-9b-7}
 {\omega}^g_{B\textbf{k}}&=&\frac{{\omega}^g_{0\textbf{k}}-
 \partial_\eta{\theta^g}_{\textbf{k}}}{\coth 2r^g_{\textbf{k}}}.
 \eea
A general solution of the classical equations can be written with the aid of a
complete set of the initial data $ b^{\pm}_{0\textbf{k}}$:
 \bea \label{cc-10g-7}
 b^{\pm}_{\textbf{k}}(\eta)=\exp\left\{\pm i \int\limits^{\eta}_{\eta_0}
 d\overline{\eta}{\,\omega}^g_{B\textbf{k}}(\overline{\eta})\right\}
 b^{\pm}_{0\textbf{k}}.
 \eea

On the other hand, quantities $ b^{+}_{0\textbf{k}}( b^{-}_{0\textbf{k}})$ can
be considered as the creation (annihilation) operators, which satisfy the
commutation relations:
 \be
 [b^{-}_{0\textbf{k}},b^{+}_{0\textbf{k}'}]=\delta_{\textbf{k},\textbf{-k}'},\quad
 [b^{-}_{0\textbf{k}},b_{0\textbf{k}'}^{-}]=0,\quad
 [b^{+}_{0\textbf{k}},b^{+}_{0\textbf{k}'}]=0, \label{bcom-7}
 \ee
if one
introduces the   vacuum state as $b^{-}_{0\textbf{k}}|0\rangle=0$. Indeed,
relations (\ref{bcom-7}) are the results of: i) the classical Poisson
bracket $\{P_{\widetilde{F}},\widetilde{F}\}=1$
which transforms into
 \be
 [\widetilde{g}^{-}_{\textbf{k}},\widetilde{g}^{+}_{-\textbf{k}}]
 =\delta_{\textbf{k},\textbf{k}'};
 \label{pcom-7}
 \ee
ii) the solution (\ref{cc-10g-7}) for the initial data;
iii) the Bogoliubov transformations (\ref{cc-6-7}), (\ref{cc-6c-7}).

With the aid of Eqs.~(\ref{cc-6-7})--(\ref{cc-6cc-7}) and
(\ref{cc-10g-7})--(\ref{pcom-7}) we are able to calculate the  vacuum expectation
value of the total energy (\ref{2-newh-7}), (\ref{3-8ac-7})
 \be
 \label{H-2ad-7}
 \langle 0|\textsf{H}_\eta^g(a)|0\rangle=
\sum\limits_{{\bf k}}^{}{\omega}^g_{0\textbf{k}}|\beta_{\textbf{k}}|^2=
 \sum\limits_{{\bf k}}^{}
 {\omega}^g_{0\textbf{k}}
 \frac{\cosh\{2r_{\bf k}^g(a)\}-1}{2}.
 \ee

The numerical analysis \cite{Arbuzov:2010fz-7}
 of Eqs.~(\ref{cc-8-7})--(\ref{cc-9-7}) for unknown variables $(r^g_{\textbf{k}},
\theta^g_{\textbf{k}})$
 with the zero boundary conditions at $a=a_{I}$ (at the beginning of creation)
 \bea \label{3-9bb0h-7}
  {r^g}_{\textbf{k}}(a_I)=0,
\qquad \theta^g_{\textbf{k}}(a_I)=0
 \eea
  enables us to suggest an
approximate analytical solution for the evolution equations.

Our approximation consists in the following. It arises, if instead of
$r_{\textbf{k}}$ one substitutes an approximate value $r_{\rm apr}$ in the
vicinity of  the soft mode of the  Bogoliubov energy (\ref{cc-9b-7})
$\omega_{0\rm appr}=\partial_\eta{\theta^g}_{\rm appr}$,
 \bea \label{c-4cn1-7}
 r_{\rm appr}&=&\frac{1}{2}
 \int\limits_{X_I=2\theta^g_{\rm appr}(a_I)}^{X=2\theta^g_{\rm appr}(a)}
 \frac{d\overline{X}}{\overline{X}}
 {\cosh \overline{X}}\simeq 2\langle D\rangle_I,
 \\ \label{c-4cn12-7}
 X(a)&=&2\theta^g_{\rm appr}(a)=
 2\int\limits_{\eta(a_I)}^{\eta(a)}d\eta \omega_{0\bf k}.
 \eea
 This  soft
mode  provides a  transition \cite{Arbuzov:2010fz-7}  at the point $a^2_{\rm relax}\simeq
2a^2_{\rm Pl}$ from the unstable state of the particle creation to the stable
state with almost a constant occupation number  during the relaxation time
 \be\label{relax-1-7}
 \eta_{\rm relax}\simeq \frac{2e^{-2\langle  D\rangle_I}}{2H_0}\equiv \frac{2a_I^2}{2H_0}.
 \ee
 At the point of the relaxation,
the determinant of Eqs.~(\ref{cc-5-7-7}) changes its sign and becomes
positive~\cite{Andreev:1996qs-7}. Finally, we obtain
 \bea
 \label{3-8v-7}
  \langle 0|{\cal H}_{\textbf{k}}^g|0\rangle\big|_{(a>a_{\rm relax})} \!=\!
\omega^g_{0\textbf{k}}\frac{\cosh[2r_{\textbf{k}}^g]-1}{2}
  \approx \frac{\omega^g_{0\textbf{k}}}{4a^4_I}.
 \eea
 We have verified that the deviation of the results obtained with the aid of
 this formula  from  the numerical
solutions of  Eqs.~(\ref{cc-8-7})--(\ref{cc-9-7}) (see Ref.~\cite{Arbuzov:2010fz-7}) does
not exceed 7\%.

In virtue of this result, we obtain the total energy
 \bea\label{H-2a-7} \!\!
 \langle 0|\textsf{H}_\eta^g|0\rangle\big|_{(a>a_{\rm relax})}\!\! \approx
 \frac{1}{2a_I^4}\sum\limits_{{\bf k}}^{}\frac{{\omega}^g_{0\textbf{k}}}{2}
 \equiv \frac{\textsf{H}^g_{\eta\; {\rm Cas}}({a})}{2a_I^4},
 \eea
 where $\textsf{H}^g_{\eta\; {\rm Cas}}({a})$
 is the  Casimir vacuum energy. 

Thus, the total energy of the created gravitons is
 \bea\label{H-2ab-7}
 \langle 0|\textsf{H}_\eta^g|0\rangle \simeq
 \frac{{\widetilde{\gamma}}H_0}{4a^2a^4_I}.
 \eea
 It appeared that the dilaton initial data $a_I=e^{-\langle D\rangle_I}$
and $H_0$ determine both the total energy  (\ref{H-2a-7}) of the created gravitons
and their occupation number $N_g$ at the relaxation time (\ref{relax-1-7}):
\bea\label{H-2abr-7}
 N_g(a_{\rm relax})\simeq \frac{\langle
0|\textsf{H}_\eta^g|0\rangle}{\langle\omega^g_k\rangle} \simeq
 \frac{\widetilde{\gamma}^{(g)}}{16 a^6_I}\simeq 10^{87},
 \eea
 where we divided the total energy by the mean one-particle energy
$$\langle\omega^g_k\rangle \approx \langle\omega^{(2)}\rangle (a_I)$$
defined in Eq.~(\ref{ce-1}).
For numerical estimations we use $\widetilde{\gamma}^{(g)}\approx 0.03$.
The number of the primordial gravitons is compatible with the number of
the CMB photons as it was predicted in Ref.~\cite{Babak:1999dc-7}.

The main result of this Section consists in the evaluation  of the primordial
graviton number~(\ref{H-2abr-7}). We suppose that the Casimir energy is defined
by the total ground state energy of created excitations, see Eq.~(\ref{H-2a-7}).

\newpage
\section{Summary}
\makeatletter
\renewcommand{\@evenhead}{\raisebox{0pt}[\headheight][0pt]
{\vbox{\hbox to\textwidth{\hfil \strut \thechapter . Quantization of gravitons in terms of Cartan forms
\quad\rm\thepage}\hrule}}}
\renewcommand{\@oddhead}{\raisebox{0pt}[\headheight][0pt]
{\vbox{\hbox to\textwidth{\strut \thesection .\quad
Summary and literature
\hfil \rm\thepage}\hrule}}}
\makeatother

We developed a Hamiltonian approach to the gravitational model, formulated
as the nonlinear realization of joint affine and conformal symmetries.
With the aid of the
Dirac -- ADM foliation, the conformal and affine symmetries provide
a natural separation of the dilaton
and gravitational dynamics in terms of the  Maurer -- Cartan forms.
As a result, the exact solution of the energy constraint
yields the diffeo-invariant evolution operator in the field space.

In the CGR, the conformal symmetry breaking happens due to the
Casimir vacuum energy.
This energy is
obtained as a result of the quantization scheme of the Hamiltonian dynamics.
The diffeo-invariant dynamics in terms of the Maurer -- Cartan
forms with application of the affine symmetry condition leads to the reduction
of the  graviton representation to the  one-component field.
The affine graviton strong wave yields
the effect of expansion (or contraction) in the hypersurface perpendicular to the
direction of the wave propagation.
We demonstrated that the Planck least action postulate applied to the
Universe limited by its horizon provides the value of the cosmological
scale factor at the Planck epoch. A hierarchy of cosmological energy
scales for the states with different conformal weights is found.
The intensive creation of primordial gravitons and
Higgs bosons is described assuming that the Casimir vacuum energy is the source
of this process.
We have calculated the total energy of the created particles, Eq.~(\ref{H-2a-7}), and
their occupation numbers, Eq.~(\ref{H-2abr-7}). 



\chapter{Mathematical principles of description of the Universe}
\renewcommand{\theequation}{8.\arabic{equation}}
\setcounter{equation}{0}
\section{The classical theory of gravitation
\label{sect_matter}}
\renewcommand{\theequation}{8.\arabic{equation}}
\setcounter{equation}{0}
\makeatletter
\renewcommand{\@evenhead}{\raisebox{0pt}[\headheight][0pt]
{\vbox{\hbox to\textwidth{\hfil \strut \thechapter . Mathematical principles of description of the Universe
\quad\rm\thepage}\hrule}}}
\renewcommand{\@oddhead}{\raisebox{0pt}[\headheight][0pt]
{\vbox{\hbox to\textwidth{\strut \thesection .\quad
The classical theory of gravitation
\hfil \rm\thepage}\hrule}}}
\makeatother

The classical theory of gravitation, presented in our monograph is
based on the following three principles:

\begin{enumerate}
\item The joint nonlinear realization of \textit{affine and conformal symmetry groups} via Cartan's forms,
described in Chapter 4.

\item 3+1 foliation of a pseudo--Riemannian space with \textit{kinemetric} subgroup of the group of
general coordinate transformations, described in Chapter 5.

\item \textit{The reduction of the phase space} by solving of all constraints.

\end{enumerate}

The solving of all constraints, including the Hamiltonian one, which were presented in Chapter 5, reveals
a diffeoinvariant physical content of the considered conformal and affine theory of gravitation.
The diffeoinvariant content of the conformal theory of gravity includes in itself:
\begin{itemize}
\item
dynamics at the surface of all the constraints, which is described by the action (\ref{1-3dgc1})
\bea\label{1-3dgc1-7}
  {W}_{C=0}=
\eea
$$
\!=\!\int\limits_{ }^{ }\!\! d^3x \biggl[\int\left(
 p_{(a)(b)}\omega_{(a)(b)}^R(d)
 +p_Q\,d{{{Q}}}
 +p_{A{(b)}}dA_{(b)}\right)\biggr]\!-\!\int P_{\langle D\rangle} d{\langle D\rangle};
$$
\item square of the geometric interval (\ref{ds-12}) as a sum of squares of the components of the Fock's frame
in terms of the observed values
\bea\label{ds-12-7}
 \widetilde{ds}^2\!=\!
 e^{-4D}\!\frac{\langle\sqrt{\widetilde{{\cal H}}}\rangle^2}
 {\widetilde{\cal H}}\!d\tau^2\!-\!
 \left( dX_{(b)}-X_{(c)}\,\omega^R_{(c)(b)}(d)
 \!-\!
 {\cal N}_{(b)} d\tau\!\right)^2;
\eea

\item geometrodynamics (such as Hubble's law)
 \bea\label{Dirac-c2-7}
 \tau &=&\int\limits_{\langle D \rangle_I}^{\langle D \rangle_0}
 \frac{d\langle D \rangle}{\left\langle \sqrt{\widetilde{{\cal H}}}\right\rangle},
 \eea
as cosmological relationship between geometry and dynamics as a function of the geometric
luminosity interval from zero harmonic of the dilaton.
\end{itemize}

The action (\ref{1-3dgc1-7}) contains an operator of evolution of the Universe
\bea \label{5-C-7}P_{\langle D\rangle}
&=&\pm {\bf E}_{\rm U},\\ \label{6-7}
{\bf E}_{\rm U}&=&2\int d^3x \sqrt{\widetilde{{\cal H}}},
\eea
determined from the exact solution of the constraint (\ref{5-C})
\be \label{5-C-7c}P^2_{\langle D\rangle}-{\bf E}^2_{\rm U}=0.
\ee
The role of the evolution parameter in the field space of events
performs a value $\langle D\rangle$ called in observational
cosmology as luminosity (or brightness), and $P_{\langle D\rangle}$ is its canonical momentum.
The value of the generator of evolution of the Universe (\ref{6-7}) on the equations of motion   is
\bea \label{7-7}
\frac{\delta{W}_{C=0}}{\delta F}=0,~~~~~~~\frac{\delta{W}_{C=0}}{\delta P_F}=0,
\eea
where $F$ are field variables, we call the \textit{energy of the Universe}
in the field space of events by analogy with the \textit{energy of a particle} in Minkowskian space in the Special Relativity.

\section{Foundations of quantum theory\\ of gravity
\label{sect_matter-8}}
\renewcommand{\theequation}{8.\arabic{equation}}
\makeatletter
\renewcommand{\@evenhead}{\raisebox{0pt}[\headheight][0pt]
{\vbox{\hbox to\textwidth{\hfil \strut \thechapter . Mathematical principles of description of the Universe
\quad\rm\thepage}\hrule}}}
\renewcommand{\@oddhead}{\raisebox{0pt}[\headheight][0pt]
{\vbox{\hbox to\textwidth{\strut \thesection .\quad
Foundations of quantum theory of gravity
\hfil \rm\thepage}\hrule}}}
\makeatother

\subsection{The irreducible unitary representation\\  of the group $A(4)\otimes C$}

The theory of gravity was presented above as a nonlinear realization
of finite-dimensional affine and conformal groups of \textit{symmetry},
that close the group of general coordinate transformations.
Therefore, as mentioned above, there is a unique opportunity to build further
classification of experimental and observational data, using the unitary irreducible
representations of these groups, without resorting to the classical \textit{laws of dynamics}
as initial statements of the physical theory, or concluding classical \textit{laws of dynamics}
from the first principles of symmetry.

In the quantum theory of the Universe at the level of operator quantization
in the field space of events $[\langle D\rangle|F]$
Hamiltonian constraint equation (\ref{5-C-7})
becomes the equation of Wheeler -- De Witt type (\ref{q11})
\be \label{5-C-7cc}\left[\hat P^2_{\langle D\rangle}-{\bf E}^2_{\rm U}\right]
{\hat \Psi}_{\langle D\rangle_I, \langle D\rangle_0}=0,
\ee
corresponding to the dimension of the kinemetric subgroup of invariance of the Hamiltonian formulation.
In the quantum theory, the canonical variables $\hat P_{\langle D\rangle}, {\langle D\rangle}$
become operators with a commutation relation
 $$[\hat P_{\langle D\rangle}, {\langle D\rangle}]=\imath.$$
The general solution of this Wheeler -- De Witt equation in the approximation of the empty Universe
with Casimir vacuum energy obtained in  Section 6.6 by the Bogoliubov transformation.

By analogy with the unitary irreducible representation of the Poincar\'e group  (see Chapter 2 (\ref{q1-13}))
in quantum field theory we get a general operator solution of the Wheeler -- De Witt equation (\ref{5-C-7cc})
for the Universe as a sum of two
Т-ordered with respect to parameter $\langle D\rangle$ exponents:
\bea \label{09-07}
{\hat \Psi}_{\langle D\rangle_I, \langle D\rangle_0}=
{\hat A}_{\langle D\rangle_I}^+
 {\hat{\mathbb{U}}}^{0}_{I}\frac{1}{{\sqrt{2{\bf E}_{0\rm U}}}} +
{\hat A}_{\langle D\rangle_I}^-
 \frac{1}{{\sqrt{2{\bf E}_{0\rm U}}}}{\hat{\mathbb{U}}}^{I\dagger}_{0},
\eea
describing the creation of the Universe at the time of $\langle D\rangle_I$,
its evolution from $\langle D\rangle_I$ till the moment $\langle D\rangle_0$
and a state at the modern epoch $\langle D\rangle_0$.
The two terms correspond to the positive and negative energy,
where
${\hat A}_{\langle D\rangle_I}^+$
can be interpreted as an operator of creation of the Universe at the moment
$\langle D\rangle_I$
from a state of vacuum,
and ${\hat A}_{\langle D\rangle_I}^-$ is an operator of annihilation of the Universe, correspondingly,
with a commutation relation
$$[{\hat A}_{\langle D\rangle_I}^-,{\hat A}_{\langle D\rangle_I}^+]=1:$$
\bea \label{08-02}
 {\hat{\mathbb{U}}}^{0}_{I}= T_{\langle D\rangle}\exp \left\{-\imath
 \int\limits_{\langle D\rangle_I}^{\langle D\rangle_0}
 d\langle D\rangle {\bf E}_{\rm U} \right\};
 ~~~~~~~~~~~\hat{\mathbb{U}}\cdot \hat{\mathbb{U}}^{\dagger}=\hat I
\eea
is an operator of evolution in the space of events, or space of measurements
$[\langle D\rangle|F]$,
relatively to the evolution parameter
$[\langle D\rangle]$.
A vacuum state
$B^-\big|_{\langle D\rangle_I}0 \rangle=0$
is set by actions of the Bogoliubov operators $B^{\pm}$, which diagonalize the evolution equations,
as was shown above in Section 6.6.
Negative energy is removed by the second quantization of the Universe and all the fields.

Thus, the reduction of the extended phase space to the subspace of physical variables gives the corresponding
reduced action (\ref{1-3dgc1-7}),
which \textit{is rejected} in the Standard Hamiltonian formulation of the General Relativity \cite{ADM-8} as trivial.
This action here is at the \textit{forefront}, as the basic element of constructing of a quantum operator of
creation and evolution of the Universe in the field space of events, by analogy with dynamic formulation
of the Special Theory of Relativity.

Using a direct correspondence of Wheeler -- De Witt between a particle in the Special Relativity and the
 Universe in the General Relativity
(see Table at the end of Section 6.6),
and the definition of irreducible unitary representations of the Poincar\'e group in the space of events
$[P_{(\alpha)}|X_{(\alpha)}]$:
\bea\label{q1-13-02}
\!\!\!\! &&\Psi[P_{(\alpha)}|X_{(\alpha)}]=\\ \nonumber
&&\!\!\!\!=
 \frac{1}{\sqrt{2\, |P_{(0)}|}}
 \left[a^+\Psi_{P_{(0)+}}\theta(X_{(0)}\!-\!{X_{I(0)}})\!+\!
 a^-\Psi_{P_{(0)-}}^*\theta({X_{I(0)}}\!-\!X_{(0)})\!\right],
 \eea
we can interpret the functional
${\hat \Psi}_{\langle D\rangle_I, \langle D\rangle_0}$ (\ref{09-07})
as an unitary representation of the group
 $A(4)\otimes C$ in the field space of events $[\langle D\rangle|F]$.
In the quantum geometrodynamics of the Universe for the relativistic theory of gravitation, we shall not
forget also the geometric interval (\ref{ds-12-7}) and the relation  (\ref{Dirac-c2-7})
between the geometric interval and the dynamic parameter of evolution.
This relation is the Hubble law in the exact theory, which includes quantum effects, such as the arrow of time,
appearing in the quantum description of the Universe as a consequence of the postulate of
the existence of the vacuum.

The unitary of representation of the (\ref{09-07})
$$\hat{\mathbb{U}}\cdot \hat{\mathbb{U}}^{\dagger}=\hat I$$
follows from the assumption of a positive definite metric in the Hilbert space of states.
In the future, we show that the theory used to describe the matter
after the constraints have been solved, indeed, contains
only self-conjugated fields with a positive probability, for which the energy of the Universe
(\ref{6-7}) is positive and has not an imaginary part
\bea \label{08-04}
{\bf E}_{\rm U}=2\int d^3x \sqrt{\widetilde{{\cal H}}}\geq 0;~~~~Im \sqrt{{\bf H}}=0.
\eea
For the construction of the irreducible representations (\ref{09-07})
we introduce a complete set of orthogonal states
\bea \label{08-05}
\langle Q|Q'\rangle=\delta_{Q,Q'},~~~~\sum_Q|Q\rangle \langle Q|=\hat I.
\eea
Here $\hat I$ is a unit operator, and $Q$ are quantum numbers, which characterize
this representation of orthonormal states of the Universe,
arisen out of the vacuum as the state with the lowest energy by an action of the operator of creation.
The set $Q$ includes numbers of occupation of particles and their one-particle energies, spins and
other quantum numbers.
All these definitions are within a \textit{framework of the axiomatic approach in Quantum Field Theory},
including the postulate of the existence of vacuum \cite{BLOT-87-Math-8} and
representations of the Poincar\'e group in the tangent space of Minkowski.

The new fact is only that we are expanding representations of the Poincar$\acute{\rm e}$ group in the tangent space of Minkowski
by zero harmonic of the dilaton in full compliance with two classes of functions
of kinemetric subgroup of diffeomorphisms of Hamiltonian describing of the evolution of the Universe.
It should  describe the physical excitations of quantum gravity by the two classes of functions.
Thus, all physical excitations in the reduced phase space can be classified by the
homogeneous dilaton (\textit{zero harmonic}) and localized \textit{field--particles},
and the Newtonian-like  \textit{potentials} with zero momenta.
Two independent variables: \textit{dilaton and graviton} are squeezed oscillators,
that allow the quantization and the Casimir vacuum energy specified in Chapter 6.

\subsection{Casimir's vacuum}

The canonical momentum of the zero harmonic of the dilaton is an evolution operator in the field
space of events. The canonical momentum of the dilaton is not equal to zero if there is a
non-zero Casimir energy of all the other fields in the empty
space, as was shown in Chapter 6.
One method of measuring of this homogeneous dilaton is the redshift of spectral lines of atoms.
Moreover, the occurrence of atoms of matter is
also described by the \textit{operator of creation of the Universe}  (\ref{09-07}).
In this regard, we will consider the homogeneous dilaton $\langle D\rangle$ as
a form of matter, along with non-homogeneous particles and their bound states, if matter is understood as
all that is measured and independent of an observer.
In any case, the separation of the dilaton from the metric of a space which was offered by Dirac, allows
us to include  the homogeneous dilaton in a field space of events as a parameter of evolution.
The key idea of creation of the Universe (and, as we shall see later, its matter)
is that the homogeneous dilaton in Conformal quantum theory of gravity
is a squeezed oscillator.
Therefore, the statement
 \textit{that at the beginning of the Universe there was the homogeneous dilaton and the
 Casimir vacuum energy of all fields}\footnote{A similar phrase in the Standard cosmology
 ``\textit{in the early Universe there was a redshift of spectral lines of atoms of
matter, and then there  appeared the very atoms of matter}'' recalls
rather a fabulous statement about the Cheshire Cat:
\textit{at first there was a smile of a Cat, and then himself}
(Carroll, Lewis: \textit{Alice's Adventures in Wonderland}. Macmillan and Co., London (1865).)}
is physically correct in the context of solving the problem of
creation of the Universe and particles of matter from the vacuum
in the early Universe by standard quantum theory methods, as it has been done in
Chapter 7 in describing  creation of gravitons from the vacuum.

\subsection{An approximation of a nearly empty Universe}

According to the conformal scenario, the Universe was empty and cold at the time of its creation from vacuum.
At this moment, the Casimir vacuum energy  dominated.
The Universe is almost empty throughout its evolution, including the modern era, according to
the latest data on the Supernovae. As was shown in Chapter 6, in the framework of the Conformal cosmology,
a cosmological scale factor of the empty Universe depends on the measured time interval as the square
root.
This is consistent with the description of the chemical evolution, which indicates
that there are only a few percents of the baryonic matter in the Universe.
 In other words, the Universe is {\it cold and almost empty}  all the time of its
existence, since the conformal temperature is a constant equal to three Kelvin.
This constant  appeared under the normal ordering of the field operators
in the Hamiltonian as a spontaneous breaking of conformal symmetry.
The operator of creation and evolution of the Universe also contains additional quantum anomalies,
such as the arrow of time.
After the procedure of normal ordering the Hamiltonian takes the form
\bea \label{08-06}
\widetilde{{\cal H}}&=&\rho_{\rm Cas}+:\widetilde{{\cal H}}:,
\eea
where
\bea \label{08-06N}
\rho_{\rm Cas}&=&\sum\limits_{f,Q}^{} \frac{\omega_{f,Q}}{2}
\eea
is the density of the Casimir energy of all the particles, as discussed in Chapter 6.
The Casimir energy is appear as a result of normal ordering the product of the field operators in the free Hamiltonian and the interaction Hamiltonian.
According to the observational data on the Supernovae, the Casimir energy
\bea \label{08-07}
{\bf E}^I_{\rm U}=2\int d^3x \sqrt{\rho_{\rm Cas}}
\eea
of the Universe is dominated. Dominance of the Casimir energy is the second cornerstone of our construction.

Let us consider further an expansion of the generator of the evolution (\ref{08-04}) regarding this vacuum expectation
$$
{\bf E}_{\rm U}=2\int d^3x
\sqrt{
\rho_{\rm Cas}+
:\widetilde{{\cal H}}:
}=
$$
\bea \label{08-08}
=2\int d^3x \sqrt{\rho_{\rm Cas}} +
\frac{\int d^3x:\widetilde{{\cal H}}:}{\sqrt{\rho_{\rm Cas}}}+ \cdots =
\eea
$$
={\bf E}^I_{\rm U}+\frac{{\bf H}_{\rm QFT}}{\sqrt{\rho_{\rm Cas}}}+\cdots
$$
where in the expression
\bea \label{08-09}
{\bf H}_{\rm QFT}=\int d^3x:\widetilde{{\cal H}}:
\eea
it is easy to recognize the Hamiltonian of all fields of matter, including the field of gravitons.
All of these fields have a positive definite metric after the explicit solving of all constraints in the
frame of reference, selected by a unit time-like vector \cite{BLOT-87-Math-8}
(see Chapter 7).

In the case of  approaching  nearly empty space, the evolution operator (\ref{08-02}) is presented as
the product of three factors
\bea \label{08-10}
\!\! \frac{\hat{\mathbb{U}}}{{\sqrt{2{\bf E}_{\rm U}}}}=
\frac{{\mathbb{U}}_0}{{\sqrt{2{\bf E}^I_{\rm U}}}} \cdot
 \left[ 1- \frac{1}{4}
 \hat \Omega_{\rm creation}\right]  \cdot
 T_{\widetilde{t}}\exp \left\{-\imath \int\limits_{\widetilde{t}_I}^{\widetilde{t}_0}
 d\widetilde{t} {\bf H}_{\rm QFT}\right\}.
\eea
In the first factor on the left you can recognize the cosmological wave function of the empty Universe
${\mathbb{U}}_0/\sqrt{2{\bf E}^I_{\rm U}}$,
earlier discussed in Section 6.6.
The second factor in the form of square brackets, containing the
operator of the relative density of matter creation in the Universe,
includes the ratio of the Hamiltonian of QFT to the vacuum energy
\bea \label{08-10o}
\hat \Omega_{\rm creation}=\frac{{\bf H}_{\rm QFT}}{V_0 \rho_{\rm Cas}}.
\eea
The third factor
\bea \label{08-11o}
 T_{\widetilde{t}}\exp \left\{-\imath \int\limits_{\widetilde{t}_I}^{\widetilde{t}_0}
 d\widetilde{t} {\bf H}_{\rm QFT}\right\}\equiv{\hat{\mathbb{U}}}^{\widetilde{t}_0}_{\widetilde{t}_I}
\eea
is a standard evolution operator in quantum field theory with respect to time
\bea \label{08-11}
d\widetilde{t}=\frac{d\langle D\rangle}{\sqrt{\rho_{\rm Cas}}},
\eea
which is given by the effective parameter of evolution in the field space of events.
We shall see later that this time (\ref{08-11}) coincides with the conformal time.
\bea \label{08-12}
 d\widetilde{t}=d\eta.
\eea
The third factor can be presented as the product of $N$ factors, breaking up the entire time interval
of evolution of the Universe on $N$ parts.
\bea \label{08-11o-3}
 {\hat{\mathbb{U}}}^{\widetilde{t}_0}_{\widetilde{t}_I}
 =\prod\limits_{n=1}^{n=N} {\hat{\mathbb{U}}}^{\widetilde{t}_0-n\triangle t}_{\widetilde{t}_I}.
\eea
Inserting an identity operator between the factors  as a
sum over a complete set of all possible states
$$\hat{I}=\sum\limits_{Q} |{Q}\rangle \langle {Q}|,$$
one can get the elements of the S - matrix in the representation of interaction
\cite{BLOT-87-Math-8}
\bea \label{08-11o1}
\left\langle Q'| T_{\widetilde{t}}\exp \left\{-\imath
\int\limits_{\widetilde{t}_0-n\triangle t}^{\widetilde{t}_0-({n-1})\triangle t}
 d\widetilde{t} {\bf H}_{\rm QFT}\right\}|Q''\right\rangle\equiv
 \left\langle {Q'}_{\rm int}|\hat S|Q''_{\rm int}\right\rangle .
\eea
A time interval
$\triangle t$
is determined by the energy resolution of physical devices and
characteristic time of the processes in high-energy physics in modern
accelerators\footnote{In this case, the interval $\triangle t$
is a moment of time life of physicists, and the lifetime of physicists is a moment of the lifetime of the Universe.}.

Thus, the Hamiltonian formulation of the theory of gravity in the \textit{reduced} phase space
leads to a modification of a well-defined theory of the S - matrix, which will be discussed in the next chapter.
The \textit{reduced} Hamiltonian approach is the primary method of learning  the theory of
gauge fields, starting with the pioneering work of Dirac \cite{d1927-8, dCan-8},
Heisenberg and Pauli \cite{hp-Math-8, hpII-Math-8} and papers of Schwinger
on the quantization of non-Abelian fields \cite{sch2-0-8}
(see details in \cite{pol-8, f1-Math-8, hpp-8, 6P-8}
and Appendix A).

\begin{center}
{\vspace{0.31cm}}
\parbox{0.5\textwidth}{
\includegraphics[height=9.truecm,
angle=-0]{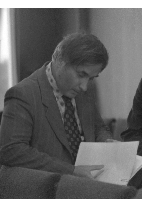}}
\parbox{0.48\textwidth}
{\small Igor V. Polubarinov (1928, Moscow — 1998, Dubna) is a Russian physicist.
He is known by his pioneering results in the field - theoretical interpretation of
gauge theories and gravity, obtained in close cooperation with V.I. Ogievetsky,
and Hamiltonian formulation of the S-matrix for physical gauge fields, left after solving of all constraints.
One of the most valuable results of I.V. Polubarinov is construction of an explicit form
relativistic transformations of the physical fields
from one frame to another. He was at the forefront of the fundamental operator quantization of gauge fields,
on which the present monograph is based.
}
{\vspace{0.31cm}}
\end{center}

In all these papers, time components of the vector field with negative contributions to the energy
are excluded, as was accepted in the Dirac approach to quantum electrodynamics  \cite{d1927-8,dCan-8}.
Dirac's Hamiltonian approach to QED in 1927 was based on a
gauge-invariant action on the surface of constraint
 \bea \label{6-2-8}
 W^{\rm Dirac}_{\rm QED}=
 W_{\rm QED} \Big|_{\frac{\delta W_{\rm QED}}{\delta A^{\ell}_0}=0},
  \eea
where a component
$A^{\ell}_0= (A\cdot {\ell})$
is defined as the scalar product of the vector $A_\mu$ and a unit time-like vector ${\ell}_\mu$.

Such an elimination of the time component results into static interactions that form simultaneous
bound states in QED, described by the Schr\"odinger equation, and in QCD, described by Salpeter equation
(see Appendix B).
It was shown that the Dirac Hamiltonian approach leads to the correct relativistic transformations of
observed quantized fields in non-Abelian gauge theories and theories of massive vector fields
 \cite{sch2-0-8, hpp-8, 6P-8}.
The Hamiltonian formulation \cite{f1-Math-8} is considered as a justification of
modern methods of quantization of gauge
theories\footnote{There is a stronger statement of Julian Schwinger
``{\it we rejected all Lorentz gauge
formulations as unsuited to the role of providing the fundamental
operator quantization}'' (see \cite{sch2-0-8} {p.324}).}, including the method of Faddeev -- Popov
\cite{fp1-Math-8},
which used to describe the Standard Model of elementary particles
\cite{db-8}.

The operator of creation of the Universe in the approximation (\ref{08-10}) describes the
three classes of processes: vacuum creation of matter, given in the previous chapter on the example of gravitons;
scattering and decay of elementary particles described by the S-matrix
and the interference of the S-matrix and vacuum creation.
Below, we describe the physical content of the operator of creation of the Universe (\ref{08-10}),
having considered the two extreme cases: when the cosmological scale factor tends to zero
(as a statement of the problem of origin of matter in the Universe), and when the cosmological scale factor
tends to unity
(as a modified description of the S-matrix elements of the scattering, decay, and formation
of bound states of elementary particles).

\section{Summary}
\renewcommand{\theequation}{8.\arabic{equation}}
\setcounter{equation}{0}
\makeatletter
\renewcommand{\@evenhead}{\raisebox{0pt}[\headheight][0pt]
{\vbox{\hbox to\textwidth{\hfil \strut \thechapter . Mathematical principles of description of the Universe
\quad\rm\thepage}\hrule}}}
\renewcommand{\@oddhead}{\raisebox{0pt}[\headheight][0pt]
{\vbox{\hbox to\textwidth{\strut \thesection .\quad
Summary and literature
\hfil \rm\thepage}\hrule}}}
\makeatother

Classical and quantum theory of the Universe appear as a result of solving of all constraint equations
in the conformal-invariant theory.
Herewith, the only source of the violation of the conformal symmetry is quantum anomalies such as the
Casimir energy type, or Casimir condensates, arising via normal ordering of
products of field operators in this theory.
Dominance of the Casimir energy, confirmed, as was shown in Chapter 6, by the present data on
Supernovae, leads to the approximation of a nearly empty Universe.
This approximation means the factorization constructed above of the operator of creation and evolution
of the Universe, on the wave function of the Universe,
describing data on the Supernova by the Casimir energy and modified by the operator of evolution of fields
of matter in QFT. As a result, we have a well-defined \textit{cosmological modification} of the
operator of evolution of fields, under their quantization in the phase
space of field variables that remain after the solution of the constraint equations in the considered theory of
gravity. Thus, the Hamiltonian approach provides an adequate formalism for unification of
the theory of gravitational field with the Standard Model of elementary particles, in which both theories are
considered at the quantum level, after solving all constraint equations in a certain frame of reference.
In the future, we will consider in detail the above mentioned \textit{cosmological modification} of the operator of
evolution of fields in QFT, and also
the creation of particle-like perturbations of these fields from the vacuum of the Universe.

{}

\chapter{Creation of matter in the Universe}
\renewcommand{\theequation}{9.\arabic{equation}}
\setcounter{equation}{0}
\section{The Big Bang or the vacuum creation?
\label{sect_matter-9}}
\makeatletter
\renewcommand{\@evenhead}{\raisebox{0pt}[\headheight][0pt]
{\vbox{\hbox to\textwidth{\hfil \strut \thechapter . Creation of matter in the Universe
\quad\rm\thepage}\hrule}}}
\renewcommand{\@oddhead}{\raisebox{0pt}[\headheight][0pt]
{\vbox{\hbox to\textwidth{\strut \thesection .\quad
The Big Bang or the vacuum creation?
\hfil \rm\thepage}\hrule}}}
\makeatother

\subsection{Statement of the problem}

Recall that quantum field theory in the Early Universe
describes a set of particles as oscillators interacting with the cosmological scale factor.
The scale factor squeezes the phase space of these oscillators. This squeezing
is a source of the cosmological creation of the particles from vacuum.
There is the following classification of these squeezed oscillators.
Massless particles (fermions and  photons) are not squeezed and created;
massive particles (fermions and transversal components of massive vector
$W$-, $Z$- bosons) are created very weakly; and strongly squeezed oscillators
(gravitons, scalar particles, and longitudinal components of electroweak massive vector bosons)
suffer the intensive cosmological creation from vacuum
due to their strong dependence on the cosmological scale factor.
The Cosmic Microwave Background radiation and the baryon matter in the
Early Universe can be the decay products of such
primordial electroweak bosons and their annihilations.

The question:  \textit{Is modern theory able to explain the origin of
observed matter in the Universe by its cosmological
production from a vacuum?} was considered at the end of the 60s -- early 70s of the last century
in the set of papers \cite{1-9, 2-9, 4-9, 5-9, 6-9}.
As it is well known,
the answer to this question is associated with the
problem of particle creation in the vicinity of a cosmological
singularity. Thus far, it has been common
practice to assume that the number of product pairs
is by far insufficient for explaining the total amount of
observed matter  \cite{7-9}.
We recall that the cosmological creation of massive
particles is calculated by going over to conformal
variables  \cite{9-9, 11-9}, for which the limit of zero scale factor
(point of a cosmic singularity) means vanishing
of masses. Massive vector and scalar bosons are the only particles of
the Standard Model that have a singularity at zero
mass  \cite{12-9, 13-9}. In this limit, the normalization of the
wave function for massive  bosons is singular in
mass  \cite{12-9, 13-9}. The absence of the massless limit in the
theory of massive vector bosons is well known \cite{14-9, 15-9}. In
calculations in the lowest order of perturbation theory,
this leads to a divergence of the number of produced
longitudinal bosons  \cite{7-9, 11-9}.
There exist two opinions concerning the removal
of this singularity. In  \cite{7-9},  the divergence of the
number of particles is removed by means of a standard
renormalization of the gravitational constant.
However, it is also indicated in the monograph of
Grib {\it et al}, \cite{7-9} that the number of produced particles
is determined by the imaginary part of loop Feynman
diagrams; since, in quantum field theory, it is the real
parts of these diagrams that are subjected to renormalization,
this means that the above mentioned divergence of
the number of particles does not belong to the class
of divergences in quantum field theory that are removed
by means of a conventional renormalization of
physical quantities. Indeed, the physical origin of this
divergence is that the problem of a cosmological creation
of particles from a vacuum is treated within an
idealized formulation. The point is that the quantum
production of particles in a finite volume for a system
featuring interaction and exchange effects may lead
to a set of Bose particles having a specific statistical
distribution with respect to energy, so that it is able
to ensure the convergence of the respective integral of
the momentum distribution.
In the present study, we analyze physical conditions
and models for which the number of product
vector bosons may be quite sufficient for explaining
the origin of matter in the Universe. Such cosmological
models include the Conformal cosmology \cite{16-9, Blaschke_04-9}, where
conformal quantities of the General Theory of Relativity
and of the Standard Model are defined as observables
 for which there are relative reference units of intervals.

The ensuing exposition of this Chapter is organized as follows.
The next Subsection 9.1.2 is devoted to discussion of the problem of an origin
of the Cosmic Microwave Background (CMB) radiation in the light of our classification
of observational data in order to determine the initial data of the CMB origin.
 Section 9.2 describes the cosmological creation of the primordial scalar bosons
 that are a source of the origin of all matter in the Unverse.
 In Section 9.3 the physical states of the matter are classified  in the context
 of the irreducible unitary representation of the Poincar$\acute{\rm e}$ group.
 In Section 9.4 we give the generation and modification of the Faddeev -- Popov path integral method
 in the light of this classification at the present day  value of the
 cosmological scale factor $a=1$.

\subsection{Observational data on the  CMB\\ radiation origin
}

%
Observational data on the  CMB testify about the first instances of creation and evolution
of the matter in the Universe. According to these data obtained by means of satellites, air-balloons,
and  in observatories, the  picture of temperature
distribution of the CMB radiation is asymmetrical.
 This asymmetry is treated in the modern cosmology \cite{Giovannini-9}
as a signal of the motion of the Earth with reference to the CMB radiation.
This motion has a velocity 368 km/sec towards the constellation Leo.
To remove this  asymmetry one should pass to a co-moving frame of reference, where the CMB radiation
takes the symmetric form.
Such a co-moving frame of reference by no means can be associated with any heavy body,
as it was accepted in the celestial mechanics. In physics a new situation arises,
when a frame  of reference is connected with the parameters of the photon gas.
In this case the choice of the co-moving frame  of reference allows us to remove the dipole component
of CMB radiation, separating out the motion of the very observer.
As it was shown in the previous chapters, the choice of the conformal etalons
allows us to separate out the cosmic evolution of massive devices of the very observer.
In terms of conformal etalons \cite{Behnke:2001nwempty-9}
observational Supernovae data \cite{SN1997ff-9, SN1997f-9} allows us to obtain
dependence of the horizon $r(z)$ on the redshift $1+z$
$$r(z)=(1+z)^{-2}\cdot H^{-1}_0 \simeq (1+z)^{-2}\times 10^{29}~\mbox {mm},$$
that is almost universal for all epoch, including the Beginning of the Universe.
We propose that this Hubble law was valid at the instances of creation of the primordial
particles and the  origin of the CMB temperature.
What does an origin of the concept of temperature mean?
And how to define the range of the validity of this concept?
According  to D.I. Blokhintsev \cite{Blokhin1949-9},
the definition of the ranges of the validity of concepts can predict new effects
and determine the values of physical magnitudes describing these effects,
as we wrote in Chapter 1.

The  concept of the temperature begins arising when the mean wave-length of a CMB photon
coincides with the horizon (\textit{id est} with the visible size of the Universe)
$$r(z_I) \simeq (1+z_I)^{-2}\times 10^{29}\mbox{mm}= 1~\mbox{mm}.$$
From this it follows that at this instance the redshift is equal to
\be
1+z_I\simeq 3 \times 10^{14}.
\ee
The second observational fact consists in the present day values of masses of the
primordial particles, the decay products of which give the CMB photons
at the Beginning
\be\label{WZ-9}
M_0=(1+z_I)\cdot T_0\sim 100~ {\rm\mbox{GeV}},
\ee
that are in the region of the electroweak scale defined by the values of $W$- $Z$- masses.
\begin{center}
\includegraphics[width=0.8\textwidth]{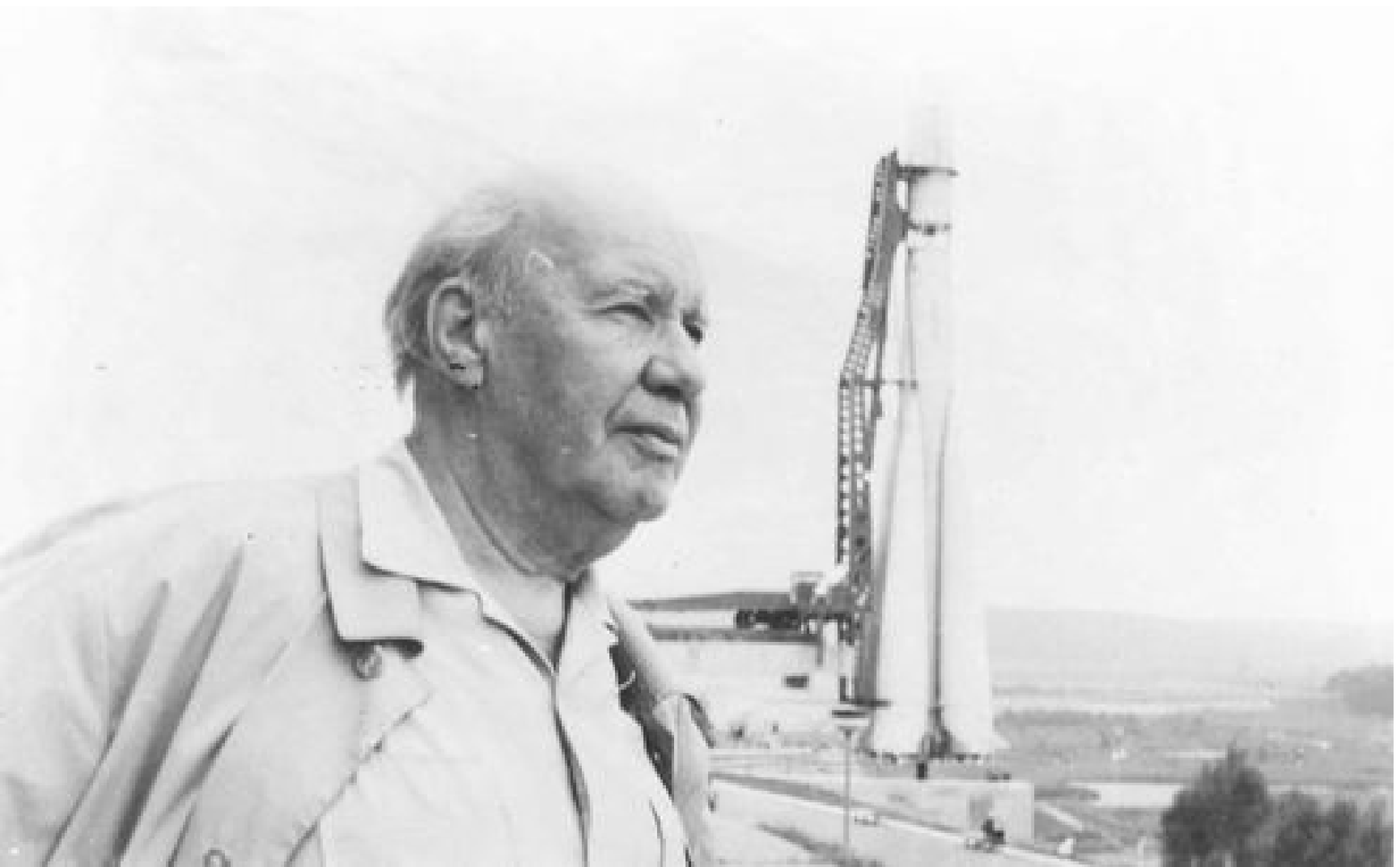}
\end{center}
{\small Dmitry I. Blokhintsev (29 December 1907 (11 January 1908), Moscow -- 27 January 1979,
Dubna), Russian physicist.
Independently studied the basics of differential and integral calculus.
Corresponded with K.E. Tsiolkovsky. From Tsiolkovsky Dmitry Ivanovich took the spirit of Russian science
of the early 20-th century, which is not expressed so much in the quest to achieve concrete results, but
how to create a holistic harmonious world.
He graduated from the Physics Department of Moscow State University (1930).
He taught there (since 1936 - professor, then head of the Department of Theoretical Nuclear Physics).
He was the founder of the Division of Nuclear Physics at the Physics Department of Moscow State University.
In 1935-1947 he worked in the Lebedev Physics Institute of the USSR Academy of Sciences.
Since 1947 he was the Director of the Research Laboratory in Obninsk,
on the basis of which under his leadership the Physics -- Energy Institute was created.
Together with I.V. Kurchatov, Blokhintsev initiated the establishment of the world's first nuclear power
electrostation.
Blokhintsev initiated the establishment of the
Joint Institute for Nuclear Research (JINR) in Dubna.
In 1956, he was elected as first director of the Institute.
In the years 1956--1965 --- JINR Director, since 1965 -- Director of the Laboratory of Theoretical Physics, JINR.
}

Here
$$T_0=2.35 \times 10^{-13} ~\mbox{GeV}$$
is the mean energy of the CMB photons
$$(1+z_I)\simeq 3 \times 10^{14}$$
is the critical value of the redshift obtained above.
The third observational fact is that the time of the CMB origin
$(1+z_I)\simeq 3 \cdot 10^{14}$ coincides with the time when the Higgs particle Compton size
becomes of order of the visible size of the Universe $H_0/a^2_I$.
The equality of such times yields the same region  for e-w boson mass values.
$$M_0=\sqrt{\frac{T_0^3}{H_0}}\sim 100 ~{\rm\mbox{GeV}}.$$

Just at this time one can introduce the concepts of these particles,
and at this instance they are created from the vacuum \cite{Blaschke_04-9, 2009-9, grg-9}.
This approximal coincidence of the instances of origin of the CMB temperature
and the creation of the primordial e-w bosons points out that the CMB is the decay product
of these bosons. In the next Section we give the direct calculation of the occupation number
of the primordial Higgs particles created from the vacuum.

\section{Vacuum creation of scalar bosons \label{sect_Higgs}}
\makeatletter
\renewcommand{\@evenhead}{\raisebox{0pt}[\headheight][0pt]
{\vbox{\hbox to\textwidth{\hfil \strut \thechapter . Creation of matter in the Universe
\quad\rm\thepage}\hrule}}}
\renewcommand{\@oddhead}{\raisebox{0pt}[\headheight][0pt]
{\vbox{\hbox to\textwidth{\strut \thesection .\quad
Vacuum creation of scalar bosons
\hfil \rm\thepage}\hrule}}}
\makeatother

In our model the interaction of scalar bosons and gravitons with the dilaton may be considered on an equal footing
\cite{Barbashov:2005perturb-9}.
Using this fact, we can consider the intensive creation of the Higgs
scalar particles from the vacuum using the results of Section 7.3 devoted to
the creation of gravitons.

To proceed we have to add the SM sector to the theory under construction.
In order to preserve the common origin of the conformal symmetry breaking by the
Casimir vacuum energy, we have to exclude the unique dimensional parameter from
the SM Lagrangian, {\it id est} the Higgs term with a negative squared tachyon mass.
However, following Kirzhnits~\cite{linde1-9}, we can include the vacuum
expectation of the Higgs field $\phi_0$, so that:
$$\phi=\phi_0+\overline{h}/[a\sqrt{2}],~~~~~~~\int d^3x \overline{h}=0.$$
The origin of  this vacuum expectation value $\phi_0$ can be associated with
the Casimir energy arising as a certain external initial data at $a=a_{\rm Pl}$.
In fact, let us apply the Planck least action postulate to the Standard Model action:
$$\phi_0\approx a_{\rm Pl}^3 H_0$$
 in agreement with its value in Table~6.1.


 The standard vacuum stability conditions at $a=1$
 \bea
<0|0>|_{\phi=\phi_0}=1, \qquad \frac{d<0|0>}{d\phi}\bigg|_{\phi= \phi_0}=0
 \eea
\bea
V_{\rm eff}(\phi_0)=0, \qquad \frac{dV_{\rm eff}(\phi_0)}{d\phi_0}=0.
\eea
It results in a zero contribution of the Higgs field vacuum expectation
into the Universe energy density. In other words, the SM
mechanism of a mass generation can be completely repeated in the framework
of our approach to the spontaneous symmetry breaking.

In particular, one obtains that the Higgs boson mass is determined from the equation
$V''_{\rm eff}(\langle \phi\rangle)=M_h^2$. Note that in our construction the Universe
evolution is provided by the dilaton, without making use of any special potential
and/or any inflaton field. In this case we have no reason to spoil the
renormalizablity of the SM by introducing the non-minimal interaction between the
Higgs boson and the gravity. 
In the middle field approximation (far from heavy masses), 
our gravitation  theory 
supplemented by the
Standard Model the Higgs bosons are described by the action
 \bea
 \label{hh-2s}
 {W}_h&=& \int\limits_{}^{}d\tau
 \sum\limits_{\textbf{k}^2\not =0}
 \dfrac{v^h_{\textbf{k}}v^h_{-\textbf{k}}\!-\!
 {h}_{\textbf{k}}{h}_{-\textbf{k}}a^2{\omega^h_{0{\bf k}}}^2}{2}
 =\sum\limits_{\textbf{k}^2\not =0}
 p_{-\textbf{k}}^hv^h_{\textbf{k}}-\textsf{H}^h_{\tau},
 \eea
where
 \bea\label{2-10hh}
 \omega^h_{0{\bf k}}({a})=\sqrt{{\bf k}^2+a^2M^2_{0\rm h}}
 \eea
is the massive one--particle energy with respect to the conformal time interval.

There are values of the scale factor $a$,  when the mass term in
one--particle energy  is less than the conformal Hubble parameter value
$$a M_{0\rm h}<H_0a^{-2}.$$
As a result, the Casimir energy for the Higgs particles  coincides
with the graviton one at the considered epoch:
\bea\nonumber \textsf{H}^h_{\rm Cas}&\simeq&
\sum\limits_{\bf k}\frac{{\sqrt{{\bf k^2}}}}{2}=\textsf{H}^g_{\rm Cas}.
\eea
In this case the calculation of the scalar particle creation energy completely  repeats
the scheme for the graviton creation discussed in Chapter 7.
Assuming thermalization in the primordial epoch, we expect that the occupation number
of the primordial Higgs bosons is of the order of the known CMB photon one
 \be\label{nu-1}
 \textsf{N}_h \sim \textsf{N}_\gamma=411 \mbox{mm}^{-3}\times
 \frac{4\pi r_h^3}{3}\simeq 10^{87}.
 \ee
 Thus, the CGR provides
a finite occupation number of the produced primordial particles. Note
that in other approaches~\cite{7-9} a subtraction is used to achieve a finite result.
Moreover, the number of produced particles happens to be of the
order of the known CMB photon number. To our opinion this coincidence
supports our model, since the number of photons can naturally inherit
the number of primordial Higgs bosons (if one considers the photons
as one of the final decay products of the bosons).
According to our model, the relativistic matter has been created very soon after
the Planck epoch at $z_{\rm Pl}\simeq 10^{15}$. Later on it cooled down and
at $z_{\rm CMB}\simeq 1000$ the CMB photons decoupled from recombined ions
and electrons as discussed by Gamow. In our model the CMB
temperature is defined directly from the Hubble parameter
and the Planck mass (related to the Universe age $a_{\rm Pl}$).

Note that the obtained occupation number 
corresponds to the thermalized system of photons with the mean wave length
$T\simeq 3^\circ$~K
in finite volume 
$V_0\sim H_0^{-3}$:
\be\label{U-1}
\left(\textsf{N}_\gamma\right)^{1/3} \simeq 10^{29} \simeq \lambda_{\rm CMB} H_0^{-1}.
\ee

As concerns vacuum creation of spinor and vector SM particles, it is known~\cite{7-9}
to be suppressed very much with respect to the one of scalars and gravitons.

The intensive creation of primordial gravitons and Higgs bosons is described assuming
that the Casimir vacuum energy is the source of this process \cite{grg-9}.

\begin{figure}[t]
\includegraphics[width=0.90\textwidth,height=0.45\textwidth]{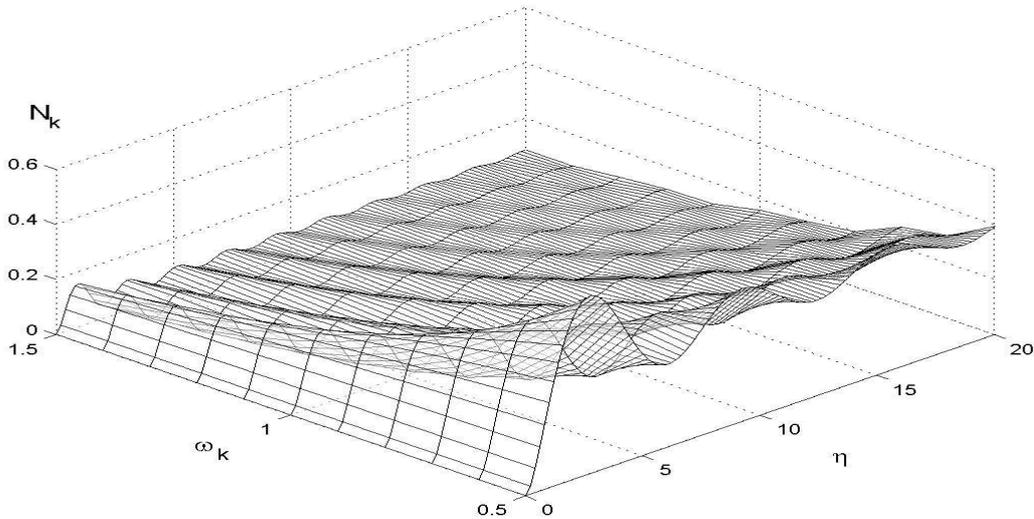}
\caption{\small The process of the vacuum creation of $N_h\sim 10^{88}$ Higgs particles
during the first $10^{-12}$ sec is shown. Here \fbox{$\nearrow$} is the time axis,
\fbox{$\uparrow$} is the number of scalar bosons
and longitudinal components of vector ones, and \fbox{$\searrow$} is the axis of their momenta \cite{grg-9}.
}
\end{figure}

The vacuum creation of massive spinors and transversal components of vector bosons \cite{7-9}
is suppressed with respect to the creation of scalar particles and gravitons.

\section{Physical states of matter
}
\label{sect_R}
\makeatletter
\renewcommand{\@evenhead}{\raisebox{0pt}[\headheight][0pt]
{\vbox{\hbox to\textwidth{\hfil \strut \thechapter . Creation of matter in the Universe
\quad\rm\thepage}\hrule}}}
\renewcommand{\@oddhead}{\raisebox{0pt}[\headheight][0pt]
{\vbox{\hbox to\textwidth{\strut \thesection .\quad
Physical states of matter
\hfil \rm\thepage}\hrule}}}
\makeatother
According to the general principles of quantum field theory (QFT), physical
states  of the lowest order of perturbation theory are completely covered  by
local fields as  particle-like representations of the Poincar$\acute{\rm e}$ group of
transformations of four-dimensional space-time.
The existence of each elementary particle is associated with a quantum field
$\psi$. These fields are operators defined in all space-time and acting on
states $|{\cal P},s\rangle$ in the Hilbert space with positively defined
scalar product. The states correspond to the wave functions
$\Psi_\alpha(x)=\langle 0|\psi_\alpha(x)|{\cal P},s\rangle $ of free
particles.

 Its algebra is formed by generators of the four translations $\hat {\cal P}_\mu=i\partial_\mu$
 and six rotations $\hat{M}^{\mu\nu} =
i[x_\mu\partial_\nu-x_\nu \partial_\mu]$. The unitary and irreducible
representations are eigen-states of the Casimir operators of mass and spin,
given by
 \bea \label{3-1} \hat{\cal P}^2|{\cal P},s\rangle
&=&m_{\psi}^2~|{\cal P},s\rangle ,\\ \label{3-1w}
 -\hat w_p^2 |{\cal P},s\rangle &=&s(s+1)|{\cal P},s\rangle ,\\ \label{3-1ww}
 \hat w_\rho&=& \frac{1}{2}
 \varepsilon_{\lambda\mu\nu\rho} \hat{\cal P}^\lambda \hat{M}^{\mu\nu}.
 \eea
The unitary irreducible Poincar$\acute{\rm e}$ representations describe wave-like
dynamical local excitations of two transverse   photons in QED
\bea \label{3-1ph}
 A^{T}_{(b)}(t,\textbf{x})=
\eea
$$
= \int
 \frac{d^3k}{(2\pi)^3}\sum\limits_{\alpha=1,2}\frac{1}{\sqrt{2\omega({\bf k})}}
 \varepsilon_{(b)\alpha}\left[
 e^{\imath
 (\omega_{\textbf{k}}t-\textbf{\textbf{k}\textbf{x}})} A^{+}_{\textbf{k},\alpha}+
 e^{-\imath
 (\omega_{\textbf{k}}t-\textbf{\textbf{k}\textbf{x}})}A^{-}_{\textbf{k},\alpha} \right].
$$
Two independent polarizations $\varepsilon_{(b)\alpha}$
are perpendicular to the wave vector and to each
other, and the photon dispersion is given by
$\omega_{\textbf{k}}=\sqrt{{\textbf{k}}^2}$.
The creation and annihilation operators of photon obey the commutation relations
$$[A^{-}_{\textbf{k},\alpha},A^{+}_{\textbf{k}',\beta}]=
\delta_{\alpha,\beta}\delta({\textbf{k}-\textbf{k}'}).$$

The bound states of elementary particles (fermions) are associated with
bilocal quantum fields formed by the instantaneous potentials (see
\cite{MarkovYukawa-9, kad-9, LukierskiOziewicz-9})
 \bea\label{set-1}
 {\cal M}(x,y)={\cal M}(z|X)
 =\sum\limits_H\int\frac{d^3{\cal
P}}{(2\pi)^{3}\sqrt{2\omega_H}}\int\frac{d^4qe^{\imath q\cdot z}}{(2\pi)^4}\times
 \eea
$$\times
 \left[
e^{\imath{\cal P}\cdot{X}} \Gamma_H(q^{\bot}|{\cal P})a^+_H(\bm{{\cal
P},q^{\bot}})+
e^{-\imath{\cal P}\cdot{X}} \bar{\Gamma}_H(q^{\bot}|{\cal
P})a^-_H(\bm{{\cal P},q^{\bot}})\right],
$$
where
$${\cal P}\cdot X=\omega_H X_0-\bm{{\cal P}} \mathbf{X},~~~~~q_{\mu}^{\bot}=q_{\mu}-
\frac{{\cal P}\cdot q}{M_H^2}{\cal P}_\mu,$$
${\cal P}_\mu=(\omega_H,\bm{{\cal P}})$ are the momentum components
on the mass shell,
$$\omega_H=\sqrt{M_H^2+\bm{{\cal P}}^2},$$
and
\bea
\label{set-1a} X=\frac{x+y}{2},~~~~~~~~~~~~~~ z=x-y,
\eea
are the total coordinate
and the relative one, respectively.
The functions $\Gamma $ belong to the
complete set of orthonormalized solutions  of the BS equation \cite{a15-9} in a
specific gauge theory, $a^{\pm}_H(\bm{{\cal P},q^{\bot}})$ are coefficients
treated in quantum theory as the creation (+) and annihilation (-) operators (see
Appendix B).

The irreducibility constraint, called the Markov -- Yukawa constraint, is imposed on
the class of instantaneous bound states \cite{MarkovYukawa-9}
 \bea \label{1set-2}
 z^\mu \hat {\cal P}_\mu{\cal M}(z|X)\equiv \imath z^\mu \frac{d}{d
X^\mu}{\cal M}(z|X)=0.
  \eea
In Ref.~\cite{BLOT-87-9} the in- and out- asymptotical states are the ``rays''
defined as a product of these irreducible representations of the Poincar$\acute{\rm e}$
group
\be \label{3-2-1} \langle {\rm out}|=\langle \prod_{J}{{\cal
P}_J,s_J}~\big|, ~~~|{\rm in}\rangle =\big|\prod_{J }{{\cal P}_J,s_J}\rangle.
\ee
This means that all particles (elementary and composite)  are  far enough
from each other to neglect their interactions in the in-, out- states. All
their asymptotical states  $\langle {\rm out}|$ and $|{\rm in}\rangle $
including the bound states are considered as the irreducible representations
of the Poincar$\acute{\rm e}$ group.

These irreducible representations form a complete set of states, and the
reference frames are distinguished by the eigenvalues of the appropriate time
operator $\hat \ell_\mu=\dfrac{\hat{\cal P}_\mu}{M_J}$
 \bea \label{3-2a}
 \hat \ell_\mu|{{\cal P},s}\rangle =\frac{{\cal P}_{J\mu}}{M_{J}}|{{\cal P}_J,s}\rangle ,
 \eea
 where the Bogoliubov -- Logunov -- Todorov  rays (\ref{3-2-1}) can include bound states.
\begin{center}
\parbox{0.5\textwidth}{
\includegraphics[height=9.truecm,angle=-0]{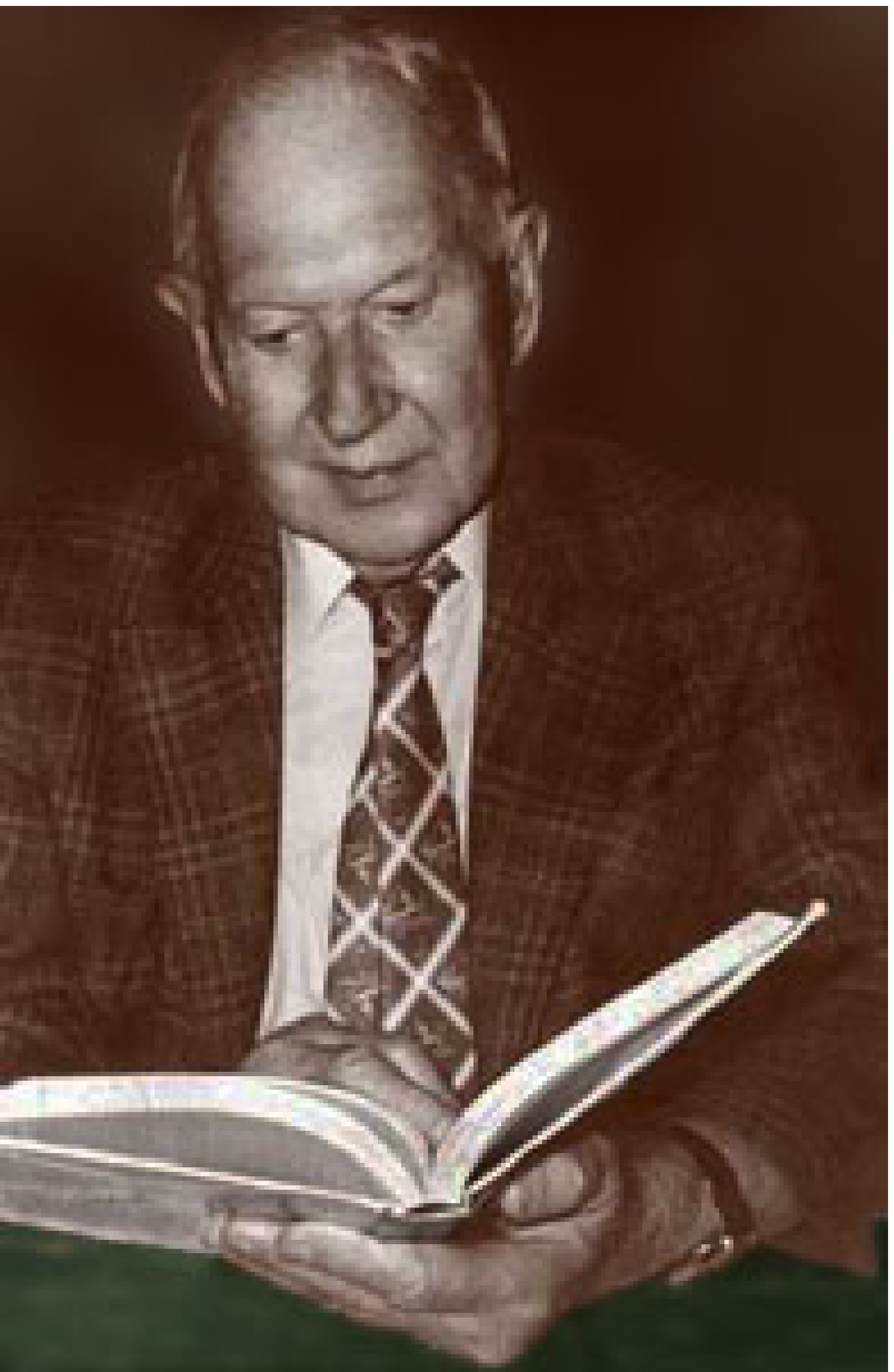}}
\parbox{0.48\textwidth}
{\small
Moisei A. Markov (30 April (13 May) 1908, Rasskazovo, Tambov province — 1 October 1994, Moscow),
Russian physicist - theorist, Academician.
Markov graduated from Moscow State University in 1930.
He is a member of the Presidium of the Academy of Sciences of the USSR.
He has been chairman of the Interdepartmental Commission for Nuclear Physics since 1971.
He was one of the organizers of the Joint Institute for Nuclear Research.
He made a pioneer contribution into the development of neutrino investigations,
demonstrated the expediency of conducting neutrino experiments at great depths underground and the possibility
of conducting such experiments in accelerators.
He studied fundamental problems
of elementary particle physics and quantum gravitation on the boundary of particle physics and cosmology.
}
\end{center}

\section{QU modification of S-matrix in QFT
}
\label{sect_R-9}
\makeatletter
\renewcommand{\@evenhead}{\raisebox{0pt}[\headheight][0pt]
{\vbox{\hbox to\textwidth{\hfil \strut \thechapter . Creation of matter in the Universe
\quad\rm\thepage}\hrule}}}
\renewcommand{\@oddhead}{\raisebox{0pt}[\headheight][0pt]
{\vbox{\hbox to\textwidth{\strut \thesection .\quad
Conformal modification of S-matrix in QFT
\hfil \rm\thepage}\hrule}}}
\makeatother

The S-matrix elements are defined as the evolution operator expectation values
between in- and out- states
 \bea \label{4-2}\underbrace{{{\cal M}_{\rm in,out}}}_{P-inv,G-inv}=
 \underbrace{\langle {\rm out}|}_{P-covariant}~
 \underbrace{\hat S[\hat \ell]}_{P-covariant,G-inv}~\underbrace{|{\rm in}\rangle }_{P-covariant},
\eea
where the abbreviation $``G-inv''$, or ``gauge-invariant'', assumes the
invariance of S-matrix with respect to the gauge transformations, and 
$``P-covariant''$ means relativistic covariance with respect to
the Poincar$\acute{\rm e}$ group transformations.
The conformal modification of $S-$ matrix, in this case, means that the conformal
symmetry can be violated by the quantization procedure with the vacuum postulate.
in the reduced phase space after resolution of all constraints.
Such reduced phase space quantization coincides with the Dirac approach to the gauge
invariant theories. The Dirac approach to gauge-invariant S-matrix was formulated at the rest
 frame $\ell_\mu^0=(1,0,0,0)$ \cite{Dirac-9, hp-9, Polubarinov-9}. The Dirac Hamiltonian
approach to QED of 1927 was based on the constraint-shell action 
\cite{Dirac-9}
 \bea \label{6-2} W^{\rm Dirac}_{\rm QED}=
 W_{\rm QED} \Big|_{\frac{\delta W_{\rm QED}}{\delta A^{\ell}_0}=0},
  \eea
where the component $A^{\ell}_0$ is
defined by the scalar product
$$A^{\ell}_0=(A\cdot{\ell})$$
of vector field
$A_\mu$ and the unit time-like vector ${\ell}_\mu$.
The gauge was established by Dirac as the first integral of the Gauss constraint
\bea \label{7-2} \int^t dt{\frac{\delta W_{\rm QED}}{\delta
A^{\ell}_0}=0},~~~~~~~~ t=(x\cdot \ell).
\eea
In this case, the S-matrix
elements (\ref{4-2}) are relativistic invariant and independent of the frame
reference provided the condition (\ref{3-2a}) is fulfilled
\cite{LukierskiOziewicz-9}. 

Dirac introduced the radiation variables
\bea\label{lc1-9}
\imath e A^{*}_k[A_j]&=&u^*[A_j]\left(\imath e A_k-\partial_k\right)(u^*)^{-1}[A_j],\\\label{lc2-9}
 \psi^*[A_j,\psi]&=&u^*[A_j]\psi,
 \eea
where the phase factors  $u^*[A_j]$ satisfy the equation
\bea
 \label{lc3-9ph}
u^*[A_j]\left(\imath e a_0[A_j]-\partial_0\right)(u^*)^{-1}[A_j]=0;
\eea
here $a_0[A_j]$ is a solution of the Gauss constraint equation
\bea
 \label{lc3-9}
 \triangle a_0[A_j]&=&\partial_j \partial_0A_j.
 \eea
One can be convinced that the radiation variables (\ref{lc1-9}) are gauge-invariant functionals.

 Then the problem
arises how to construct a gauge-invariant S-matrix in an arbitrary frame of
reference. It was  Heisenberg and Pauli's question   to von Neumann \cite{hp-9}: ``How to
generalize the Dirac Hamiltonian approach to QED of 1927 \cite{Dirac-9} to any
frame?'' \cite{hp-9,Polubarinov-9,Pervu2-9,Pervu-9}. The reply of von Neumann was to go
back to the initial Lorentz-invariant formulation
\bea\label{lc5-9}
\imath e A_k&=&(u^*)^{-1}[A_j]\left(\imath e A^{*}_k[A_j]-\partial_k\right)u^*[A_j],\\\label{lc6-9}
 \psi&=&(u^*)^{-1}[A_j]\psi^*[A_j,\psi],
 \eea
and to choose the co-moving
frame
\be\label{vN-1} \ell_\mu^0=(1,0,0,0) \to \ell_\mu^{\rm
co-moving}=\ell_\mu,~~~~~\ell_\mu \ell^\mu=(\ell\cdot \ell)=1
\ee
and to repeat
the gauge-invariant Dirac scheme in this frame for calculation of the spectrum and $S-$ matrix
elements(\ref{4-2}).
In the following we call this gauge-invariant approach  the von Neumann -- Polubarinov
formulation because Polubarinov constructed the corresponding gauge transformations (\ref{vN-1})
 in the manifest  form \cite{Polubarinov-9, Pervu2-9, Pervu-9}.
 In this approach the $S-$ matrix elements (\ref{4-2}) are relativistic invariant and
do not contain nonphysical states with indefinite metrics provided by the constraint (\ref{3-2a})
 \cite{LukierskiOziewicz-9, Kalinovsky-chapter9}.
Therefore,  relativistic  bound states can be successfully included in the
relativistic covariant unitary perturbation theory \cite{Kalinovsky-chapter9}. They
satisfy the Markov -- Yukawa constraint (\ref{1set-2}), where the time axis  $\ell^{0}$
is the eigenvalue of the total momentum operator of instantaneous bound states.

In QED this  framework yields the observational  spectrum of bound states \cite{Salpeter-9}, and leads to
the Schr\"odinger equation (see Appendix B),
and paves the way for constructing a bound state generating functional in QCD. The
functional construction is based on the Poincar$\acute{\rm e}$ group representations
with $\ell^{0}$ being the eigenvalue of the total momentum
operator of instantaneous bound states.
In order to demonstrate the Lorentz-invariant version of the Dirac method \cite{Dirac-9}
 given by Eq.~(\ref{6-2})
 in a non-Abelian theory, we consider
 the simplest example of the Lorentz-invariant  formulation of the naive
 path integral without any ghost fields and FP-determinant.
 Recall that
 the simplest example of the Lorentz  formulation that uses the naive
 path integral without any ghost fields and FP-determinant is described by the
 generating functional
 \bea\label{fp-1c}
 Z[J,\eta,\overline{\eta}]&=&\int\left[\prod\limits_{\mu,a}^{}d A^a_\mu\right]
 d\psi d\overline{\psi} e^{\imath W[A,\psi,
 \overline{\psi}]+\imath S[J,\eta,\overline{\eta}]}.
 \eea
 We use standard the QCD action $W[A,\psi, \overline{\psi}]$ and the source terms
\bea \label{fp-2}
W&=&\int d^4x
\left[-\frac{1}{4}F^a_{\mu\nu}{F^a}^{\mu\nu}
-\overline{\psi}(\imath\gamma^\mu(\partial_\mu +\hat A_\mu) -m)\psi\right],\\
\label{fp-4} F^a_{0k}&=&\partial_0 A^a_k -\partial_0
A^a_k\partial+gf^{abc}A^b_0 A^c_k\equiv
\dot A^a_k-\nabla^{ab}_k A^b_0,\\
\label{fp-42} S[A_\mu]&=&\int d^4x \left[A_\mu
J^\mu+\overline{\eta}\psi+\overline{\psi}\eta\right], ~~\hat
A_\mu= g\frac{\lambda^aA^a_\mu}{2\imath}.
\eea
There are a lot of drawbacks of this path integral (\ref{fp-1c}) from the point of view of QFT and
the Faddeev -- Popov functional \cite{fp1-9}. They are the following:
\begin{enumerate}
\item The  time component $A_0$ has indefinite metric.
\item The integral (\ref{fp-1c}) contains the infinite gauge factor.
\item The bound state spectrum contains tachyons.
\item The analytical properties of  field propagators are gauge dependent.
\item Operator foundation is absent \cite{sch-9}.
\item Low-energy region of hadronization is not separated from the high-energy one.
\end{enumerate}
All these defects can be
removed by the integration over the indefinite metric  time component $A_\mu \ell^\mu\equiv (A\cdot \ell)$,
 where $\ell^\mu$ is an arbitrary
unit time-like vector: $\ell^2=1$. If $\ell^{0}=(1,0,0,0)$ then $A_\mu
\ell^\mu= A_0$. In this case the generational functional (\ref{fp-1c}) takes the form
\bea \nonumber 
 Z[{{\ell^0}}]&=&\!\int\!\!  \left[\prod\limits_{x,j,a}^{}dA^{a*}_j(x)\!\right]\!e^{\imath W^*_{\rm YM}}
~ \delta\left(L^a\right) \left[\det{(\nabla_j(A^*))^2}\right]^{-1/2}Z_{\psi},\\\nonumber
L^a&=&\int\limits_{}^{t} d\overline{t}\nabla^{ab}_i(A^*)\dot A^{*b}_i=0,\\\nonumber
W^*_{\rm YM}&=&\int d^4x \frac{(\dot {A_j^a}^*)^2-(B^a_j)^2}{2},\\
 \label{9-2-3}
Z_{\psi}[J^*,\eta^*, \overline{\eta^*}]&=& \int d\psi^*
d\overline{\psi^*}e^{-\frac{\imath}{2}\left(\psi^*\overline{\psi^*},{\cal
K}\psi^*\overline{\psi^*}\right)-\left(\psi^* \overline{\psi^*},G^{-1}_{A^*}\right)
+\imath S[J^*,\eta^*, \overline{\eta^*}]}\\\nonumber
\left(\psi^* \overline{\psi^*},G^{-1}_{A^*}\right)&=&
\int d^4x\overline{\psi^*}\left[\imath\gamma_0\partial_0\!-
\!\gamma_j(\partial_j+\hat A^*_j)\!\!-m\!\right]{\psi^*},\\\nonumber
\left(\psi^*\overline{\psi^*},{\cal K}\psi^*\overline{\psi^*}\right)&=&
\int d^4x d^4y
j^a_0(x)\left[\frac{1}{(\nabla_j(A^*))^2}\delta^4(x-y)\right]^{ab} j^b_0(y).
\eea

The infinite factor is removed by the gauge fixing (\ref{7-2}) treated
as an antiderivative function of the Gauss constraint. $A^{*a}_i$ denotes
fields $A^{a}_i$ under gauge fixing condition (\ref{7-2}).
It becomes homogeneous
$$\nabla^{ab}_i(A^{*})\dot A^{*b}_i=0,$$
because $A^*_0$ is
determined by the interactions of currents. 
 It is just the non-Abelian
generalization \cite{Pervu2-9,Pervu-9} of the Dirac approach to QED \cite{Dirac-9}.

In the case of QCD there is a possibility to include the nonzero condensate
of transverse gluons
$$\langle A^{*a}_jA^{*b}_i\rangle=2C_{\rm gluon}\delta_{ij}\delta^{ab}$$
as a source of the conformal symmetry breaking.
The Lorentz-invariant bound state matrix elements can be obtained, if we choose
the time-axis $\ell$ of the Dirac Hamiltonian dynamics  as the operator
acting in the complete set of bound states (\ref{3-2a})
and given by  Eqs.~(\ref{set-1a}) and (\ref{1set-2}).
 This scheme enables us to
restore the Lorentz-invariance, if  the time-axis is taken as the operator $\hat\ell$
 proper frame of reference of each bound state.
 This means the von Neumann substitution (\ref{vN-1}) given in \cite{hp-9}
\begin{equation}\label{3-2c}
Z_{}[\ell^{0}\,]\to Z[\,\ell\,]\to Z[\,{\hat \ell}].
\end{equation}
Thus, we shown how to use the Dirac Hamiltonian formulation in order to describe
bound states by a relativistic-invariant manner with the Feynman rules
depending on the Markov -- Yukawa frame of reference

In the modern literature such dependence is treated as a defect
that complicates the perturbation theory. In order to remove this dependence,
in the Faddeev -- Popov approach \cite{fp1-9}  one passes to the variables of the type of
gauge transformation,
\bea\label{lc1-9-1}
\hat A^{*}_k[A^{L b}_j]&=&u^*[A^{L b}_j]\left(\hat A^L_k+\partial_k\right){u^*}^{-1}[A^{L b}_j],\\\label{lc2-9-1}
 \psi^*[A^{L b}_j,\psi^L]&=&u^*[A^{L b}_j]\psi^L,
 \eea
where $A^L_\mu$  obeys to condition that does not depend on a frame of reference,
in particular, the Lorentz constraint $\partial_\mu  A^L_\mu=0$,
and the phase factors $u^*[A^L_j]$ satisfy the equation
\bea
 \label{lc3-9phn}
u^*[A^L_j]\left(\hat a_0[A^L_j]+\partial_0\right)(u^*)^{-1}[A^L_j]=0.
\eea
Here $a^c_0[A_j]$ is a solution of the Gauss constraint
\bea\label{lc4-9-1}
 [((\nabla_j(A^L))^2]^{cb} a^b_0=
\nabla^{cb}_i(A^L)\dot A^{L b}_i.
\eea
A solution of Eq.~(\ref{lc3-9phn}) takes the form
\bea
\label{lc3-9-1a}
 u^*[A^{L b}_j]&=& v({\bf x}) T\exp\left\{\int\limits_{}^{t} d\overline{t}\hat a_0[A^{L b}_j]\right\},
 \eea
where the symbol  $T$ means the time ordering  of matrix under the exponent sign,
$v({\bf x})$ are initial data of Eq.~(\ref{lc3-9phn}).
These gauge transformations keep the action due to its gauge invariance.
\bea\label{lc5-9a}
W^*=W[A^L_\mu].
 \eea
  Such variable change is known as the \textit{choice of a gauge} \cite{fp1-9}.
 The choice of gauge changes the Feynman rules. One can choose a \textit{gauge} for which
 the Feynman rules completely do not depend on initial data \cite{f1-9}.
 However, in the generating functional  (\ref{9-2-3}),
 there are sources of the transversal fields $S[J^*,\eta^*, \overline{\eta^*}]$
 that depend on initial data $v({\bf x})$ of Eq. (\ref{lc3-9phn}).

The assertion about the \textit{gauge}-independence of the physical contents of the generating functional
that removes any initial data is called the Faddeev theorem \cite{f1-9}.
To prove this theorem one has to be convinced also that the sources of the radiation
variables can be replaced by the  sources of the fields in the Lorentz \textit{gauge}
formulation
\bea\nonumber
\!\!S[J^*,\eta^*, \overline{\eta^*}]&\!=\!&\int d^4x \left[J^{*c}_k A^{*c}_k[A^{L b}_j]
+\overline{\eta^*}\psi^*[A^{L b}_j,\psi^L] + \overline{\psi^*}[A^{L b}_j,\psi^L]\eta^*\right]\\\label{lc5-9b}
~&\!\to\!& ~
S[A^L_\mu,\psi^L,\overline{\psi}^L],
 \eea
where $S[A^L_\mu]$ is given by Eq.~(\ref{fp-42}).
Actually,  this theorem was proved in paper \cite{f1-9} only for the scattering
processes of the elementary particles in QED.

However, as we saw above, this \textit{change of a gauge} can disturb  the bound state
spectrum. In any case, questions arise about the range of validity
of  the Faddeev theorem \cite{f1-9}.
In the next chapters we shall return to these questions
and discuss the status of the primordial initial data and their physical
effects.

\section{Summary}
\makeatletter
\renewcommand{\@evenhead}{\raisebox{0pt}[\headheight][0pt]
{\vbox{\hbox to\textwidth{\hfil \strut \thechapter . Creation of matter in the Universe \quad\rm\thepage}\hrule}}}
\renewcommand{\@oddhead}{\raisebox{0pt}[\headheight][0pt]
{\vbox{\hbox to\textwidth{\strut \thesection .\,
Summary and literature
\hfil \rm\thepage}\hrule}}}
\makeatother

The gravitation theory treated as the joint nonlinear realization of affine and conformal symmetry groups
in terms of the diffeoinvariant Cartan forms
allows us to quantize these forms.  This quantum gravitation theory is unified with
the Standard Model of the elementary particles
on the equal footing in the form of the wave function of the Universe as
a joint irreducible unitary representation (IUR) of the
affine and conformal symmetry groups constructed in the previous Chapter 8. Results given
in Chapter 9 testify that this wave function of the Universe yields the classification
of real physical processes in the Universe, including its creation from vacuum together with
its matter content in agreement with observational and experimental data.

In this Chapter we present theoretical and observational arguments in favor of the fact that there was the Beginning
and there arose from vacuum $10^{88}$ primordial Higgs particles at the Beginning
 during the first $10^{-12}$ sec. All matter content of the present day Universe can be
 the decay products of these primordial bosons. In this case, one can expect that
 the wave function of the Universe yields classification of the real processes in the
 Universe at present time. In any case,
we reminded the Poincar$\acute{\rm e}$ classification of the particle states including the bound ones
and obtained relation of our approach to QFT with the accepted Faddeev -- Popov path integral.
The next Chapter will be devoted to the status of the initial data in QCD.

\chapter{Reduced phase space  QCD}
\renewcommand{\theequation}{10.\arabic{equation}}
\setcounter{equation}{0}
\section{Topological confinement}
\makeatletter
\renewcommand{\@evenhead}{\raisebox{0pt}[\headheight][0pt]
{\vbox{\hbox to\textwidth{\hfil \strut \thechapter . Reduced phase space  QCD
\quad\rm\thepage}\hrule}}}
\renewcommand{\@oddhead}{\raisebox{0pt}[\headheight][0pt]
{\vbox{\hbox to\textwidth{\strut \thesection .\quad
Topological confinement
\hfil \rm\thepage}\hrule}}}
\makeatother

At the beginning of the sixties of the twentieth century Feynman found that
the naive generalization of his path integral method (\ref{fp-1c}) of  construction of QED fails in the
non-Abelian theories. The unitary S-matrix in the non-Abelian theory was
obtained in the form of the Faddeev -- Popov (FP) path integral by the brilliant application of
the theory of connections in vector bundle (see \cite{fp1-9} in the previous Chapter). Many physicists are of
opinion that the FP path integral is the highest level of quantum description
of the gauge constrained systems. Really, the FP integral at least allows us
to prove  both renormalizability of the unified theory of electroweak
interactions and asymptotic freedom of the non-Abelian theory.
However, the FP integral still remains  serious and challenging problems
 of the quark confinement (in the form of the Feynman quark-hadron duality),
spontaneous chiral symmetry breaking and hadronization. These problems require
generalization of the FP path integral to the bound states in the non-Abelian
theories in their frame of reference, including corresponding initial data.

In this Chapter, we consider these problems in the context of the Quantum Universe (QU) modification of QCD
described in the previous Chapter.
The QU modification of QCD means  QCD in the reduced phase space obtained by
the manifest resolution of all
constraints in a concrete frame of reference with initial data.
 The choice of the vacuum
and normal ordering of the field products leads to  a source of spontaneous conformal symmetry
breaking in the quantum theory.

The role of constraints is played by the equations of the time component.
In the YM theory (\cite{Gitman-10}, \S 16) the time component
of a Yang-Mills field occupies a particular place, since it has no canonical
momentum.  Dirac \cite{Dir-10} and other
authors of the first classical theories on quantization of gauge fields
\cite{Heisenberg-10,Fermi-10}, who followed him, removed the time component of gauge
fields by gauge transformation. In our case, the similar transformation is
\bea\label{lc1-10-1}
\hat A^{*}_k[A^b_j]&=&u^*[A^b_j]\left(\hat A_k+\partial_k\right){u^*}^{-1}[A^b_j],\\\label{lc2-10-1}
 \psi^*[A^b_j,\psi]&=&u^*[A^b_j]\psi,\\\label{lc3-9-1}
 u^*[A^b_j]&=& v({\bf x}) T\exp\left\{\int\limits_{}^{t} d\overline{t}\hat a_0[A^b_j]\right\},
 \eea
 where the symbol $T$ denotes the time ordering of the matrices under the exponent sign.
It determines a non-Abelian analogue of the Dirac variables
(see also \cite{Pervush1-10}, Section 2.2)  within arbitrary stationary matrices $v({\bf x}$
considered as initial data of the solution to the equation (\ref{lc3-9phn})
\bea\label{lc4-10-1}
 [((\nabla_j(A^*))^2]^{cb} a^b_0=
\nabla^{cb}_i(A^*)\dot A^{*b}_i=0
\eea
 at the time instant $t_0$.  Here $\hat a_0[A^b_j]$ is the solution of the Gauss constraint.
One can see that at the level of Dirac variables the Lorentz
transformations of initial fields become nonlinear (formula (28) in \cite{Pervush1-10}),
and the group of gauge transformations reduces to the group of
stationary transformations, which set the  degeneration of the initial data of physical
fields (including the classical vacuum $A_0=A_i=0$, determined as a state of
zero energy).
 We imply as a gauge fixation, in the given case, the putting
of the initial data  in perturbation theory as a transversality condition
\cite{Pervush1-10, Zumino-10}. \par
In non-Abelian theories a set of
stationary gauge transformations is the set of three-dimensional paths in
the group space of  the Lie group $SU_c(2)$ subdivided into  topological manifolds.
These manifolds are determined by integers, {\it degrees of the map}  \cite{Belavin-1975-10}:
\bea   \label{degree}
n=
- \frac {1}{24\pi ^2}\int d^3 x \epsilon ^{ijk}\times
\eea
$$\!\!\times
Tr [v^{(n)}(
{\bf x})\partial_i v^{(n)}({\bf x})^{-1} v^{(n)}({\bf x})\partial_j
v^{(n)}({\bf x})^{-1} v^{(n)}({\bf x})\partial_k v^{(n)}({\bf x}
)^{-1}].
$$
The degree of the map shows how many times a  three-dimensional path $v(\mathbf{x})$
 turns around the $SU_c(3)$ manifold when the coordinate $x_i$
runs over the space where it is given. The condition (\ref{degree}) means
that all the sets of three-dimensional paths have the homotopy group
 $$\pi_3(SU_c(3)) ={\bf Z},$$ and all the fields $v^{(n)}\partial_i
v^{(n)-1}$ are given in the class of functions for which the
integral (\ref{degree}) has a finite (or countable) value
\be
\label{degeneration1}
{\hat A}^{(n)}_i= v^{(n)} ({\hat A}_i^{ (0)}+
\partial _i)v^{(n)-1},\qquad v^{(n)}({\bf x})=
\exp [n\Phi _0({\bf x})]~.
\ee
Due to the gauge invariance of the action,
the phase factors of the topological degeneration
 disappear. However,
these phase factors remain at the sources of physical
fields in the generating functional. 
The theory with the topological degeneration
of the initial data, where the sources contain the phase
factors of the topological degeneration, $$tr [\hat J^i v^{(n)}\bar{\hat A}_i^{(0)}v^{(n)-1}],$$
differs from the theory without  the degeneration and
with the sources $tr[\hat J^i\bar{\hat A}^{(0)}_i]$. In
the theory with  the  degeneration of the initial data
it is necessary to average the amplitudes over the
degeneration parameters. Such averaging can lead \it to the
disappearance of  a series of physical states\rm. \par

 It has been shown in  \cite{Pervush4-10,Nguen2-10} that the topology can be the origin of
 color confinement as complete destructive interference of
 the phase factors of the topological degeneration of initial data.

 The mechanical analogy of the topological degeneration of initial data
 is the free rotator $N(t)$ with the action of free particle
 \be\label{rot}
 W(N_{out},N_{in}|t_1)=\int\limits_{0 }^{t_1 } dt \frac{\dot N^2}{2}I,
 ~~p=\dot N I,~~H_0=\frac{p^2}{2I}
 \ee
 given on a ring where the points $N(t)+n$
 ($n$ is integer) are physically equivalent.
 Instead of a initial date $N(t=0)=N_{in}$ in the mechanics in the
 space with the trivial topology, the observer of the rotator
 has the manifold of initial data
 $$N^{(n)}(t=0)=N_{in}+n;~~~~~~n=0,\pm 1,\pm  2,\ldots .$$

 An observer does not know where is the rotator. It can be at points
 $$N_{in},~~~~N_{in}\pm 1,~~~~N_{in}\pm 2,~~~~N_{in}\pm 3,\ldots .$$
 Therefore, he
 should to average a wave function
 $$
 \Psi (N)=e^{\imath pN}~.
 $$
 over all values of the topological degeneration
 with the $\theta$-angle measure $\exp (\imath\theta n)$. In the result
 we obtain the wave function
 \be\label{wf}
 \Psi (N)_{\rm observable}=\lim\limits_{L \to \infty}\frac{1}{2L}
 \sum\limits_{n=-L }^{n=+L }
 e^{\imath\theta n}\Psi (N+n)=\exp\{\imath (2\pi k+\theta)N\}~,
 \ee
 where $k$ is integer. In the opposite case $p\not = 2\pi k + \theta$
 the corresponding wave function ({\it id est}, the probability amplitude) disappears
  $$\Psi(N)_{\rm observable}=0$$
  due to the complete destructive interference.

 The consequence of this topological degeneration is that
 a part of values of the momentum spectrum becomes unobservable in
 the comparison with the trivial topology.

 This fact can be treated as confinement of those values which do
 not coincide with the discreet ones
 \be\label{spc}
 p_k=2\pi k + \theta, ~~~~~ 0\leq \theta \leq \pi~.
 \ee
 The observable spectrum follows also
 from the constraint of the equivalence
 of the point $N$ and $N+1$
 \be\label{wf1}
 \Psi (N)=e^{\imath\theta}\Psi (N+1), ~~~~~~\Psi (N)=e^{\imath pN}~.
 \ee
 In the result we obtain the spectral decomposition
 of the Green function of the free rotator~(\ref{rot})
 (as the probability amplitude of transition from the point $N_{in}$
 to $N_{out}$) over the observable values of spectrum~(\ref{spc})
 \bea \label{rotgf}
 G(N_{out},N_{in}|t_1)&\equiv& <N_{out}|\exp(-\imath\hat Ht_1)|N_{in}>=\\\nonumber
 &=& \frac{1}{2\pi}\sum\limits_{k=-\infty }^{k=+\infty }\exp\left[
 -\imath \frac{p_k^2}{2I}t_1+\imath p_k(N_{out}-N_{in})\right]~.
 \eea
 Using the connection with the Jacobian theta-functions~\cite{poll-10}
 $$
 \Theta_3 (Z|\tau)=\sum\limits_{k=-\infty }^{k=+\infty }\exp[\imath\pi k^2\tau
 +2\imath kZ]=(-\imath\tau)^{-1/2}\exp[\frac{Z^2}{\imath\pi\tau}]\Theta_3\left(
 \frac{Z}{\tau|-\frac{1}{\tau}}\right)
 $$
 we can represent expression~(\ref{rotgf}) in a form of the sum over
 all paths
 \be \label{rotgfp}
 G(N_{out},N_{in}|t_1)
 = \sqrt{\frac{I}{(\imath 4\pi t_1)}}\sum\limits_{n=-\infty }^{n=+\infty }
 \exp[\imath\theta n]\exp\left[+\imath W(N_{out},N_{in}+n|t_1)\right]~,
 \ee
 where
 $$
 W(N_{out}+n,N_{in}|t_1)=\frac{(N_{out}+n-N_{in})^2I}{2t_1}
 $$
 is the rotator action~(\ref{rot}).

\section{Quark--hadron duality}
\makeatletter
\renewcommand{\@evenhead}{\raisebox{0pt}[\headheight][0pt]
{\vbox{\hbox to\textwidth{\hfil \strut \thechapter . Reduced phase space  QCD
\quad\rm\thepage}\hrule}}}
\renewcommand{\@oddhead}{\raisebox{0pt}[\headheight][0pt]
{\vbox{\hbox to\textwidth{\strut \thesection .\quad
Quark--hadron duality
\hfil \rm\thepage}\hrule}}}
\makeatother

 All physical states and the Green functions should be averaged over
 all topological  copies  in the
 group space. Averaging over all parameters of the degenerations
 can lead to a complete destructive interference of all color
 amplitudes~\cite{Pervush4-10,Nguen2-10}. In this case, only colorless
 (``hadron'') states have to form a complete set of physical states.
 Using the example of a free rotator,
 we have seen that the disappearance of a part of physical states
 due to the topological degeneration (confinement) does not violate
 the composition law for Green functions
  \be \label{compos}
 G_{ij}(t_1,t_3) =\sum\limits_{h }^{ } G_{ih}(t_1,t_2)G_{hj}(t_2,t_3)
 \ee
 defined as the amplitude of the probability to find a system with the
 Hamiltonian $H$
 in a state $j$ at the time $t_3$, if at the time $t_1$  this system was
 in a state $i$, where $(i,j)$ belongs to a complete set of all states
 $\{h\}$:
 $$G_{ij}(t_1,t_3)=<i|\exp\left(-\imath\int\limits_{t_1 }^{t_3 }H\, dt\right)|j>
 $$
 The particular case of this composition law~(\ref{compos}) is the unitarity
 of S-matrix
 $$
 SS^+=I~\Rightarrow~\sum\limits_{h }^{ }<i|S|h><h|S^+|j>=<i|j>~
 $$
 known as the law of probability conservation for S-matrix elements
$$S=I+iT:$$
 \be \label{qhd}
 \sum\limits_{h }^{ } <i|T|h> <h|T^*|j> =2 {\rm Im}<i|T|j>~.
 \ee
 The left side of this law is the analogy of the spectral
 series of the free rotator~(\ref{rotgf}).
 The destructive interference keeps only colorless ``hadron'' states.
 Whereas, the right side of this law
 far from resonances can presented  in a form of the perturbation
 series over the Feynman diagrams that follow from the Hamiltonian.
 Due to gauge invariance  $H[A^{(n)},q^{(n)}]=H[A^{(0)},q^{(0)}]$
 this Hamiltonian  does not depend on the topological
  phase factors and it contains the perturbation series in terms
 of only the zero - map  fields ({id est}, in terms of constituent color
 particles) that can be identified with the Feynman partons.
 The Feynman path integral as the generation functional
 of this perturbation series is the analogy of the sum over
 all path of the free rotator~(\ref{rotgfp}).

 Therefore,  confinement in the spirit of the
 complete destructive interference of color
 amplitudes \cite{Pervush4-10,Nguen2-10} and
 the law of probability conservation for S-matrix elements~(\ref{qhd})
 leads to
 Feynman quark-hadron duality, that is foundation
 of all the parton model~\cite{feynman-10}
 and QCD application~\cite{efremov-10}. The quark-parton
 duality gives the method of direct experimental measurement
 of the quark and gluon quantum numbers from the deep inelastic
 scattering cross-section~\cite{feynman-10}.
 For example, according to Particle Data Group
 the ratio of the sum of the probabilities of $\tau$-decay
 hadron modes to the probability of $\tau$-decay muon mode is
 $$
 \frac{\sum\limits_{h }^{ }w_{\tau \to h}}{w_{\tau \to \mu}}=3.3\pm 0.3~.
 $$
 This is the left-hand side of Eq.~(\ref{qhd}) normalized to the value of the
 lepton mode probability of $\tau$-decay. On the right-hand side of
 Eq.~(\ref{qhd}) we have the ratio of the imaginary part of the sum of
 quark-gluon diagrams (in terms of constituent fields free from the topological
 phase factors) to the one of the lepton diagram. In the lowest order of QCD
 perturbation on the right-hand side we get the number of colors $N_c$ and,
 therefore,
 $$
 3.3\pm 0.3=N_c~.
 $$
 Thus in the constraint-shell QCD we can understand not only
 ``why we do not see quarks'', but also ``why we can measure their
 quantum numbers''.
 This mechanism
 of confinement due to the quantum interference of phase factors (revealed
 by the explicit resolving the Gauss law constraint~\cite{Pervush4-10,Nguen2-10})
 disappears after the change of ``physical'' sources
 $A^*J^*~\Rightarrow~A J$ that
 is called the transition to another gauge in the gauge-fixing method.

\section{Chiral symmetry breaking in QCD
}
\makeatletter
\renewcommand{\@evenhead}{\raisebox{0pt}[\headheight][0pt]
{\vbox{\hbox to\textwidth{\hfil \strut \thechapter . Reduced phase space  QCD
\quad\rm\thepage}\hrule}}}
\renewcommand{\@oddhead}{\raisebox{0pt}[\headheight][0pt]
{\vbox{\hbox to\textwidth{\strut \thesection .\quad
Chiral symmetry breaking in QCD
\hfil \rm\thepage}\hrule}}}
\makeatother

Instantaneous QCD interactions are described by the non-Abelian generalization
of the Dirac gauge in QED
The relativistic invariant bilocal effective action obtained in \cite{Pervush1-10},
takes  the form for the quark sector, in the color singlet channel
\bea\label{1-1-7} W_{\rm instant}  &=& \int d^{4} x  \bar {q}(x) \; ( \imath
\rlap/\partial - {\hat {m}}^{0} ) {q}(x)-
\eea
$$
-{ 1 \over {2}}
\int d^4x d^4y j^a_0(x)\left[\frac{1}{(\nabla_j(A^*))^2}\delta^4(x-y)\right]^{ab} j^b_0(y),
$$
where
$$j^a_0(x)=\bar q(x)\dfrac{\lambda^a}{2}\gamma_0 q(x)$$
is the 4-th component of the quark current,
with the Gell-Mann color matrices $\lambda^a$
(see the notations in Appendix A).
For  simplicity indices $(1,1'|2,2')$ denote in (1) all spinor, color and flavor
ones.
The symbol
$$
{\hat {m}}^{0} = \text{diag}(m_{u}^{0}, m_{d}^{0},m_{s}^{0})
$$
denotes the bare quark mass matrix.
The normal ordering of the transverse gluons in the nonlinear action  (\ref{1-1-7}) 
$$\nabla^{db}A^b_0\nabla^{dc}A^c_0$$
leads to the condensate of gluons
\bea\label{c-1g}
\!\! g^2f^{ba_1d}f^{da_2c}\langle
 A^{a_1*}_iA^{a_2*}_j\rangle =2g^2 N_c \delta^{bc} \delta_{ij}C_{\rm gluon}=M_g^2
 \delta^{bc}\delta_{ij},\eea
where
 \bea\label{c-1gc}
\langle A^{*a}_jA^{*b}_i\rangle&=&2C_{\rm gluon}\delta_{ij}\delta^{ab}.
 \eea
This condensate yields the squared effective gluon mass in  the squared covariant derivative
$$\nabla^{db}A^b_0\nabla^{dc}A^c_0=:\nabla^{db}A^b_0\nabla^{dc}A^c_0: + M_g^2 A^d_0A^d_0.$$
The constant
$$C_{\rm gluon}=\int\dfrac{ d^3k}{(2\pi)^32\sqrt{{\bf k}^2+M_g^2}}$$
is finite after substraction of the infinite volume contribution,
and its value is determined by the hadron size like
the Casimir vacuum energy. 
Finally, in the lowest order of perturbation theory,
this gluon condensation  yields  the effective Yukawa potential in
the colorless meson sector \cite{Belkov-94-10}
\be\label{Y-1}
\underline{V}({\mathbf{k}})=\frac{4}{3}
\frac{g^2}{\mathbf{k}^2+M_g^2}
\ee
and the NJL type model with the effective gluon mass $M_g^2$.
While deriving the last equation, we use the relation
\begin{eqnarray*}
\left[\sum\limits_{a=1}^{a=N_c^2-1}  \frac{\lambda^{a}_{1,1'}}{2}
\frac{\lambda^{a}_{2,2'}}{2}\right]=\frac{1}{2}\delta_{1,2'}\delta_{2,1'}-
\frac{1}{6}\delta_{1,1'}\delta_{2',2'}.
\end{eqnarray*}
The product of this expression by the unit matrix $\delta_{1,2'}$ and summation yield
the coefficient $3/2-1/6=4/3$ in front of the Yukawa potential (\ref{Y-1}).
Below we consider the potential model (\ref{Y-1}) in the form
\bea\label{1-1q}
W_{\rm instant}[q, \bar{q}]  &=& \int d^{4} x  \bar {q}(x) \; ( \imath
\rlap/\partial - {\hat {m}}^{0} ) {q}(x)-
\eea
$$
-{ 1 \over {2}}
\int d^4x d^4y j^a_{\ell}(x)V(x^{\bot}-y^{\bot}) \delta ((x-y)\cdot \ell) j^a_{\ell}(y)\equiv
$$
$$\equiv
\imath \left(q \bar {q}, G_0^{-1}\right)-\frac{1}{2}\left(q \bar{q}, K q \bar{q}\right),
$$
 with the  choice of the time axis as the eigenvalues of the bound state total momentum. This model can
 formulate  the effective action in terms of bound state bilocal fields given in Appendix B.
 In this case the semiclassical approach repeats   the ladder approximation.

 In particular, the equation  of stationarity (\ref{8}) coincides
  with the   Schwinger -- Dyson (SD) equation
 \begin{eqnarray}\label{sd}
 \Sigma(x-y) = m^{0} \delta^{(4)} (x-y) + \imath {\cal K}(x,y)
 G_{\Sigma}(x-y)~.
 \end{eqnarray}
It describes the spectrum of Dirac particles in bound states. In the momentum space  with
$$
 \underline{\Sigma}(k) = \int d^{4}x \Sigma(x) e^{\imath k\cdot x}
$$
for the Coulomb type kernel, we obtain the following equation for  the mass operator
 $ \underline {\Sigma} $
 \begin{eqnarray} \label{sdm}
 \underline{\Sigma}(k) = m^{0} - {\imath} \int { d^{4}q \over { (2\pi)^{4}
 }} \underline {V} ( k^{\perp} - q^{\perp} ) \rlap/\ell
 \underline{G}_{\Sigma}(q) \rlap/\ell ,
 \end{eqnarray}
where
$$ \underline{G}_{\Sigma}(q) \equiv ( \rlap/q - \underline{\Sigma}(q))^{-1}$$
is the Fourier representation of the  potential,
$$ k^{\perp}_{\mu} = k_{\mu} - \ell_{\mu} ( k \cdot \ell) $$
is the relative transverse momentum.
The quantity $\underline{\Sigma}$  depends only on the transverse momentum
$\underline{\Sigma}(k) = \underline{\Sigma}(k^{\perp})$,
because of the instantaneous form of the potential $\underline{V}(k^{\perp} )$. We can put
\begin{eqnarray} \label{sd21}
 \underline{\Sigma}_{\mathrm{a}}(q) = 
   E_{\mathrm{a}} ({\bf q})\cos 2 {\upsilon}_{\mathrm{a}}({\bf q})\equiv M_{{\mathrm{a}}}({\bf q}).
 \end{eqnarray}
Here
 $M_{\mathrm{a}}({\bf q})$ is the constituent quark mass and
 \begin{eqnarray} \label{sd3}
  \cos 2 {\upsilon}_{\mathrm{a}}({\bf q})&=&\frac{M_{\rm a}(\bf q)}{\sqrt{M^2_{\rm a}(\bf q)+{\bf q}^2}}
 \end{eqnarray}
determines  the Foldy -- Wouthuysen type  matrix
 \bea\label{s-1}
\!\!\!\! S_{\mathrm{a}}({\bf q})=\exp [\imath({\bf q}\bm{\gamma}/q){\upsilon}_{\mathrm{a}}({\bf q})]=
 \cos {\upsilon}_{\mathrm{a}}({\bf q})+ \imath({\bf q}\bm{\gamma}/q)\sin {\upsilon}_{\mathrm{a}}({\bf q}),
 \eea
with  the vector of Dirac matrices
$\bm{\gamma}=(\gamma_1,\gamma_2,\gamma_3)$
 lying in the range $0\leqslant {\upsilon}_{\mathrm{a}}(q)\leqslant \pi/2$,
  and  ${\upsilon}_{\mathrm{a}}({\bf q})$ is the  Foldy -- Wouthuysen angle.
The fermion spectrum can be obtained by solving the SD equation (\ref{sdm}).
It can be integrated over the longitudinal momentum $q_{0}= (q \cdot  \ell) $
in the reference frame $\ell^{0}=(1,0,0,0)$, where $q^{\perp}=(0,\mathbf{q})$.
By using Eq.~(\ref{s-1}), the quark Green function can be presented in the form
 \begin{eqnarray} \label{sd4}
 \underline{G}_{\Sigma_{\mathrm{a}}} &=&
 [ q_{0} \rlap/\ell - E_{\mathrm{a}}(q^{\perp}) S_{\mathrm{a}}^{-2}(q^{\perp})]^{-1} = \nonumber \\
 & = & \left[ { \Lambda^{(\ell)}_{(+)\mathrm{a}} (q^{\perp}) \over { q_{0} -
  E_{\mathrm{a}}(q^{\perp}) +\imath \epsilon} } +
 { \Lambda^{(\ell)}_{(-)\mathrm{a}} (q^{\perp}) \over
 { q_{0} + E_{\mathrm{a}}(q^{\perp}) +\imath \epsilon} } \right] \rlap/\ell,
 \end{eqnarray}
where
\begin{eqnarray} \label{ope1}
\!\!  \Lambda^{(\ell)}_{(\pm)\mathrm{a}}(q^{\perp})= S_{\mathrm{a}}(q^{\perp})
 \Lambda^{(\ell)}_{(\pm)}(0) S_{\mathrm{a}}^{-1}(q^{\perp}) , \,\,\,
 \Lambda^{(\ell)}_{(\pm)}(0)= ( 1 \pm \rlap/\ell ) / 2
  \end{eqnarray}
are the operators separating the states with positive
($ + E_{\mathrm{a}} $) and negative ($ - E_{\mathrm{a}} $) energies.
As a result, we obtain the following equations for the one-particle energy $ E $
  and the angle $ {\upsilon} $ (\ref{sd3}) with the potential given by Eq.~(\ref{Y-1})
 \bea\label{sd1}\nonumber
E_{\mathrm{a}}(k^{\perp}) \cos 2 {\upsilon}_{\mathrm{a}}(k^{\perp}) = m^{0}_{\mathrm{a}} +
 {1\over2}\int { d^{3}q^{\perp}\over (2\pi)^{3} }
 \underline{V}(k^{\perp}-q^{\perp}) \cos 2{\upsilon}_{\mathrm{a}}(q^{\perp}).
\eea
In the rest frame $\ell^{0}=(1,0,0,0)$ this equation takes the form
\bea\label{sd2}
M_{\mathrm{a}}({\bf k})&=&
m^{0}_{\mathrm{a}} +
 {1\over2}\int { d^{3}q\over (2\pi)^{3} }
 \underline{V}({\bf k}-{\bf q}) \cos 2{\upsilon}_{\mathrm{a}}({\bf q}).
  \eea
By using the integral over the solid angle
$$\int_{0}^{\pi} d\vartheta \sin\vartheta \dfrac{2\pi}{M_g^2+({\bf k}-{\bf q})^2}
=\int\limits_{-1}^{+1} d\xi\dfrac{2\pi}{M_g^2+{k}^2+{q}^2-2{k}{q}\xi}=
$$
$$
=\frac{\pi}{kq}\ln \frac{M_g^2+(k+q)^2}{M_g^2+(k-q)^2}$$
and the definition of the QCD coupling constant $\alpha_s=4\pi g^2$, it can be rewritten as
 \bea\label{sd-2a}
\!\!M_{\mathrm{a}}(k)=m^{0}_{\mathrm{a}} + \frac{\alpha_s}{3\pi k}\int\limits_{0}^{\infty}
 d q \frac{q M_{\mathrm{a}}(q)}{\sqrt{M^2_{\mathrm{a}}(q)+q^2}}\ln
 \frac{M_g^2+(k+q)^2}{M_g^2+(k-q)^2}.
 \eea
The suggested scheme allows us to consider the SD equation (\ref{sd2}) in the limit when
 the bare current mass $m^{0}_{\mathrm{a}}$ equals to zero \cite{5856-10}.
 Then the ultraviolet divergence is absent,
and, hence, the renormalization procedure can be successfully avoided.
This kind of nonlinear integral equations was considered in the paper \cite{puz-10} numerically.

The solution for
$M_{\mathrm{a}}(q)$  in the separable approximation \cite{a20-10}
in the form of the step function was used for the estimation of
 the
quark and meson spectra in agreement with the experimental data.
Currently, numerical solutions of the nonlinear  equation (\ref{sd-2a}) are under way, and the details of
computations will be published elsewhere.

As discussed in the Appendix B, the SD equation (\ref{sd2}) can be rewritten in the form (\ref{sd-2a}).
 Once we know the solution of Eq.~(\ref{sd-2a}) for $M_{\mathrm{a}}(q)$, we can determine
 the Foldy -- Wouthuysen angles $\upsilon_{\mathrm{a}},(\mathrm{a}=u,d)$ for u-,d- quarks with
 the help of relation (\ref{sd3}). Then the BS equations in the form 
\be\label{BS-1}
 M_{\pi}L^{\pi}_{2}(\textbf{p})=[E_u({\bf p})+E_d({\bf p})]L^{\pi}_{1}(\textbf{p})-
\ee
$$
\int\!\frac{d^3q}{(2\pi)^3}\underline{V}(\textbf{p}\!-\!\textbf{q})
 L^{\pi}_{1}(\textbf{q})[c^{-}({\bf p})c^{-}({\bf q})\!+\!\xi s^{-}({\bf p})s^{-}({\bf q})],
$$
$$
 M_\pi L^{\pi}_{1}(\textbf{p})=[E_u({\bf p})+E_d({\bf p})]L^{\pi}_{2}(\textbf{p})-
$$
$$
 \int\!\frac{d^3q}{(2\pi)^3}\underline{V}(\textbf{p}-
 \textbf{q})L^{\pi}_{2}(\textbf{q})[c^{+}({\bf p})c^{+}({\bf q})\!+\!\xi s^{+}({\bf p})s^{+}({\bf q})]
$$
yield the pion mass $M_\pi$ and wave functions $L^{\pi}_{1}(\textbf{p})$ and $L^{\pi}_{2}(\textbf{p})$.
Here $m_u,m_d$ are the current quark masses,
$$E_{\mathrm{a}}=\sqrt{{\bf p}^2+M^2_{\mathrm{a}}({\bf p})},~~~~~~~ (\mathrm{a}=u,d)$$
are the u-,d- quark energies,
$\xi=(\textbf{p\,q})/pq$, and we use the notations
 \bea
E({\bf p})&=&E_{\mathrm{a}}({\bf p})+E_{\mathrm{b}}({\bf p})~,\label{E-1a}\\
{\tt c}^{\pm}({\bf p})&=&\cos[\upsilon_{\mathrm{a}}({\bf p}) \pm \upsilon_{\mathrm{b}}({\bf p})]~,\label{E-2a}\\
{\tt s}^{\pm}({\bf p})&=&\sin[\upsilon_{\mathrm{a}}({\bf p}) \pm \upsilon_{\mathrm{b}}({\bf p})]~.\label{E-3a}
\eea
The model is simplified in some limiting cases. Once the quark masses
$m_u$ and $m_d$ are small and approximately equal, Eqs.~(\ref{sd2}) and
(\ref{BS-1}) take the form
\begin{align}\label{sd-3u}
 m_{\mathrm{a}}&=M_{\mathrm{a}}({\bf p})-\frac{1}{2}\int\frac{d^3q}{(2\pi)^3}\underline{V}(\textbf{p}
 -\textbf{q})
 \cos 2{\upsilon}_u({\bf q}),\\
\label{bs-3u}
 \frac{M_{\pi}L^{\pi}_{2}(\textbf{p})}{2}&=E_u({\bf p})L^{\pi}_{1}(\textbf{p})\!-\!
 \frac{1}{2}\!\int\!\frac{d^3q}{(2\pi)^3}\underline{V}(\textbf{p}\!-\!\textbf{q})
 L^{\pi}_{1}(\textbf{q}).
\end{align}
Solutions of equations of this type are  considered in the numerous
papers~\cite{a4-10, a5-10, a6-10, a7-10, a10-10}
(see also review~\cite{puz-10}) for different potentials.
 One of the main results of these papers was the pure quantum effect of spontaneous chiral symmetry breaking.
  In this case, the instantaneous interaction leads to rearrangement of the perturbation series and strongly
  changes the spectrum of elementary excitations and bound states in contrast to the naive perturbation theory.

In the limit of massless quarks $m_u=0$ the left-hand side of Eq.~(\ref{sd-3u})
is equal to zero. The nonzero solution of Eq.~(\ref{sd-3u}) implies that
  there exists a mode with zero pion mass $M_{\pi}=0$
 in accordance with the Goldstone theorem.
 This means that the BS equation (\ref{bs-3u}), being the equation for the wave function of the Goldstone pion,
 coincides
with the the SD equation (\ref{sd-3u}) in the case of $m_u=M_{\pi}=0$.
Comparison of the equations yields
 \bea\label{bs-4a}
 L^{\pi}_{1}({\bf p})=\frac{M_u({\bf p})}{\sqrt{2}F_\pi E_u({\bf p})}=
 \frac{\cos2\upsilon_u({\bf p})}{\sqrt{2}F_\pi},
 \eea
where the constant of the proportionality $F_\pi$ in Eq. (\ref{bs-4a}) is
called the  weak decay constant. In the more general case of massive quark
$m_u\neq 0,\, M_{\pi}\neq0$, this constant is determined from the normalization condition
(\ref{nakap})
\bea
  \label{1fpi2}
\!\!\!\!  1 &=&\frac{4 N_c}{M_\pi} \int\limits_{ }^{ } \frac{d^3q}{(2\pi)^3} L_2 L_1=
 \frac{4 N_c}{M_\pi} \int\limits_{ }^{ } \frac{d^3q}{(2\pi)^3} L_2 \frac{\cos2\upsilon_u({\bf p})}{F_\pi}
  \eea
with $N_c=3$.
In this case the wave function $L^{\pi}_{1}(p)$ is proportional
to the  Fourier component of the quark condensate
 \bea\label{1fpi2q}
 C_{\rm quark}=\sum_{n=1}^{n=N_c}\langle q_n(t,{\bf x})\overline{q}_n(t,{\bf y})\rangle=
\eea
$$=
 4N_c\int \frac{d^3 p}{(2\pi)^3}\frac{M_u({\bf p})}{2\sqrt{{\bf p}^2+M^2_u({\bf p})}}\,.
$$
Using Eqs.~(\ref{sd3}) and (\ref{bs-4a}), one can rewrite the definition of the quark condensate (\ref{1fpi2q})
 in the form
\bea \label{gmor1}
 C_{\rm quark}&=&{4 N_c} \int\frac{d^3q}{2(2\pi)^3}  \cos2\upsilon_u({\bf q}).
\eea
Let us assume that the representation for the wave function $L_1$ (\ref{bs-4a})  is still valid
for non-zero but small quark masses. Then
the subtraction of the BS equation (\ref{bs-3u}) from the SD one (\ref{sd-3u})
multiplied by the factor $1/F_\pi$ determines the second meson wave function
$L_2$
 \bea\label{bs-u}
 \frac{M_{\pi}}{2}L^{\pi}_{2}({\bf p})&=&
\frac{m_u}{\sqrt{2}\,F_\pi}.
 \eea
The wave function $L^{\pi}_{2}(p)$ is independent of the
momentum in this approximation. Substituting the equation
$$L_2=\frac{2m_u}{\sqrt{2}\,M_\pi F_\pi}=\text{const}$$
 into the normalization condition (\ref{1fpi2}), and using
Eqs.~(\ref{bs-4a}) and (\ref{gmor1}), we arrive at the Gell-Mann -- Oakes -- Renner (GMOR)
relation \cite{gmor-10}
  \bea \label{gmor}
 M^2_{\pi}F^2_{\pi}&=&{2m_u} C_{\rm quark}\,.
\eea
Thus, in  the framework of instantaneous interaction we prove the Goldstone theorem in the bilocal variant,
and the GMOR relation directly results from the existence of the gluon and quark condensates.
Strictly speaking,
the postulate that the finiteness of the  quark condensate is finite implies that
in QCD the ultraviolet  divergence can be removed. 
Here one can remind that the ultraviolet  divergence in the
observational Casimir energy of photons between two metal plates
in one dimensional space is removed by the  Bose - distribution function
with the temperature as the inverse free length \cite{Kaz-77-10}.
In this case, it is well known that the spectrum of
the vacuum oscillations coincides with the spectrum of   an  absolute black body
with the energy density
\bea\label{gmor-2}
\rho^{1}_{\rm Cas}= \frac{1}{2L}\sum\limits_{n=0}^{\infty}\omega_n=
\frac{1}{\pi}\int\limits_{0}^{\infty} \frac{d\omega \omega}{1-e^{2L\omega}}=-\frac{\pi}{24L^2}
\eea
where $\omega_n=\pi n/ L, \, n=1, 2,...$ is the complete set of the one-photon energy
in the  space between two metal plates of a size $L$.

We suppose that the energy of the vacuum oscillations of fermions in the Dirac sea
is suppressed by the  Fermi - distribution function (see Appendix C)
\be\label{pich-1}
f_{(+)}(q)=\frac{1}{\exp\{(\omega(q)-1)L\}+1},
\ee
$$\omega(q)=\sqrt{1+(q^2/M^2(0))},$$
where $L$ is the inverse effective temperature in the unit of the constituent mass.
The substitution of this  Fermi - distribution function under the sign of the integral in Eq.~(\ref{gmor1})
and the  change of variable of
integration
$$dp\, p^2= d\omega \, \omega \, \sqrt{\omega^2-1}$$
leads to the expression of the Casimir condensate in the units of the mass $M(0)=1$
\bea \label{C-Cas-1}
 {C}_{\rm quark}(L)&=& \frac{3}{\pi^2} \int\limits_{1}^{\infty}
 \frac{d\omega\sqrt{\omega^2-1}}{1+ e^{(\omega-1)L}}~\Bigg|_{L=1}\simeq 0.39,
\eea
where $L$ is the inverse effective temperature and the constituent mass is the unit.
In the chiral massless limit ($m^0 \to 0$) the solution of the Schwinger--Dyson equation
and the Salpeter one yield a meson 
spectrum via the constituent quark masses $M_{\rm const}\simeq 320$ GeV \cite{Kalinovsky-10}.
Using the GMOR relation (\ref{gmor}) and the constituent quark mass value $\sim 320$ MeV
we can define a {\it conformal invariant} as
the ratio of the condensate value to the cubed constituent mass
\be
\label{gmor-10}
{\frac{<u\bar u>}{M_u^3}=\frac{M_\pi^2F_{\pi}^2}
{2 m_u M_u^3}\simeq 0.41\pm 0.08}.
\ee
The comparison of the theoretical value (\ref{C-Cas-1}) with
the experimental one (\ref{gmor-10}) shows us that the  inverse effective temperature $L$
coincides with the Compton length $(L=M_u^{-1}=1)$.

Thus, the breakdown of the chiral symmetry in QCD can be characterized by the quark condensate
(\ref{gmor-10}).
It is clear that the substitution of the
 effective vacuum temperature factor (\ref{pich-1}) $f_{(+)}(q)$ at $(L=M_u^{-1}=1)$
 in the integral (\ref{sd2}) at the large momenta
allows us to neglect in the potential its  dependence on the momentum of integration ${\bf q}$
\bea\nonumber
M_{\mathrm{a}}({\bf k})&=&
m^{0}_{\mathrm{a}} +
 {1\over2}\int { d^{3}q\over (2\pi)^{3} }
  \underline{V}({\bf k}-{\bf q})\cos 2{\upsilon}_{\mathrm{a}}({\bf q})f_{(+)}(q)=\\\nonumber
&\simeq&m^{0}_{\mathrm{a}} +
 {\underline{V}({\bf k})}{1\over2}\int { d^{3}q\over (2\pi)^{3} }
  \cos 2{\upsilon}_{\mathrm{a}}({\bf q})f_{(+)}(q)=\\\label{sd2p}
  &=&m^{0}_{\mathrm{a}} +\frac{g^2}{3(\mathbf{k}^2+M_g^2)}<u\bar u>,
  \eea
where $\underline{V}({\bf k})$ is given by Eq.~(\ref{Y-1}). This expression is in agreement with
the short distance operator product expansion applied to quark fields \cite{politzer-10}.

\newpage
\section{Summary}
\makeatletter
\renewcommand{\@evenhead}{\raisebox{0pt}[\headheight][0pt]
{\vbox{\hbox to\textwidth{\hfil \strut \thechapter .
Reduced phase space  QCD \quad\rm\thepage}\hrule}}}
\renewcommand{\@oddhead}{\raisebox{0pt}[\headheight][0pt]
{\vbox{\hbox to\textwidth{\strut \thesection .\,
Summary and literature
\hfil \rm\thepage}\hrule}}}
\makeatother

We considered the status of the initial data in QCD in the context of the
Quantum Universe wave function classification of physical processes in the Universe given in Chapters 8 and 9.
This classification proposes the reduced space phase QCD Hamiltonian.
 One of the main results is the topological degeneration of initial data
 that leads to
 the color confinement in the form of quark-hadron duality.
 In the constraint-shell QCD we can understand not only
 ``why we do not see quarks'', but also ``why we can measure their
 quantum numbers''.
 The normal ordering of the field products in the reduced space phase QCD Hamiltonian
 leads to both the  QCD inspired model
 of the low energy meson physics \cite{a20-10} with
  the Gell-Mann -- Oakes -- Renner (GMOR) relation \cite{gmor-10} and the short distance operator
   product expansion applied to quark fields \cite{politzer-10}.

\chapter{QU modification of the Standard Model}
\renewcommand{\theequation}{11.\arabic{equation}}
\setcounter{equation}{0}

\section{SM Lagrangian}
\makeatletter
\renewcommand{\@evenhead}{\raisebox{0pt}[\headheight][0pt]
{\vbox{\hbox to\textwidth{\hfil \strut \thechapter . QU modification of the Standard Model
\quad\rm\thepage}\hrule}}}
\renewcommand{\@oddhead}{\raisebox{0pt}[\headheight][0pt]
{\vbox{\hbox to\textwidth{\strut \thesection .\quad
SM Lagrangian
\hfil \rm\thepage}\hrule}}}
\makeatother

The Standard Model (SM) known as the Weinberg -- Salam -- Glashow
minimal electroweak theory was constructed on the basis of the Yang -- Mills theory \cite{ym-11}
with the symmetry group  ${SU(2)}\otimes{U(1)}$ \cite{STZhB-11} in two steps.
 The first step was the choice of the Lagrangian  ${\cal L}_{\rm G}$ and physical variables.
 The second step was the choice of a mechanism of the mass generation.
 Let us consider the gauge-invariant Lagrangian
 \bea\label{M-11}
&&{\cal L}_{\rm G}=-\frac{1}{4}G^a_{\mu\nu}
G^{\mu\nu}_{a}-\frac{1}{4}F_{\mu\nu} F^{\mu\nu}\\\nonumber
&+&\sum_s\bar{s}_1^{R}\imath \gamma^{\mu}\left(D_{\mu}^{(-)}
 +\imath
g^{\prime}B_{\mu}\right)s_1^{R}+\sum_s\bar{L}_s\imath
\gamma^{\mu}D^{(+)}_{\mu}L_s,
\eea
where
$$G^a_{\mu\nu}=\partial_{\mu}A^{a}_{\nu}-\partial_{\nu}A^{a}_{\mu}
 +g\varepsilon_{abc}A^{b}_{\mu}A^{c}_{\nu}$$
is the tension of  the  non-Abelian $SU(2)$ fields and
$$F_{\mu\nu}=\partial_{\mu}B_{\nu}-\partial_{\nu} B_{\mu}$$
is the tension of  the Abelian  $U(1)$ field,
$$D_{\mu}^{(\pm)}=\partial_{\mu}-\imath {g}\frac{\tau_{a}}{2}A^a_{\mu}\pm\frac{\imath}{2}g^{\prime}B_{\mu}$$
are covariant derivatives, and $\bar L_s=(\bar
s_1^{L}\bar s_2^{L})$ are the fermion duplets,  $g$ and $g'$ are the Weinberg coupling constants.

The physical variables as measurable bosons
$W^+_{\mu},~W^-_{\mu},~Z_{\mu}$
are determined by the relation
\bea
W_{\mu}^{\pm}&\equiv&{A}_{\mu}^1\pm{A}_{\mu}^2={W}_{\mu}^1\pm{W}_{\mu}^2,\\
Z_{\mu}&\equiv&-B_{\mu}\sin\theta_{W}+A_{\mu}^3\cos\theta_{W},\\
\tan\theta_{W}&=&\frac{g'}{g},\eea
where $\theta_{W}$ is the Weinberg angle.
In terms of these variables the Lagrangian  (\ref{M-11}) takes the form
\bea \nonumber
{\cal L}_G&=&-\frac{1}{4}(\partial_\mu A_\nu-\partial_\nu A_\mu)^2-
\frac{1}{4}(\partial_\mu Z_\nu-\partial_\nu Z_\mu)^2-\\\nonumber
&-&\frac{1}{2}|D_\mu W_\nu^+-D_\nu W_\mu^+|^2-\imath e(\partial_\mu A_\nu-\partial_\nu A_\mu)W^{+\mu}W^{-\nu}-
\\ \nonumber
&-&g^2\cos^2\theta_W[Z^2(W^+W^-)-(W^+Z)(W^-Z)]+
\\ \nonumber
&+&\imath g\cos\theta_W(\partial_\mu Z_\nu-\partial_\nu Z_\mu)W^{+\mu}W^{-\nu}+\\ \nonumber
&+&\frac{1}{2}\imath g\cos\theta_W[(D_\mu W_\nu^+-D_\nu W_\mu^+)(W^{-\mu}Z^\nu-W^{-\nu}Z^\mu)-h.c.],\eea
where
$D_\mu=\partial_\mu+\imath e A_\mu$ is covariant derivative,
$A_\mu$ is a photon field, $e$ is the electromagnetic interaction coupling.
According to the \textit{principles of quantum Universe}, the conformal symmetry
of the theory can be broken by only the normal ordering of the quantum field products
in the reduced phase space after resolution of all constraint equations.
In the previous Chapter 10, it was shown that the non-Abelian fields have the topological
degeneration of initial data. This degeneration can be removed by the interaction
of non-Abelian fields with an elementary scalar field.
 The conformal-invariant Lagrangian of the scalar field $h$ interacting with vector bosons and fermions
 $f$ is chosen in the form
\bea \nonumber
{\cal L}_h&=&\frac{1}{2}(\partial_\mu h)^2-\frac{\lambda^2}{8}h^4+
\sum_{f=s_1,s_2}\bar f \left[\imath\gamma\partial-g_f h\right]f
\\ \nonumber
&+&\frac{1}{8}h^2g^2\left[(W^+W^-+W^-W^+)+Z_\mu^2/\cos^2\theta_W\right]
\eea
where
$W^{\pm}$-, $Z$- are vector fields with the Weinberg coupling $g=0.645$;
 $\theta_W$ is the Weinberg angle,
and $\sin^2\theta_W=0.22$.
Masses of the vector bosons arise if the scalar field $h$ has the zero harmonics
\bea \label{M-11-2} h=v+H, ~~~~~~~~\int d^3x H=0.
\eea
The  value of the zero harmonics $v$ is determined by the Casimir condensate.
They automatically arise after the normal ordering procedure for all field operators.
The normal ordering of the interaction Hamiltonian of the scalar field yields the condensate
density
${\langle H H \rangle}_{\rm Cas}$
\be\label{C-1}
{\langle  H H  \rangle}_{\rm Cas}
=\!\frac{1}{V_0}
 \sum_p\frac{1}{2\sqrt{p^2+m^2}}\,.
\ee
This magnitude is connected with the Casimir energy \cite{Kirsten-11, condensate-11}
\be
E_{\rm Cas}=\frac{1}{2}\sum_{\bf p} \sqrt{{p}^2+m^2}
\ee
by the relation
\be\label{C-2}
{\langle  H H  \rangle}_{\rm Cas}=\frac{2}{V_0}\dfrac{\partial }{\partial m^2}E_{\rm Cas}\,.
\ee
In the continual limit of the QFT one has
\bea\label{universality}
&&\frac{1}{V_0}\sum_{\bf p}\frac{1}{2\sqrt{p^2+m_t^2}}\Rightarrow
\int \frac{d^3p}{(2\pi)^3}\frac{1}{2\sqrt{p^2+m^2}}=\nonumber\\
&&=m^2\int \frac{d^3x}{(2\pi)^3}\frac{1}{2\sqrt{x^2+1}}\equiv\gamma_0 \cdot m^{2}.
\eea
Thus, the Casimir condensate density
of a massive scalar field in the absence of any additional scale
is proportional to its squared mass
\be\label{Cas-2}
\langle H H \rangle_{\rm Cas}=\gamma_0 \cdot m^{2} \Rightarrow
\frac{\langle H H \rangle_{\rm Cas}}{m^{2}}\equiv \gamma_0\, ,
\ee
where $\gamma_0$ is {\it a dimensionless conformal parameter}
with a zero conformal weight (see discussion on conformal weights in \cite{grg-11}).
The normal ordering of a fermion pair (we intentionally interchange the order
of fermion fields to deal with positive condensates)
$$f \bar f=:f \bar f:+ \langle f \bar f \rangle$$
yields the condensate density of the fermion field $\langle f\bar f \rangle$
in the  Yukawa interaction term in Eq.~(\ref{L_int}).
In virtue of the above given results, we have for the top quark Casimir condensate density
\be\label{11}
 {\langle t\bar t \rangle}_{\rm Cas}= 4 N_c \frac{m_t}{V_0}\sum_{\bf p}
 \frac{1}{2\sqrt{p^2+m_t^2}}=4N_c\cdot \gamma_0 \cdot m_t^{3},
 \ee
where $N_c=3$ is the color number.
\section{The condensate mechanism of Higgs boson mass }
\makeatletter
\renewcommand{\@evenhead}{\raisebox{0pt}[\headheight][0pt]
{\vbox{\hbox to\textwidth{\hfil \strut \thechapter . QU modification of the Standard Model
\quad\rm\thepage}\hrule}}}
\renewcommand{\@oddhead}{\raisebox{0pt}[\headheight][0pt]
{\vbox{\hbox to\textwidth{\strut \thesection .\quad
The condensate mechanism of Higgs boson mass
\hfil \rm\thepage}\hrule}}}
\makeatother



Recently a few research groups have reported upon the discovery of scalar
particles with almost similar masses around $125 - 126$ GeV
\cite{:2012gk-11, :2012gu-11, Aaltonen:2012qt-11}. The experimentalists express extreme
caution in the identification of these particles with the long-waiting Higgs boson.
Indeed, the literature contains a plethora of predictions on lower and upper limits
of the Higgs mass based
on many different ideas, models and numerical techniques which are close
to the observed values. The question on a genuine mechanism which provides
an unambiguous answer about the Higgs mass is a real challenge to high energy
physics and is crucial for the base of the Standard Model (SM)
\footnote{\textit{Rencontres de Moriond}. La Thuile, Italy (2013).}.
It is  especially noteworthy that in the SM the Higgs mass is introduced {\it ad hoc}.


According to the general wisdom, all SM particles (may be except neutrinos) own
masses due to the spontaneous symmetry breaking (SSB)
of the electroweak gauge symmetry \cite{Englert-11, higgs-11, Kibble-11}.
In particular, one deals with the potential (in notation of Ref.~\cite{Beringer:1900zz-11}):
\be\label{5a}
 V_{\rm Higgs}(\phi)=\frac{\lambda^2}{2}(\phi^\dagger\phi)^2 + \mu^2\phi^\dagger\phi,
 \ee
where one component of the complex scalar doublet field
$$\phi=\left(\begin{array}{c}\phi^+\\ \phi^0\end{array}\right)$$
acquires a non-zero vacuum expectation value
$$\langle\phi^0 \rangle = \frac{v}{\sqrt{2}}.$$

Note that the tachyon-like mass term in the potential is critical for this construction.
In contrast to the SSB, it breaks the conformal symmetry explicitly being
the only {\em fundamental} dimension-full parameter in the SM.
We recall that the explicit conformal symmetry breaking in the
Higgs sector gives rise to the unsolved problem of fine tuning in the renormalization
of the Higgs boson mass.
That is certainly one of the most unpleasant features of the SM.
In the classical approximation, from the condition of the potential minimum one obtains
the relation between the vacuum expectation value and the primary parameters
$\mu$ and $\lambda$ in the form $v=\sqrt{-2{\mu^2}}\,/\lambda$. This quantity can be
defined as well with the aid of the Fermi coupling constant
derived from the muon life time measurements:
$$v=(\sqrt{2}G_{\mathrm{Fermi}})^{-1/2}\approx 246.22~\mbox{\rm GeV}.$$
The experimental studies at the LHC~\cite{:2012gk-11,:2012gu-11}
and Tevatron~\cite{Aaltonen:2012qt-11} observe an excess of events in the data
compared with the background in the mass range around $\sim 126$ GeV.
Taking into account radiative corrections, such a mass value makes the
SM being stable up to the Planck mass energy scale~\cite{Bezrukov:2012sa-11}.
Nevertheless, the status of the SM and the problem of the mechanism of
elementary particle mass generation are still unclear.


The idea on dynamical breaking of the electroweak gauge symmetry with the
aid of the top quark condensate has been continuously discussed in literature since
the pioneering papers~\cite{Nambu:1990dx-11,Nambu:1990hj-11,Bardeen:1989ds-11}
(see also review~\cite{Cvetic:1997eb-11} and references therein).
Such approaches suffer, however, from formal quadratic divergences in
tadpole loop diagrams leading, in particular, to the
naturalness problem (or fine tuning) in the renormalization of the Higgs boson mass.
All mentioned facts suggest that it might be good to examine if the
SSB is also responsible for the Higgs field.
To begin with, we suppose that there is a general mechanism of the SSB,
which is responsible for the appearance
of all SM field condensates.

\begin{center}
\includegraphics[width=0.8\textwidth]{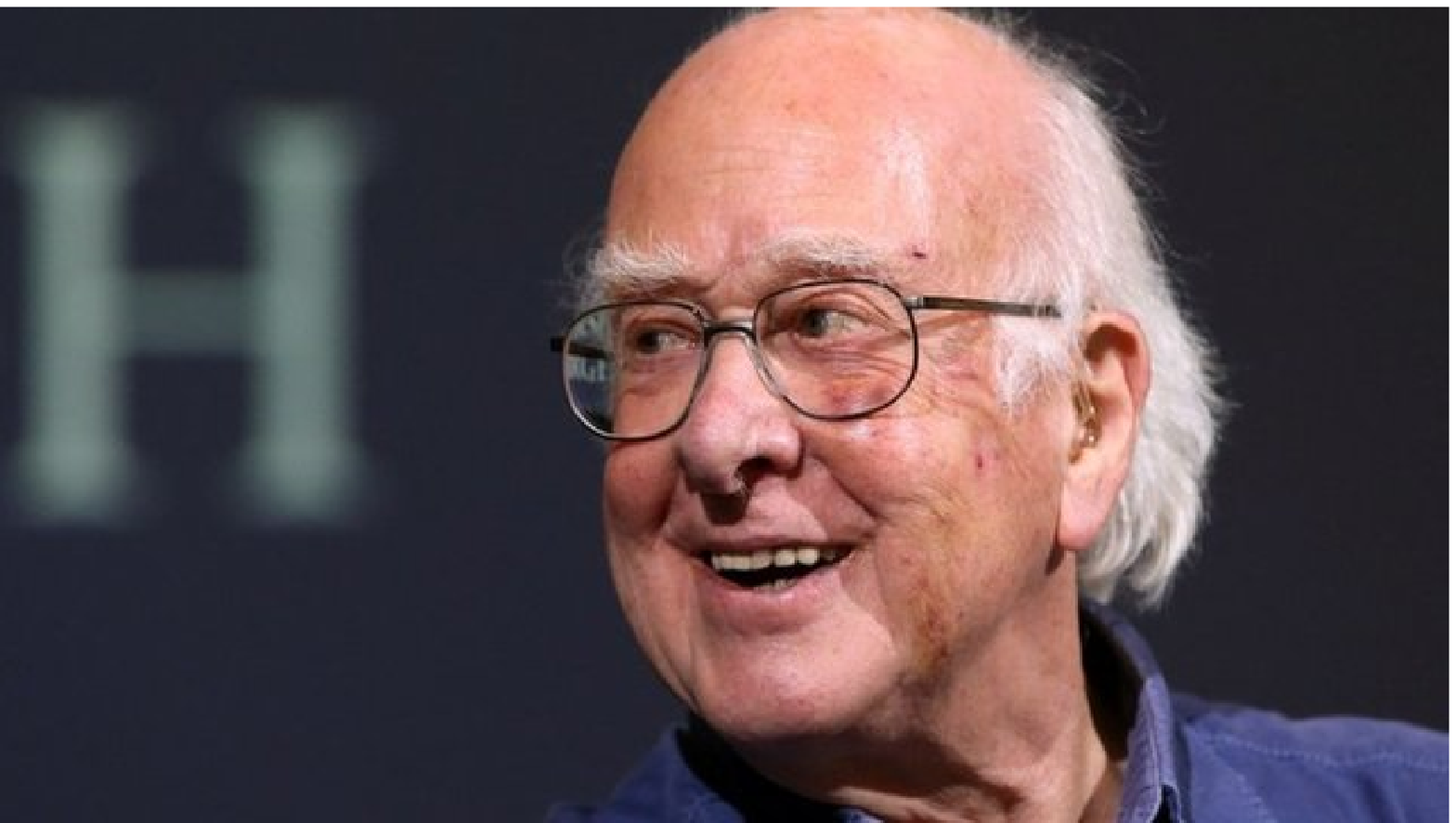}
\end{center}
{\small
Peter Ware Higgs, (born 29 May 1929, Newcastle Upon Tyne, England)
is a British theoretical physicist, Nobel Prize laureate (shared with Francois, Baron Englert)
 and emeritus professor at the University of Edinburgh.
He is best known for his 1960s proposal of broken symmetry in electroweak theory,
explaining the origin of mass of elementary particles in general and of the W- and Z- bosons in particular.
This so-called Higgs mechanism, which was proposed by several physicists besides Higgs at about the same time,
predicts the existence of a new particle, the Higgs boson.
CERN announced on 4 July 2012 that they had experimentally established the existence of a Higgs-like boson,
but further work is needed to analyze its properties and see if it has the properties expected from the Standard Model Higgs boson.
Higgs himself said on this occasion, that he
did not expect the experimental confirmation of his theory in his life.
On 14 March 2013, the newly discovered particle was tentatively confirmed to be ``+'' parity and zero spin,
two fundamental criteria of a Higgs boson, making it the first known scalar particle to be discovered in nature.
The Higgs mechanism is generally accepted as an important ingredient in the Standard Model of particle physics,
without which certain particles would have no mass.
}

\begin{center}
{\vspace{0.31cm}}
\parbox{0.5\textwidth}{
\includegraphics[height=9.truecm,
angle=-0]{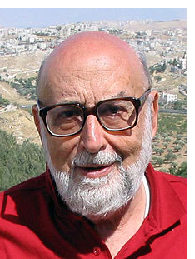}}
\parbox{0.48\textwidth}
{\small
Francois, Baron Englert (born 6 November 1932) is a Belgian theoretical physicist and 2013 Nobel prize laureate (shared with Peter Higgs).
He is Professor emeritus at the Universit$\acute{\rm e}$ libre de Bruxelles where he is member of the Service de Physique Th$\acute{\rm e}$orique.
He is also a Sackler Professor by Special Appointment in the School of Physics and Astronomy at Tel Aviv University and a member of the
Institute for Quantum Studies at Chapman University in California. He was awarded the 2010 J. J. Sakurai Prize for Theoretical Particle Physics
(with Gerry Guralnik, C. R. Hagen, Tom Kibble, Peter Higgs, and Robert Brout), the Wolf Prize in Physics in 2004 (with Brout and Higgs)
and the High Energy and Particle Prize of the European Physical Society (with Brout and Higgs) in 1997 for the mechanism which unifies
short and long range interactions by generating massive gauge vector bosons.
He is the recipient of the 2013 Prince of Asturias Award in technical and scientific research,
together with Peter Higgs and the CERN.
}
{\vspace{0.31cm}}
\end{center}

The main feature of our approach
is the assumption about the underlying
(softly broken) conformal symmetry which protects the jump of the Higgs
boson mass to a cut-off scale.
We will call this mechanism the spontaneous conformal symmetry breaking (SCSB) \cite{Baldin-11}.
Evidently, in this case one should require the conservation
of the conformal symmetry of the genuine theory fundamental Lagrangian.
It will be shown that the SCSB provides
the breaking of the gauge, chiral, and conformal symmetries on equal footing.
Therefore, it allows one also to introduce the universal relation
between different condensates determined relative to the corresponding
mass power depending on quantum statistics (see Eq.~(\ref{13})). Our basic assumption is
that this relation is not violated by the SCSB.

Following the ideas of Nambu \cite{Nambu:1990dx-11,Nambu:1990hj-11},
we generate the SCSB of the Higgs potential, using the top quark condensate.
It is assumed that the general construction of the SM should remain unchanged.
Let us start with the conformal invariant Lagrangian
of Higgs boson interactions (\ref{5a})
\be \label{L_int}
{\cal L}_{\mathrm{int}} = - \frac{\lambda^2}{8}h^4 - g_t h~\bar{t}t.
\ee
Here, for the sake of discussion, we consider only the most intensive terms:
the self-interaction and the Yukawa ones of the top quark coupling
constant $g_t$. From the beginning, we assume that the $O(4)$ symmetry of the
Higgs sector is spontaneously broken to the $O(3)$ symmetry.
Contributions of other interaction terms will be considered below as well.


Keeping in mind all these results, we are ready to treat the contribution
of  the top quarks to the effective potential \cite{condensate-11} generated by the term  Eq.~(\ref{L_int}):
\bea\label{V_cond}
V_{\rm cond}(h) &=& \frac{\lambda^2}{8}h^4 -g_t\langle t\bar t\,\rangle h.
\eea

The extremum condition for the potential
$$\frac{dV_{\rm cond}}{dh}\Biggl|_{h=v}=0$$
yields the relation
\be \label{lambda}
{v^3}\frac{\lambda^2}{2} = {g_t\langle t\bar t\,\rangle}.
\ee
This relation follows from the fact that the Higgs field has a zero
harmonic $v$ in the standard decomposition
of the field $h$ over harmonics  $$h = v  + H,$$
 where $H$ is the sum of all nonzero
harmonics with a condition
$$\int d^3x H=0.$$
Here, the Yukawa coupling of
the top quark $g_t \approx 1/\sqrt{2}$ is known from the experimental
value of top quark mass
$$m_t=v g_t\simeq 173.4~\mbox{\rm GeV}.$$

The spontaneous symmetry breaking yields the potential minimum which results in the
nonzero vacuum expectation value $v$ and Higgs boson mass.
The substitution $h = v  + H$ into the potential~(\ref{V_cond}) leads to the result
\be
V_{\rm cond}(h)=V_{\rm cond}(v)+\frac{m_H^2}{2}H^2+
\frac{\lambda^2 v}{2} H^3+\frac{\lambda^2}{8}H^4,
\ee
which defines the scalar particle mass as
\be
\label{mh}
m_H^2\equiv\frac{\lambda^2}{2}3v^2 .
\ee
We stress that this relation is different from the one $(m_H=\lambda v)$
which emerges in the SM with the Higgs potential~(\ref{5a}).

With the aid of Eqs.~(\ref{lambda}), (\ref{mh}), the squared scalar particle
mass can be expressed in terms of the $t$ quark condensate:
\be\label{h_mass}
m^2_H = \frac{3 g_t \langle t\bar t\,\rangle}{v}.
\ee

The hypothesis on universality of the conformal invariant ratios of the field
condensate densities and the corresponding mass powers (see Eqs.~(\ref{universality})--(\ref{11}))
allows us to determine the $t$ quark condensate density with the aid
of the light quark one.

\section{Estimation of the Higgs boson mass}
\makeatletter
\renewcommand{\@evenhead}{\raisebox{0pt}[\headheight][0pt]
{\vbox{\hbox to\textwidth{\hfil \strut \thechapter . QU modification of the Standard Model
\quad\rm\thepage}\hrule}}}
\renewcommand{\@oddhead}{\raisebox{0pt}[\headheight][0pt]
{\vbox{\hbox to\textwidth{\strut \thesection .\quad
Estimation of the Higgs boson mass
\hfil \rm\thepage}\hrule}}}
\makeatother
The supposition about the Casimir condensate  density of the light quarks in Chapter 10
(see  (\ref{gmor-10}), (\ref{universality})--(\ref{11})) allows us to determine
the Casimir condensate  density of the $t$-quark
\be
\label{13}
\frac{\langle t\bar t \rangle}{m^3_{t }}=\frac{\langle q\bar q \rangle}{m^3_{q}},
 \ee
and estimate the Higgs boson mass.
We consider the left and right hand sides as scale invariants. However, their numerators
and denominators are scale variables. Therefore, we have to choose the proper scales.
For the left hand side, the scale is naturally defined by the known $t$ quark mass.
We define the scale of the right hand side by the light quark condensate  density
$\langle q\bar q \rangle$. It is quite accurately determined in the chiral limit of the
QCD low-energy phenomenology \cite{Beringer:1900zz-11}:
\be
\label{13c}
\langle q \bar{q} \rangle\simeq (250~{\rm \mbox{МeV}})^3.
\ee
At this scale, the light quark possesses the constituent mass
$$m_{q}\approx 330~\mbox{MeV}$$
estimated in the QCD-inspired model~\cite{gmor1-11}. With the aid of Eq.~(\ref{13})
one determines the top quark condensate value
\be
\langle t\bar t\,\rangle \approx (126\ {\rm\mbox{GeV}})^3.
\ee
Such a large value of the top quark condensate does not affect the low energy QCD phenomenology,
since its contribution is very much suppressed
by the ratio of the corresponding energy scales (squared).

By means of Eqs.~(\ref{13}), (\ref{13c}),  in the tree approximation we obtain for the scalar particle mass
\be\label{h_mass1}
(m_H^0)^2 = (130\pm 15~{\rm\mbox{GeV}})^2\,.
\ee
Here, we have assigned 10\% uncertainty into the ratio light quark condensate and its constituent mass.

The tentative estimate of the Higgs boson mass given above is rather preliminary.
In order to improve this value we consider below the contributions
of other condensates at the tree level. The mass can be also affected by
radiative corrections which will be analyzed elsewhere.
Under the assumption of $\gamma_0$ universality, the normal ordering of the field operators
$$HH = :HH: + \langle HH\rangle$$
yields
\be
\label{27a}
\frac{\langle HH\rangle}{{m^2_H}}
=\gamma_0.
\ee
The normal ordering of the vector fields $V_iV_j$
defines the vector field condensates normalized on each degree of freedom
\be\label{27} \langle VV\rangle
={M^2_V}\cdot \gamma_0,~~~V=W^\pm,Z\,,
\ee
calculated in the gauge $V_0=0$. Here, $M_V$ is a corresponding mass of the vector field.
Transverse and longitudinal components are considered
on equal footing  in the reduced phase space quantization of the massive vector theory~\cite{h-pp-11}.
As a result, one obtains the upper limit of the vector field condensate contributions
for the mass formula (\ref{h_mass1}) at the tree level for the SM
\be\label{28}
\Delta m_H^2=\frac{3\lambda^2}{4}\langle HH\rangle+\frac{3}{8}g^2
\left(2\langle WW\rangle+\frac{\langle ZZ\rangle}{\cos^2\theta_W}\right),
\ee
where $g$ and $\theta_W$ are the Weinberg coupling constant and the mixing angle.
In Eq. (\ref{28}) the first term is a contribution to the square mass due to
the very scalar field condensate $\langle HH\rangle$. Taking into account
the values of coupling constants, mixing angle, masses,
and condensates, we arrive to the following result
\be
m_H=m^0_H{\left[1+4\frac{\Delta m_H^2}{v^2}\right]}^{1/2}\approx
m^0_H\times \left(1+0.02\right),
\ee
where $m^0_H$ is given by Eq. (\ref{h_mass1}).
If there exist additional heavy fields interacting with the SM Higgs boson,
their condensates would contribute to the Higgs boson mass.

\newpage
\section{Summary}
\makeatletter
\renewcommand{\@evenhead}{\raisebox{0pt}[\headheight][0pt]
{\vbox{\hbox to\textwidth{\hfil \strut \thechapter .  QU modification of the Standard Model
\quad\rm\thepage}\hrule}}}
\renewcommand{\@oddhead}{\raisebox{0pt}[\headheight][0pt]
{\vbox{\hbox to\textwidth{\strut \thesection .\quad
Summary and literature
\hfil \rm\thepage}\hrule}}}
\makeatother

In conclusion, we suggest the condensate mechanism of Spontaneous
Conformal Symmetry Breaking in the Standard Model of strong and electroweak interactions.
We suppose that this mechanism is related to the vacuum Casimir energy in the Standard Model.
Our key assumption is that the condensates of all fields normalized to their
masses and degrees of freedom represent a conformal invariant. This idea enables
us to avoid efficiently  the problem of the regularization of the divergent
tadpole loop integral. The top quark condensate supersedes the phenomenological
negative square mass term in the Higgs potential.
In contrast to the standard Higgs mechanism, the condensate mechanism allows one
to establish relations between all condensates and masses including the Higgs one.
According to our results, the latter is of the order $130\pm 15$ GeV,
if the universality relation (\ref{13}) holds true.

%

\newpage

\chapter{Electroweak vector bosons
\label{S-12a}}
\renewcommand{\theequation}{12.\arabic{equation}}
\setcounter{equation}{0}
\section{Cosmological creation of electroweak\\ vector bosons}
\makeatletter
\renewcommand{\@evenhead}{\raisebox{0pt}[\headheight][0pt]
{\vbox{\hbox to\textwidth{\hfil \strut \thechapter . Electroweak vector bosons
\quad\rm\thepage}\hrule}}}
\renewcommand{\@oddhead}{\raisebox{0pt}[\headheight][0pt]
{\vbox{\hbox to\textwidth{\strut \thesection .\quad
Cosmological creation of electroweak vector bosons
\hfil \rm\thepage}\hrule}}}
\makeatother

Let us consider the vector massive particles in the conformal flat metrics
 \be \label{cst}
 {\widetilde{ds}}^2=\widetilde{g}_{\mu\nu}dx^\mu dx^\nu
  = (d\eta)^2-(dx^i)^2,
 \ee
in the cosmological model, with the time interval
 \be\label{cst1}  d\eta=\bar{N}_0(x^0)dx^0
 \ee
that follows from the homogeneous approximation of our gravitation theory developed in
the previous chapters.
The field motion equations including the dilaton one are derived from the action
\be\label{totl}
W=W_{\rm Cas}+W_{ v}~.
 \ee
Here
\be\label{uncol1}
 W_{\rm Cas}=- V_0
 \int\limits_{x^{0}_{1}}^{x^{0}_{2}} dx^{0}
 \bar N_0\left[ \left(\frac{da}{\bar N_0dx^0}\right)^2
 +\rho_{\rm Cas}(a) \right]
\ee
is the action of the cosmological scale factor in the supposition of the Casimir energy dominance
$$\rho_{\rm Cas}(a)=\dfrac{H_0^2}{a^2};$$
\begin{equation}
W_{  v}=\int d^4x \sqrt{-g}
\left[-{1\over 4}F_{\mu\nu}F^{\mu\nu}
-{1\over 2}M_v^2v_\mu v^\mu\right]~
\label{Lem20}
\end{equation}
is the vector boson action. 

The classification of observational data in the framework of the considered model
of the quantum Universe as representations of the $A(4)\otimes C$ group
supposes that the concept of a particle in QFT can be associated with the fields
that have positive energy and the positive probability.
The negative energies are removed by resolution of constraints and
the causal quantization in the reduced phase space. According to the causal quantization
the creation operator of a particle with a negative energy is replaced by
the annihilation  operator of a particle with a positive energy.
The results of such quantization in the metrics (\ref{cst}) are given in Appendix A.

The Quantum Universe model supposes the identification of real observational magnitudes
with the conformal variables. This identification yields the Universe evolution
that differs from the Standard model. In the following we shall use the Hamiltonian
form of the field Fourier harmonics
$${\bf \rm v}_k^{I}=\int\limits_{}
d^3xe^{\imath\bf k\cdot x}{\bf \rm v}^{I}({\bf x}).$$
The action takes the form 
\begin{eqnarray}
\label{grad}
W&=&\int\limits_{x^0_1 }^{x^0_2 }dx^0
\sum\limits_{k}
 \left[{\bf p}_{k}^{\bot}\partial_0{\bf \rm v}_{k}^{\bot}
 + {\bf p}_{k}^{||}\partial_0{\bf \rm v}_{k}^{||}\right]+\\\nonumber
&+&\int\limits_{x^0_1 }^{x^0_2 }dx^0\left(
- P_{a} \frac{da}{dx^0} +{N}_{0}\left[\frac{P_{a}^2}{4V_0}-
(H^{\bot} + H^{||})
\right]\right),
\end{eqnarray}
where 
${\bf p}_{k}^{\bot},{\bf p}_{k}^{||}$ are transversal and longitudinal
momenta and 
\begin{eqnarray}
&H^{\bot} = \sum\limits_{k} \dfrac{1}{2}\left[{\bf p}_k^{\bot}{}^2 +
\omega^2 {\bf \rm v}_k^{\bot}{}^2\right]~,
\nonumber\\ [-8mm]&\label{grad1}
\\[.2cm]&\nonumber
H^{||} = \sum\limits_{k} \dfrac{1}{2}\left
[\left(\dfrac{\omega(a,k)}{M_{\rm v} a}\right)^2{\bf p}_{k}^{||}{}^2 +
(M_{\rm v} a)^2 {\bf \rm v}_{k}^{||}{}^2 \right]~
\end{eqnarray}
are the free Hamiltonian with one-particle energy 
$$\omega(a,k) = \sqrt{{\bf k^2} + (M_{\rm v} a)^2};$$
here we introduced the notions 
$${\bf p}_{k}^{||}{}^2\equiv ({\bf p}_{k}^{||}{}\cdot{\bf p}_{-k}^{||}){}.$$

Let us consider the case of the rigid equation state with initial data
$$a(\eta)=a_I\sqrt{1+2H_I (\eta-\eta_I)},~~~~~~~~(a^2_IH_I=H_0),$$
$$a_I=a(\eta=\eta_I):$$
\be\label{g-12}
 \tau=2\eta H_I=\frac{\eta}{\eta_I},~~~~~~~~~ x=\frac{q}{M_{I}},
 ~~~~~~~~\gamma_{\rm v}=\frac{M_{I}}{H_I},
\ee
$$M_{I}=M_{\rm v}(\eta=\eta_I).$$

In terms of the conformal variables the one particle energy takes the form
$$\omega_{\rm v}=H_I\gamma_{\rm v}\sqrt{1+\tau+x^2}.$$

\begin{figure}[t]
 \includegraphics[width=0.37\textwidth,angle=-90]{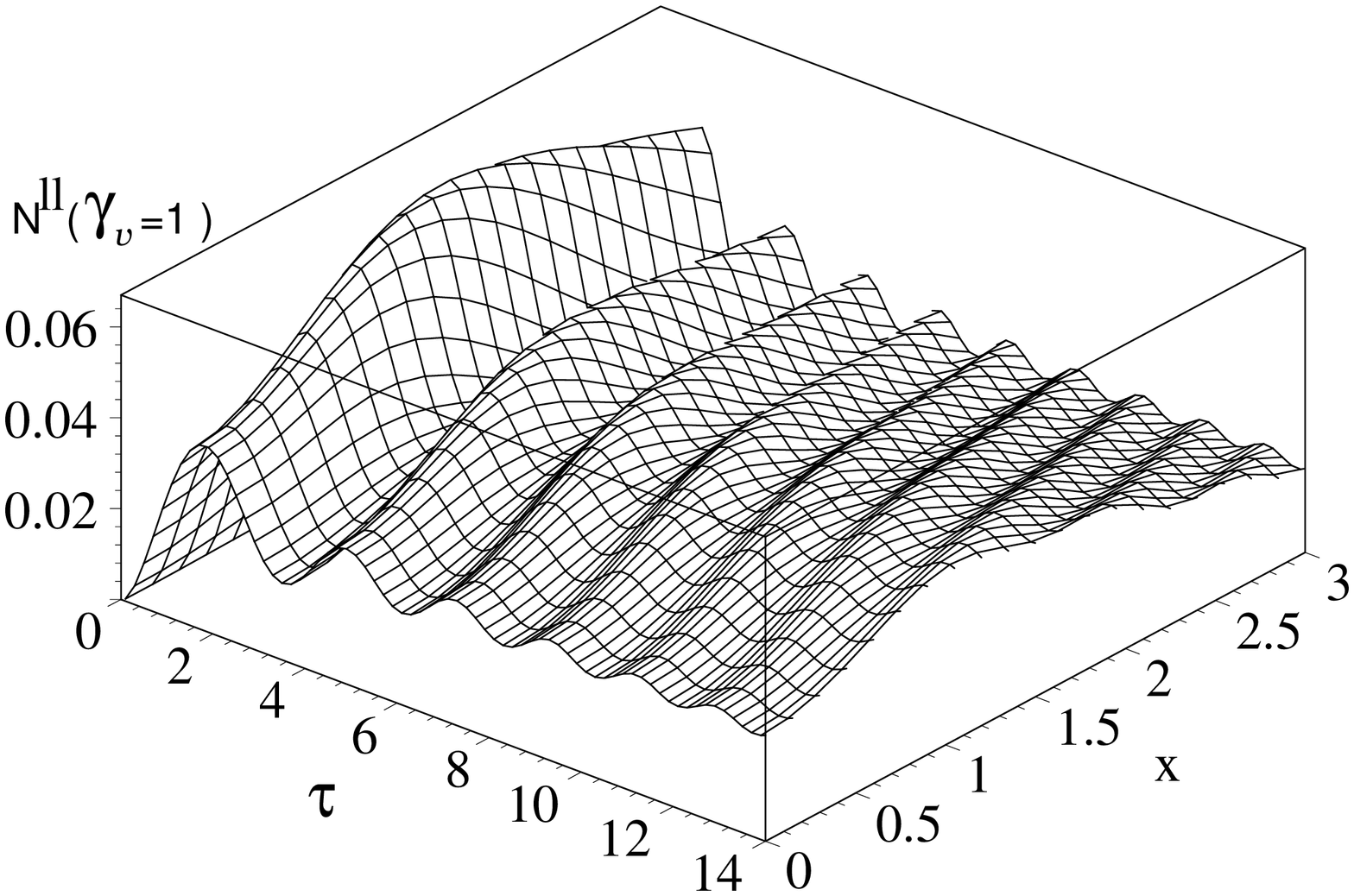}
 \includegraphics[width=0.37\textwidth,angle=-90]{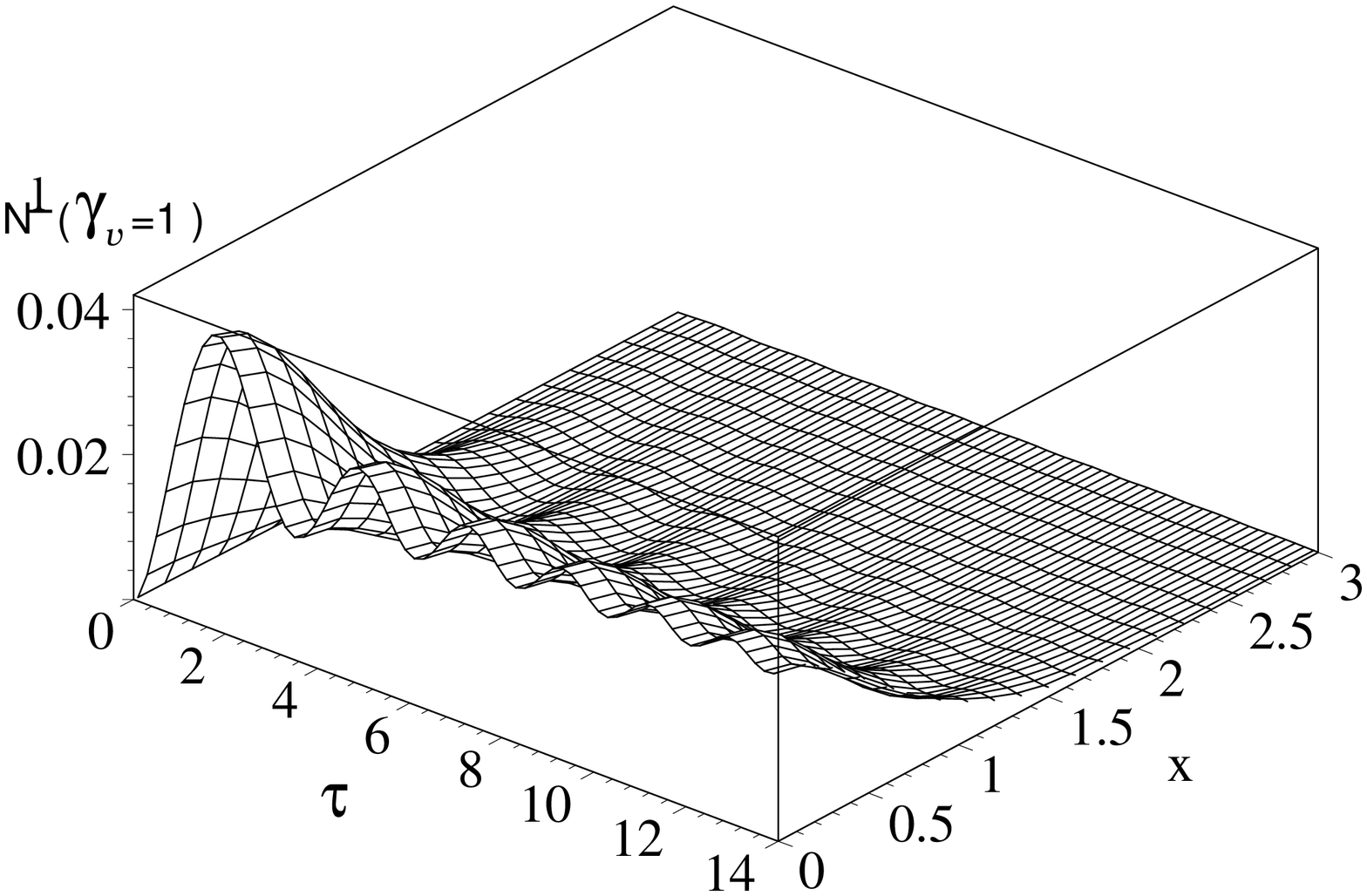}
\caption{\small
Longitudinal (${\cal N}^{||}(q,\eta)$) and transverse (${\cal N}^{\bot}(q,\eta)$)
 components of the boson distribution versus the dimensionless time  $\tau= 2\eta H_I$ and the
dimensionless momentum $x = q/M_I$ at the initial data MI = HI $M_I = H_I$ ($\gamma_v = 1$)
\cite{ppgc-12, 114:a-12, vin-12}.
 }
\label{distbosons}
\end{figure}

Then the Bogoliubov equations are 
\bea\label{1g12}
&&\left[\frac{\gamma_{\rm v}}{2}\sqrt{(1+\tau)+x^2} -
\frac{d\theta^{||}_{\rm v}}{d\tau}\right] \tanh(2r^{||}_{\rm v})\\\nonumber
 &&=
-\left[\frac{1}{2(1+\tau)}-\frac{1}{4\left[ (1+\tau)+x^2\right]}\right]
\sin(2\theta^{||}_{\rm v}),
\eea
$$
\frac{d}{d\tau}r^{||}_{\rm v} =
\left[\frac{1}{2(1+\tau)}-\frac{1}{4\left[ (1+\tau)+x^2\right]}\right]
\cos(2\theta^{||}_{\rm v}),
$$
$$
\left[\frac{\gamma_{\rm v}}{2}\sqrt{(1+\tau)+x^2} -
\frac{d}{d\tau}\theta^{\bot}_{\rm v}\right]  \tanh(2r^{\bot}_{\rm v}) =
-\left[\frac{1}{4\left[ (1+\tau)+x^2\right]}\right]
\sin(2\theta^{\bot}_{\rm v}),
$$
\be\label{1g}
\frac{d}{d\tau}r^{\bot}_{\rm v} =
\left[\frac{1}{4\left[ (1+\tau)+x^2\right]}\right]
\cos(2\theta^{\bot}_{\rm v}).
\ee
We solved these equations numerically \cite{ppgc-12, 114:a-12, vin-12}  at positive values
of the momentum $x=q/M_I$, considering that 
 the asymptotic behavior of the solutions is
given by
$$r(\tau)\to{\rm const}\cdot\tau,\qquad \theta(\tau)=O(\tau),\qquad \tau\to+0.$$
The distributions
of longitudinal ${\cal N}^{||}(x,\tau)$ and transverse ${\cal N}^{\bot}(x,\tau)$
vector bosons are given in Fig. \ref{distbosons} for
the initial data $H_I=M_I,~\gamma_{\rm v}=1$.

From the figure, it can be seen that for $x > 1$ the
longitudinal component of the boson distribution is
everywhere much greater than the transverse component,
demonstrating a more copious cosmological
creation of longitudinal bosons in relation to
transverse bosons. A slow decrease in the longitudinal
component as a function of momentum leads to
a divergence of the integral for the density of product particles
\be\label{nb-12}
 n_{\rm v}(\eta)=\frac{1}{2\pi^2}
\int\limits_{0 }^{\infty }
d q q^2
\left[ {\cal N}^{||}(q,\eta) + 2{\cal N}^{\bot}(q,\eta)\right]\to\infty.
\ee

The divergence of the integral in  (\ref{nb-12}) stems from
idealizing the problem of the production of a pair
of particles in a finite volume for a system where
there are simultaneous interactions associated with
the removal of fields having a negative probability and
where identical particles affect one another (so-called
exchange effects). It is well-known \cite{kittel-69-12, Muller-07-12} that in this
case one deals with the production of a set (rather
that a pair) of particles, which acquires, owing to the
aforementioned interactions, the properties of a statistical
system. As a model of such a statistical system,
we consider here a degenerate Bose -- Einstein gas with the Boltzmann -- Chernikov
 distribution function that has the form
\be \label{bose}
{\cal F}\left(T_{\rm v},q,M_{\rm v}(\eta),\eta\right)=
 \left\{\exp\left[\frac{\omega_{\rm v}(\eta)- M_{\rm v}(\eta)}{ T_{\rm v}}\right]
 -1\right\}^{-1},
 \ee
(we use
the system of units where the constant is $k_{\rm B}=1$),
where $T_{\rm v}$ is the boson temperature. We set apart the
problem of theoretically validating such a statistical
system and its thermodynamic exchange, only assuming
fulfillment of specific conditions ensuring its
existence. In particular, we can introduce the notion
of the temperature $T_{\rm v}$ only in an equilibrium system.
A thermal equilibrium is thought to be stable if the
time within which the vector-boson temperature Tv is
established, that is, the relaxation time \cite{Ignatyev123-12,ber-12}
\be \label{rel01}
 \eta_{\mbox{\small  rel}} =
 \left[{n(T_{\rm v})\sigma_{\mbox{\small scat}}}\right]^{-1}
 \ee
[as expressed in terms of their density $n(T_{\rm v})$ and
the scattering cross section $\sigma_{\mbox{\small  scat}} \sim 1/ M_{I}^2$], does not
exceed the time of vector-boson-density formation
owing to cosmological creation, the latter time being
controlled by the primordial Hubble parameter, $\eta_{\rm v}=1/ H_I$.
 From (\ref{rel01}), it follows that the particle-number
density is proportional to the product of the Hubble
parameter and the mass squared (this product being
an integral of the motion in the present example); that is,
\be \label{bose11}
 n(T_{\rm v})=n(T_{\rm v},\eta_{\rm v})\simeq
  C_H H_IM_I^2, 
\ee
where $C_{H}$ is a constant. The expression for the density
$n(T_{\rm v},\eta)$ in Eq.~(\ref{bose11}) assumes the form
\be\label{n1}
 n_{{\rm v}}(T_{\rm v},\eta)=\frac{1}{2\pi^2}
\int\limits_{0 }^{\infty }
d q q^2{\cal F}\left(T_{\rm v},q,M(\eta),\eta\right)
\left[ {\cal N}^{||}(q,\eta) + 2{\cal N}^{\bot}(q,\eta)\right].
\ee
Here, the probability of the production of a longitudinal
and a transverse boson with a specific momentum
in an ensemble featuring exchange interaction
is given (in accordance with the multiplication law
for probabilities) by the product of two probabilities,
the probability of their cosmological creation, ${\cal N}^{||,\bot}$,
and the probability of a single-particle state of vector
bosons obeying the Bose -- Einstein distribution
in  (\ref{bose}).
A dominant contribution to the integral in  (\ref{n1})
from the region of high momenta (in the above idealized
analysis disregarding the Boltzmann factor
this resulted in a divergence) implies the relativistic
temperature dependence of the density,
\be \label{nc1}
 n(T_{\rm v},\eta_{\rm v}) = C_T T_{\rm v}^3,
 \ee
where $C_T$ is a coefficient. A numerical calculation
of the integral in (\ref{n1}) for the values $T_v = M_I = H_I$,
which result from the assumption about the choice
of initial data ($C_T = C_H$), reveals that this integral
is weakly dependent on time in the region $\eta \geq \eta_{\rm v}=H_I^{-1}$ and,
for the constant $C_T$, yields the value
\be \label{nc}
C_T = \frac{ n_{\rm v}}{ T_{\rm v}^3} = \frac{1}{2\pi^2}
\left\{ [1.877]^{||}+2 [0.277]^{\bot}=2.431   \right\}~,
\ee
where the contributions of longitudinal and transverse
bosons are labeled with the superscripts $(||)$ and $\bot)$,
 respectively.

On the other hand, the lifetime $\eta_L$ of
boson product in the early Universe in dimensionless units,
$\tau_L=\eta_L/\eta_I$, where $\eta_I=(2H_I)^{-1}$, can be estimated
by using the equation of state 
and
the W-boson lifetime within the Standard Model.
Specifically, we have
\be \label{life}
1+\tau_L=
\frac{2H_I\sin^2 \theta_{(W)}}{\alpha_{\rm QED} M_W(\eta_L)}=
\frac{2\sin^2 \theta_{(W)}}{\alpha_{\rm QED}\gamma_{\rm v}\sqrt{1+\tau_L}},
\ee
where $\theta_{(W)}$ is the Weinberg angle, $\alpha_{\rm QED}=1/137$ and
$\gamma_{\rm v}=M_{I}/ H_I\geq 1$.
From the solution to Eq.~(\ref{life}),
\be \label{lifes}
\tau_L+1=
\left(\frac{2\sin^2\theta_{(W)}}{\gamma_{\rm v}\alpha_{\rm QED}}\right)^{2/3}
\simeq \frac{16}{\gamma_{\rm v}^{2/3}}~
\ee
it follows that, at $\gamma_{\rm v}=1$, the lifetime of
bosons product is an order of magnitude longer than the
Universe relaxation time:
\be \label{lv}
\tau_L =\frac{\eta_L}{\eta_I}\simeq \frac{16}{\gamma_{\rm v}^{2/3}}-1=15.
\ee
Therefore, we can introduce the notion of the
vector-boson temperature $T_v$, which is inherited by
the final vector-boson-decay products (photons).
According to currently prevalent concepts, these
photons form Cosmic Microwave Background radiation
in the Universe. Indeed, suppose that one
photon comes from the annihilation of the products
of $W^{\pm}$-boson decay and that the other comes from
$Z$ bosons. In view of the fact that the volume of the
Universe is constant within the evolution model being
considered, it is then natural to expect that the photon
density coincides with the boson density \cite{ppgc-12}
\be \label{1nce}
n_\gamma={T_\gamma^3}\frac{1}{\pi^2}\times
 2.404
\simeq n_{\rm v}.
\ee
On the basis of (\ref{bose11}), (\ref{nc1}), (\ref{nc}), and  (\ref{1nce}), we can
estimate the temperature $T_{\gamma}$ of Cosmic Microwave Background radiation arising upon the annihilation
and decay of $W-$ and $Z-$ bosons. This yields
\be \label{1nce1}
T_\gamma\simeq \left[\frac{ 2.431}{2.404 \times 2 }\right]^{1/3}T_{\rm v}=0.8\times T_{\rm v},
\ee
where the vector-boson temperature
$$T_{\rm v} = [H_I M^2_I ]^{1/3}$$
is an invariant quantity within the model being considered.
This invariant can be estimated at
\be\label{cmb-12-1}
T_{\rm v}= [H_IM^{2}_I]^{1/3}=[H_0 M_W^2]^{1/3}=2.73/0.8 ~K=3.41 ~K ,
\ee
which is a value that is astonishingly close to the observed
temperature of Cosmic Microwave Background
radiation. In the present case, this directly follows,
as is seen from the above analysis of our numerical
calculations, from the dominance of longitudinal
vector bosons with high momenta and from the fact
that the relaxation time is equal to the inverse Hubble
parameter. The inclusion of physical processes, like
the heating of photons owing to electron–positron
annihilation  \cite{ee-12}, amounts to multiplying the photon
temperature (\ref{1nce1}) by $(11/4)^{1/3}=1.4$; therefore, we
have
\be \label{1nce11}
T_\gamma(e^+~e^-)\simeq (11/4)^{1/3}\times 0.8~ T_{\rm v}=2.77 ~K~.
\ee
We note that in other models \cite{mar-12} the fluctuations
of the product-particle density are related to primary
fluctuations of Cosmic Microwave Background radiation
\cite{33-12}.

One can find the relation of the energy density of the vector bosons
$$ \rho_{\rm v}(\eta_I)\sim  T^4\sim H_I^4\sim M^4_{I}$$
to the Universe density:
\be \label{F}
\frac{\rho_{\rm v}(\eta_I)}{\rho_{\rm tot}(\eta_I)}= \frac{  M^2_{I}}{(M_{\rm Pl}a_I)^2}
=\frac{M^2_{W}}{M_{\rm Pl}^2}=10^{-34}.
\ee
This value indicates that the inverse effect of product
particles on the evolution of the Universe is negligible.

Thus, the quantum version of General Relativity and the
Standard Model, 
considered as the result
of a spontaneous breakdown of the scale symmetry
of a conformal-invariant theory in a specific frame and initial data
can explain the origin of the Universe and its matter from vacuum.


\section{Sources of CMB radiation anisotropy
\label{sect_AppF-12}}
\makeatletter
\renewcommand{\@evenhead}{\raisebox{0pt}[\headheight][0pt]
{\vbox{\hbox to\textwidth{\hfil \strut \thechapter . Electroweak vector bosons
\quad\rm\thepage}\hrule}}}
\renewcommand{\@oddhead}{\raisebox{0pt}[\headheight][0pt]
{\vbox{\hbox to\textwidth{\strut \thesection .\quad
Sources of CMB radiation anisotropy
\hfil \rm\thepage}\hrule}}}
\makeatother
In Sections 7.3 and 9.2 we discussed two different methods of the estimation of
the primordial boson number. The first of them is the direct solution of the equations for
Bogoliubov transformation coefficients with initial data determined by the Planck principle of
minimal action for the Universe filled  in the Casimir vacuum energy.
The second is the cut of the momentum integral by means of the Boltzmann -- Chernikov distribution function
 \cite{Muller-07-12, CH-63-12}, where the temperature parameter value (\ref{cmb-12-1}), (\ref{1nce11})
is determined from the quantum uncertainty principle, and this conformal temperature
is the cosmological motion integral of the vacuum equation of state with the dominance
of the same Casimir energy.

The coincidence of these two different methods of calculation of the primordial boson number
testifies to the early thermalization of the primordial bosons before the instance
of formation of their decay products that include the Cosmic Microwave Background radiation.

Really, the arguments considered before,  mean  that the CMB photons can inform us about the parameters
of electroweak interactions and masses, including the Higgs particle
mass~\cite{2009-12}, and a possibility to estimate the magnitude of the CMB anisotropy.


Its observational value  about $\alpha_{\rm QED}^2\sim 10^{-5}$  \cite{WMAP-8-12}
testifies  to the dominance of the two photon processes. Therefore,
the CMB anisotropy revealed  in the region of the
three peaks
$$220\pm 20,\quad 546\pm 50,\qquad 800\pm 80$$
can reflect parameters of the primordial bosons and their
decay processes, in particular
$$h\to \gamma\gamma,\qquad W^+W^-\to \gamma\gamma,\quad \mbox{and}\quad ZZ \to \gamma\gamma.$$
These values were known until quite recently with accuracy as minimum 10\%.
For their description by means of  the Standard cosmological model   one uses
the metric scalar perturbations  with the negative probability
\cite{WeinbergCosmology-8-12}, that are forbidden in the considered Conformal General Relativity
with the vacuum postulate \cite{Barbashov_06-8-12}, as we have seen above in Chapter 8.

In this case,
two-photon decays of primordial Higgs particles and
annihilation processes of primordial  $W$, and $Z$  bosons
can explain three clear peaks in the CMB power spectrum
without any acoustic waves with negative energy \cite{Giovannini-12},
if the values for multipole momenta $\ell_P$ of their processes are proportional
to number of emitters at the horizon length
$$\ell_P=\widetilde{H}^{-1}(z_{\rm P})\widetilde{M}(z_{\rm P})\sim (1+z_{\rm P})^{-3}$$
(in accordance with the new CC analysis of
Supernovae type Ia data)
and the energy of any photon is proportional to the mean photon energy in CMB multiplied by
the effective scale factor
$$(1+z_P)^{-1}\sim\ell_P^{1/3}$$
in accordance with the experience
of description of the recombination epoch and  the primordial helium abundance \cite{Behnke_04,Wein_73}.

One can be convinced that first three peaks
$$\ell_1 = {220},\qquad \ell_2 ={546},\qquad \ell_3 =800$$
reflect
the ratio of $W$ and $Z$ masses
\be\label{80}
{M_Z}/{M_W} = 1.134 \approx \left({800}/{546}\right)^{1/3}= 1.136 ~\to
\ee
$$\to
~~[\,\sin^2\theta_W \approx 0.225\,]$$
and  the value of Higgs particle mass as
\be\label{118}
m_h=2M_W\left({\ell_1}/{\ell_2}\right)^{1/3}=
\ee
$$=2M_W\left({220\pm 20}/{546\pm50}\right)^{1/3}
\simeq
120\pm8 \,\mbox{\rm GeV}.$$

The Higgs boson mass is close to the present fit of the LEP
experimental data supporting rather low values just above the experimental
limit
$$114.4<m_h< 128 ~ {\rm GeV}.$$
To get a more accurate estimate of the Higgs mass and a better description of
the CMB power spectrum within the model under consideration, one has to
perform an involved analysis of the kinetic equations for
nonequilibrium Universe~\cite{Ignatyev123-12} with primordial particle creation and
subsequent decays.

For a reader who wishes to describe these peaks as acoustic disturbances
according to the formulas given in the book \cite{WeinbergCosmology-8-12},
we will make the following remarks. The first is the choice of \textit{conformal} magnitudes,
as the real observables, instead of \textit{worlds}.
The second is the choice of the rigid state equation, instead of dominance of radiation.
The third is the history of evolution of masses, instead of temperature history of the Universe.
In this case, all the characteristic resonance processes of the
history of evolution of masses (like the transition from plasma to the atoms)
occur at the same values of the redshift
$z\simeq 1100$,
as in the temperature history of the Universe.
However, in the Conformal cosmology the constant temperature
(\ref{cmb-12-1}), (\ref{1nce11})
plays the role of a fundamental parameter, determined from the microscopic quantum theory.


\section{Baryon asymmetry of the Universe
\label{sect_AppF}}
\makeatletter
\renewcommand{\@evenhead}{\raisebox{0pt}[\headheight][0pt]
{\vbox{\hbox to\textwidth{\hfil \strut \thechapter . Electroweak vector bosons
\quad\rm\thepage}\hrule}}}
\renewcommand{\@oddhead}{\raisebox{0pt}[\headheight][0pt]
{\vbox{\hbox to\textwidth{\strut \thesection .\quad
Baryon asymmetry of the Universe
\hfil \rm\thepage}\hrule}}}
\makeatother
It is well known that, because of a triangle
anomaly, $W-$ and $Z-$ boson interaction with lefthanded
fermion doublets  $\psi_L^{(i)}$
$i=1,2,...n_L$, leads to
a nonconservation of the number of fermions of each
type ${(i)}$ ~\cite{bj-12, th-12, ufn-12, 039a-12},
%
\bea \label{rub}
\partial_\mu j^{(i)}_{L\mu}=\frac{1}{32\pi^2}
{\rm Tr}\hat F_{\mu\nu}{}^*{\hat F_{\mu\nu}},
\eea
where
$$\hat F_{\mu\nu}=-\imath F^a_{\mu\nu}g_W\tau_a/2$$
is the strength of the vector fields
$$F^a_{\mu\nu}=
\partial_\mu A_\nu^a-\partial_\nu A_\mu^a+g\epsilon^{abc}A_\mu^bA_\nu^c.$$
In each of three generations of leptons (е, $\mu$, $\tau$) and color quarks, we have four fermion
doublets, in all there are: $n_L=12$ of them. Each of the 12 fermions doublets interacts with the triplet
of non-Abelian fields
$$A^1=(W^{(-)}+W^{(+)})/\sqrt{2},~~~~
A^2=\imath (W^{(-)}-W^{(+)})/\sqrt{2},~~~~A^3=Z/\cos\theta_{(W)}$$
with constant
$g=e/\sin\theta_{(W)}.$

Taking the integral of the quality in (\ref{rub}) with respect to conformal time and three-dimensional
variables $d^4x$, we can find a relation between the change
$$\Delta F^{(i)}=\int d^4x \partial_\mu j^{(i)}_{L\mu}$$
of the fermionic number
$$ F^{(i)}=\int d^3x j_0^{(i)}$$
and the Chern class functional:
$$N_{CS}=\frac{1}{32\pi^2}\int d^4x {\rm Tr}\hat F_{\mu\nu}{}^*{\hat F_{\mu\nu}}.$$
The difference is equal to
\be
\label{rub2} \Delta F^{(i)}= N_{CS} \not = 0, ~~~i=1,2,...,n_L.
\ee
The equality (\ref{rub2}) is considered as a selection rule: that, the fermionic number changes
identically for all fermion types \cite{ufn-12}:
$$N_{CS}=\Delta L^e=\Delta L^\mu=\Delta L^\tau=\Delta B/3,$$
at the same time, the change in the baryon charge $B$
and the change in the lepton charge
$$L=L^e+L^\mu+L^\tau$$
are related to each other in such a way that $B-L$ is conserved, while $B+L$ is not invariant.
Upon taking the sum of the equalities in (\ref{rub2}) over all doublets, one can obtain
$$\Delta (B+ L)=12 N_{CS}.$$

%
One can evaluate the expectation value of the Chern functional (\ref{rub2}) (in the lowest order of perturbation theory
in the coupling constant) in the Bogoliubov vacuum
 $b|0>_{\rm sq}=0$. Specifically, we have
\be
 N_{CS}=N_{\rm W}\equiv
 -\frac{1}{32\pi^2}\int\limits_0^{\eta_{L}} d\eta \int {d^3 x} \;
 \langle 0|{\rm Tr}\hat F^{\rm W}_{\mu\nu}
 {}^*\!{\hat F^{\rm W}_{\mu\nu}}|0\rangle ,
 \ee
%
where $\eta_{L_W}$ is the W- boson lifetime, and $N_{\rm W}$ is the contribution of the primordial W--bosons.
$\eta=0$ and $\eta =\eta_L$ is given by
 $$N_{\rm W} ={2}{\alpha_W}V_0
 \int\limits_{0}^{{\eta_{L}}} d\eta \int\limits_{0 }^{\infty }dk
 |k|^3 R_{\rm W}(k,\eta),
 $$
where
 $${\alpha_W}={{\alpha}_{\rm QED}}/{\sin^{2}\theta_{W}}$$
and
 $$R_{\rm W}=\frac{\imath}{2}{}_b<0|b^+b^+-b^-b^-|0>_b=-\sinh(2r(\eta_L))\sin(2\theta(\eta_L))
 $$
is the Bogoliubov condensate, that is specified by relevant solutions to the Bogoliubov equations.

By calculating the integral, with time values of life of vector bosons
$$\tau_{L_W}= 15, \qquad n_\gamma\simeq n_{\rm v},$$
we obtain the estimate of the average value of the functional of Chern -- Simons on states of the primary
bosons \cite{039a-12}
 \be \frac{N_{CS}}{V_{0}}=\frac{( N_W)}{V_{0}}
 =\frac{{\alpha}_{\rm QED}}{\sin^{2}\theta_{(W)}}
  T^3  4\times 1.44= 0.8 ~ n_{\gamma}.
\ee
Hence we obtain the following estimate of the value of violation of density of
fermion number in the considered cosmological model \cite{039a-12}
\begin{eqnarray}
\frac{\Delta F^{(i)}}{V_{0}}&=&\frac{N_{CS}}{V_{0}}
  = 0.8~ n_{\gamma},
\end{eqnarray}
where
$$n_{\gamma}={ 2.402  \times T^3 }/{\pi^2}$$
is the density of number of relic photons.
According to Sakharov \cite{sufn-12}, the violation of the fermion number is
frozen by ${\rm CP}$ - non-conservation, which leads to the density of baryon numbers
\be
\label{X} n_{\rm b}= X_{\rm CP}\frac{\Delta
F^{(i)}}{V_{(r)}}\simeq X_{\rm CP}n_{\gamma}~.
\ee
where a multiplier
$X_{\rm CP}$ is defined by electroweak interaction between $d$ and $s$ quarks $(d+s~\rightarrow ~s+d)$,
responsible for $CP$-violation, experimentally observed in the decay
of $K$-mesons \cite{o-12}.

From the ratio of the number of baryons to number of photons it is possible to estimate the
constant of ultra-weak interaction:  $X_{\rm CP}\sim 10^{-9}$.
Thus, the evolution of the Universe, the primary vector bosons and
ultra-weak interaction~\cite{o-12},
responsible for the violation of $CP$-symmetry with the constant $X_{CP}\sim 10^{-9}$,
lead to the baryonic asymmetry of the Universe with a density
\be\label{data6}
 \rho_{\rm b}(\eta=\eta_{L})
 \simeq 10^{-9} \times 10^{-34}\rho_{\rm cr}(\eta=\eta_{L}),
 \ee
where for $\eta_{L}$ it is possible to assess the future evolution of the density of
baryons, select the lifetime of the W-boson.

After the decay of bosons their temperature is inherited by the
Cosmic Microwave Background radiation. All subsequent evolution of matter in a constant cold
Universe scenario repeats the known hot Universe one \cite{three-12},
because this evolution is defined as a
conformally invariant relation of mass and temperature $m/T$.

Formulae (\ref{life}), (\ref{F}), and (\ref{data6}) provide an opportunity to assess
the ratio of the present values of the baryon density and Casimir energy density
played the role of primary quintessence of the considered model:
\be
\Omega_{\rm b}(\eta_0)=\frac{\rho_{\rm b}(\eta_{0})}{\rho_{\rm cr}(\eta_{0})}=
\left[\frac{a_0}{a_L}\right]^3=\left[\frac{a_0}{a_I}\right]^3
\left[\frac{a_I}{a_L}\right]^3.
\ee
It is take into account that the baryon density increases as mass, and the density of Casimir energy decreases
as the inverse square of the mass. Recall the value of relationships
$${\left[\frac{a_0}{a_I}\right]}^3\sim 10^{43},$$
and relation
$[{a_I}/{a_L}]^3$ given by a lifetime of bosons
(\ref{lifes}) and by the equation of state of matter $a(\eta)\sim
\sqrt{\eta}$, hence we obtain $\Omega_{\rm b}(\eta_0)$:
taking into account the delay of baryon production for the life of the vector bosons
\be\label{data7}
\Omega_{\rm b}(\eta_0)
=\left[\frac{a_0}{a_L}\right]^3 10^{-43}\sim 10^{43}
\left[\frac{\eta_I}{\eta_L}\right]^{3/2}10^{-43} \sim
\left[\frac{\alpha_{QED}}{\sin^2 \theta_{(W)}}\right] \sim 0.03 , \ee
which is consistent with observations \cite{fuk-12}.

\section{Summary}

\makeatletter
\renewcommand{\@evenhead}{\raisebox{0pt}[\headheight][0pt]
{\vbox{\hbox to\textwidth{\hfil \strut \thechapter . Electroweak vector bosons
\quad\rm\thepage}\hrule}}}
\renewcommand{\@oddhead}{\raisebox{0pt}[\headheight][0pt]
{\vbox{\hbox to\textwidth{\strut \thesection .\quad
Summary and literature
\hfil \rm\thepage}\hrule}}}
\makeatother

General Relativity and Standard Model are considered as a theory of dynamical scale symmetry with
definite initial data compatible with the accepted Higgs mechanism. In this theory the Early Universe
behaves like a factory of electroweak bosons and Higgs scalars, and it gives a possibility to identify three
peaks in the Cosmic Microwave Background power spectrum with the contributions of photonic decays
and annihilation processes of primordial Higgs, W and Z bosons in agreement with the QED coupling
constant,Weinberg’s angle, and Higgs’ particle mass of about $120\pm 6$ GeV.

The key points of our construction are the choice of the conformal variables and conformal interval.
Thus the Standard Model supplemented with the Casimir vacuum energy
does not contradict with the following scenario of the evolution of the
Universe within the Conformal cosmology~\cite{Blaschke_04}:\\
 $\eta \sim 10^{-12}s,$ {creation of vector bosons from a
 ``vacuum''};\\ [1.5mm] $10^{-12}s < \eta <
 10^{-10} s,$ {formation of baryon-antibaryon asymmetry;}\\
 $\eta \sim 10^{-10}s,$ {decay of vector bosons;}\\
 $10^{-10}s <\eta < 10^{11}s,$ { primordial chemical
evolution of matter;}\\ 
$\eta \sim 10^{11}s,$ {recombination or separation of Cosmic Microwave Background radiation;}\\
 $\eta \sim  10^{15}s,$ {formation of galaxies;}\\
  $\eta > 10^{17}s,$ { terrestrial experiments and evolution
of Supernovae.}

Our description is not complete, but it gives us a clear consistent statement of
the problems in the framework of the well established principles of
classification of observational and experimental facts in physics and astrophysics.
There are tendencies in modern cosmology \cite{Giovannini-12,MFB} to ignore these principles,
in particular, to  replace the initial data with the fundamental parameters of
the equations of motion.
However, this replacement  leads to contradictions in modern models in
applying mathematical tools of the type of the classification of
relativistic states, or the Hamiltonian method because the latter
were developed especially to solve equations with initial data.
The scale-invariant cosmological model has several features quite different
from those in  the widely accepted $\Lambda$CDM model.

In particular, the  model considered here does not need

1) considering acoustic waves $P_{D}\not =0$ and their creation from \textit{vacuum},
when the \textit{vacuum} postulate excludes these waves by the Dirac
constraint $P_{D} =0$;

2) proposing ``a dynamical inflation''
$\textsf{V}_{0}\not = \textsf{V}_{I}$ and $\textsf{K}\not = 0$, when
the inflation equation
$\rho=\textsf{K}+\textsf{V}=-p=-\textsf{K}+\textsf{V}$
is valid only if
$\textsf{K} = 0$;

3) proclaiming the dominance of the potential term
$\textsf{V}_{I}/\textsf{K}_{I}\sim a^{-6}$
at the Planck epoch, when $a^{-1}_I\sim 10^{61}$
is valid, and the kinetic term $\textsf{K}$ has a huge enhancement factor
of $a^{-6}_I\sim 10^{366}$ times with respect to the potential
one $\textsf{V}$;

4) use of the Planck epoch initial data in the dynamical theory,
where these initial data should be free from any fundamental parameters of
equations including the Planck one.

Thus, here we propose to construct Cosmology as a theory of {\it a quantum relativistic Universe}
in a direct analogy to the quantum field theory.
In the model the  Universe evolves  in the field super-space of events \cite{WdW}
with respect to   four-dimension interval in the   Minkowskian  space-time.
The latter is  defined as the tangent space-time of
invisible  Riemannian manifold  as an object of general coordinate transformations
that lead to constraints including the energy one.

Within the conformal formulation of the General Theory of Relativity and the Standard Model, we have
investigated conditions under which the origin of
matter can be explained by its cosmological creation
from a vacuum. We have presented some arguments
in support to the statement that the number of product
vector-boson pairs is sufficient for explaining the total
amount of observed matter and its content, provided
that the Universe is considered as a conventional
physical object that is characterized by a finite volume
and a finite lifetime and which is described by a
conformal invariant version of the General Theory of Relativity and the Standard Model featuring scale invariant
equations where all masses, including the
Planck mass, are replaced by the dilaton variable and
where the spatial volume is replaced by a constant.
In this case, the energy of the entire Universe in the
field space of events is described by analogy with the
description of the energy of a relativistic quantum
particle in the Minkowski space: one of the variables
(dilaton in the case being considered) becomes an
evolution parameter, while the corresponding canonically conjugate momentum assumes the role of
energy. This means that measured quantities
are identified with conformal variables that are used
in observational cosmology and in quantum field
theory in calculating cosmological particle creation
from a vacuum. Within the errors of
observation, this identification of conformal variables
with observables is compatible with data on the
chemical evolution of matter and data on Supernovae,
provided that cosmic evolution proceeds via the
regime dominated by the density of the vacuum.
Thus, the identification of conformal coordinates
and variables used in observational cosmology and in
quantum field theory with measured quantities is the
first condition under which the origin of matter can
be explained by its cosmological creation from the vacuum.
This is possible within a conformal invariant
unified theory, where the Planck mass, which is an
absolute quantity in the General Theory of Relativity,
becomes an ordinary present-day value of the dilaton
and where the Planck era loses its absolute meaning.
The construction of a stable vacuum of perturbation
theory by eliminating (through the choice of
gauge-invariant variables) unphysical fields whose
quantization leads to a negative normalization of the
wave function in this reference frame is the second
condition.

Finally, the elimination of divergences in summing
the probabilities of product particles over their momenta
by thermalizing these particles in the region
where the Boltzmann H-theorem is applicable is the
third condition.
Under these conditions, it has been found in
the present study that, in describing the creation
of vector bosons from a vacuum in terms of conformal
variables, one arrives at the temperature
of Cosmic Microwave Background
radiation as an integral of the motion of the
Universe and at the baryon antibaryon asymmetry
of the Universe with the superweak-interaction coupling
constant and the baryon density
of these results being
in satisfactory agreement with the corresponding
observed values and being compatible with the most
recent data on Supernovae and nucleosynthesis.

In Chapter 9, it was shown that only gravitons,
Higgs bosons (and the corresponding longitudinal components of the vector bosons) can be created from the vacuum
in an empty Universe.
One of the central results is the creation from the vacuum of $10^{88}$ particles of boson Higgs field,
decays of which form
all the material content of the Universe in accordance with modern observational facts
on Supernovae and primordial nucleosynthesis.
We have presented arguments in favor of the fact that the data on
Supernovae, primordial nucleosynthesis, and cosmological particle production converted to units of
relative standard length, can be described in conformal evolution scenario of the Universe by a
single regime of the Casimir vacuum energy dominance.
This chapter shows that transversal bosons in the course of their lifetimes, form baryon asymmetry of the Universe
as a consequence of ``polarization'' of these bosons the vacuum Dirac sea of left fermions according to the selection rules
of the Standard Model \cite{ufn-12}. In this case the difference between the number of baryons and leptons remains,
and their sum is not preserved. The experimentally observed ultra-weak interaction~\cite{o-12},
responsible for the violation of CP-symmetry with the constant
$X_{ CP}\sim 10^{-9}$,
freezes the baryon asymmetry of the Universe with a density (\ref{data6}).
After the decay of bosons, their temperature is inherited by Cosmic Microwave Background radiation.

All subsequent evolution of matter in the constant cold Universe follows the known scenario of the hot
Universe \cite{three-12}, because this evolution is defined as
conformally invariant relations of mass and temperature $m/T$.
Baryon density increases as the mass, and density of the Casimir energy
decreases as the inverse square of the mass. As a result, the present value of the relative baryon density is
equal, up to a factor of order unity, to the Weinberg constant
$$\Omega_{\rm b}(\eta_0)\simeq \dfrac{\alpha_{QED}}{\sin^2 \theta_{W}} \sim 0.03$$
in agreement with observations.

In this Chapter we have shown that the standard definition of the temperature of the primary boson
(a fundamental constant in the conformal evolution of the Universe)
as the square root of the product of the density of the particles and their scattering cross sections
is not contrary to the direct calculation of
number of produced particles with a constant temperature.
In other words,
the temperature of the primary vector W-, Z-bosons can be estimated from the same formula
which is used to describe the chemical evolution of matter \cite{three-12}: time of setting of temperature,
equal to the inverse of the product of the particle density on the cross section of scattering can not
exceed the lifetime of the Universe, which is proportional to the inverse of the Hubble parameter.
The temperature history of the hot Universe, rewritten in conformal variables, looks like the history of
evolution of the masses of elementary particles in the Universe at a constant cold
temperature of the CMB.

Independence of the temperature of the background radiation $T_{\rm CMB}\sim 2.725$  K
from the redshift $z$, at first glance, is in contrast to the observation
6. 0 K < $T_{\rm CMB}$ ($z$ =
2.3371) < 14 K.
The relative occupancy of the various energy
levels, where the temperature was brought to this observation,
follows from the Boltzmann statistics.
However, the argument of the Boltzmann factor
as the weight ratio of the temperature has the same dependence on
factor z
and in the cold Universe.
Therefore, this ratio can be interpreted as the dependence of the energy levels, {\it id est} mass, of
redshift at constant temperature.
Distribution of chemical elements as well,
determined mainly by Boltzmann factors, depend
on conformal invariant relationship of temperature to mass.
\chapter{Conformal cosmological perturbation theory
\label{sect_13cosm}}
\renewcommand{\theequation}{13.\arabic{equation}}
\setcounter{equation}{0}

\makeatletter
\renewcommand{\@evenhead}{\raisebox{0pt}[\headheight][0pt]
{\vbox{\hbox to\textwidth{\hfil \strut \thechapter . Conformal cosmological perturbation theory
\quad\rm\thepage}\hrule}}}
\renewcommand{\@oddhead}{\raisebox{0pt}[\headheight][0pt]
{\vbox{\hbox to\textwidth{\strut \thesection .\quad
The equations of the theory of perturbations
\hfil \rm\thepage}\hrule}}}
\makeatother

\section{The equations of the theory of\\ perturbations}
In this Chapter, the conformal cosmological theory of
perturbations will be considered for calculation of the lapse function ${\cal N}$
and nonzeroth harmonics of dilaton $\overline{D}$,
with a certain geometric interval (\ref{ds-12})
\bea\label{ds-12-13}
 \widetilde{ds}^2
&=& e^{-4\overline{D}}{\cal N}^2 d\eta^2-
\\\nonumber
&-&\left(
dX_{(b)}-X_{(c)}[\omega^R_{(c)(b)}(d)+\omega^L_{(c)(b)}(d)] -
 {\cal N}_{(b)} d\tau\!\right)^2,
\eea
where
$$
 {\cal N}=\frac{\langle\sqrt{\widetilde{{\cal H}}}\rangle^2}{\widetilde{\cal H}}.
$$
Recall that in the general case the local energy density (\ref{CC-3}) is
 \bea\label{CC-1a}
  \widetilde{{\cal H}}=
  -\frac{4}{3}{e}^{-7D/2}
 \triangle  {e}^{-D/2}+ \!\!\!\!
  \sum\limits_{J=0,2,3,4,6} {e}^{-JD}{\cal T}_J(\widetilde{F}),
\eea
$$\triangle =\partial_i[{\bf e}^i_{(b)}{\bf e}^j_{(b)}\partial_j]$$
is the Beltrami -- Laplace operator. The sum is over of the densities of states:
rigid radiation $(J=2)$, matter $(J=3)$, curvature $(J=4)$, $\Lambda$-type term $(J=6)$,
respectively, in terms of the conformal fields
\be\label{ratio-1b}
 \widetilde{F}^{(n)}=e^{nD}{F}^{(n)},
\ee
where  is the conformal weight.

In this case, the equation of the nonzeroth harmonics (\ref{e2t2}) and (\ref{e2tc2}) takes the form
\cite{Barbashov:2005perturb-13}
 \bea \label{CC-4}
  T_D-{\langle T_D\rangle}=0,
 \eea
where
 \bea \label{1-D}
 {T}_{D} =\frac{2}{3}
 \left\{7{\cal N}{e}^{-7D/2}
 \triangle {e}^{-D/2}+{e}^{-D/2} \triangle
 \left[{\cal N}{e}^{-7D/2}\right]\right\}+
 \eea
$$ +{\cal N}\sum\limits_{J=0,2,3,4,6}J {e}^{-JD}{\cal T}_J.$$

One can solve all Hamiltonian equations (\ref{CC-1a}), and (\ref{CC-4})
to define simplex components
 \bea \label{dg-2aab}
 &&{\widetilde{\omega}}_{(0)}=e^{-2{D}}{\cal N} d\tau,~~~
{\cal N}=\dfrac{\langle\sqrt{\widetilde{{\cal
H}}}\rangle}{\sqrt{\widetilde{{\cal H}}}},
\\
 \label{dg-3ab}
&&{\widetilde{\omega}}_{(b)}=dX_{(b)}-X_{(c)}\omega^R_{(c)(b)}
+{\cal N}_{(b)}d\tau.
 \eea
Recall that in the lowest order of perturbation
theory, $\omega^R_{(c)(b)}$ describes the free
one-component transverse strong gravitational wave considered in Section~3. The longitudinal
component of the shift-vector ${\cal N}_{(b)}$, unambiguously determined by the  constraint (\ref{h-2g1}),
 becomes equal to
\be \label{h-2g13}
\partial_\eta e^{-3\overline{D}}+
\partial_{(b)}\left(e^{-3\overline{D}}{\cal N}_{(b)}\right)=0.
\ee

\section{The solution of the equations for\\ small fluctuations}
\makeatletter
\renewcommand{\@evenhead}{\raisebox{0pt}[\headheight][0pt]
{\vbox{\hbox to\textwidth{\hfil \strut \thechapter . Conformal cosmological perturbation theory
\quad\rm\thepage}\hrule}}}
\renewcommand{\@oddhead}{\raisebox{0pt}[\headheight][0pt]
{\vbox{\hbox to\textwidth{\strut \thesection .\quad
The solution of the equations for small fluctuations
\hfil \rm\thepage}\hrule}}}
\makeatother

For small fluctuations
 \bea\label{e1-2-13}
 {\cal N}e^{-7\overline{D}/2}=1-{\nu}_1,\qquad
e^{-\overline{D}/2}=1+{\mu}_1+\cdots
\eea
the first order Eqs. of (\ref{CC-1a}) and (\ref{1-D}) take the form
\bea\nonumber
[-\hat{\triangle}+14\rho_{(0)}-\rho_{(1)}]\mu_{1} +
   2\rho_{(0)}\nu_1=\overline{\cal T}_{(0)},
\eea
\bea\nonumber
[7\cdot 14\rho_{(0)}-14\rho_{(1)}+\rho_{(2)}]\mu_1
 +[-\hat{\triangle}+
14\rho_{(0)}-\rho_{(1)}]\nu_1 =
7\overline{\cal T}_{(0)}-\overline{\cal T}_{(1)},
\eea
where
 \bea\label{ec1-3}
  \rho_{(n)}=\langle{\cal T}_{(n)}\rangle \equiv
 \sum\limits_{J=0,2,3,4,6}  (2J)^n(1+z)^{2-J}
  \langle{\cal T}_{J}\rangle ,
  \eea
  \bea
\label{ec1-4}
 {\cal T}_{(n)}= \sum\limits_{J=0,2,3,4,6}(2J)^n(1+z)^{2-J}
{\cal T}_{J}.
 \eea

 In the first order
of perturbation with respect to the Newton coupling constant the lapse function and the dilaton
takes the form \cite{Barbashov:2005perturb-13}
  \bea \label{12-17}
\!\!\!\!\!\!e^{-\overline{D}/2}
\!=\!1\!+\!\frac{1}{2}\int d^3y\Bigg{[}G_{(+)}(x,y)
\overline{T}_{(+)}^{(D)}(y)\!+\!G_{(-)}(x,y) \overline{T}^{(D)}_{(-)}(y)\Bigg{]},
 \\ \label{12-18}
\!\!\!\!\!\!{\cal N}e^{-7\overline{D}/2}
\!=\!1\!-\!\frac{1}{2}\int d^3y\Bigg{[}G_{(+)}(x,y)
 \overline{T}^{(N)}_{(+)}(y)\!+\!G_{(-)}(x,y) \overline{T}^{(N)}_{(-)}(y)\Bigg{]},
  \eea
where $G_{(\pm)}(x,y)$ are the Green functions satisfying the equations
 \be
 \label{2-19}\nonumber
 [\pm  m^2_{(\pm)}- \triangle ]G_{(\pm)}(x,y)=\delta^3(x-y).
 \ee
Here
 $$m^2_{(\pm)}= H_0^2 \dfrac{3(1\!+\!z)^2}{4}\Bigg{[}\!{14(\beta\pm 1)}
\Omega_{(0)}(a)\mp \Omega_{(1)}(a)\Bigg{]},$$
$$\beta=\sqrt{1+[\Omega_{(2)}(a)-14\Omega_{(1)}(a)]/[98\Omega_{(0)}(a)]},$$
and
 \bea
  \label{1cur1}
 \overline{T}^{(D)}_{(\pm)} = \overline{{\cal T}}_{(0)}\mp7\beta
  [7\overline{{\cal T}}_{(0)}-\overline{{\cal T}}_{(1)}],
  \eea
 \bea\label{1curv2}
 \overline{T}^{(N)}_{(\pm)} = [7\overline{\cal T}_{(0)}
 -\overline{\cal T}_{(1)}]
 \pm(14\beta)^{-1}\overline{\cal T}_{(0)},
 \eea
 are the local currents, and
 \bea
 \label{1cur3}
 \Omega_{(n)}(a)=\sum\limits_{J=0,2,3,4,6}(2J)^n(1+z)^{2-J}\Omega_{J},
 \eea
where
$$\Omega_{J=0,2,3,4,6}= \frac{\langle{\cal T}_J\rangle}{H_0^2}$$
are partial densities of states: rigid, radiation, matter, curvature, $\Lambda$-term, respectively;
$$\Omega_{(0)}(a=1)=1,\qquad 1+z=a^{-1}$$
 is the Hubble parameter.

In the context of  these definitions, a full  family of solutions  (\ref{12-17}), (\ref{12-18}) for
the lapse function and the nonzeroth dilaton harmonics of the Hamiltonian constraints
(\ref{1-fbb3}) -- (\ref{Dirac-c2}) yields a Newton-type potential. In particular, for a point mass
distribution in a finite volume which corresponds to the nonzero terms with

a)$J=0,3$ in Eq. (\ref{ec1-3});

b)$J=3$ in Eq. (\ref{ec1-4});

c)$J=0,3$ in Eq. (\ref{1cur3})\\
(otherwise zero), we have
 \be \label{T-1}
 \overline{{\cal T}}_{(0)}(x)=\frac{\overline{{\cal T}}_{(1)}(x)}{6}
 \equiv \frac{3}{4a^2} M\left[\delta^3(x-y)-\frac{1}{V_0}\right].
 \ee

As a
result, solutions (\ref{12-17}) and (\ref{12-18}) are transformed to the Schwarzschild-type form
 \bea
 \label{N-1}
\!\!\!\!  e^{-\overline{D}/2}=1+
  \frac{r_{g}}{4r}\Bigg{[}\frac{1+7\beta}{2}e^{-m_{(+)}(a)r}
 \!+\! \frac{1-7\beta}{2}\cos{m_{(-)}(a) r}\Bigg{]},\\
 \label{N-2}
\!\!\!\! {\cal N}e^{-7\overline{D}/2}=1-
 \frac{r_{g}}{4r}\Bigg{[}\frac{14\beta+1}{28\beta}e^{-m_{(+)}(a)r}
 \!+\!\frac{14\beta-1}{28\beta}\cos{m_{(-)}(a) r}\Bigg{]},
 \eea
where
$$r_{g}=M/M^2_{\rm Pl},~~ \beta=5/7,~~ m_{(+)}=3m_{(-)},$$
$$m_{(-)}=H_0\sqrt{3(1+z)\Omega_{\rm Matter} /2}.$$
These solutions describe the
Jeans-like spatial oscillations of
 the scalar potentials (\ref{N-1}) and (\ref{N-2}) even for the
 case of zero pressure.
 These spatial oscillations can determine the clustering of matter  in the recombination epoch,
 when  the redshift is close to the value
  $z_{\rm recomb.}\simeq 1100$.
  Indeed, if we use for the matter clustering parameter
(that follows from spatial oscillations of the modified Newton law
(\ref{N-1}), (\ref{N-2}))
  the observational value  \cite{Bajan-14-8-13}
\begin{equation}\label{cl-1}
r_{\rm clustering} \simeq 130\, \mbox{\rm Mps} \simeq\dfrac{1}{m_{(-)} }
=\dfrac{1}{ H_0[\Omega_{\rm Matter} (1+z_{\rm recomb})]^{1/2}},
 \end{equation}
 one obtains  $\Omega_{\rm Matter}\sim  0.2$.
This estimation is in agreement with the one
recently discovered in the quest of the large scale periodicity
distribution (see for details in~\cite{Zakharov:2010perturb-13}).

Constraint  (\ref{h-2g1}) yields
the shift of the coordinate origin in the process of the evolution
 \be \label{2-23}
 {\cal N}^i=\left(\frac{x^i}{r}\right)
 \left(\frac{\partial_\eta V}{\partial_r V}\right), \qquad
 V(\eta,r)=\int\limits_{}^{r}d\widetilde{r}\;
 \widetilde{r}^2e^{-3\overline{D}(\eta,\widetilde{r})}.
 \ee
In the limit $H_0=0$ at $a_0=1$, the solutions (\ref{N-1}) and (\ref{N-2}) coincide
with the isotropic Schwarzschild solutions:
 $$e^{-\overline{D}/2}=1+\frac{r_g}{4r},~~~~
 {\cal N}e^{-7\overline{D}/2}=1-\frac{r_g}{4r},~~~~
 {\cal N}_{i}=0.$$
 Solution~(\ref{N-1})
doubles the angle of the photon beam deflection by the Sun field. Thus, the CGR  provides also the
Newtonian limit in our variables.

\section{Summary}
\makeatletter
\renewcommand{\@evenhead}{\raisebox{0pt}[\headheight][0pt]
{\vbox{\hbox to\textwidth{\hfil \strut \thechapter . Conformal cosmological perturbation theory
\quad\rm\thepage}\hrule}}}
\renewcommand{\@oddhead}{\raisebox{0pt}[\headheight][0pt]
{\vbox{\hbox to\textwidth{\strut \thesection .\quad
Summary and literature
\hfil \rm\thepage}\hrule}}}
\makeatother

Chapter 13 was developed to the conformal diffeoinvariant version of the cosmological perturbation theory.
We have received a modification of the Schwarzschild solutions for the evolution of the Universe.
It was shown that the non zero harmonics of the dilaton lead to the Jeans oscillations even in the case of a massive dust.

\chapter{Cosmological modification of Newtonian dynamics}
\renewcommand{\theequation}{14.\arabic{equation}}
\setcounter{equation}{0}
\section{Free motion in conformally flat metric}
\makeatletter
\renewcommand{\@evenhead}{\raisebox{0pt}[\headheight][0pt]
{\vbox{\hbox to\textwidth{\hfil \strut \thechapter . Cosmological modification of Newtonian dynamics
\quad\rm\thepage}\hrule}}}
\renewcommand{\@oddhead}{\raisebox{0pt}[\headheight][0pt]
{\vbox{\hbox to\textwidth{\strut \thesection .\quad
Free motion in conformally flat metric
\hfil \rm\thepage}\hrule}}}
\makeatother

The considered above
joint unitary irreducible representations of the affine and conformal groups added by the SM fields contain
both the Newtonian dynamics of a massive classical particle and  the Friedmann cosmological metrics.
The problem of the validity of the Newtonian dynamics arises when the
Newtonian  velocity value  of a cosmic object becomes the order of the Hubble velocity value of this
object \cite{Einstein:1945-14}.
Another problem is the choice of a frame of reference where initial data are given.

In the cosmological models considered above,
we faced with the three classes of frames of reference. First of them is the world time-space
interval in
the Friedmann -- Lem\^aitre -- Robertson -- Walker (FLRW) metrics
\be\label{ds:2w}
{ds}^2= dt^2 -a^2(t)dx_i^2=a^2(\eta)\widetilde{ds}^2=a^2(\eta)\left[d\eta^2-dx_i^2\right],
\ee
associated with heavy massive body.
Here $dt$ is the world time,
$\eta$ is the conformal time, $x_1$, $x_2$, $x_3$ are the conformal coordinates, and $a(\eta)$ is
the conformal scale factor.
The second class differs from the first one by  the conformal long interval
\be\label{ds:2}
\widetilde{ds}^2=\left[d\eta^2-dx_i^2\right].
\ee
 and varying mass
\be\label{ds:2m}
m(\eta)=m_0a(\eta).
\ee
 The third class of frames of reference is associated with luminosity interval
 \be\label{ds:2L}
{ds_L}^2= a^{-6}(\eta){ds}^2=a^{-4}\widetilde{ds}^2=
a^{-4}\left[d\eta^2-dx_i^2\right].
\ee
This class of frames of reference comoves the void  local volume element.

In this Chapter we consider the dynamics of a classical particle in both the world interval (\ref{ds:2w})
  and the conformal one (\ref{ds:2})
  \cite{BehnkeSuper-14, Zorin-3-14, Zorin-5-14, Zorin-6-14}.

The one-particle energy $E=p_0$ is defined by the constraint
\be\label{qft1}
p_\mu p_\nu-m^2(\eta) 
=0,
\ee
which implies that
 \be\label{solution:p_0}
p_0=\sqrt{p^2+m^2(\eta)}\simeq m(\eta) + \frac{p^2}{2m(\eta)},
\ee
where $m(\eta)=m_0a(\eta)$ is a running mass (\ref{ds:2m}).

The action of a relativistic particle in a conformally flat metric  (\ref{ds:2}) in the
non-relativistic limit leads to the classical action for a particle
 \be\label{qft2} S_0=\int\limits_{\eta_I}^{\eta_0} d\eta
\left[p_ix'_i-p_0+m\right],
\ee
where $x'_i=dx^i/d\eta$, а $p_0$
is given by expression (\ref{solution:p_0}).

Let us consider action (\ref{qft2}) for the radial motion in the
non-relativistic limit:
\be\label{S:N}
S_{N}=\int\limits_{\eta_I}^{\eta_0} d\eta
\frac{{r'}^2m(\eta)}{2};
\ee
here $r=\sqrt{x_ix^i}$ and
$r'=dr/d\eta$.
In this case, the equation of motion is
\be
\label{qft3}[r'(\eta)m(\eta)]'=0
\ee
with initial data
$$r_I=r(\eta_I),\qquad r'_I=p_I/m_0,\qquad m_I=m(\eta_I).$$
This equation has the following solution:
\be
\label{qft4}
r(\eta)=r_I+p_I\int\limits_{\eta_I}^{\eta}\frac{d \bar
\eta}{m(\bar \eta)}.
\ee

The Friedmann world time $dt=d\eta \, a(\eta)$ and the absolute coordinate
\be
R(t)=a(\eta)r(\eta)
\ee
are determined by the conformal transformation with the scale factor
$$a(\eta(t))=a(t),$$
which is usually chosen to be unity for the modern age
$$\eta=\eta_0: \qquad a(\eta_0)=1,$$
the scale factor in the initial
time moment
$\eta=\eta_I$ is defined by the $z$-factor:
$$a(\eta_I)=a_I=\frac{1}{(1+z_I)},$$
where $z(\eta_I)=z_I$.
Since Friedmannian variables are tied to the modern era
$\eta=\eta_0$, the time $\eta_I$ is convenient to replace by $\eta_0$.
Then the world spatial interval
$$R(t)=a(\eta)r(\eta)$$
is set by the expression
\be\label{qft5}
R(t)=a(t)\left[r_0+\frac{p_I}{m_0}\int\limits_{t_I}^{t} \frac{d
\bar t}{a^2(\bar t)}\right]
\ee
and satisfies the equation of motion
\be\label{qft6} \ddot
R(t)-(H^2+\dot H)R=0,
\ee
where
$H(t)=\dot a(t)/a(t)$ is the Hubble parameter.
The equation of motion follows from the action
\be\label{qft7}
S_{N}(t)=\int\limits_{t_I}^{t_0}dt\frac{\left(\dot R-HR
\right)^2m_0}{2}.
\ee
The same action can be obtained geometrically using the definitions of the measured intervals
in the Standard cosmology
$$dl=a(t)dr=d [ra(t)]-r\dot a(t)dt=[\dot R-HR]dt,$$
including the world space interval
$R(t)=r\,a(t)$
in a space-time with the Friedmann -- Lem\^aitre -- Robertson -- Walker (FLRW) metric
\bea\label{gr5}
(ds^2)=(dt)^2-a^2(t)(dx^i)^2.
\eea
The observable coordinates
$X^i$
of the expanding Universe can be written as
\bea\label{fc}
X^i=a(t)x^i, \quad dX^i=a(t)dx^i+x^ida(t),
\eea
and instead of the Euclidean differentials $dX^i$ the covariant ones are used
\be\label{A0}
a(t)dx^i=d[a(t)x^i]-x^ida(t)=dX^i-X^i\frac{da(t)}{a(t)}.
\ee
In the Standard cosmology a particle mass is constant.

The interval (\ref{gr5}) in terms of variables (\ref{fc}) becomes equal to
 \be\label{A}
 (ds^2)=(dt)^2-\sum_{i=1,2,3}\left(dX^i-H(t)X^idt\right)^2,
 \ee
where $H(t)$ is the world Hubble parameter. All these equations by
conformal transformations are reduced to the equations given in the book of Peebles \cite{Peebles-14}.

\section{The motion of a test particle\\ in a central field}
\makeatletter
\renewcommand{\@evenhead}{\raisebox{0pt}[\headheight][0pt]
{\vbox{\hbox to\textwidth{\hfil \strut \thechapter . Cosmological modification of Newtonian dynamics
\quad\rm\thepage}\hrule}}}
\renewcommand{\@oddhead}{\raisebox{0pt}[\headheight][0pt]
{\vbox{\hbox to\textwidth{\strut \thesection .\quad
The motion of a test particle in a central field
\hfil \rm\thepage}\hrule}}}
\makeatother

The energy of a particle that moves along a geodesic line in a space with a given metric can be found by solving the equation
of the mass shell. Equating the square of 4-momentum
$p_\mu p^\mu$
to the square of the mass in the metric (\ref{ds:2}):
\be p^2=g^{\mu\nu}
p_\mu p_\nu = m^2
\ee
we find an expression for the energy $p_0$
\be\label{eq:approx} p_0 \approx
\pm\left[\left(1-\frac{r_g}{2r}\right)m + \frac{p_r^2}{2m}
+\frac{p_\theta^2}{2mr^2} \right].
\ee
From the condition of positive energy $p_0>0$ in the right-hand side (\ref{eq:approx})
we choose a positive sign; as a result, in the non-relativistic
limit, we arrive to the action\footnote{The equations of motion for a
free particle, taking into account the expansion of the Universe, do not differ from the ones
given in the monograph of Peebles \cite{Peebles-14}.}
\bea\label{S:classic}
S_{\rm classic} &=& \int\limits_{\eta_I}^{\eta_0} d\eta \left[p_r r'
+p_\theta \theta' - E_{\rm classic}\right],
\eea
where
\be\label{E:classic}
E_{\rm classic}=\frac{p_r^2}{2m}+\frac{p_\theta^2}{2mr^2}-\frac{r_gm}{2r},
\ee
and $m=m(\eta)$ is the conformal mass of a test body, which depends on the time (evolution)
 and is determined by (\ref{ds:2m}).
The product $r_gm$ is a conformal invariant independent of time. For a constant mass $ m = m_0 $ one obtains
the classical action.

In the case of a particle with a constant mass moving along a circle ($r=r_0$) the Newtonian speed
$$w_0=\sqrt{\frac{r_g}{2r_0}}$$
coincides with the orbital one
$$v_0=\frac{p_\theta}{m_0r_0}.$$
The equality $w_0=v_0$ is the basis of the analysis of observational data on dark matter in the Universe \cite{ch-14}.

To determine the region of applicability of the Newton theory with a constant mass and the status of the circular trajectories,
we shall investigate the Kepler problem for variable masses (\ref{def:m}), the dependence on time of
which is determined by astrophysical data on the Supernovae \cite{BehnkeSuper-14}.

\section{The Kepler problem in the\\ Conformal theory}\label{exact:solution}
\makeatletter
\renewcommand{\@evenhead}{\raisebox{0pt}[\headheight][0pt]
{\vbox{\hbox to\textwidth{\hfil \strut \thechapter . Cosmological modification of Newtonian dynamics
\quad\rm\thepage}\hrule}}}
\renewcommand{\@oddhead}{\raisebox{0pt}[\headheight][0pt]
{\vbox{\hbox to\textwidth{\strut \thesection .\quad
The Kepler problem in the Conformal theory
\hfil \rm\thepage}\hrule}}}
\makeatother

Taking into consideration the dependence of the coordinate distance from the conformal time
(\ref{ds:2}) and cosmic
evolution in a rigid state condition
one can move from the evolution parameter $\eta$ to a
monotonically increasing function $a(\eta)$
\be\label{def:a}
a(\eta)=\sqrt{1+2H_0(\eta-\eta_0)}.
\ee
Then from the equation of motion for the Newtonian action (\ref{S:classic}) taking into account the dependence
of the mass from the conformal time (\ref{def:m}) and relation
(\ref{def:a}) we obtain an explicit parametric solution $a(\tau)$ and
$r(\tau)$ with a parameter $\tau$ introduced in \cite{Zorin-5-14, Zorin-6-14, ZP-14}:
\bea\label{14-28}
a(\tau) = c_1 \frac{N_1(\tau)}{\tau^{2/3}N(\tau)}, ~~~~~~~~
\frac{r(\tau)}{r_0} = c_2\tau^{2/3}N(\tau),
\eea
where
 \bea
N(\tau)=\alpha_1 U(\tau)^2+\beta_1 U(\tau) V(\tau) + \gamma_1 V(\tau)^2,
\eea
\bea
N_1(\tau)=\left(\tau\frac{dN(\tau)}{d\tau}+\frac
23 N(\tau)\right)^2\pm 4\tau^2N(\tau)^2 +\omega^2 \Delta,
\label{def:N2}
\eea
\bea
\Delta = 4 \alpha_1 \gamma_1 - \beta_1^2 > 0, ~~~~
c_1,~c_2,~\alpha_1,~\beta_1, ~\gamma_1={\rm const},
\label{sol:delta}
\eea
\bea
c_1=\left(\frac{3w_0^2}{4c_0^2}\right)^{1/3}
\frac{c_0v_0}{2w^2|\omega|\Delta^{1/2}},~~~~
c_2=\left(\frac{4c_0^2}{3w_0^2}\right)^{1/3} \frac{v_0}{
|\omega|\Delta^{1/2}}.
 \eea
Here
\bea\label{def:vwc}
w_0^2=\frac{r_g}{2r_0}, \qquad v_0=\frac{p_\theta}{m_0r_0}, \qquad
c_0=H_0r_0
\eea
are the Newtonian, orbital and cosmic velocity, respectively.

For the upper sign in (\ref{def:N2})
\be\label{solution:restricted}
U(\tau)=J_{1/3}(\tau),~~~~V(\tau)=Y_{1/3}(\tau),~~~~ \omega=\frac
2\pi,
\ee
where $J_{1/3}(\tau)$ and $Y_{1/3}(\tau)$ are the Bessel functions of
the first and the second kind. For the lower sign
\be\label{solution:unrestricted}
U(\tau)=I_{1/3}(\tau),~~~~V(\tau)=K_{1/3}(\tau),~~~~ \omega=-1,
\ee
where $I_{1/3}(\tau)$ and $K_{1/3}(\tau)$ are the modified Bessel functions
of the first and the second kind.

The solution (\ref{14-28}) -- (\ref{solution:unrestricted}) includes
five independent constants which can be found from the following
algebraic system of equations:
\bea
 \left.\frac{r}{r_0}\right|_{\tau=\tau_0}&=&1,\quad
 \left.\frac{dr}{da}\right|_{\tau=\tau_0}=0,\quad
 \left.a\right|_{\tau=\tau_0}=1,\nonumber\\ &&\label{eq:ConstIntegration}\\
 \frac 9{64}\left(\frac{c_2^2}{c_1}\right)^2\omega^2\Delta&=&\frac{v_0^2}{c_0^2},\qquad\qquad
 \frac 9{128}\left(\frac{c_2}{c_1}\right)^3 =  \frac{w_0^2}{c_0^2} \nonumber
\eea
in the region of their solvability. For example, for
$$\tau_0=1,\qquad v_0^2=0,25,\qquad w_0^2=0,05,\qquad c_0=1$$
the system has the following solution:
$$c_1=-0,48,\quad c_2=-0,32,\quad \alpha_1=-0,78,\quad \beta_1=0,\quad \gamma_1=-0,48,$$
with the condition $\Delta>0$.

The solution corresponding to the lower sign in (\ref{def:N2})
is restricted to zero and is not limited in infinity because of the properties of the
function $K_{1/3}(\tau)$. The solution corresponding to the upper sign in (\ref{def:N2})
is limited in infinity and describes a finite motion along an ellipse.
A character of the motion at short times can be regarded as the determination of a periodic regime after some initial
perturbation.

These two types of solutions correspond to two different signs of energy
(\ref{E:classic}): positive energy corresponds to a free motion of a particle and its
negative energy corresponds to a bound state.

\section{Capture of a particle by a central field}
\makeatletter
\renewcommand{\@evenhead}{\raisebox{0pt}[\headheight][0pt]
{\vbox{\hbox to\textwidth{\hfil \strut \thechapter . Cosmological modification of Newtonian dynamics
\quad\rm\thepage}\hrule}}}
\renewcommand{\@oddhead}{\raisebox{0pt}[\headheight][0pt]
{\vbox{\hbox to\textwidth{\strut \thesection .\quad
Capture of a particle by a central field
\hfil \rm\thepage}\hrule}}}
\makeatother

\begin{figure}[htb]
\begin{center}
\includegraphics[width=0.7\textwidth]{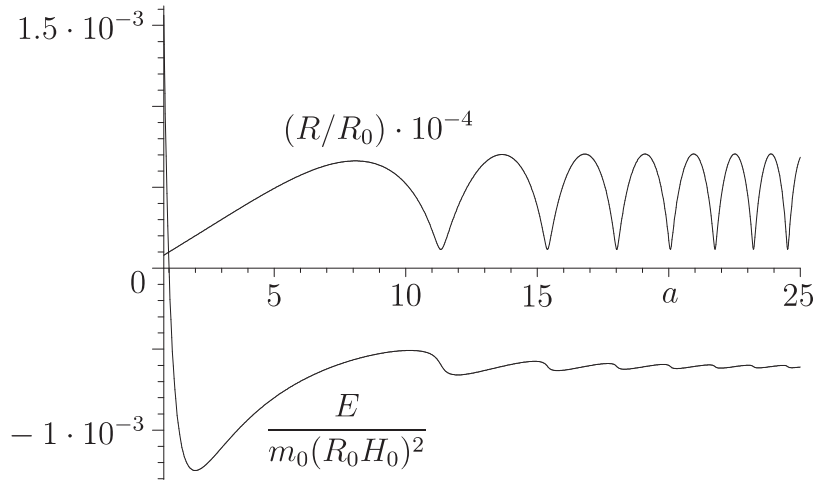}
\end{center}
\caption{\small The upper part of the graph shows the solution of the equations for action (\ref{S:classic})
in dimensionless variables $y(x)=R/R_I$ and $x=H_I(t-t_I)$ with the boundary conditions $y(x=0)=1$ and $y'(x=0)=0$.
The curve in the lower part of the graph shows the evolution of the total
energy (\ref{E:classic}) in variables $R=ar$ and $P=p/a$.}
\label{fig:h-x}\end{figure}

From the equations of motion resulting from
(\ref{S:classic}) and determination of the energy (\ref{E:classic}) one can
find the rate of the energy change of the object:
\be
\label{eq:E-eta}\frac{d E_{\rm
classic}}{d\eta}=-H(\eta)\left[\frac{p_r^2+p_\theta^2/r^2}{2m}\right],
\ee
where $H(\eta)=da/d\eta/a$ is the Hubble parameter. From (\ref{eq:E-eta}) it follows that the derivative of the energy is always
negative and tends to zero, so the energy itself asymptotically decreases to a negative value, and the reason
of the non-conservation of energy is the cosmic evolution of the masses
(see Fig. \ref{fig:h-x}) \cite{Zorin-2-14, Zorin-9-14}.

Thus, the cosmic evolution of the mass reduces the energy of the test particle to negative
asymptotic values under the condition $E=0$ which, in particular, takes place under the initial data
$v_I^2=2w_I^2$; particle transfers to a bound state and its trajectory is an ellipse.
The described mechanism of capturing particles can be applied to the dynamics of stars and galaxies
and should lead to the formation of galaxies and clusters with an anisotropic
distribution of the Hubble flows in the Local Group, which is supported by the
observations \cite{Karach1-14, Longair-14}.

\section{The problem of dark matter \\ in Superclusters}
\label{Chapter:Bound}
\makeatletter
\renewcommand{\@evenhead}{\raisebox{0pt}[\headheight][0pt]
{\vbox{\hbox to\textwidth{\hfil \strut \thechapter . Cosmological modification of Newtonian dynamics
\quad\rm\thepage}\hrule}}}
\renewcommand{\@oddhead}{\raisebox{0pt}[\headheight][0pt]
{\vbox{\hbox to\textwidth{\strut \thesection .\quad
The problem of dark matter in Superclusters
\hfil \rm\thepage}\hrule}}}
\makeatother

In modern cosmological investigations, the effects of dark matter are analyzed by using characteristics of the
Newtonian motion in the gravitational fields or clusters of galaxies
\cite{ch-14, E:1-14, E:2-14, Rubin:83-14, pri-14}; however, the following
discrepancy occurs: The Newtonian motion of galaxies is described in a flat space-time
$$ds^2=dt^2-\sum_i(dx^i)^2,$$
and analysis of the observational data is carried out in terms of the metric FLRW (\ref{gr5}).

Let us consider the Newtonian motion of a particle in a gravitational field of a space
with the FLRW metric in which for the observed coordinates
in an expanding Universe the coordinates (\ref{fc}) are adopted and instead of
differentials of the Euclidean space $dX^i$ the
covariant differentials of the FLRW space are used (\ref{A0}).
In this case, the Kepler problem is defined by the equation
\be\label{eq:Kepler}
\ddot R(t)-(H^2+\dot
H)R-\frac{(m_0R^2\dot\theta)^2}{m_0^2R^3}+\frac{r_g}{2R^2}=0.
\ee
This equation reduces to the equation that has been solved by transition to conformal variables.

The law of conservation of energy in the flat space ($H(t) = 0$)
leads to the next dependence of the radius from the orbital speed:
\be \label{1cr} R \dot \theta =
\sqrt{\frac{r_g}{2R}},
\ee
where
$r_g=2\alpha/m_I \simeq 3 \cdot 10^5 M $ cm  is the gravitational radius of the object, $M$ is its mass
expressed in solar masses. In the considered case of the rigid state equation
(\ref{eq:Kepler}) in the class of solutions $R=R_I$, $\dot R_I=0$ one has the expression
\be \label{2cr} R_I \dot
\theta=\sqrt{\frac{r_g}{2R_I}+2(H_IR_I)^2},
\ee
or
$$v_I=\sqrt{w_I^2+2c_I^2},$$
where
\be v_I=R_I\dot \theta,\qquad w_I^2
=\frac{r_g}{2R_I}, \qquad c_I=R_IH_I.
\ee

In a more general case, for the metric (\ref{ds:2})
$$(ds^2)=a^2(\eta)\left[d\eta^2-(dx^i)^2\right]$$
with the equation of state (\ref{1uncol})
$$\frac{1}{H_0^2}{\left(\frac{da}{d\eta}\right)}^2=\Omega(a),$$
where $\Omega(a)$ is defined in (\ref{def:Omega}) one has
$$ \dot H + H^2 = -H^2\left(1-\frac{a}{2\Omega}\frac{d\Omega}{da}\right),$$
or, substituting it into (\ref{eq:Kepler}), we get
$$ \ddot
R+H^2\left(1-\frac{a}{2\Omega}\frac{d\Omega}{da}\right)R -
\frac{(m_0R^2\dot \theta)^2}{m_0^2R^3} + \frac{r_g}{2R^2} = 0.
$$
In this case, the equality $v_I^2=w_I^2$ becomes the following relation:
\be
v_I^2=w_I^2+(HR_I)^2\left(1-\frac{a}{2\Omega}\frac{d\Omega}{da}\right).
\nonumber\ee

\begin{center}
\begin{tabular}[b]{|l|ccc|ccc|ccc|ccc|}
  \hline &&&&&&&&&&&&\\
  $\Omega(a)=$
  &&$\Omega_{\rm rigid}a^{-2}$&
  &&$\Omega_{\rm rad}$&
  &&$\Omega_{M}a$&
  && $\Omega_{\rm \Lambda}a^4$&
  \\ \hline    &&&&&&&&&&&&\\
  $\displaystyle\left(1-\frac{a}{2\Omega}\frac{d\Omega}{da}\right)=$
  && 2&
  && 1&
  && 1/2&
  && -1&
  \\   &&&&&&&&&&&&\\ \hline
\end{tabular}
\end{center}
The Table shows that the lowest deficit of the dark matter of the four ``pure'' states is related to the
rigid state $\Omega_{\rm rigid}$ which corresponds to
the case under consideration (\ref{2cr}).
Note that in the Standard cosmology, the cosmic evolution increases the deficit of dark matter:
\be \label{2cr1}
R_I \dot\theta=\sqrt{\frac{r_g}{2R_I}-\frac{(H_IR_I)^2}{2}}.
\ee

From (\ref{2cr1}) it follows that the conventional Newtonian characteristics to describe the behavior of the orbital
velocities are not applicable to radial distances when the double square of cosmic speed is comparable in magnitude to
square Newtonian
velocities\footnote{This fact was known to Einstein and Straus \cite{Einstein:1945-14}
(see also \cite{Zorin-3-14, Zorin-5-14, Zorin-2-14, Zorin-9-14}).}:
$2c_I^2\geq w_I^2$.

For evaluation of the radial distance, in this case one can get the distance (we will call it a critical distance)
$R_{\rm cr}$, wherein $2c_I^2 = w_I^2$, hence
\be
\label{cr} R_{\rm cr}=\left(\frac{r_g}{2H_I^2}\right)^{1/3}.
\ee
The current value of the Hubble parameter $H^{-1}_0\simeq 10^{28}$ cm
leads to the value of the critical distance
\be R_{\rm cr}\simeq
10^{20}\left(\frac{M}{M_\odot}\right)^{1/3}~\mbox{\rm cm}.
\ee
The critical radius for the Coma cluster
($M\simeq 10^{15}M_\odot$
\cite{ch-14}) is comparable with the size of the cluster:
\be
R_{\rm size}\sim 3
\cdot 10^{25}~ \mbox{\rm cm}>R_{\rm cr}\sim 10^{25}~ \mbox{\rm cm},
\ee
and our arguments are applicable.
For our galaxy, ($M \! \simeq \! 10^{12}M_\odot$)
the corresponding estimate gives
\be
R_{\rm size}\sim 10^{23}~
\mbox{\rm cm}<R_{\rm cr}\sim 10^{24}~ \mbox{\rm cm,}
\ee
that is the critical radius of our galaxy, an order of magnitude larger than its size.
\begin{figure}[htb!]
\begin{center}
\includegraphics[width=0.5\textwidth]{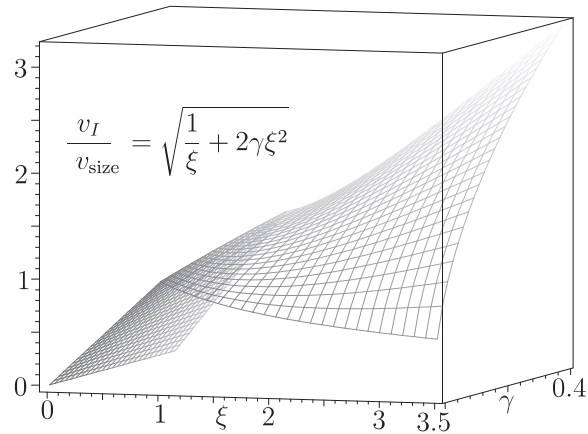}
\end{center}
\caption{\small
The dependence of the orbital velocity of the ``particle'' $v_I$ on its radius
{\it id est} the distance from the center of the object, $\xi=R/R_{\rm size}$, where $R_{\rm size}$
is a radius of the object, $\gamma=(R_{\rm size}/R_{\rm cr})^3$ and
$R_{\rm cr} =[r_g/H^2]^{1/3} =10^{20} M^{1/3}$ cm is a value of radius for which Newtonian
velocity coincides with the Hubble one, $M$ is a mass of the object in units of solar masses.
Under $\gamma =0$ a rotation curve coincides with the curve obtained in Newtonian mechanics.}
\label{halo}
\end{figure}

It is convenient to consider
the rotational curve of the circular speed $v_I=R_I \dot \theta$
(\ref{2cr}) in dimensionless terms
$\xi=R/R_{\rm size}$ and $\gamma$:
\be
 \label{abc}
 \frac{v_I}{v_{\rm size}}=\sqrt{\frac{1}{\xi}+2\gamma\xi^2},
 \ee
where
$$v_{\rm size} = \sqrt{\frac{r_g}{2R_{\rm size}}},~~~~~~~~~~~
\gamma={\left(\frac{R_{\rm size}}{R_{\rm cr}}\right)}^3,$$
$R_{\rm size}$ is a size of the object and
$$R_{\rm cr} = {\left(\frac{r_g}{H^2}\right)}^{1/3} = 10^{20} M^{1/3} cm $$
is the value of radius in cm, for which the Newtonian velocity coincides with the
Hubble one, $M$ is the mass of the object in units of solar masses
(Fig. \ref{halo}).
The dependence (\ref{abc}) at $\gamma=0$ corresponds to the Newtonian case, and the curve at $\gamma\not =0$
deviates from the Newtonian curve. This deviation cannot be explained by the introduction of the halo of
dark matter \cite{E:1-14, E:2-14, Rubin:83-14, pri-14}, but rather a cosmological modification of the Newtonian
dynamics described in this monograph.
Therefore, the violation of the virial theorem for $R\geq R_{\rm cr}$, found in clusters of galaxies and
interpreted as evidence for the existence of dark matter,
in the Conformal cosmology is considered as the result of evolution of the Universe
\cite{Zorin-3-14, Zorin-5-14, Zorin-2-14, Zorin-9-14}, as was predicted by
Einstein and Strauss in \cite{Einstein:1945-14}.

\section{The Kepler problem in the generalized \\ Schwarzschild field}\label{Chapter:Schw}
\makeatletter
\renewcommand{\@evenhead}{\raisebox{0pt}[\headheight][0pt]
{\vbox{\hbox to\textwidth{\hfil \strut \thechapter . Cosmological modification of Newtonian dynamics
\quad\rm\thepage}\hrule}}}
\renewcommand{\@oddhead}{\raisebox{0pt}[\headheight][0pt]
{\vbox{\hbox to\textwidth{\strut \thesection .\quad
The Kepler problem in the generalized Schwarzschild field
\hfil \rm\thepage}\hrule}}}
\makeatother

\begin{figure}[hp]
\begin{center}
\begin{minipage}[h]{\textwidth}
  \centering
\hbox to \textwidth{
\includegraphics[width=0.43\textwidth]{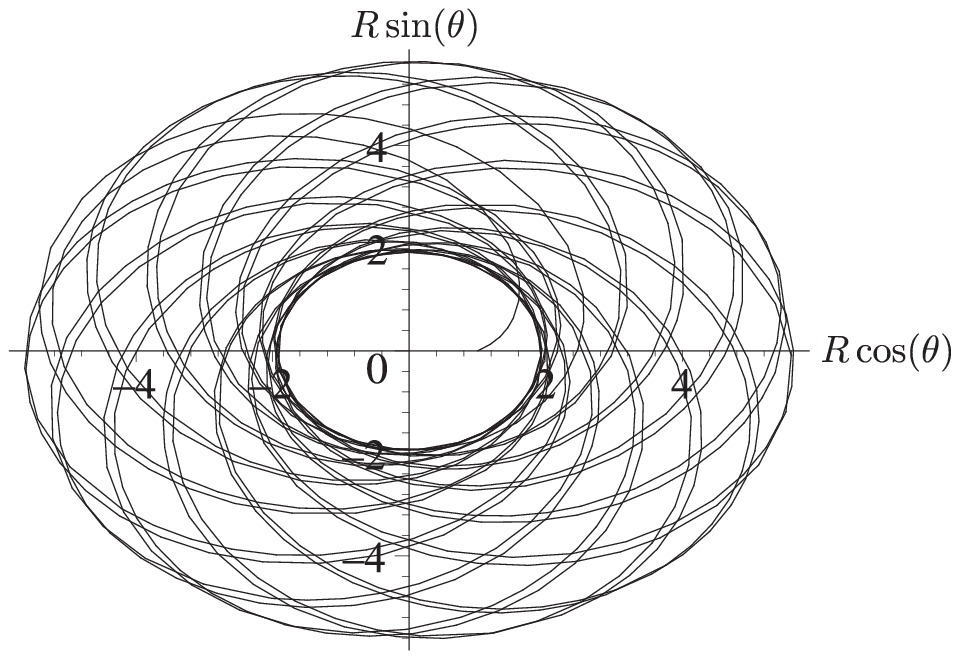}
\hfil
\includegraphics[width=0.43\textwidth]{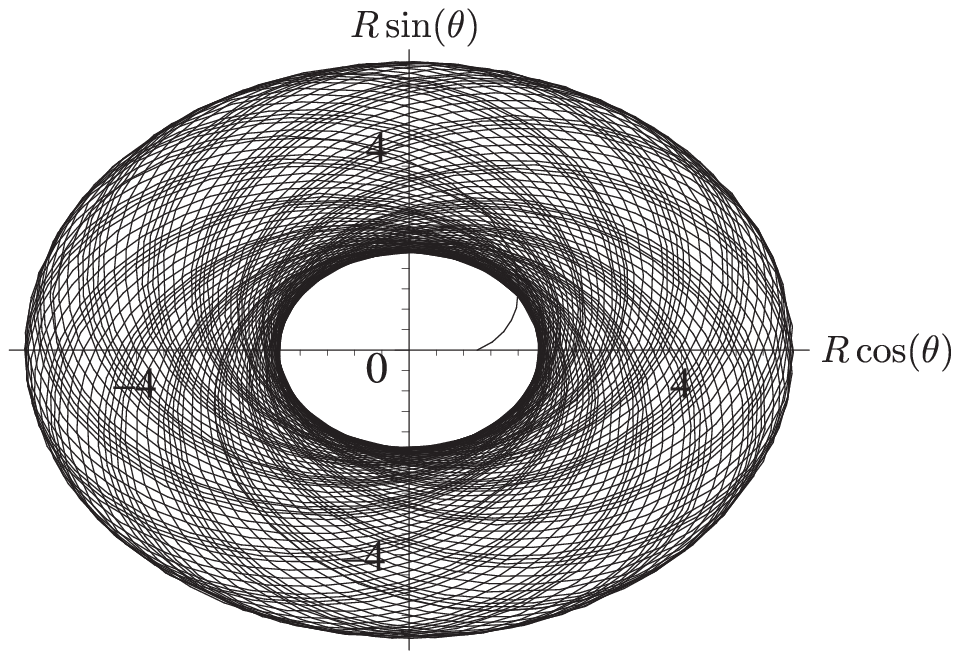}
 }
\caption{\small
The solution of the equations of motion for the action (\ref{S:schw:eta}) at
$c_I=1$, $v_I=1$, and $w_I^2=0,25$. In both figures there is shown the trajectory of the same object from
the starting point $(1,0)$ for different intervals of time in the
generalized Schwarzschild field (\ref{def:metrics:schw}).}
\label{fig:bh:1}
\end{minipage}

\vskip0.01\textwidth

\begin{minipage}[h]{.43\textwidth}
  \centering
\includegraphics[width=\textwidth]{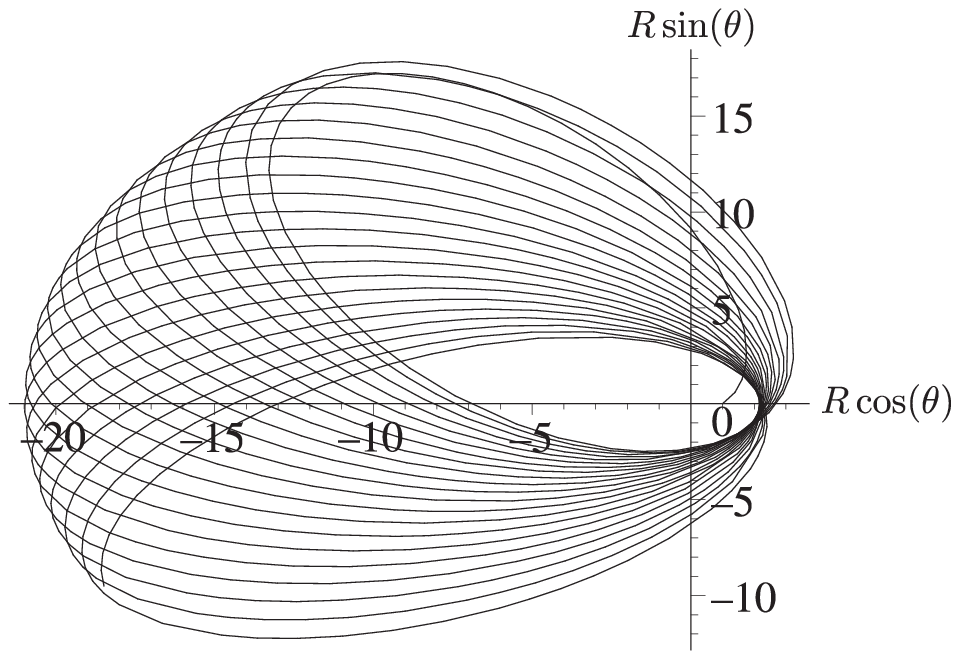}
\caption{\small
The solution of the equations of motion for (\ref{S:schw:eta}) at
$c_I$ $=0,25$, $v_I=0,25$, and $w_I^2=0,015625$. These values of parameters correspond to the
relativistic limit of equations for (\ref{S:schw:eta}) in which the classical ellipse
begins to turn counterclockwise.}\label{fig:bh:4}
\end{minipage}
\hfil
\begin{minipage}[h]{.43\textwidth}
  \centering
\includegraphics[width=\textwidth]{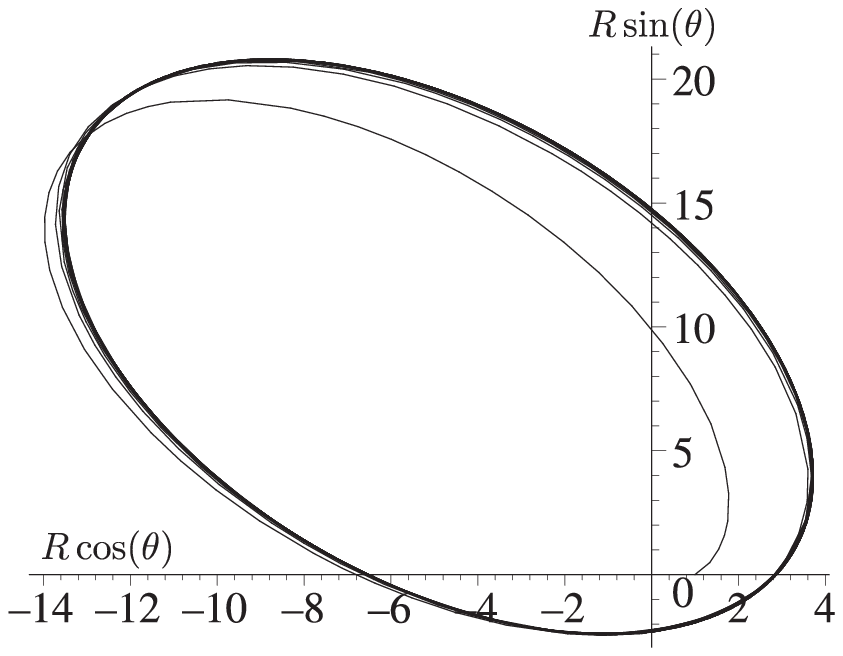}
\caption{\small
The solution of the equations of motion for (\ref{S:schw:eta}) at
$c_I=0,01$, $v_I=0,01$, and $w_I^2=2,5\cdot 10^{-5}$. These
values of parameters correspond to the classical limit and the classical
ellipse on relatively large times since the beginning of motion.
Just as in the original case $c_I = 1$ (generalized Schwarzschild field), the particle at small times
is ``captured'' by an ellipse.}\label{fig:bh:100}
\end{minipage}
\end{center}
\end{figure}

Let us consider the general case of the motion of a test body or a particle in a spherically-symmetric gravitational
field of the heavy mass. We generalize the Schwarzschild metric in the synchronous reference frame by replacing the ordinary
mass $m_0$ by its conformal analogue $m_0 a(\eta) = m(\eta)$:
\bea\label{def:metrics:schw}
ds^2= \left(1-\frac{2\alpha}{m r}\right) dt^2
-\frac{dr^2}{1-{2\alpha}/{(m r)}}-r^2\sin(\theta)^2d\theta^2,
 \eea
where
$$m=m(\eta),~~~~~~~~r=\sqrt{x_ix^i},~~~~~~~~~a(\eta)=\sqrt{1+2H_I(\eta-\eta_I)},$$
and consider the motion in the cylindrical coordinates
 \be
 X^1=R\cos\Theta, \quad X^2=R\sin\Theta, \quad R=ar.
 \ee
Here $H_I$ is the initial value of the Hubble velocity in the space with the rigid state equation of matter \cite{BehnkeSuper-14}
when the density of energy and pressure are equal. In terms of the conformal time $d\eta=dt/a$ and conformal values $r=R/a$
let us write the action for a particle in the form
\begin{eqnarray} \label{S:schw:eta} S_{\rm Schw}&=&\int\limits_{\eta_I}^{\eta_0} d\eta
\left[ P_r\frac{dr}{d\eta} + P_\theta\frac{d\theta}{d\eta} -
E_{\rm Schw} \right], \end{eqnarray}
where
$$\displaystyle Q_{\rm Schw}=\left(1-\frac{r_g}{r}\frac{m_I}{m}\right)^{1/2},\qquad
r_g={M_{\rm O}G},
$$
$P_r$, $P_\theta$ are conjugated momenta of the corresponding coordinates and $E_{\rm Schw}$ is the
energy of the system
\be\label{def:E:schw}
E_{\rm Schw}=Q_{\rm Schw}\sqrt{P_r^2Q_{\rm
Schw}^2+P_\theta^2/r^2+m^2}- m.
\ee
The trajectory of the test particle is shown in Fig. \ref{fig:bh:1}, and the Newtonian limit of the action
(\ref{S:schw:eta}) takes the form
\be
 S_A=\int\limits_{\eta_I}^{\eta_0}d\eta\left[P_r\frac{dr}{d\eta}
 +P_\theta\frac{d\theta}{d\eta}-\frac{P_r^2+{P_\theta^2}/{r^2}}{2m}-\frac{\alpha}{r}\right],
\ee
where
$$\alpha={M_{\rm O} m_I G}$$
is the Newton constant of interaction of the galaxy with a mass $m_I$ in the central
gravitational field with a central mass ${M_{\rm O}}$.

Let us consider three velocities:
\be w_I=\sqrt{\frac{r_g}{2r_I}}, \quad
v_I=\frac{P_\theta}{m_Ir_I}, \quad c_I=H_Ir_I
\ee
Newtonian, orbital and cosmic, respectively. The limit of small velocities
$w_I$, $v_I$, $c_I$
$\longrightarrow$ $0$
\label{limit:classic} corresponds to the classical approximation (see Fig. \ref{fig:bh:100}) ---
classical Kepler's problem with the expansion of the Universe.
In this limit, we obtain the action (\ref{S:schw:eta}), where instead of the Schwarzschild
Hamiltonian (\ref{def:E:schw}) its Newtonian limit is:
 \be \label{def:H:n}
 E_{\rm Schw}\sim E_{\rm
 classic}=\frac{P_r^2}{2am_I}+\frac{P_\theta^2}{2am_Ir^2}-\frac{r_gm_I}{2r}.
 \ee

It is convenient to study the solution of the problem in terms of dimensionless magnitudes
\be\label{def1} x=H_I (\eta-\eta_I),~~~~~~ r = r_I y,~~~~~P_r=m_I
p,
\ee
in terms of which the effective action for the radial motion takes the form
\bea
S_{\rm eff} = r_Im_I\int\limits_{x_I}^{x_0}  dx  \left(
p\frac{dy}{dx}  - \frac{1}{c_I}E_{\rm eff}\right),
\eea
where
$$
E_{\rm eff}=\frac{E_{\rm Schw}}{m_I} =
$$
$$= \sqrt{1-{2 w_I^2}/{(a  {} y)}} \sqrt{a^2 {} +
 \left(1  -  {2 w_I^2}/{(a  {} y)}\right) p^2+v_I^2/y^2}-  a \simeq
$$
\bea
 \simeq  \frac{p^2+v^2_I/y^2}{2a}-\frac{w_I^2}{y},
\eea
$a=\sqrt{1+2x}$.
The approximate equality holds here for small velocities, whereas if we put $a = 1$, we obtain the classical
orbital motion $ y = 1, ~ p = 0 $, where the Newtonian velocity $w_I$
coincides with the orbital $v_I$. This equality, rather its violation, is the basis for theoretical analysis of observational
data on dark matter in the Universe \cite{ch-14, E:1-14, E:2-14, pri-14}.

In Fig. \ref{fig:bh:1}, there is shown a numerical solution in dimensionless magnitudes (\ref{def1}) of the
Schwarzschild equations of motion
which begins in the state of zero energy (\ref{def:E:schw})
and zero radial velocity $P_I=0$. It can be seen that the particle is being trapped in a bound state and this is true for all
space velocities.
In Figs. \ref{fig:bh:1}, \ref{fig:bh:4}, and \ref{fig:bh:100}, there are shown solutions of equations (\ref{S:schw:eta})
following under the initial conditions
$$y(0)=1,~~~~~~~\frac{dy}{dx}(0)=0$$
and parameters
$$v_I=c_I,\qquad w_I^2=0,25c_I^2,\qquad (c_I=1,~0,25,~0,01).$$
In all the figures a trajectory starts from the point $(1,~0)$. It can be seen that
a trajectory of a test object is removed at some distance from the starting point and then becomes a periodic
(``capture'' of an object) in both time and space (Fig.\ref{fig:bh:100}).
In decreasing velocities of particles, their trajectories gradually
pass into the classical ellipses of the Kepler problem.
Thus, the exact solution of a modified Kepler problem with Hamiltonian (\ref{def:H:n}) and numerical solutions in the case of
Hamiltonian (\ref{def:E:schw}) show that the cosmic evolution of mass reduces the energy of a test particle
(stars and galaxies). Cosmic evolution reduces the energy of free stars and galaxies causing them to form
bound states such as galaxies or their clusters, respectively.

\section{Quantum mechanics of a particle \\ in Conformal cosmology}
\makeatletter
\renewcommand{\@evenhead}{\raisebox{0pt}[\headheight][0pt]
{\vbox{\hbox to\textwidth{\hfil \strut \thechapter . Cosmological modification of Newtonian dynamics
\quad\rm\thepage}\hrule}}}
\renewcommand{\@oddhead}{\raisebox{0pt}[\headheight][0pt]
{\vbox{\hbox to\textwidth{\strut \thesection .\quad
Quantum mechanics of a particle in Conformal cosmology
\hfil \rm\thepage}\hrule}}}
\makeatother

Let us consider a quantum mechanics of a particle in the Conformal cosmology,
where the masses of elementary particles also become dynamic \cite{Zorin-5-14}
\be\label{def:m}
 m(\eta)=m_0 \cdot \tilde a(\eta).
\ee
These masses determine the emission spectrum of atoms at time moment $\eta$; their change
$m'/m={a}'/{a}\sim 10^{-42}$ GeV is significantly less than the energy levels of the atom for
$\widetilde{a}(\eta_0)=1$ with quantum number $k$
\be\label{b1}
E^0_{k}=-\frac{{m\alpha^2}}{2{k^2}} \sim 10^{-8} ~ \mbox{\rm GeV},
\ee
being the eigenvalues of the Schr\"odinger stationary equation
\be\label{b2}  {\bf \hat E}(p,x)
\Psi^0\equiv\left[\frac{\hat p^2}{2m_0} -\frac{\alpha}{r}
\right]\Psi^0=E^0_{k} \Psi^0.
 \ee
The spectrum of the hydrogen atom with a mass-dependent time (\ref{def:m}), at any other instant $\eta=\eta_0-r$
can be found by solving the quasistationary Schr\"odinger equation
 \be\label{1Sch1}  {\bf \hat E}_c(p,x) \Psi\equiv\left[\frac{\hat p^2}{2m_0 \tilde
a(\eta)}-\frac{\alpha}{r} \right]\Psi=E_{k}(\eta)\Psi.
 \ee
Its solution is the spectrum
 \be\label{1Sch2}
 E_k(\eta)=\tilde{a}(\eta)E^0_{k},
 \ee
where $E^0_{k}$ are the levels of the atom with constant mass (\ref{b1}).
A rigorous derivation of (\ref{1Sch2}) is based on the canonical transformation
to the Friedmann variables \cite{Zorin-5-14, Zorin-6-14}
$$(p,x)\longrightarrow (P=p/a,X=xa),$$
and the non-stationary Schr\"odinger equation with the variable mass
$${\bf \hat E}_c(p,x)\Psi_c=-\imath\frac{\partial}{\partial\eta}\Psi_c$$
results into the Schr\"odinger equation
\be
{\bf \hat E}(P,X)\Psi=-\imath\frac{\partial}{\partial t} \Psi - H(t)PX\Psi
\ee
with a constant mass and an additional term disappearing at $H\rightarrow 0$, where
$H(t)$ is the Hubble parameter.

From (\ref{1Sch2}) follows the definition of the  redshift $z(r)$
\be\label{1gr} z(r)+1=\frac{E_{k}(\eta_0)}{E_{k}(\eta_0-r)}
=\frac{1}{\widetilde{a}(\eta_0-r) }
\ee
of the spectral lines on a space object at {\it coordinate}
distance $r$ from the Earth, relative to the spectral lines of the Earth atoms
$$E_k^0=E_k(\eta_0)$$
when photons are being detected under the condition $\widetilde{a}(\eta_0)=1$.
\newpage
\section{Summary}
\makeatletter
\renewcommand{\@evenhead}{\raisebox{0pt}[\headheight][0pt]
{\vbox{\hbox to\textwidth{\hfil \strut \thechapter . Cosmological modification of Newtonian dynamics
\quad\rm\thepage}\hrule}}}
\renewcommand{\@oddhead}{\raisebox{0pt}[\headheight][0pt]
{\vbox{\hbox to\textwidth{\strut \thesection .\quad
Summary and literature
\hfil \rm\thepage}\hrule}}}
\makeatother

In this Chapter, there were considered equations of dynamics of a test particle in a central gravitational
field taking into account the evolution of the Universe and found the exact analytical solution for the Conformal
cosmological model, compatible with the latest data on the Supernovae.
These equations were used to describe the effect of the capture of a test particle in a gravitational field
of the expanding Universe. It was shown that the capture effect can lead to the formation of
galaxies and their clusters with anisotropic radial vector field of velocities. Such a velocity field
could explain the anisotropy of the Hubble flow of velocities in the Local Group of galaxies
observed by I. Karachentsev with the colleagues.

In the framework of the model the limits of applicability of the Newtonian approximation
commonly used in the literature to describe the dark matter were assessed. The formula to describe the
orbital speeds with the cosmological evolution of the Universe predicted by
Einstein and Straus in 1945 \cite{Einstein:1945-14} is obtained.
According to this formula, the evolution of the Universe can imitate the effect of
dark matter for Superclusters of galaxies.
\chapter{Afterword}

\section{Questions of Genesis}
\makeatletter
\renewcommand{\@evenhead}{\raisebox{0pt}[\headheight][0pt]
{\vbox{\hbox to\textwidth{\hfil \strut \thechapter . Afterword
\quad\rm\thepage}\hrule}}}
\renewcommand{\@oddhead}{\raisebox{0pt}[\headheight][0pt]
{\vbox{\hbox to\textwidth{\strut \thesection .\quad
Questions of the Universe
\hfil \rm\thepage}\hrule}}}
\makeatother

In the far away future descendants will certainly define our time as the time of great astrophysical discoveries
comparable in importance to the era of great geographical discoveries of the late fifteenth and early
sixteenth centuries. On unknown vastness of the Universe, scientists have discovered traces of hitherto unseen
physical objects: neutron stars, quasars, pulsars, almost homogeneous Cosmic Microwave Background radiation
with temperature about three kelvins, filling the entire visible Universe, and much more. Astrophysicists
found a redshift of spectral lines of atoms that emit photons at distant space objects obeying the Hubble
law: The farther the object -- the greater the redshift. Modern researchers,
as once brave explorers of past centuries, have realized that they can already reach the limits of the visible
Universe, those distances that a light beam flies over the lifetime of the Universe.
Astrophysicists can see space objects remote from us at distances of the order of the size of the Universe, and thus can
determine the dependence of large redshift of the distances comparable to the size of the visible part of the Universe.
New data for large values of the redshift suggest that our Universe is
filled with basically not massive dust of distant galaxies, but invisible and mysterious substance of completely
different nature, with a different equation of state, called dark energy.
The results of measuring of the distribution of chemical elements in the Universe indicate the predominance of the photons
in the era of the primary chemical synthesis of elements and negligibly small contribution of the visible baryonic matter
(about 3 \%) to the cosmic evolution. On the other hand, the speed of rotation of stars in spiral galaxies and
the speed of rotation of galaxies in all giant superclusters, according to Newtonian mechanics, shows that
apart from the baryonic matter of which we are composed dark matter is present in galaxies, the mass of which
is ten times larger the mass of the visible baryonic matter.
As a result of these recent discoveries, the following most pressing \textit{questions of the Universe} face
the Standard cosmology:

\begin{enumerate}

\item How was our Universe created?
\item What had been in the Universe before its appearance?
\item What is the Universe made?
\item What is the nature of the dark energy and the dark matter?
\item Why did background radiation flash in the early Universe?
\item How was the matter created?
\item How can the baryon asymmetry of the Universe in which one baryon corresponds to billion photons be explained?
\item
Is it possible to build a physical theory that would not only answer all these questions, but could
predict the evolution of matter in the Universe, just as Newton's celestial mechanics
was able not only to explain to his contemporaries problems of the Universe current at that time,
but also to calculate the movement of the planets and predict the existence of new planets really discovered later?
\end{enumerate}
Modern physicists have to answer these questions and to \textit{explain} them on the basis of the first
principles. According to Wigner, there are three levels of ``\textit{explanation}'':

\begin{enumerate}

\item New empirical phenomena and mechanisms of the type of mechanism of inflation.

\item New laws of dynamics.

\item New additional symmetry principles of theories of gravitation and elementary particles.
\end{enumerate}

Recall that in the Standard cosmology to explain the data on the Supernova there is a mechanism of
inflationary expansion of the Universe, just the one that was proposed and developed by modern
physicists to solve the problems of the Standard cosmology.
However, the initial energy density of the inflation of the Universe differs $10^{57}$ times from the present energy density
of the inflation.
This huge difference has not yet found a convincing explanation at the level of new laws of dynamics in the framework
of the \textit{dynamic model}
of inflation with a single scalar field (inflanton).
On the other hand, the mechanism of \textit{primary inflation} absolutizes the current values of the constants of the Newton and
Hubble parameter measured by contemporary observers, just as Ptolemy's system absolutizes
the position and velocity of the Earth observer in celestial mechanics.

To explain the data on the Supernova, the authors of this book have preferred the third level, choosing as
new principles of symmetry the groups of affine and conformal transformations and the corresponding conformal
Dirac's variational principle.

\section{General discussion of results
}
\makeatletter
\renewcommand{\@evenhead}{\raisebox{0pt}[\headheight][0pt]
{\vbox{\hbox to\textwidth{\hfil \strut \thechapter . Postface
\quad\rm\thepage}\hrule}}}
\renewcommand{\@oddhead}{\raisebox{0pt}[\headheight][0pt]
{\vbox{\hbox to\textwidth{\strut \thesection .\quad
General discussion of results
\hfil \rm\thepage}\hrule}}}
\makeatother

\subsection{Results of the work}

Let us briefly list the results and conclusions of this monograph. For the formulation of the problem of
classification of data on physical measurements and astrophysical observations we trace the evolution of
ideas and mathematical methods of theoretical physics in the last five centuries of its development from
Copernicus' principle of relativity to Einstein's principles of relativity, Poincar$\acute{\rm e}$'s group and gauge theories.
Irreducible unitary representations of Poincar$\acute{\rm e}$'s group underlie the classification of quantum
relativistic particles and quantum field theory, which describes the creation of particles, their decays and interactions.

The book is devoted to the construction of the quantum wave function of the Universe as a joint
\textit{unitary irreducible representation of affine and conformal groups}
with the inclusion of the Standard Model of electroweak interactions and QCD.
This arrangement consists of the following stages.
\begin{enumerate}

\item Derivation of the theory of gravitation as joint nonlinear realization of \textit{affine and conformal groups}
(Ogievetsky and Borisov) in the tangent Fock space in terms of Cartan's forms.

\item Choice of conformal measurement standards (Dirac), which allow us to separate
the cosmic evolution of the devices of observation from the evolution of cosmic objects.

\item Choice of the reference frame of the Universe (Markov and Yukawa).

\item Solution of constraints in the event space (Dirac).

\item Primary and secondary quantization with the postulate of the vacuum (Fock).

\item The definition of the initial data according to the uncertainty principle (Blokhintsev).

\item  Diagonalization of the operators of creation and annihilation of particles and the Universe (Bogoliubov).

\end{enumerate}

It is shown that such a construction leads to the following results.

\begin{enumerate}
\item
  The postulate of the vacuum gives the arrow of the proper time interval and its beginning as a quantum anomaly.

\item
  The source of violation of conformal symmetry and the origin of the masses of elementary particles
  is the only normal ordering of field operators leading to Casimir's energies and condensates.

\item
  Planck's quantum of action leads to a hierarchy of cosmological scales for the matter
  fields, in accordance with their conformal weights $(n)$. These scales include the current value of Hubble's
  parameter $(n=0)$, temperature of the Cosmic Microwave Background radiation $(n=2)$, the mass scale
  of the electroweak bosons $(n=3)$, and the Planck mass $(n=4)$.

\item
  Observational data on the Supernova (1998 -- 2013), recalculated for Dirac's conformal
  long space-time intervals testify that our Universe is cold and almost empty.

\item
  The wave function of the quantum Universe as a joint irreducible unitary representation of affine
  and conformal symmetry groups is
  factorized by the wave function of the empty Universe and the S-matrix used in high energy physics to describe the
  processes of creation and interaction of particles.

\item
  The resulting S-matrix corresponds to the quantization of gauge fields in the reduced phase space of the field variables.
  Thus, the Hamiltonian formulation is a basis for unification of the theory of the gravitational field
  with the Standard Model of elementary particles,
  in which both the theories are considered at the quantum level in a certain frame of reference after solving of
  constraint equations.

\item
  The physical consequences of reduced quantization for QED, QCD, SM and the theory of gravity were considered.
  The solution of the Gaussian constraint in QED coincides with Dirac's approach to the quantization of electrodynamics in 1927.
  It is shown that the quantization of massive vector field is consistent with the principles of
  axiomatic approach to quantum field theory, in particular, with the postulate of the existence of the vacuum.

\item
  By solving the Gaussian constraint for the non-Abelian theory, in particular QCD, the corresponding
  generating functional was obtained in terms of gauge-invariant observables of color fields including their bound states.
  The method of generating functional is compared with the standard Faddeev -- Popov path integral one.
  The difference of the modified QCD from the standard Faddeev -- Popov approach consists in that the initial data of
  gauge-invariant observables of color fields are topologically degenerated.
  Destructive interference of these phase factors of the topological degeneration leads to zero amplitudes of creation
  of all color states. Therefore, this destructive interference can be interpreted as a purely
  kinematic confinement of colored particles and conditions.
  As a result of the kinematic confinement, there arises a quark-hadron duality, widely
  used in high-energy QCD to describe the deep inelastic interactions, where free
  quarks and gluons are treated as free partons with the standard propagators.

\item
  The normal ordering of field operators is a source of the
  spontaneous symmetry breaking and the appearance of dimensional parameters such as local two-particle correlation
  functions (Casimir's condensates) of quarks and gluons.
  The reduced QCD gives a possibility
  to obtain relations of these condensates with the parameters of low-energy interactions of quarks and mesons.
  In particular, we obtain the Gell-Mann -- Oakes -- Renner relation.

\item
  In the Standard Model of electroweak interactions the topological degeneration of
  non-Abelian fields is removed by their interactions with the Higgs field.
  After the normal ordering of the electroweak bosons and fermions in SM there appear quantum anomalies
  as two-particle correlations ({\it id est} the Casimir condensates).
  Assuming the universality of the ratio of
  the Casimir condensates of fields to their masses in power of their conformal weight,
  we estimated the mass spectrum of these bosons.
  The obtained mass of the Higgs boson $\sim 130 \pm $ 15 GeV is in agreement with recent experimental data.

\item
  The affine gravitons (in terms of linear Cartan's forms) are described
  by the free action in the approximation of zero Newtonian interaction.
  We consider some arguments testified that the affine gravitons can play a role of the
  dark matter in spiral galaxies.

\item
  It is shown that in the empty Universe there is a possibility of the intensive vacuum creation of only gravitons and Higgs particles
  (and the corresponding longitudinal components of the vector bosons).
  One of the main results is the creation from the vacuum of an order of $10^{88}$ of the Higgs boson particles and longitudinal
  components of the vector fields, decays of which form all the matter content of the
  Universe, in accordance with current observational facts.
  The values of the baryon density, the ratio of the number of photons to the number of baryons, and the CMB temperature
  are in agreement with observations.

\item
  The data on the Supernovae, primordial nucleosynthesis and cosmological particle creation, re-calculated to units of
  relative standard length, can be described by a single regime of vacuum dominance of the Casimir energy.

\item
  On the basis of the developed Hamiltonian method the theory of cosmological perturbations is formulated.
  The Schwarzschild-type solutions are obtained with Jeans-like spatial oscillations.

\item
  Cosmic evolution of the masses leads to the capture of cosmic objects by a central field and provides a mechanism
  of formation of galaxies and clusters of galaxies; in this case there is a
  class of essentially ellipsoidal trajectories of galaxies. The reality of existence of this type of trajectories is confirmed
  by recent observations of the anisotropy of the Hubble flow velocities in the
  Local Group of galaxies by I.D. Karachentsev's group. Finally, the chaos of freely moving particles is organized by a cosmic motion
  in the observational structures of matter.

\item
  According to the cosmological modified Newtonian dynamics, the square of Newton's velocity of the galaxy in the
  COMA-type superclasters is replaced by
  the sum of squares of two velocities (Hubble and Newton ones).
  Therefore, violation of the virial theorem found in clusters of galaxies and interpreted as evidence
  for the existence of dark matter,
  in the Conformal cosmology, can be explained by the Hubble velocities of these galaxies, without the dark matter halo,
  as was predicted by Einstein and Strauss in 1945 yet.
\end{enumerate}

Thus, on the basis of the principles of the quantum Universe, formulated before 1974, it is possible to
explain different properties of the world.

\subsection{Discussion}

One can send, imaginatively, an observer to the beginning of the Universe, just as Copernicus put his
observer, imaginatively, on the Sun. Our observer knows that there was the beginning, and at the beginning
there was an empty Universe. The observer finds that in this Universe there is only the dilaton zero mode and
Casimir's energy. A method of measuring the dilaton is the redshift.
There are systems of reference, co-moving to an empty element of the space, where the initial data of the dilaton
are constants of motion. These constants of motion define its position (the present value of Newton's constant)
and speed (the present value of the Hubble parameter).
The principles of symmetry, of which the action of Dirac is derived, and the postulate of vacuum lead to a hierarchy
of cosmological scales and the wave function of the Universe. The wave function of the Universe allows us to unify modern
observational data on the Supernova with the latest experimental value of the Higgs particle mass.

The level of mathematical description of the nature, claimed in modern physics, formulates a lot
of other questions. How can the affine group extend to include all fields of the Standard Model in the Goldstone fields?
Why does the world have these symmetries, and not others?
Why is there such a precise fit of the initial data and the dimensionless coupling constants under the anthropic principle?

Maybe in the future someone will find the answers to these questions in harmony of the very principles of symmetry, like
Copernicus and Kepler found their answers to the questions of the Universe in harmony of epicycles,
and Einstein, Weyl and other researchers in the 20th century found their answers in harmony of the laws of nature.

Steven Weinberg fatefully wrote in the Introduction of his book\footnote{Weinberg, S.:
\textit{The First Three Minutes: A Modern View of
the Origin of the Universe.} Basic Books, New York (1977).} about the future of the Standard Cosmological model:
``{\it Can we really be sure of the standard model? Will new discoveries overthrow it and replace the present standard model
with some other cosmogony, or even revive the steady-state model? Perhaps. I cannot deny a feeling of unreality in
writing about the thirst three minutes as if we really know what we are talking about''.}

In this book, we gave theoretical and observational arguments in favour of Steven Weinberg's predictions.
New present-day discoveries, in particular, the recent LHC experimental data on a small value of the Higgs particle mass
confirm the conformal symmetry principle on which the Standard Model of elementary particles is based.
The origin of elementary particle masses is explained by the quantum anomaly of the vacuum postulate, not by the phenomenological
Higgs potential. The same quantum anomaly in the form of the vacuum Casimir energy explained the present-day Supernovae
observational data without the lambda-term in the framework of the Conformal cosmology.

\appendix
\chapter{Reduced Abelian field theory
\label{sect_AppA}}
\renewcommand{\theequation}{A.\arabic{equation}}
\setcounter{equation}{0}
\makeatletter
\renewcommand{\@evenhead}{\raisebox{0pt}[\headheight][0pt]
{\vbox{\hbox to\textwidth{\hfil \strut \thechapter . Appendix
\quad\rm\thepage}\hrule}}}
\renewcommand{\@oddhead}{\raisebox{0pt}[\headheight][0pt]
{\vbox{\hbox to\textwidth{\strut \thechapter .\quad
 Reduced Abelian field theory
\hfil \rm\thepage}\hrule}}}
\makeatother
\section{Reduced QED}
\makeatletter
\renewcommand{\@evenhead}{\raisebox{0pt}[\headheight][0pt]
{\vbox{\hbox to\textwidth{\hfil \strut \thechapter .  Reduced Abelian field theory
\quad\rm\thepage}\hrule}}}
\renewcommand{\@oddhead}{\raisebox{0pt}[\headheight][0pt]
{\vbox{\hbox to\textwidth{\strut \thesection .\quad
Quantum electrodynamics
\hfil \rm\thepage}\hrule}}}
\makeatother

\subsection{Action and frame of reference}

 Let us recall the Dirac approach to QED \cite{Dirac-A, hp-A, Polubarinov-A}. The theory is given by
 the well-known action
 \be\label{1e}
 S=\int d^4x\bigg(-\frac{1}{4}F_{\mu\nu}F^{\mu\nu}+
 \bar \psi [\imath\rlap/\partial -m]\psi+A_\mu j^{\,\mu}\bigg),
 \ee
where $F_{\mu\nu}=\partial_\mu A_\nu -\partial_\nu A_\mu$ is
tension, $A_{\mu}$ is a vector potential, $\psi$ is the
 Dirac electron-positron bispinor field, and
 $j_{\mu}=e  \bar {\psi}  \gamma_{\mu} \psi$ is the charge
 current,
 $\rlap/\partial \equiv \partial^{\mu} \gamma_{\mu}$.
 This action is invariant with respect to the collection of gauge
transformations
 \bea
 \label{3e1}
 A^{\lambda}_\mu=A_\mu+\partial_\mu\lambda,~~~
\psi^{\lambda}=e^{+\imath e\lambda}\psi.
 \eea
 The variational principle used for the action (\ref{1e}) gives the Euler -- Lagrange equations of motion known as the Maxwell equations
 \be\label{vp}\partial_\nu F^{\mu\nu}+j^\mu=0,
 \ee
Physical solutions of the Maxwell equations are obtained in a fixed
{\it inertial reference frame} distinguished by a unit time-like
vector \mbox{$n_{\mu}$}. This vector splits the
 gauge field $A_\mu$ into the time-like $A_0=A_\mu n_{\mu}$
 and space-like
$$A^{\bot}_\nu=A_\nu - n_{\nu}(A_\mu n_{\mu})$$
 components. Now we rewrite the Maxwell equations in terms of
  components\bea\label{c1}
 \Delta A_0-\partial_0\partial_{k}A_k &=&j_{0},\\\label{jc1}
\Box A_k-\partial_k[\partial_{0}A_0-\partial_iA_i ]&=&-j_{k}.
  \eea
The field component $A_0$
  cannot be a {\it degree of freedom}
   because its canonical conjugate momentum vanishes.
The Gauss constraints (\ref{c1}) have the solution \bea\label{c2}
  A_0+\partial_0\Lambda=-\frac{1}{4\pi}\int d^3y\frac{j_0(x_0,y_k)}{|\mathbf{x}-\mathbf{y}|},
  \eea
 where
  \bea\label{lc2}
  \Lambda=-\frac{1}{\Delta}\partial_{k}A_{k}=\frac{1}{4\pi}\int{d^3y}\frac{\partial_{k}A_{k}}{|\mathbf{x}-\mathbf{y}|}
  \eea
  is a longitudinal component.
  The result (\ref{c2}) is treated as the {\it Coulomb potential field} leading to
 the {\it static} interaction.

\subsection{Elimination of time component}

 Dirac \cite{Dirac-A} proposed to eliminate the time component by
 substituting  the manifest resolution of the Gauss constraints given by
(\ref{c2}) into the
 initial action (\ref{1e}). This substitution -- known as the reduction
 procedure --
  allows us to eliminate nonphysical pure gauge degrees of
  freedom. After this step the action (\ref{1e}) takes
the form
\bea\label{2.3-1}S\!=\!\int d^4x
\left(\frac{1}{2}(\partial_{\mu}A^{\rm T}_k)^2
 \!+\!
 \bar \psi
 [\imath\rlap/\partial\!-\!m]\psi\!-\!j_0\partial_0 \Lambda\!-\!
 A^{\rm T}_kj_{k}\!+\!\frac{1}{2}
 j_0\frac{1}{\triangle}j_0\right),
 \eea
where
 \be\label{lc1}
 A^{\rm T}_k=\left(\delta_{ij}-
\frac{\partial_i\partial_j}{\triangle}\right)A_j.
 \ee
This substitution leaves the longitudinal component
  $\Lambda$
 given by Eq.~(\ref{lc2}) without any kinetic term.

 There are two possibilities. The first one is to treat
 $\Lambda$ as the Lagrange factor that leads to
 the conservation law (\ref{vp}).  In this approach, the longitudinal component
is treated
 as an independent variable. This treatment violates gauge invariance because
 this component is gauge-variant and it  cannot  be measurable.
 Moreover, the time derivative of the longitudinal  component in Eq.~(\ref{c2}) looks like
 a physical source of the Coulomb potential. By these reasons we will not consider this
approach in this paper.

 In the second possibility, a measurable
 potential stress is identified with the gauge-invariant quantity  (\ref{c2})
 \be\label{2-3-4}
 A_0^{\rm R}=A_0-\frac{\partial_0\partial_k }{\triangle}A_k~,
 \ee
 This approach  is consistent with    the
 principle of gauge invariance
  that identifies
 observables  with gauge-invariant quantities.
 Therefore, according to the gauge-invariance,
   the longitudinal
 component should be eliminated from the set of degrees of freedom of QED, too.

\subsection{Elimination of longitudinal  component}

 This elimination is fulfilled by the choice of the ``radiation variables''
 as gauge invariant functionals
 of the initial fields, \textit{id est} ``dressed fields'' \cite{Dirac-A}
 \bea
 A^{\rm R}_\mu=A_\mu+\partial_\mu\Lambda,~~~~~~~~\psi^{\rm R}=e^{\imath e\Lambda}\psi,\label{d2}
 \eea
 In this case, the linear term $\partial_k A_k$ disappears
 in the Gauss law (\ref{c1})
 \be\label{1c1}
 \Delta A^{\rm R}_0=j^{\rm R}_{0}\equiv e
 \bar \psi^{\rm R}\gamma_0\psi^{\rm R}.
  \ee
 The source of the gauge-invariant {\it potential
field} $A^{\rm R}_0$
 can be only an
 electric  current $j^{\rm R}_0$ whereas
 the spatial components  of the vector field  $A^{\rm R}_k$
 coincide with the  transversal one
\be\label{kc1}
 \partial_k A^{\rm R}_k=\partial_k A^{\rm T}_k\equiv {0}.
  \ee
In this manner the frame-fixing $A_\mu=(A_0,A_k)$ is compatible with
understanding of $A_0$ as a classical field, and the use of
  the Dirac  dressed fields (\ref{d2}) of the Gauss constraints (\ref{c1})
leads to understanding of the  variables (\ref{d2}) as
gauge-invariant functionals of the initial fields.

\subsection{Static interaction}

 Substitution of the manifest resolution
 of the Gauss constraints (\ref{c1})
  into the initial action (\ref{1e})
  calculated on constraints leads to that the initial action
 can be expressed
 in terms of the gauge-invariant radiation variables
  (\ref{d2})
 \cite{Dirac-A, Polubarinov-A}
 \bea\label{14-2}
\!\!S=\!\!
 \int d^4x \left(\frac{1}{2}(\partial_{\mu}A^{\rm R}_k)^2
 +
 \bar \psi^{\rm R}
 [\imath\rlap/\partial -m]\psi^{\rm R}-
 A^{\rm R}_kj^{\rm R}_{k}+\frac{1}{2}
 j_0^{\rm R}\frac{1}{\triangle}j_0^{\rm R}\right).
 \eea
The Hamiltonian, which corresponds to this action, has the form
 \bea\label{2-5-2}
 {\cal  H}=
 \frac{(\Pi_k^{\rm R})^2+(\partial_j A_k^{\rm R})^2}{2}+p^{R}_{\psi}\gamma_{0}
[\imath\gamma_k\partial_k+m]\psi^{\rm R}+
\eea
 $$
 +A^{\rm R}_kj^{\rm R}_{k}-\frac{1}{2}
 j_0^{\rm R}\frac{1}{\triangle}j_0^{\rm R},
 $$
  where $\Pi_{k}^{R}$, $p_\psi^{\rm R}$
  are the canonical conjugate momentum fields of the theory calculated
  in a standard way. Hence, the
  vacuum can be defined as a state with minimal energy obtained
  as the value of the Hamiltonian for the equations of motion.
 Relativistic covariant transformations of the
 gauge-invariant fields are proved at the level of the fundamental
 operator quantization in the form of
 the Poincar\'e
 algebra generators \cite{sch-A}.
 The status of the theorem of equivalence between the
 Dirac radiation variables and the Lorentz gauge formulation
 is considered in \cite{Pervu-A,fp1-A, f1-A,h-pp-A}.

\subsection{Comparison of radiation variables with the Lorentz gauge ones}

 The static interaction and the corresponding bound states
  are lost in any frame free formulation including the
  Lorentz gauge one. The action (\ref{2.3-1}) transforms into
\bea\label{2-6-1} S=\int d^4x
\left(-\frac{1}{2}(\partial_{\mu}A^{\rm L}_\nu)^2
 +
 \bar \psi^{\rm L}
 [\imath\rlap/\partial -m]\psi^{\rm L}+
 A^{\rm L}_\mu j^{\rm L\mu}\right) ,
 \eea
 where
\bea\label{2-6-2}
A^{\rm L}_\mu=A_\mu+\partial_\mu\Lambda^{\rm
L},~~~~\psi^{\rm L}=e^{ie\Lambda^{\rm L}}\psi,~~~~\Lambda^{\rm
L}=-\frac{1}{\Box}\partial^\mu A^{\rm L}_\mu
 \eea
 are the manifest gauge-invariant functionals satisfying
 the equations of motion
 \be\label{2-6-3}
 \Box A^{\rm L}_\mu=-j^{\rm L}_\mu,
 \ee
 with the current
 $$j^{\rm L}_\mu=-e\bar{\psi}^{\rm L}\gamma_{\mu}\psi^{\rm L}$$
 and
 the gauge constraints
 \be\label{2-6-4}
\partial_\mu A^{\rm L\mu}\equiv 0.
 \ee
 Really, instead of the potential, satisfying the Gauss constraints
$$\triangle A^{\rm R}_0= j^{\rm R}_0,$$
 and two transverse variables
 in QED in terms of the
 radiation variables (\ref{d2}) we have here  three independent  dynamic
 variables,
 one of which $A^{\rm L}_0$ satisfies the equation
 \be\label{2-6-5}
\Box A_0^{\rm L}= -j_0,
 \ee
   and
 gives a negative contribution
 to the energy.

 We can see that there are two distinctions of the
 ``Lorentz gauge formulation'' from the
 radiation variables.
 The first is the loss of Coulomb poles (\textit{id est} static
 interactions). The second is the treatment of the time component
 $A_0$ as an independent variable with the negative contribution
 to the energy; therefore, in this case,
 the vacuum as the state with the
 minimal energy is absent.  In other words,
 one can say that the static interaction
 is the consequence of the vacuum postulate,
 too. The inequivalence between the radiation variables
  and the Lorentz ones
 does not mean
violation
 of the gauge invariance,
 because both the variables  can be defined as
 the gauge-invariant functionals of the initial gauge fields
   (\ref{d2}) and (\ref{2-6-2}).

In order to demonstrate the inequivalence between the radiation
variables
  and the Lorentz ones, let us consider
 the electron-positron scattering amplitude
$$T^R=\langle e^+,e^-|\hat S|e^+,e^-\rangle.$$
One can see that the
Feynman rules in the radiation gauge give the amplitude in terms of
the current $j_\nu=\!\bar e \gamma_\nu e$
 \bea\label{1wr}
 T^R&=&\frac{j^2_0}{\mathbf{q}^2}\!+\!
 \left(\!\delta_{ik}-\frac{q_iq_k}{\mathbf{q}^2}\!\right)
 \frac{j_ij_k}{q^2+\imath\varepsilon}\!\\\nonumber
 &\equiv&\frac{-j^2}{q^2+\imath\varepsilon}+
 \fbox{$\dfrac{(q_0j_0)^2-
 (\mathbf{q}\cdot\mathbf{j})^2}{\mathbf{q}^2[q^2+\imath\varepsilon]}$}~.
 \eea
 This amplitude coincides with the Lorentz gauge one
 \be
 \label{2wr}
 T^L =
 -\frac{1}{q^2+\imath\varepsilon}
 \left[j^2-\fbox{$\dfrac{(q_0j_0-
 \mathbf{q}\cdot\mathbf{j})^2}{q^2+\imath\varepsilon}$}\,\,\right]~
 \ee
 when the box terms  in Eq.~(\ref{1wr}) can be
 eliminated. Thus, the Faddeev equivalence theorem \cite{fp1-A, f1-A} is valid
  if the currents
 are conserved
 \be \label{3wr}
 q_0{j}_0-\mathbf{q}\cdot\mathbf{j}=qj=0,
 \ee
 However, for the action with the external sources   the currents
 are not conserved. Instead of the classical conservation laws
 we have the Ward -- Takahashi identities for
 Green functions, where the currents are not conserved
\be \label{3wr1}
 q_0{j}_0-\mathbf{q}\cdot\mathbf{j}\not =0.
 \ee
 In particular, the
Lorentz gauge perturbation theory (where the propagator has only the
light cone singularity $q_{\mu}q^{\mu}=0$) cannot describe
instantaneous Coulomb atoms; this perturbation theory contains only
the Wick -- Cutkosky bound states whose spectrum is not observed in
the Nature.

Thus, we can give a response to the question: What are new physical
results that follow from the
 Hamiltonian approach to QED in comparison with the
 frame-free Lorentz gauge formulation? In the framework of the perturbation
 theory, the Hamiltonian presentation of  QED contains the static
 Coulomb interaction  (\ref{1wr})  forming
 instantaneous bound states observed in the Nature, whereas
 all  frame free formulations lose this static interaction
 together with instantaneous bound states in the lowest order of
 perturbation theory on retarded interactions
   called the radiation correction. Nobody has proved that
   the sum of these retarded radiation corrections with the
   light-cone singularity    propagators (\ref{2wr}) can restore
   the Coulomb interaction that was removed from
   propagators (\ref{1wr})
   by  hand
   on the level of the action.
\section{Reduced vector boson theory}
\makeatletter
\renewcommand{\@evenhead}{\raisebox{0pt}[\headheight][0pt]
{\vbox{\hbox to\textwidth{\hfil \strut \thechapter .  Reduced Abelian field theory
\quad\rm\thepage}\hrule}}}
\renewcommand{\@oddhead}{\raisebox{0pt}[\headheight][0pt]
{\vbox{\hbox to\textwidth{\strut \thesection .\quad
 Reduced vector boson theory
\hfil \rm\thepage}\hrule}}}
\makeatother

\subsection{Lagrangian and reference frame}

The classical Lagrangian of massive QED is
\begin{equation}
\label{3-1-1} {\cal L}=
-\frac{1}{4}F_{\mu\nu}F^{\mu\nu}+\frac{1}{2}M^2V_\mu^2+
\bar\psi(\imath\rlap/\partial-m)\psi +V_{\mu}j^{\mu}~,
\end{equation}
In a fixed reference frame this Lagrangian takes the form
\be \label{3-1-2}
 \mathcal{L}=\frac{(\dot V_k-\partial_kV_0)^2-
 (\partial_jV_k^{\rm T})^2\!+\!M^2(V_0^2\!-
 \!V_k^2)}{2}\!+
\ee
$$
+\bar\psi(\imath\rlap/\partial-m)\psi\!+\!V_{0}j_{0}-V_{k}j_{k},
$$
where $\dot V=\partial_0V$ and $V_k^{\rm T}$ is the transverse
component defined by the action of the projection operator given in
Eq.~(\ref{lc1}). In contrast to QED this action is not invariant
with respect to gauge transformations. Nevertheless,
 from the Hamiltonian viewpoint the massive theory
has the same problem as QED. The time component of the massive boson
has a vanishing canonical momentum.

\subsection{Elimination of time component}

In \cite{h-pp-A}, one supposed to eliminate the time component from the
set of degrees of freedom like the Dirac approach to QED,
\textit{id est}, using the action principle. In the massive case it
produces the equation of motion
\begin{equation}\label{3-1-3}
(\triangle-M^2)V_0=\partial_i\dot{V}_i+j_0.
\end{equation}
which is understood as constraints and has the solution
\bea
\label{3-1-4} V_0
=\left(\frac{1}{\triangle-M^2}\,\partial_i{V}_i\right)^{\cdot}
+\frac{1}{\triangle-M^2}\,j_0.
\eea
In order to eliminate the time component, let us insert (\ref{3-1-4}) into the Lagrangian
(\ref{3-1-2}) \cite{Dirac-A,h-pp-A}
$$
 {\cal L}=\frac{1}{2}\left[(\dot V_k^{\rm
 T})^2+V_k^{\rm T}(\triangle-M^2)V^{\rm T}_k
 +j_0\frac{1}{\triangle-M^2}j_0\right]+
$$
\bea\label{3-2-7}
+\bar\psi(\imath\not{\partial}-m)\psi-V^{\rm T}_k j_k+
\eea
$$+ \frac{1}{2}\left[\dot V_k^{\rm ||}
 M^2\frac{1}{\triangle-M^2}\dot V_k^{\rm ||}-M^2
 (V_k^{\rm ||})^2
 \right]-V^{\rm
 ||}_kj_k+
$$
$$
+j_0\frac{1}{\triangle-M^2}\partial_k \dot V_k^{\rm ||},
$$
where we decomposed the vector field
$$V_k=V_k^{\rm T}+V_k^{\rm ||}$$
by means of the  projection
 operator by analogy with (\ref{lc1}). The last two terms are the contributions of the longitudinal
component only. This Lagrangian contains the longitudinal component
  which is  the dynamical variable described by the bilinear term.
  Now we propose the following transformation:
 \be\label{3-2-8b}
\bar\psi(\imath\not{\partial}-\!m)\psi-V^{\rm ||}_kj_k+j_0
\frac{1}{\triangle-M^2}\partial_k \dot V_k^{\rm ||}=
 \ee
$$=\bar\psi^{\rm R}(\imath\not{\partial}-m)\psi^{\rm R}-V^{\rm R ||}_kj_k,
$$
where
  \be\label{3-2-9}
  V^{\rm R||}_k=V^{\rm||}_k-\partial_k\frac{1}{\triangle-M^2}\,\partial_i{V}_i=
-M^2\frac{1}{\triangle-M^2}V^{\rm ||}_k~,
\ee
\be
 \psi^{\rm R}=\exp\left\{-\imath e\frac{1}{\triangle-M^2}\,\partial_i{V}_i\right\}\psi
 \ee
are the radiation-type variables. It removes the linear term
$\partial_i\dot{V}_i$ in the Gauss law
 (\ref{3-1-3}). If the mass $M\not = 0$, one can
 pass from the initial variables $V^{\rm ||}_k$
 to the radiation ones $V^{\rm R ||}_k$ by the change
 \be\label{3-2-9a} V^{\rm ||}_k= \hat{Z}V^{\rm R ||}_k,~~\hat{Z}=\frac{M^{2}-\triangle}{M^{2}}
  \ee
Now the Lagrangian (\ref{3-2-7}) goes into
$$
 {\cal L}=\frac{1}{2}\left[(\dot V_k^{\rm
 T})^2+V_k^{\rm T}(\triangle-M^2)V^{\rm T}_k
 +j_0\frac{1}{\triangle-M^2}j_0\right]+\bar\psi^{\rm R}(\imath\not{\partial}-m)\psi^{\rm R}+
$$
\bea\label{3-2-10}
+\frac{1}{2}\left[\dot V_k^{\rm R\rm ||}
 \hat{Z}\dot V_k^{\rm R\rm ||}+
 V_k^{\rm R\rm ||}(\triangle-M^{2})\hat{Z}V_k^{\rm R\rm ||}
 \right]-V^{\rm T}_k j_k-V^{\rm R ||}_k j_k.
\eea
The Hamiltonian corresponding to this Lagrangian can be constructed
in the standard canonical way. Using the rules of the Legendre
transformation and canonical conjugate momenta
$$\Pi_{V^{\rm T}_k},\quad \Pi_{V^{\rm R ||}_k},\quad \Pi_{\psi^{R}}$$
we obtain
$$
{\cal{H}}=\frac{1}{2}\left[\Pi_{V^{\rm T}_k}^{2}+V_k^{\rm
T}(M^2-\triangle)V^{\rm T}_k
 +j_0\frac{1}{M^2-\triangle}j_0\right]-
$$
\bea\label{3-2-14}
-\Pi_{\psi^{R}}\gamma_0(\imath\gamma_{k}\partial_{k}+m)\psi^{\rm R}
\eea
$$
+\frac{1}{2}\left[\Pi_{V^{\rm R ||}_k}\hat{Z}^{-1}\Pi_{V^{\rm R
 ||}_k}+
 V_k^{\rm R\rm ||}(M^{2}-\triangle)\hat{Z}V_k^{\rm R\rm
 ||}\right]+
$$
$$+V^{\rm T}_kj_k+V^{\rm R
 ||}_kj_k.
$$
One can be convinced \cite{h-pp-A} that the corresponding
 quantum system has a vacuum as a state with
 minimal energy and correct relativistic transformation
 properties.

\subsection{Quantization}

We start the quantization procedure from the canonical quantization
by using the following equal time canonical commutation relations
(ETCCRs):
\bea\label{3-4-1}
 \left[\hat{\Pi}_{V^{\rm T}_k},\hat{V}^{\rm
 T}_k\right]=\imath\delta_{ij}^{\rm T}\delta^3(\mathbf{x}-\mathbf{y}),
\eea
\bea
\left[\hat{{\Pi}}_{V^{\rm R ||}_k}, \hat{V}^{\rm R ||}_k\right]=
 \imath\delta_{ij}^{||}\delta^3(\mathbf{x}-\mathbf{y}).
\eea
The Fock space of the theory is built by the ETCCRs
\bea
\left[{a^{-}_{(\lambda)}\left({\pm
k}\right),a_{(\lambda')}^{+}\left({\pm k'}\right)}\right]=\delta
^{3}\left({{\bf k}-{\bf k'}}\right)\delta_{(\lambda)(\lambda')};
\eea
\bea
\left\{b^{-}_\alpha\left({\pm k}\right),b_{\alpha'}^{+}\left({\pm
k'}\right)\right\}=\delta^{3}\left({{\bf k}-{\bf
k'}}\right)\delta_{\alpha\alpha'};
\eea
\bea
\left\{{c^{-}_\alpha\left({\pm k}\right),c_{\alpha'}^{+}\left({\pm
k'}\right)}\right\}=\delta^{3}\left({{\bf k}-{\bf
k'}}\right)\delta_{\alpha\alpha'}.
\eea
with the vacuum state $|0\rangle$ defined by the relations
 \be \label{3-4-7}
 a_{(\lambda)}^-|0\rangle=b_\alpha^-|0\rangle=c_\alpha^-|0\rangle=0.
 \ee
The field operators have the Fourier decompositions in the plane
wave basis
$$
 V_j\left(x\right)=
 \!\int[d\mathbf{k}]_{v}
 \epsilon_{j}^{(\lambda)}{\left[{a_{(\lambda)}^{+}\left({\omega,{\bf k}}
 \right)e^{-\imath\omega t + \imath{\bf kx}}+a_{(\lambda)}^{-}
 \left({\omega,-{\bf k}}\right)e^{\imath\omega t -\imath{\bf kx}}}\right]}
$$
$$
 \psi\left(x\right)=\sqrt{2m_{s}}\int[d\mathbf{k}]_{s}
 {\left[{b^+_\alpha\left(k\right)u_\alpha
 e^{-\imath\omega t + \imath{\bf kx}}+c^-_\alpha\left({-k}\right)\nu _\alpha
 e^{\imath\omega t -\imath{\bf kx}}}\right]}
$$
$$
 \psi^{+}\left(x\right)=\sqrt{2m_{s}}\int[d\mathbf{k}]_{s}
 {\left[{b^-_\alpha\left(k\right)u_\alpha^{+}e^{\imath\omega t -\imath{\bf kx}}+c^+_\alpha
 \left({-k}\right)\nu _\alpha^{+}e^{-\imath\omega t +\imath{\bf
 kx}}}\right]}
$$
with the integral measure
$$[d\mathbf{k}]_{v,s}=\dfrac{1}{\left({2\pi}\right)^{3/2}}\dfrac{{d^3\bf
k}}{{\sqrt{2\omega_{v,s}(\bf k)}}}$$
and the frequency of
oscillations
$$\omega_{v,s}(\mathbf{k})=\sqrt{\,\mathbf{k}^2+m^2_{v,s}}.$$
One can
define the vacuum expectation values of the
 instantaneous products of the field operators
\bea \label{3-4-8}
 V_i(t,\vec x)V_j(t,\vec y)=:V_i(t,\vec x)V_j(t,\vec y):
 +\langle V_i(t,\vec x)V_j(t,\vec y) \rangle,
\eea
\bea
 \overline{\psi}_\alpha(t,\vec x) \psi_\beta(t,\vec y)=:
 \overline{\psi}_\alpha(t,\vec x) \psi_\beta(t,\vec y):
 +\langle \overline{\psi}_\alpha(t,\vec x) \psi_\beta(t,\vec y),
\eea
where
\bea \label{3-4-9}
 \langle V_i(t,\vec x)V_j(t,\vec y) \rangle=\frac{1}
 {(2\pi)^3}\int\frac{d^3\bf{k}}{2\omega_v(\bf{k})}
 \sum\limits_{(\lambda)}^{}
 \epsilon_{i}^{(\lambda)}\epsilon_{j}^{(\lambda)}e^{-\imath\bf{k}(\bf{x}-\bf{y})},
\eea
\bea\label{3-4-9a}
 \langle \overline{\psi}_\alpha(t,\vec x) \psi_\beta(t,\vec y) \rangle=
 \frac{1}{(2\pi)^3}\int\frac{d^3\bf{k}}{2\omega_s(\bf{k})}
 (\mathbf{k}\vec{\gamma}+m)_{\alpha\beta}\,e^{-\imath\bf{k}(\bf{x}-\bf{y})}
\eea
are the  Pauli -- Jordan functions.

\subsection{Propagators and condensates}

The vector field in the Lagrangian (\ref{3-2-10}) is given by the
formula
\begin{equation}\label{3-5-1}
V^{\rm R}_i= \left[\delta_{ij}^{\rm T}+\hat Z^{-1}\delta_{ij}^{\rm
||}\right]V_j=V_i^{\rm T}+\hat Z^{-1}V_i^{\rm ||}.
\end{equation}
Hence, the propagator of the massive vector field in radiative
variables is
\begin{equation}\label{3-5-2}
D^R_{ij}(x-y)=\langle
0|TV^R_{i}(x)V^R_{j}(y)|0\rangle=
\end{equation}
$$=-\imath\int
\frac{d^4q}{(2\pi)^4}\frac{e^{-\imath q\cdot
(x-y)}}{q^2-M^2+\imath\epsilon}\left(\delta_{ij}-\frac{q_{i}q_{j}}{\mathbf{q}^2+M^2}\right)~.
$$
 Together with the instantaneous  interaction described by
  the current--current term in
 the Lagrangian (\ref{3-2-10})  this propagator
 leads to the amplitude
\bea\label{3-2-11}
T^{\rm R}&=& D^{\rm
R}_{\mu\nu}(q)\widetilde{j}^\mu \widetilde{j}^\nu =\\\nonumber
\frac{\widetilde{j}_0^2}{\mathbf{q}^2+M^2}&+&\left(\delta_{ij}-\frac{q_i
q_j}{\mathbf{q}^2+M^2}\right) \frac{\widetilde{j}_i
\widetilde{j}_j}{{q}^2-M^2+\imath\epsilon}~
\eea
 of the current-current interaction which differs from the acceptable one
\begin{equation}\label{3-1-8}
 T^{\rm L}=\widetilde{j}^{\mu}D^{\rm L}_{\mu\nu}(q)\widetilde{j}^{\nu}=
-\widetilde{j}^{\mu}\frac{g_{\mu\nu}-\dfrac{q_\mu q_\nu}{M^2}
}{q^2-M^2+\imath\epsilon}\widetilde{j}^{\nu}.
\end{equation}
The amplitude given by Eq.~(\ref{3-2-11})  is the generalization of
the  radiation amplitude  in QED. As it was shown in  \cite{h-pp-A},
the Lorentz transformations of classical radiation variables
coincide with the  quantum ones  and they both (quantum and
classical) correspond to the transition to another Lorentz frame of
reference distinguished by another time-axis, where the relativistic
covariant propagator takes the form
\begin{equation}\label{3-5-4}
D^{R}_{\mu\nu}(q|n)=
\end{equation}
$$=\frac{-g_{\mu\nu}}{q^2-M^2+\imath\epsilon}+
\frac{n_{\mu}n_{\nu}(qn)^2-[q_{\mu}-n_{\mu}(qn)][q_{\nu}-n_{\nu}(qn)]}
 {(q^2-M^2+\imath\epsilon)(M^2+|q_{\mu}-n_{\mu}(qn)|^2)},
$$
where $n_{\mu}$ is determined by the external states. Remember that
 the conventional local field massive vector propagator
 takes the form  (\ref{3-1-8})
 \begin{equation}\label{3-5-5}
 D^L_{\mu\nu}(q)=-
\dfrac{g_{\mu\nu}-\dfrac{q_\mu q_\nu}{M^2}}{q^2-M^2+\imath\epsilon}~.
\end{equation}
In contrast to this conventional  massive vector propagator the
 radiation-type propagator (\ref{3-5-4})  is regular in the
limit $M\rightarrow 0$ and is well behaved for large momenta,
whereas the propagator (\ref{3-5-5}) is singular. The radiation
amplitude (\ref{3-2-11}) can be rewritten   in the alternative form
\begin{equation}
T^{\rm R}=-\frac{1}{q^2-M^2+\imath\epsilon}\left[\widetilde{j}_{\nu}^2
+\frac{(\widetilde{j}_iq_i)^2-(\widetilde{j}_0q_0)^2}{\vec{q}^2+M^2}\right]~,
\label{mvecprop2}
\end{equation}
for comparison with the conventional amplitude defined by the
propagator (\ref{3-5-5}). One can find that for a massive vector
field coupled to a conserved current
$(q_{\mu}\widetilde{j}^{\mu}=0)$ the collective current-current
interactions mediated by the radiation propagator  (\ref{3-5-4}) and
by the conventional propagator (\ref{3-5-5}) coincide
\begin{equation}\label{3-5-52}
\widetilde{j}^{\mu}D^{\rm R}_{\mu\nu}\widetilde{j}^{\nu}=
\widetilde{j}^{\mu}D^{\rm L}_{\mu\nu}\widetilde{j}^{\nu}=T^{\rm L}~.
\end{equation}
 If the  current is not conserved
 $$\widetilde{j}_0q_0\not =\widetilde{j}_kq_k,$$
 the collective
 radiation field variables
 with the propagator (\ref{3-5-4})
 are inequivalent to   the initial local  variables
 with the propagator  (\ref{3-5-5})  and the amplitude
 (\ref{3-2-11}). The amplitude (\ref{3-5-52}) in the Feynman gauge is
 \be\label{3-5-52a}
 T^{\rm L} =
 -\frac{j^2}{q^2-M^2+\imath\varepsilon},
 \ee
and corresponds to the Lagrangian
 \be\label{brst}
\mathcal{L}_{F}=\frac{1}{2}(\partial_{\mu}V_{\mu})^{2}-j_{\mu}V_{\mu}+
\frac{1}{2}M^2 V_\mu^2 \ee
 In this theory the time component has a negative contribution
 to the energy. According to  this, a correctly defined
 vacuum state could not exist. Nevertheless, the vacuum expectation
 value
 $$\langle V_\mu(x)V_\mu(x) \rangle$$
 coincides with the values
 for two  propagators (\ref{3-5-4}) and
 (\ref{3-5-5})
 because in both these propagators the longitudinal part
 does not give a contribution if one treats them as derivatives of constant
 like
$$\langle \partial V_\mu(x)V_\mu(x) \rangle
 =\partial \langle  V_\mu(x)V_\mu(x) \rangle=0.$$
 In this case, we have
\bea\label{v3-4-8}
 \langle V_\mu(x)V_\mu(x) \rangle =-\frac{2}
 {(2\pi)^3}\int\frac{d^3\bf{k}}{\omega_v(\bf{k})},
\eea
\bea\label{s3-4-8}
 \langle \overline{\psi}_\alpha(x)  \psi_\alpha(x)\rangle=-
 \frac{m_s}{(2\pi)^3}\int\frac{d^3\bf{k}}{\omega_s(\bf{k})},
 \eea
where $m_s$, $M_{v}$ are masses of the spinor and vector fields.
\chapter{Quantum field theory for bound states
\label{sect_AppB}}
\renewcommand{\theequation}{B.\arabic{equation}}
\setcounter{equation}{0}
\section{Ladder approximation}\label{subsect_App-B1}
\makeatletter
\renewcommand{\@evenhead}{\raisebox{0pt}[\headheight][0pt]
{\vbox{\hbox to\textwidth{\hfil \strut \thechapter . Appendix
\quad\rm\thepage}\hrule}}}
\renewcommand{\@oddhead}{\raisebox{0pt}[\headheight][0pt]
{\vbox{\hbox to\textwidth{\strut \thechapter .\quad
Quantum field theory for bound states
\hfil \rm\thepage}\hrule}}}
\makeatother

The generating functional of quantum field theory for bound states 
can be presented by means of
the relativistic generalization of the
 Hubbard -- Stratonovich (HS) transformation \cite{Eb-B, pre-1-B}.
The Hubbard -- Stratonovich
 transformation is an exact mathematical transformation
\bea\label{eff-1a}
\exp[-ax^2/2]=[2\pi a]^{-1/2}\int_{-\infty}^{+\infty} dy\exp[-\imath xy -y^2/(2a)].
\eea
%
The basic idea of the HS transformation is to reformulate a system of
particles interacting through two-body potentials  (\ref{Y-1}) in the theory (\ref{1-1q}) into a system of independent
particles interacting with a bilocal  auxiliary field ${\cal M}^{\mathrm{a}\mathrm{b}}(x,y)$.
The HS transformation was invented by the Russian physicist Ruslan L. Stratonovich and popularized by the
British physicist John Hubbard.
\bea\label{eff-1-B}     
Z_{\psi}&=& \int d\psi
d\overline{\psi}e^{{\imath}W_{instant}[\psi\overline{\psi}]+\imath S[J^*,\eta^*,\overline{\eta}^*]}=
\\ &=&
\label{eff-2-B}
\int d\psi
d\overline{\psi}
e^{-\frac{\imath}{2}\left(\psi\overline{\psi},
{\cal K}\psi\overline{\psi}\right)-\left(\psi \overline{\psi},G^{-1}_{0}\right)+\imath S[J^*,\eta^*]}=\\\label{eff-3-B}
&=&
\int\left[\prod\limits_{x,y,a,b}^{} d {\cal
M}^{\mathrm{a}\mathrm{b}}(x,y)\right]\exp\{\imath W_{\rm eff}[{\cal M}]
+\imath(\eta\overline{\eta},G_{{\cal M}})\}.
\eea
 The effective action in Eq.~(\ref{eff-3-B})
 can be decomposed in the form
\begin{eqnarray}
W_{eff}[{\cal M}] &=& - {1 \over 2} N_{c} ( {\cal M}, {{\cal K}}^{-1} {\cal
M} ) + \imath N_{c} {\rm Tr} \ln (1+\Phi),\\
{\rm Tr} \ln (1+\Phi) &=&
 \sum_{n=1}^{\infty} {1 \over n} \Phi^{n}.
\end{eqnarray}
Here
$ \Phi \equiv G_{0} {\cal M} , \Phi^{2} , \Phi^{3}, $
{\it etc.} mean the following expressions
\begin{eqnarray} \label{3-12a}
\Phi (x,y) \equiv G_{0} {\cal M} = \int d^{4}z G_{0} (x,z) {\cal M}(z,y), \nonumber \\
\Phi^{2} = \int d^{4} x d^{4}y \Phi(x,y) \Phi(y,x), \\
\Phi^{3} = \int d^{4} x d^{4}y d^{4}z \Phi(x,y) \Phi(y,z) \Phi(z,x) \,\,  ,
\mbox{\it etc.} \nonumber
\end{eqnarray}

The first step to the semi-classical quantization of this construction
\cite{Eb-B} is the determination of its minimum of the effective action
\begin{eqnarray}\label{8}
N_{c}^{-1}{ \delta W_{eff} ({\cal M}) \over { \delta {\cal M}} } \equiv
 - {\cal K}^{-1} {\cal M} + {\imath\over { G_{0}^{-1} - {\cal M} } } = 0.
\end{eqnarray}
This equation is known as the Schwinger -- Dyson one.
We denote the corresponding classical solution for the bilocal field by $\Sigma (x-y) $.
It depends only on the difference $x-y$ at $A^*=0$ because of translation invariance of vacuum solutions.

The next step is the expansion of the effective action around the point of minimum
$ {\cal M} = \Sigma + {\cal M}^{\prime} $,
\bea\label{7-B}
W_{eff} ( \Sigma + {\cal M}^{\prime} ) =  W_{eff}^{(2)}+W_{int};
\eea
\bea\label{9-B}
W_{eff}^{(2)} ({\cal M}^{\prime} ) = W_{Q}(\Sigma) + N_{c}\left[ - {1\over2}
{\cal M}^{\prime} {\cal K}^{-1} {\cal M}^{\prime}
 + { \imath \over 2} ( G_{\Sigma} {\cal M}^{\prime} )^{2} \right],
\eea
\bea\label{10-B}
W_{int}=\sum_{n=3}^{\infty}W^{(n)} = \imath N_{c} \sum_{n=3}^{\infty} {1\over n}
( G_{\Sigma} {\cal M}^{\prime} )^{n}, \,
\eea
\bea\label{11-B}
 G_{\Sigma} = (G_{0}^{-1} - \Sigma)^{-1} .
\eea
The bilocal function ${\cal M}^{\prime} (x,y)$ in terms of the Jacobi -- type variables
$$z=x-y,\qquad X=\frac{x+y}{2}$$ can be decomposed
over the complete set of orthonormalized solutions $ \Gamma $ of the classical equation
\begin{eqnarray}\label{10}
{ \delta^{2}W_{eff} ( \Sigma + {\cal M}^{\prime} ) \over { \delta {{\cal
M}^{\prime}}^{2}} }
\cdot \Gamma = 0.
\end{eqnarray}
This series takes the form:
\bea\label{set-1b-B}
{\cal M}^{\prime}(x,y)={\cal M}^{\prime}(z|X)=
\eea
$$
=\sum\limits_H\int\frac{d^3{\cal
P}}{(2\pi)^{3}\sqrt{2\omega_H}}\int\frac{d^4qe^{\imath q\cdot
z}}{(2\pi)^4}\times
$$
$$\times
\left[ e^{\imath {\cal P}\cdot{X}}
\Gamma_H(q^{\bot}|{\cal P})a^+_H(\bm{{\cal P}})+
 e^{-\imath {\cal P}\cdot{X}}
\bar{\Gamma}_H(q^{\bot}|{\cal P})a^-_H(\bm{{\cal P}})\right],
$$
%
with a set of quantum numbers ($H$) including masses $$M_H=\sqrt{{\cal P}_\mu^2}$$
and energies
$$\omega_H=\sqrt{{\bm{{\cal P}}}^2+M_H^2}.$$
The bound state creation and annihilation operators obey the commutation relations
\begin{eqnarray} \label{comrel}
\biggl[ a^{-}_{H'}(\bm{{\cal P}}'), a^{+}_{H}(\bm{{\cal P}}) \biggr] =
\delta_{H'H} \delta^3 ( \bm{{\cal P}}' - \bm{{\cal P}} ) \,\,\, .
\end{eqnarray}

The  corresponding Green function takes the form
\be\label{green-B} {\cal
G}(q^{\bot},p^{\bot}|{\cal P})=
\ee
$$
=\sum\limits_H
\left(\frac{\Gamma_H(q^{\bot}|{\cal P})\bar{\Gamma}_H(p^{\bot}|-{\cal P})}
{({\cal P}_0-\omega_H-\imath\varepsilon)2\omega_H}-
\frac{\bar\Gamma_H(p^{\bot}|{\cal P})){\Gamma}_H(p^{\bot}|-{\cal P})} {({\cal
P}_0-\omega_H-\imath\varepsilon)2\omega_H} \right)~.
$$

To normalize vertex functions  $\Gamma$, we can use the ''free'' part of the effective action
(\ref{9-B}) for the quantum bilocal meson ${\cal M}^{\prime}$ with the
commutation relations~(\ref{comrel}). The substitution of the off-shell
$$\sqrt{{\cal P}^2} \neq M_H$$
decomposition (\ref{set-1}) into the ``free'' part of
effective action defines the reverse Green function of the bilocal field
${\cal G}({\cal P}_0)$
\begin{eqnarray}
\!\!\!\!W_{eff}^{(0)}[{\cal M}]\!=\!2\pi\delta({\cal P}_0-{\cal
P'}_0)\sum_H\int\frac{d^3{\cal P}}{\sqrt{2\omega_H}}a^{+}_{H}(\bm{{\cal P}})
a^{-}_{H}(\bm{{\cal P}}){\cal G}^{-1}_H ({\cal P}_0)
\end{eqnarray}
where ${\cal G}^{-1}_H ({\cal P}_0) $ is the reverse Green function which can be represented as a difference of two terms
\begin{eqnarray}
{\cal G}_H^{-1}({\cal P}_0) =  I( \sqrt{ {\cal P}^2 } ) -I(M_H^{\mathrm{a}\mathrm{b}}(\omega))
\end{eqnarray}
where $M_H^{ab}(\omega)$ is the eigenvalue of the equation for small fluctuations~(\ref{10-B}) and
\begin{eqnarray}
I(\sqrt{{\cal P}^2}) &=&\imath N_c \int \frac{d^4 q}{(2\pi)^4} \times \nonumber \\
&&\mbox{tr} \biggl[ G_{\Sigma_{\mathrm{b}}}\left(q-\frac{{\cal P}}{2} \right) \bar{\Gamma}_{\mathrm{a}\mathrm{b}}^H
(q^\perp | -{\cal P}) G_{\Sigma_{\mathrm{a}}}\left( q+ \frac{{\cal P}}{2} \right) {\Gamma}_{\mathrm{a}\mathrm{b}}^H
(q^\perp | {\cal P}) \biggr],\nonumber
\end{eqnarray}
where
\be\label{nak} \underline{G}_\Sigma(q)=\frac{1}{\not
q-\underline{\Sigma}(q^{\bot})},~~~~~~ \underline{\Sigma}(q) = \int d^{4}x
\Sigma(x) e^{\imath qx}
\ee
is the fermion Green function. The normalization condition is defined by the formula
\begin{equation}
2 \omega = \frac{\partial{\cal G}^{-1}({\cal P}_0)}{\partial{\cal
P}_0}|_{{\cal P}_0=\omega({\cal P}_1)} = \frac{dM({\cal P}_0)}{d{\cal
P}_0}\frac{dI(M)}{dM}|_{{\cal P}_0=\omega}~.
\end{equation}
Finally, we get that solutions of equation (\ref{10}) satisfy the normalization condition~\cite{nakanishi-B}
$$
\imath N_c\frac{d}{d
{\cal P}_0}\int \frac{d^4q}{(2\pi)^{4}} \mbox{tr} \left[
\underline{G}_\Sigma\left(q-\frac{{\cal P}}{2}\right)\bar{\Gamma}_H(q^{\bot}|-{\cal P})
\underline{G}_\Sigma\left(q+\frac{{\cal P}}{2}\right){\Gamma}_H(q^{\bot}|{\cal P})
\right]
$$
\be\label{nakanishi}
=2\omega_H.
\ee
The achievement of the relativistic covariant constraint-shell quantization of
gauge theories is the description of both the spectrum of bound states and their S-matrix elements.

It is convenient to write the relativistic-invariant matrix elements for the
action~~ (\ref{7-B}) in terms of the field operator
$$
\Phi^{\prime}(x,y)=\int d^4x_1 G_{\Sigma}(x-x_1){\cal
M}'(x_1,y)=\Phi^{\prime}(z|X).
$$
Using the decomposition over the bound state quantum numbers $(H)$
  \bea\label{set1}
 {\Phi}^{\prime}(z|X)&=&\sum\limits_H\int\frac{d^3{\cal P}}{(2\pi)^{3/2}
 \sqrt{2\omega_H}}\int\frac{d^4q}{(2\pi)^4}\times\\
 &&\left( e^{\imath{\cal P}\cdot{X}} \Phi_H(q^{\bot}|{\cal P})a^+_H(\bm{{\cal P}})
 +e^{-\imath{\cal P}\cdot{X}} \bar{\Phi}_H(q^{\bot}|-{\cal P})a^-_H(\bm{{\cal P}})\right)~,\nonumber
 \eea
where
\be \Phi_{H(\mathrm{a}\mathrm{b})}(q^{\bot}|{\cal P})=G_{\Sigma a}(q+{\cal
P}/2)\Gamma_{H(\mathrm{a}\mathrm{b})}(q^{\bot}|{\cal P})~,
\ee
we can write the matrix elements  $W^{(n)}$~(\ref{9-B})
for the interaction  between the vacuum and the n-bound state \cite{yaf-B}
\begin{multline}
\label{S-matrix}
\langle H_1{\cal P}_1, ...,H_n{\cal P}_n|\imath W^{(n)}|0\rangle =\\
=-\imath (2\pi)^4 \delta^4 \left( \sum\limits_{i=1 }^{n }{\cal P}_i\right)
\prod\limits_{j=1 }^{n } \left[\frac{1}{(2\pi)^32\omega_j}\right]^{1/2}
M^{(n)}({\cal P}_1,...,{\cal P}_n),
\end{multline}

\begin{multline} \label{S-matrix-1}
 M^{(n)}=\int \frac{\imath d^4q}{(2\pi)^4 n} \sum\limits_{\{i_k\} }^{ }
 \Phi_{H_{i_{1}}}^{\mathrm{a}_1,\mathrm{a}_2}(q| {\cal P}_{i_1})\times \\
 \Phi_{H_{i_{2}}}^{\mathrm{a}_2,\mathrm{a}_3}(q-\frac{{\cal P}_{i_1}+{\cal P}_{i_2}}{2}| {\cal
 P}_{i_2}) \Phi_{H_{i_{3}}}^{\mathrm{a}_3,\mathrm{a}_4}\left(q-\frac{2{\cal P}_{i_2}+
 {\cal P}_{i_1}+{\cal P}_{i_3}}{2}| {\cal P}_{i_3}\right)\times \\
 ...\Phi_{H_{i_{n}}}^{\mathrm{a}_n,\mathrm{a}_1} \left(q-\frac{2({\cal P}_{i_2}+...+{\cal
 P}_{i_{n-1}})+{\cal P}_{i_1}+ {\cal P}_{i_n} }{2}| {\cal P}_{i_n}\right),
\end{multline}
where $\{i_k\}$ denotes permutations over $i_k$).

Expressions  (\ref{green-B}), and
~(\ref{S-matrix-1}) represent Feynman rules for the construction of a quantum
field theory with the action~(\ref{9-B}) in terms of bilocal fields.

\section{Bethe -- Salpeter Equations}\label{subsect_App-B2}
\makeatletter
\renewcommand{\@evenhead}{\raisebox{0pt}[\headheight][0pt]
{\vbox{\hbox to\textwidth{\hfil \strut \thechapter . Appendix
\quad\rm\thepage}\hrule}}}
\renewcommand{\@oddhead}{\raisebox{0pt}[\headheight][0pt]
{\vbox{\hbox to\textwidth{\strut \thechapter .\quad
Quantum Field Theory for bound states
\hfil \rm\thepage}\hrule}}}
\makeatother

 Equations for the spectrum of the bound states (\ref{10})  can be
 rewritten in the form of the  Bethe -- Salpeter (BS) one~\cite{Salpeter-B,a15-B}
 \begin{eqnarray}\label{bs}
 \Gamma = \imath {\cal K}(x,y) \int d^{4}z_{1} d^{4}z_{2}
 G_{\Sigma}(x-z_{1}) \Gamma(z_{1},z_{2}) G_{\Sigma}(z_{2}-y)~.
 \end{eqnarray}
 In the momentum space with
 $$\underline{\Gamma}(q \vert {\cal P}) = \int d^{4}x d^{4}y
 e^{\imath{x+y \over2} {\cal P}} e^{\imath(x-y)q} \Gamma(x,y)
 $$
  the Coulomb type kernel
   we obtain the following equation for the vertex function
( $ \underline {\Gamma} $ ):
 \begin{eqnarray} \label{bs0-B}
 \underline{\Gamma}(k, {\cal P}) =
\end{eqnarray}
$$=
 \imath \int {d^{4}q \over (2\pi)^{4}}
 \underline {V} ( k^{\perp} - q^{\perp} ) \rlap/\ell \left[
 \underline{G}_{\Sigma}\left(q+{ {\cal P} \over 2 }\right) \Gamma(q \vert
 {\cal P} ) \underline{G}_{\Sigma}\left(q-{ {\cal P} \over 2 }\right) \right]
 \rlap/\ell
$$
where
 $ \underline {V} ( k^{\perp} ) $
means the Fourier transform of the potential,
$$ k^{\perp}_{\mu} = k_{\mu} - \ell_{\mu} ( k \cdot \ell) $$
 is the relative  momentum transversal with respect to
 $ \ell_{\mu} $, and ${\cal P}_{\mu}$
is the total momentum.

 The quantity  $ \underline {\Gamma} $  depends only on the transversal momentum
 \begin{eqnarray*}
 \underline{\Gamma}(k \vert {\cal P}) =
 \underline{\Gamma}(k^{\perp} \vert {\cal P}) ,
 \end{eqnarray*}
 because of the instantaneous form of the potential
 $ \underline{V}(k^{\perp} )$
 in any frame.
The Bethe -- Salpeter equation (\ref{bs}) for potential independent of the longitudinal momentum
allows to make integration over it
which at the rest frame is equal to $q_{0}$\footnote{This integral has poles of the product of two Green
functions of the parton-quarks (or leptons in QED)
 $$\dfrac{\imath}{2\pi}\int dq^0
\frac{1}{(q_0-a - \imath\varepsilon)(q_0+b + \imath\varepsilon)}=\dfrac{1}{a+b}\,.$$}.
We consider the Bethe -- Salpeter equation~(\ref{bs0-B}) after integration over the longitudinal momentum.
The vertex function takes the form
 \begin{eqnarray} \label{bs1}
 \Gamma_{\mathrm{a}\mathrm{b}}(k^{\perp} \vert {\cal P}) = \int { d^{3}q^{\perp}
 \over (2\pi)^{3} } \underline{V} ( k^{\perp}-q^{\perp} )
 \rlap/\ell \Psi_{\mathrm{a}\mathrm{b}}(q^{\perp}) \rlap/\ell,
 \end{eqnarray}
where the bound state wave function
$ \Psi_{\mathrm{a}\mathrm{b}} $
is given by
 \begin{eqnarray} \label{bs3}
 \Psi_{\mathrm{a}\mathrm{b}}(q^{\perp})=
\end{eqnarray}
$$= \rlap/\ell \left[ {
 \bar{\Lambda}_{(+)\mathrm{a}}(q^{\perp} \Gamma_{\mathrm{a}\mathrm{b}}(q^{\perp} \vert {\cal
 P} ) {\Lambda}_{(-)\mathrm{b}}(q^{\perp}) \over { E_{T} - \sqrt{ {\cal
 P}^{2} } + \imath \epsilon } } + { \bar{\Lambda}_{(-)\mathrm{a}}(q^{\perp}
 \Gamma_{\mathrm{a}\mathrm{b}}(q^{\perp} \vert {\cal P} ) {\Lambda}_{(+)\mathrm{b}}(q^{\perp})
 \over { E_{T} + \sqrt{ {\cal P}^{2} } - \imath \epsilon } }\right ]
 \rlap/\ell.
$$
Here
the sum of one-particle energies of the two particles ($a$)
 and ($b$ )
$$E_{T} = E_{\mathrm{a}} + E_{\mathrm{b}}$$
defined by~(\ref{sd1})
  and the notation~(\ref{ope1})
 \begin{eqnarray}
 \label{ope2}
 \bar{\Lambda}_{(\pm)}(q^{\perp}) = S^{-1}(q^{\perp})
 \Lambda_{(\pm)}(0) S(q^{\perp}) = {\Lambda}_{(\pm)}( - q^{\perp}).
 \end{eqnarray}
 has been introduced.

 Acting with the operators  (\ref{ope2}) on  equation (\ref{bs1}) one gets
 the equations for the wave function $ \psi $ in an arbitrary moving reference frame
\bea\label{bs2}
(E_{T}(k^{\perp}) \mp \sqrt { {\cal P}^{2}})
 \Lambda^{(\ell)}_{(\pm)\mathrm{a}}(k^{\perp}) \Psi_{\mathrm{a}\mathrm{b}}(k^{\perp})
 {\Lambda}^{(\ell)}_{(\mp)\mathrm{b}}( - k^{\perp}) =
\eea
$$=
 \Lambda^{(\ell)}_{(\pm)a}(k^{\perp})  \int { d^{3}q^{\perp} \over
 (2\pi)^{3} } \underline{V} (k^{\perp}-q^{\perp})
 \Psi_{\mathrm{a}\mathrm{b}}(q^{\perp}) ] {\Lambda}^{(\ell)}_{(\mp)b}( - k^{\perp}) .
$$
 All these equations ~(\ref{bs1}) and ~(\ref{bs2}) have been derived without any
 assumption about the smallness of the relative momentum
 $ \vert  k^{\perp} \vert $
 and for an arbitrary total momentum
\be\label{P-B}
 {\cal P}_{\mu} = \left( \sqrt { M_{A}^{2} + { {{\cal P}}}^{2} } ,\quad {\bm{{\cal P}}} \neq 0 \right)~.
\ee
 We expand the function $\Psi$ on the projection operators
 \be  \label{bs4}
 \Psi=\Psi_++\Psi_-,~~~~~\Psi_{\pm}=
 \Lambda^{(\ell)}_{\pm}\Psi
 \Lambda^{(\ell)}_{\mp}~.
 \ee
 According to Eq.~(\ref{bs3}), $\Psi$ satisfies the identities
 \be  \label{bs5}
 \Lambda^{(\ell)}_{+}\Psi\Lambda^{(\ell)}_{+}=
 \Lambda^{(\ell)}_{-}\Psi\Lambda^{(\ell)}_{-}\equiv 0~,
 \ee
 which permit the determination of unambiguous expansion of $\Psi$  in terms of the Lorentz structures:
 \be \label{bs6}
  \Psi_{\mathrm{a,b}\pm}=S^{-1}_{\mathrm{a}}\left( \gamma_5 L_{\mathrm{a,b}\pm}(q^{\bot})+
  (\gamma_{\mu}-\ell_{\mu}\not \ell)N^{\mu}_{\mathrm{a,b}\pm} \right)
 \Lambda^{(\ell)}_{\mp}(0)S^{-1}_{\mathrm{b}}~,
 \ee
where
$$L_{\pm}=L_1\pm L_2,\qquad N_{\pm}=N_1\pm N_2.$$
In the rest  frame $\ell_{\mu}=(1,0,0,0)$ we get
 $$
 N^{\mu}=(0,N^i)~;
~~~~~~N^i(q)=\sum\limits_{a=1,2 }^{ }N_{\alpha}(q)e^i_{\alpha}(q)+
 \Sigma(q) \hat q^i~.
 $$
 The wave functions $L,N^{\alpha},\Sigma$  satisfy the following equations.

\begin{center}
\underline {1. Pseudoscalar particles.
}
\end{center}

$$ M_{L}  \stackrel{0}{L}_{2}({\bf p}) =
E  \stackrel{0}{L}_{1}({\bf p})
-
\int \frac{d^3 {q}}{(2\pi)^{3}}
V({\bf p}-{\bf q})
(  {\tt c}^{-}_{p} {\tt c}^{-}_{q}
- \xi {\tt s}^{-}_{p} {\tt s}^{-}_{q})
\stackrel{0}{L}_{1}({\bf q}) \,\,\, ;
$$
$$
M_L  \stackrel{0}{L}_{1}({\bf p}) =
E \stackrel{0}{L}_{2}({\bf p})
-
\int \frac{d^3 {q}}{(2\pi)^{3}}
V({\bf p}-{\bf q})
(  {\tt c}^{+}_{p} {\tt c}^{+}_{q}
- \xi {\tt s}^{+}_{p} {\tt s}^{+}_{q})
\stackrel{0}{L}_{2} ({\bf q})     \,\,\, .
$$
 Here, in all equations, we use the following definitions
 \bea\label{E-0}
 E({\bf p})&=&E_{\mathrm{a}}({\bf p})+E_{\mathrm{b}}({\bf p})~,\\\label{E-1}
 {\tt c}^{\pm}_p&=&\cos[v_{\mathrm{a}}({\bf p}) \pm v_{\mathrm{b}}({\bf p})]~,\\\label{E-2}
 {\tt s}^{\pm}_p&=&\sin[v_{\mathrm{a}}({\bf p}) \pm v_{\mathrm{b}}({\bf p})]~,\\\label{E-3}
 \xi &=& {\hat p}_{i}
\cdot {\hat q}_{i} \,\,\, , \eea
where
$E_{\mathrm{a}},E_{\mathrm{b}}$
are  one-particle energies and
$v_{\mathrm{a}}, v_{\mathrm{b}}$
are the Foldy -- Wouthuysen angles of particles (a,b)
given by Eqs.~(\ref{sd2}) and (\ref{sd-2a}).

\begin{center}
\underline {2. Vector particles.}
\end{center}

$$
M_{\rm N} \stackrel{0}{N}_{2}{}^{\alpha} =
E \stackrel{0}{N}_{1}{}^{\alpha}  -
$$
$$
-
\int \frac{d^3 {q}}{(2\pi)^{3}}
V({\bf p}-{\bf q})
\{
(  {\tt c}^{-}_{p} {\tt c}^{-}_{q}
{\underline \delta}^{\alpha\beta}
+ {\tt s}^{-}_{p} {\tt s}^{-}_{q}
( {\underline \delta}^{\alpha\beta} \xi -
\eta^{\alpha}{\underline \eta}^{\beta} ) )
\stackrel{0}{N}_{1}^{\beta}
+
(\eta^{\alpha} {\tt c}^{-}_{p} {\tt c}^{+}_{q})
\stackrel{0}{\Sigma}_{1} \}   \,\,\,  ;
$$
$$
M_{\rm N} \stackrel{0}{N}_{1}{}^{\alpha} =
E \stackrel{0}{N}_{2}{}^{\alpha}  -
$$
$$
-\int \frac{d^3 {q}}{(2\pi)^{3}}
V({\bf p}-{\bf q})
\{
(  {\tt c}^{+}_{p} {\tt c}^{+}_{q}
{\underline \delta}^{\alpha\beta}
+ {\tt s}^{+}_{p} {\tt s}^{+}_{q}
( {\underline \delta}^{\alpha\beta} \xi -
\eta^{\alpha}{\underline \eta}^{\beta} ) )
\stackrel{0}{N}_{2}^{\beta}
+
(\eta^{\alpha} {\tt c}^{+}_{p} {\tt c}^{-}_{q})
\stackrel{0}{\Sigma}_{2}  \} \,\,\, .
$$
\begin{eqnarray*}
\eta^{\alpha} &=& {\hat q}_{i} {\hat e}^{\alpha}_{i}(p), \\\nonumber
{\underline \eta}^{\alpha} &=& {\hat p}_{i} {\hat e}^{\alpha}_{i}(q), \\\nonumber
{\underline \delta}^{\alpha\beta}
&=& {\hat e}^{\alpha}_{i}(q){\hat e}^{\beta}_{i}(p) .
\end{eqnarray*}

\begin{center}
\underline {3. Scalar particles.}
\end{center}

$$
M_{\Sigma} \stackrel{0}{\Sigma}_{2}  =
E \stackrel{0}{\Sigma}_{1}  -
$$
$$
-\int \frac{d^3 {q}}{(2\pi)^{3}}
V({\bf p}-{\bf q})
\{
( \xi {\tt c}^{+}_{p} {\tt c}^{+}_{q} + {\tt s}^{+}_{p} {\tt s}^{+}_{q})
\stackrel{0}{\Sigma}_{1}
+
({\underline \eta}^{\beta}
{\tt c}^{-}_{p} {\tt c}^{+}_{q}) \stackrel{0}{N}_{1}{}^{\beta}  \} \,\,\, ;
$$
$$
M_{\Sigma} \stackrel{0}{\Sigma}_{1}  =
E \stackrel{0}{\Sigma}_{2}   -
$$
$$
-\int \frac{d^3 {q}}{(2\pi)^{3}}
V({\bf p}-{\bf q})
\{
( \xi {\tt c}^{-}_{p} {\tt c}^{-}_{q} + {\tt s}^{-}_{p} {\tt s}^{-}_{q})
\stackrel{0}{\Sigma}_{2}
+                          ( {\underline \eta}^{\beta}
{\tt c}^{+}_{p} {\tt c}^{-}_{q}) \stackrel{0}{N_{2}}^{\beta} \} \,\,\, .
$$

 The normalization of these solutions is uniquely determined by equation~(\ref{nakanishi})
\be\label{nakap}
\frac{2N_c}{M_L}\int
\frac{d^3q}{(2\pi)^3}\left\{ L_1({\bf q})L_2^*({\bf q})+L_2({\bf q})L_1^*({\bf q})\right\}=1~, \ee \be\label{nakav}
\frac{2N_c}{M_N}\int \frac{d^3q}{(2\pi)^3}
\left\{N^{\mu}_1({\bf q})N^{\mu*}_2({\bf q})+N^{\mu}_2({\bf q})N^{\mu*}_1({\bf q})\right\}=1~, \ee \be\label{nakas}
\frac{2N_c}{M_{\Sigma}}\int \frac{d^3q}{(2\pi)^3}
\left\{\Sigma_1({\bf q})\Sigma_2^*({\bf q})+\Sigma_2({\bf q})\Sigma_1^*({\bf q})\right\}=1~.
\ee

 If the atom is at rest ( $ {\cal P}_{\mu} = ( M_{A},0,0,0 ) $ )
 equation (\ref{bs2}) coincides with the Salpeter equation \cite{a15-B}. If
 one assumes that the current mass $ m^{0} $ is much larger than
 the relative momentum, then the coupled
 equations ~(\ref{bs1}) and ~(\ref{bs2}) turn into the Schr\"odinger
 equation. In the rest frame ( $ {\cal P}_{0} = M_{A} $) equation
 ~(\ref{sd1}) 
  for a large mass  $ ( m^{0} / \vert q^{\perp} \vert
 \rightarrow \infty ) $ describes a nonrelativistic particle
 \begin{eqnarray*}
 E_{\mathrm{a}}( {\bf k} ) = \sqrt { ( m_{\mathrm{a}}^{0})^{2} + {\bf k}^{2} } \simeq
 m^{0}_{\mathrm{a}} + {1 \over 2} { {\bf k}^{2} \over m^{0}_{\mathrm{a}} }, \\
 \tan 2 \upsilon = { k \over m^{0} } \rightarrow 0 ; \,~~~~~~~~\, S({\bf
 k}) \simeq 1 ; \,~~~~~~~~\, \Lambda_{ (\pm)} \simeq { {1 \pm \gamma_{0} }
 \over 2 } .
 \end{eqnarray*}
 Then, in equation (\ref{bs2}) only the state with positive energy remains
 \begin{eqnarray}\label{dinge}
  \Psi_P^{\alpha\beta}  \simeq \Psi^{\alpha\beta}_{(+)}= [\Lambda_{(+)}\gamma_5]^{\alpha\beta}\sqrt{4\mu}\psi_{\mathrm{Sch}}, ~~~~~~
 \Lambda_{(-)} \Psi_P^{\alpha\beta} \Lambda_{(+)} \simeq 0 ,
 \end{eqnarray}
where
$$ \mu\equiv \frac{m_{\mathrm{a}} \cdot m_{\mathrm{b}}}{( m_{\mathrm{a}}+m_{\mathrm{b}})} .$$
 And finally the Schr\"odinger equation results in
\begin{eqnarray}
\label{E-4}
 \left[ {1 \over 2\mu} {\bf k}^{-2} + ( m^{0}_{\mathrm{a}} + m^{0}_{\mathrm{b}} - M_{A} )
 \right] \psi_{\mathrm{Sch}}({\bf k}) =
\end{eqnarray}
$$=
 \int { d^3{q} \over (2\pi)^{3} }
 \underline{V} ({\bf k}-{\bf q}) \psi_{\mathrm{Sch}}({\bf q}),
$$
 with the normalization
$$ \frac{1}{(2\pi)^3}\int d^3q|\psi_{\mathrm{Sch}}|^2=1.$$

 For an arbitrary total momentum $ {\cal P}_{\mu}$  (\ref{P-B}) equation~(\ref{E-4})  takes the form
 \begin{eqnarray} \label{dinger}
 &&\left[- {1 \over 2\mu}
 (k_{\nu}^{\perp})^{-2} + ( m^{0}_{\mathrm{a}} + m^{0}_{\mathrm{b}} -
 \sqrt{ {\cal P}^{2}}  ) \right] \psi_{\mathrm{Sch}}( k^{\perp})
  =
\end{eqnarray}
$$
= \int { d^{3} q^{\perp}
 \over (2\pi)^{3} } \underline{V} ( k^{\perp}- q^{\perp})
 \psi_{\mathrm{Sch}}( q^{\perp}),
$$
where
$$
k_{\mu}^{\perp}=k_{\mu}-\frac{{{\cal P}}{\bf k}}{M_H^2}{\cal P}_\mu,
$$
 and describes a  relativistic atom with nonrelativistic relative momentum
 $ \vert  k^{\perp} \vert \ll m^{0}_{a,b} $.
 In the framework of such a  derivation of the Schr\"odinger equation it is sufficient to
 define the total coordinate as  $ X=(x+y)/2$,  independently of the magnitude of the masses of the
 two particles  forming an atom.

 In particular, the Coulomb interaction leads to a positronium at rest with the bilocal wave function (\ref{dinge})
 \bea\label{posi}
 \Psi_P^{\alpha\beta}({\bf z})&=&\left(\frac{1+\gamma_0}{2}
 \gamma_5\right)^{\alpha\beta}\underline{\psi}_{\mathrm{Sch}}({\bf z})
 \sqrt{\frac {m_e} 2}~,\\\label{po-sch}
 \underline{\psi}_{\mathrm{Sch}}({\bf z})&=&
 \int\limits_{ }^{ }\frac{d^3p}{(2\pi)^3}e^{(\imath {\bf p} \cdot {\bf z})} \psi_{\mathrm{Sch}}({\bf z});
 \eea
where $\underline{\psi}_{Sch}({\bf z})$ is the Schr\"odinger normalizable wave function of the relative motion
 \be \label{relative}
 \left(-\frac 1{m_e}\frac {d^2}{d{{\bf z}}^2}
 -\frac {\alpha}{|{\bf z}|}\right)\underline{\psi}_{\mathrm{Sch}}({\bf z})=
 \epsilon\underline{\psi}_{\mathrm{Sch}}({\bf z})
 \ee
with the normalization
$$\int d^3z{\parallel\underline{\psi}_{\mathrm{Sch}}({\bf z})\parallel}^2=1$$
where $M_P=(2m_e-\epsilon)$  is the mass of a positronium,
 $(1+\gamma_0)/2$
is the projection operator on the state with positive energies of an electron and positron.

\chapter{Abel -- Plana formula
\label{sect_AppC}}
\renewcommand{\theequation}{C.\arabic{equation}}
\setcounter{equation}{0}
\makeatletter
\renewcommand{\@evenhead}{\raisebox{0pt}[\headheight][0pt]
{\vbox{\hbox to\textwidth{\hfil \strut \thechapter . Appendix
\quad\rm\thepage}\hrule}}}
\renewcommand{\@oddhead}{\raisebox{0pt}[\headheight][0pt]
{\vbox{\hbox to\textwidth{\strut \thechapter .\quad
Abel -- Plana formula
\hfil \rm\thepage}\hrule}}}
\makeatother

Only a few series in mathematics can be calculated in an exact form. Therefore, it is very important to express the sums
of series in terms of contour integrals. One of the popular methods is based on the following theorem \cite{Evgrafov}.

{\it
Let a function $f(z)$ be holomorphic in the strip $a< \Re z <b$ and satisfy the inequality
\be\label{in}
|f(x+\imath y)|\le M e^{a|y|},\qquad a< 2\pi.
\ee
Then at $k\ge a+1, n\le b-1, n>k,$ and for any $0< \theta <1$
\bea\label{APlan}
\sum\limits_{s=k}^n f(s)&=&\int\limits_{k+\theta-1}^{n+\theta} f(x)\, dx+\\\nonumber
&+&\frac{1}{2\imath}\int\limits_0^{\theta+\imath\infty}
[f(n+z)-f(k-1+z)](\cot \pi z+\imath)\, dz+\\\nonumber
&+&\frac{1}{2\imath}\int\limits_0^{\theta-\imath\infty}
[f(k-1+z)-f(n+z)](\cot \pi z-\imath)\, dz.
\eea
}

{\it Proof.}

We denote by $C_h$ a rectangle
$$k-1+\theta < \Re z < n+\theta,\qquad |\Im z|< h,$$
which in view of the conditions on $k$ and $n$ is in the band $a< \Re z< b,$ and by $J$ an integral of
$f(z)\cot \pi z$ along $C_h.$ According to the residue theorem, we have for $z=s$:
$$J=2\pi\imath\sum\limits_k^n res f(z)\cot\pi z=2\imath\sum\limits_k^n f(s).$$

We denote by $C_h^+$ the upper half of $C_h$, and by $C_h^-$ the lower half $C_h$, and the direction of
$C_h^+$ and $C_h^-$ we assume to be
the direction of a point $z=k-1+\theta$ to a point $z=n+\theta.$ Then we have
$$J=\int\limits_{C_h^-} f(z)\cot\pi z\, dz - \int\limits_{C_h^+} f(z)\cot\pi z\, dz$$
and
\bea\nonumber
J&=&\int\limits_{C_h^-} f(z)(\cot\pi z-\imath)\, dz+\imath\int\limits_{C_h^-} f(z)\, dz-\\\nonumber
&-&\int\limits_{C_h^+} f(z)(\cot\pi z+\imath)\, dz+\imath\int\limits_{C_h^+} f(z)\, dz.
\eea
The integral of $f(z)$ depends only on the ends of the loop, so the integrals of $f(z)$ along $C_h^+$ and $C_h^-$ can
be replaced by the integral over the interval $(k-1+\theta, n+\theta).$ Hence,
\bea\nonumber
J&=&2\imath\int\limits_{k-1+\theta}^{n+\theta} f(x)\, dx+\\\nonumber
&+&\int\limits_{C_h^-} f(z)(\cot\pi z-\imath)\, dz-\int\limits_{C_h^+} f(z)(\cot\pi z+\imath)\, dz.
\eea
Further,
$$\int\limits_{C_h^+} f(z)(\cot\pi z+\imath)\, dz =$$
\bea\nonumber
&=&\int\limits_0^{\theta+\imath h}[f(k-1+z)-f(n+z)](\cot\pi z+\imath)\, dz+\\\nonumber
&+&\int\limits_{h-1+\theta+\imath h}^{n+\theta+\imath h} f(z)(\cot\pi z+\imath)\, dz.
\eea
Since for ${k-1+\theta\le x \le n+\theta}$
$$\left|
\int\limits_{k-1+\theta+\imath h}^{n+\theta+\imath h} f(z)(\cot\pi z+\imath)\, dz
\right|\le$$
$$
\le (n-k+1)
|f(x+\imath h)||\cot\pi (x+\imath h)+\imath|.
$$

Since
$$|\cot\pi (x+\imath h)+\imath|<\frac{2}{e^{2\pi h}-1}$$
at $h>0,$ then, in force (\ref{in}), we have at $h\to+\infty$:
$$\left|
\int\limits_{k-1+\theta+\imath h}^{n+\theta+\imath h} f(z)(\cot\pi z+\imath)\, dz
\right|\le
(n-k+1)Me^{ah}\frac{2}{e^{2\pi h}-1}\to 0$$
and
\bea\nonumber
\int\limits_{C_h^+}(\cot\pi z+\imath)\, dz&=&\\\nonumber
&=&\int\limits_\theta^{\theta+\imath\infty}[f(k-1+z)-f(n+z)](\cot\pi z+\imath)\, dz.
\eea
Analogously, for the integral $C_h^-.$ Comparing the two expressions derived for $J,$ we arrive at the formula
(\ref{APlan}).

{\it Corollary.}

Let a function $f(z)$ be holomorphic in the half-plane $\Re z>0$ and satisfy the inequality
$$|f(x+\imath y)|<\varepsilon (x) e^{a|y|},\qquad 0< a< 2\pi,$$
where $\varepsilon (x)\to 0$ under $x\to +\infty.$ Then for any $0<\theta <1$
\be\label{APcorollary}
\lim_{n\to\infty}\left(\sum\limits_1^n f(s)-\int\limits_\theta^{n+\theta} f(x)\, dx\right)=
\ee
$$=\frac{1}{2\imath}\int\limits_\theta^{\theta-\imath\infty} f(z)(\cot\pi z-\imath )\, dz-
\frac{1}{2\imath}\int\limits_\theta^{\theta+\imath\infty} f(z)(\cot\pi z+\imath )\, dz.$$
This formula is called the {\it Abel -- Plana formula.}
It results from (\ref{APlan}), so that
$$\left|
\int\limits_{\theta}^{\theta\pm\imath\infty} f(n+z)(\cot\pi z\mp\imath)\, dz\right|\le
\varepsilon (n+\theta)\int\limits_0^\infty
Me^{-(2\pi-a)y}\, dy\to 0$$
as $n\to\infty$.

In quantum field theory an infinite number of degrees of freedom leads to zero vacuum fluctuations that
give a divergent contribution to the physical values.
To calculate the quantum values of the energy-momentum tensor the Abel -- Plana formula (\ref{APcorollary})
reduces to the form \cite{Bateman}:
\be\label{Abel-Plana}
\sum\limits_{n=0}^\infty F(n)=\int\limits_0^\infty F(x)\, dx+\frac{1}{2} F(0)+\imath\int\limits_0^\infty
\frac{F(\imath t)-F(-\imath t)}{\exp (2\pi t)-1}\, dt.
\ee
The first term on the right-hand side is the energy-momentum tensor of unlimited space.
Regularization is reduced to its subtraction.
For the regularized sum of the divergent series (\ref{Abel-Plana}) we obtain the formula \cite{GribAppendix}:
\be\label{Abel-Plana-boson}
reg\sum\limits_{n=0}^\infty F(n)=\frac{1}{2} F(0)+\imath\int\limits_0^\infty
\frac{F(\imath t)-F(-\imath t)}{\exp (2\pi t)-1}\, dt.
\ee
A modification of the derivation of formula (\ref{Abel-Plana}) allows one to get a regularization
analogous to (\ref{Abel-Plana-boson})
for divergent series in which the sum is over a half-integer values of the argument \cite{GribAppendix}:
\be\label{Abel-Plana-fermion}
reg\sum\limits_{n=0}^\infty F\left(n+\frac{1}{2}\right)=-\imath\int\limits_0^\infty
\frac{F(\imath t)-F(-\imath t)}{\exp (2\pi t)+1}\, dt.
\ee
This formula is used for carrying out calculations with a fermion field.

The Casimir effect consists of a polarization of the vacuum of quantized fields which arises as a result of a change
in the spectrum of vacuum oscillations. Calculations of the effect for manifolds of various configurations and for
fields with various spins with using the Abel--Plana formulae (\ref{Abel-Plana-boson}),
and (\ref{Abel-Plana-fermion})
for the regularization are presented in
paper \cite{MostepanenkoTrunov}.
The distribution functions for bosons and fermions over energy
$$f_\mp(\epsilon)=\frac{1}{\exp{(\epsilon-\mu)/k_B T}\mp 1}$$
are similar to the expressions under the integral signs in
(\ref{Abel-Plana-boson}) and (\ref{Abel-Plana-fermion}).



\chapter{Functional Cartan forms \label{sect_AppD}}
\renewcommand{\theequation}{D.\arabic{equation}}
\setcounter{equation}{0}

\makeatletter
\renewcommand{\@evenhead}{\raisebox{0pt}[\headheight][0pt]
{\vbox{\hbox to\textwidth{\hfil \strut \thechapter . Appendix
\quad\rm\thepage}\hrule}}}
\renewcommand{\@oddhead}{\raisebox{0pt}[\headheight][0pt]
{\vbox{\hbox to\textwidth{\strut \thechapter .\quad
Functional Cartan forms
\hfil \rm\thepage}\hrule}}}
\makeatother
\section{Dynamical model with high derivatives}

Albert Einstein in fifth edition of his book ``The Meaning of Relativity'' added his paper
``Relativistic theory of the non-symmetric field'', written with collaboration with B. Kaufman
\cite{EinsteinKaufman}. A system of differential equations of motion does not determine the field completely.
There still remail certain free data. The smaller the number of free data, the ``stronger'' is the system.
Einstein introduced a notion of ``strength'' of a system of field equations.

How can we determine the degree of freedom of the functions? This problem is studied by Whittaker under considering
spherical harmonics \cite{Whittaker}.
How can we set the initial conditions for systems of differential equations if there are gauge degrees of freedom or identities?
The problem of identifying the dynamical variables is associated with the formulation of the Cauchy problem \cite{Burlankov}.
There is an interesting problem without dynamical degrees of freedom \cite{BurlankovPavlov}.

Let us consider here an instructive example of a system with constraints: a string theory whose
Lagrangian is the $n$th power of the Gauss curvature of a space--time
$(n\in \mathbb{N},n>1)$ \cite{Pavlov7}.
Insofar the Hilbert functional of gravitation in (1+1)-dimensional
space-time gives the Gauss--Bonnet topological invariant, we take as a Lagrangian the Gauss
curvature in the $n$th power. The theory keeps its covariance. Although the calculations are
cumbersome, the problem can be solved fully. It turns out to be a useful example of using the
generalized Maurer--Cartan forms.

We analyze the dynamics of a space--time metric taken in the ADM form:
\be
(g_{\mu\nu})=\left(\begin{array}{cc}
\alpha^2+\beta^2&\gamma\beta\\
\gamma\beta&\gamma^2\end{array}\right),\qquad \sqrt{g}=\alpha\gamma,
\ee
where the metric functions $\alpha(t,x)$ and $\beta(t,x)$ have the meaning of
Lagrange multipliers. The Gauss curvature can be expressed by \cite{Pogorelov}
\be
R=-\frac{1}{2\alpha^3\gamma^2}\det\left(\begin{array}{ccc}
\alpha&\beta&\gamma\\
\dot\alpha&\dot\beta&\dot\gamma\\
\alpha'&\beta'&\gamma'\end{array}\right)- \ee
$$-\frac{1}{2\alpha\gamma}
 \left({\left[
 \frac{{(\gamma^2)}^{.}-{(\gamma\beta)}^{'}}{\alpha\gamma}\right]}^{.}-
 {\left[
 \frac{{(\gamma\beta)}^{.}-{(\alpha^2+\beta^2)}^{'}}{\alpha\gamma}\right]}^{'}
 \right).
$$
The functional of action is in the form
\be\label{AS} S=\frac{1}{2}\int_{t,x}R^n\alpha\gamma=
\ee
$$=\frac{1}{2}\int_{t,x}{(\alpha\gamma)}^{1-n}{\left[
{\left(\frac{\beta'-\dot\gamma}{\alpha}\right)}^{.}+
{\left(\frac{\beta\dot\gamma}{\alpha\gamma}\right)}^{'}-
{\left(\frac{{(\alpha^2+\beta^2)}^{'}}{2\alpha\gamma}\right)}^{'}
\right]}^n,
$$
where $\int_{t,x}$ denotes integration over the space-time. Varying $S$
by the metric one gets the Euler---Lagrange equations:
$$\frac{\partial L}{\partial\alpha}-\frac{\partial}{\partial
t}\frac{\partial L}{\partial\dot\alpha}-\frac{\partial}{\partial
x}\frac{\partial L}{\partial\alpha^{'}}+\frac{\partial^2}{\partial
x^2}\frac{\partial L}{\partial\alpha^{''}}=0,$$
\be\label{Aextremals} \frac{\partial
L}{\partial\beta}-\frac{\partial}{\partial x}\frac{\partial
L}{\partial\alpha{\beta}^{'}}+\frac{\partial^2}{\partial t\partial
x}\frac{\partial L}{\partial {\dot{\beta}^{'}}}+
\frac{\partial^2}{\partial x^2}\frac{\partial
L}{\partial\beta^{''}}=0, \ee
$$\frac{\partial L}{\partial\gamma}-\frac{\partial}{\partial t}
\frac{\partial L}{\partial\dot\gamma}-\frac{\partial}{\partial x} \frac{\partial L}{\partial
\gamma^{'}}+ \frac{\partial^2}{\partial t^2}\frac{\partial
L}{\partial\ddot\gamma}+\frac{\partial^2}{\partial t\partial x}\frac{\partial L}{\partial
{\dot\gamma}^{'}}=0,$$
where $L$ is the density of the Lagrange function.

The differential equations of the extremals (\ref{Aextremals}) are very complicated. The
matter is clearer in the Hamiltonian formulation since we deal with a nondegenerated theory with
higher derivatives. So we use a slightly modified version of the Ostrogradski method
\cite{Dubrovin}. It is relevant to introduce, along with generalized coordinates $(\alpha, \beta,
\gamma)$, the new variable
\be\label{Au} u\equiv\frac{\beta^{'}-\dot\gamma}{\alpha}.
\ee
Then the action in the coordinates $(\alpha, \beta, \gamma, u)$ takes the form
\be\label{AS1} S=\frac{1}{2}\int_{t,x}{(\alpha\gamma)}^{1-n}{\left[\dot{u}-{\left(\frac{\beta
u+\alpha^{'}}{\gamma} \right)}^{'}\right]}^n. \ee
Momentum densities are calculated, using the functional derivatives
$$\pi_u\equiv\frac{\delta S}{\delta\dot u}=\frac{\partial
L}{\partial\dot u}=\frac{n}{2}{\alpha\gamma}^{1-n}
{\left[\dot{u}-{\left(\frac{\beta u+\alpha^{'}}{\gamma}
\right)}^{'}\right]}^{n-1},$$
$$\pi_\alpha\equiv\frac{\delta S}{\delta\dot\alpha}=\frac{\partial
L}{\partial\dot\alpha},$$
$$\pi_\beta\equiv\frac{\delta S}{\delta\dot\beta}=-\frac{\partial}{\partial x}\frac{\partial
L}{\partial\dot\beta^{'}},$$
$$\pi_\gamma\equiv\frac{\delta S}{\delta\dot\gamma}=\frac{\partial
L}{\partial\dot\gamma}-\frac{\partial}{\partial t}\frac{\partial
L}{\partial\ddot\gamma}-\frac{\partial}{\partial x}\frac{\partial L}{\partial\dot\gamma^{'}}.$$
Taking into consideration (\ref{Au}), the Hamiltonian
$$H=\int_x(\pi_u\dot{u}+\pi_\alpha\dot\alpha+\pi_\beta\dot\beta+\pi_\gamma\dot\gamma-
L[\alpha,\dot\alpha,\alpha^{'},\alpha^{''}; \beta,\beta^{'},\dot\beta^{'},\beta^{''};
\gamma,\dot\gamma,\gamma^{'},\ddot\gamma,\dot\gamma^{'}])$$
becomes
$$H=\int_x(\pi_u\dot{u}+\pi_\alpha\dot\alpha+\pi_\beta\dot\beta+\pi_\gamma
(\beta^{'}-\alpha u)-L[\alpha,\alpha^{'},\alpha^{''}; \beta,\beta^{'}; \gamma,\gamma^{'};
u,\dot{u},u^{'}]).$$
Ignoring boundary terms one has the following expression for the Hamiltonian:
$$H=$$
$$=\int_x\left(\pi_\alpha\dot\alpha+\pi_\beta\dot\beta+\alpha\left[(n-1)
{\left(\frac{2}{n^n}\right)}^{1/(n-1)}\gamma\pi_u^{n/(n-1)}-u\pi_\gamma+
{\left(\frac{\pi_u^{'}}{\gamma}\right)}^{'}\right]\right.+$$
$$+\left.\beta\left[-u\left(\frac{\pi_u^{'}}{\gamma}\right)-\pi_\gamma^{'}\right]\right).$$
Along with the equations of motion obtained by varying the Hamiltonian by variables $u(t,x)$,
$\gamma(t,x)$, there are two differential constraints:
$$(n-1){\left(\frac{2}{n^n}\right)}^{1/(n-1)}\gamma\pi_u^{n/(n-1)}-u\pi_\gamma+
{\left(\frac{\pi_u^{'}}{\gamma}\right)}^{'}=0,$$
$$u\left(\frac{\pi_u^{'}}{\gamma}\right)+\pi_\gamma^{'}=0.$$

The system can be integrated once and then takes the form
$$\frac{1}{2n-1}{\left(\frac{2}{n}\right)}^{n/(n-1)}\pi_u^{(2n-1)/(n-1)}+
{\left(\frac{\pi_u^{'}}{\gamma}\right)}^2+\pi_\gamma^2=c(t),$$
$$u\pi_u^{'}+\gamma\pi_\gamma^{'}=0,$$
where $c(t)$ is an arbitrary function of time.

The Hamiltonian formulation is defined
by the Poisson structure $\hat{J}$ on the functional phase space. Its nonzero brackets are
$$\{\gamma (t,x),\pi_\gamma(t',x')\}=\delta(t-t')\delta(x-x'),$$
$$\{u(t,x),\pi_u(t',x')\}=\delta(t-t')\delta(x-x').$$
In geometrodynamics, $U$ could be considered as the phase space of the Wheeler---De Witt space.

Then we construct on the basis of constraints the functionals
$$\Phi [\phi]=$$
$$=\int_{t,x}\left(\frac{1}{2n-1}{\left(\frac{2}{n}\right)}^{n/(n-1)}
\pi_u^{(2n-1)/(n-1)}+{\left(\frac{\pi_u^{'}}{\gamma}\right)}^2+\pi_\gamma^2-c(t)\right)\phi(t,x),$$
$$\Xi[\xi]=\int_{t,x}(u\pi_u^{'}+\gamma\pi_\gamma^{'})\chi(t,x)$$
 and calculate their Poisson bracket
\be \{\Phi,\Xi\}=\int_{t,x;t',x'}\frac{\delta\Phi}{\delta z}\hat{J}\frac{\delta\Phi}{\delta z}.\ee
The result of the calculation is
$$\{\Phi[\phi],\Xi[\xi]\}=\Phi[{(\phi\chi)}^{'}]+\int c(t)\phi(t,x).$$

So the differential constraints form a
closed algebra (there are no other constraints in the theory) and they do not annihilate the Poisson
bracket. We can express the variables $\pi_\gamma$ and $u$ from the constraints as
$$\pi_\gamma^2=c(t)-\frac{1}{2n-1}{\left(\frac{2}{n}\right)}^{n/(n-1)}
\pi_u^{(2n-1)/(n-1)} -{\left(\frac{\pi_u^{'}}{\gamma}\right)}^2,$$
$$u=-\gamma\left(\frac{\pi_\gamma^{'}}{\pi_u^{'}}\right).$$

\section{Variational De Rham complex}

For the investigation of covariant theories, mathematical tools of the theory of variational
complexes \cite{Olver} that are generalization of the De Rham complexes of differential forms prove
to be useful. The variational complexes are decomposed into two components. The first part is
obtained by reformulation of the De Rham complex onto spaces of a set of differential functions set on
$V\subset X\times D$, where $X$ is a space of independent variables and $U$ is a space of dependent
variables. A differential $r-$form is given by
\be \omega^r=\sum_J P_J[u]\,dx^j, \ee
where $P_J$ are differential functions and
$$dx^J=dx^{j_1}\wedge\ldots\wedge dx^{j_r},\quad 1\le j_1<\ldots<j_r\le p$$
constitute the basis of a space of differential $r-$forms $\wedge_r T^* X.$

Since
for relativistic theories a consequence of covariance of the description is that the Hamiltonian is
zero, we will be interested here only in the second part of the variational complex. Let us suppose
the Hamiltonian constraint to be resolved. Differential forms are active on ``horizontal'' variables
$X$ from $M$, and vertical forms are constructed analogously---they are active on ``vertical'' variables
$u$ and their derivatives. A vertical $k$-form is a finite sum
\be \hat\omega^k=\sum P_J^\alpha[u]\, du_{J_1}^{\alpha_1}\wedge\ldots\wedge du_{J_k}^{\alpha_k},
\ee
where $P_J^\alpha$ are differential functions. Here independent variables are like parameters.

Insofar the vertical form $\hat\omega$ is built on a space of finite jets
$M^{(n)}$, a vertical differential has properties of bilinearity, antidifferentiation, and closure
like an ordinary differential. Here we use functional forms connected with the introduced vertical
forms, as functionals connected with differential functions.

Let $\omega^k=\int_x\hat\omega^k$ be a functional $k-$form corresponding to a vertical $k-$form
$\hat\omega^k.$ A variational differential of a form $\omega^k$ is a functional $(k+1)$-form
corresponding to a vertical differential of a form $\hat\omega^k$. The basic properties are deduced
from the properties of the vertical differential, so we get a variational complex. A variational
differential defines an exact complex
\be 0\stackrel{\delta}{\longrightarrow}\Lambda_*^0\stackrel{\delta}{\longrightarrow}
\Lambda_*^1\stackrel{\delta}{\longrightarrow}\Lambda_*^2\stackrel{\delta}{\longrightarrow}
\Lambda_*^3\stackrel{\delta}{\longrightarrow}\cdots
\ee
on spaces of functional forms on $M.$

 Of particular interest in theoretical
physics problems are functional forms: $\omega^0, \omega^1, \omega^2.$ In the present problem,
after the constraints are utilized, we get a functional 1--form as a generalization of  a
differential Cartan form for dynamical systems:
\be \omega^1=\int_{t,x}\left[\pi_\gamma\left(t,\pi_u,\left(\frac{\pi_u^{'}}{\gamma}\right)\right)
d\gamma-u\left(t,\pi_u,\left(\frac{\pi_u^{'}}{\gamma}\right),{\left(\frac{\pi_u^{'}}{\gamma}\right)}^{'}\right)
d\pi_u\right] \ee
Equations of motion
are obtained as a condition of the closedness of the 1-form: $\delta\omega^1=0.$
But, as we demonstrate below,
there is a 0-form $\omega^0$:
\be\label{Aomega0} \omega^0=\int_{t,x}\hat\omega^0(t,\gamma,\pi_u)\ee
so that
$\delta\omega^0=\omega^1$, {\it id est} $\omega^1$
is not only a closed form, it is an exact one.

Acting by the operator of the variational
differential $\delta$ on the form (\ref{Aomega0}), we get
\be
\delta\omega^0=\int_{t,x}\left[\frac{\delta\omega^0}{\delta\gamma}d\gamma+\frac{\delta\omega^0}{\delta\pi_u}d\pi_u\right].
\ee
From this we find conditions on $\hat\omega^0(t,\gamma,\pi_u)$:
\be\label{D14}
\frac{\partial\hat\omega^0}{\partial\gamma}=\pi_\gamma\left(t,\pi_u,\left(\frac{\pi_u^{'}}{\gamma}\right)\right),
\ee
\be\label{D15}
\frac{\partial\hat\omega^0}{\partial\pi_u}=\frac{\partial}{\partial
x}\left(\frac{\partial\hat\omega^0}{\partial\pi_u^{'}}\right)= -u
\left(t,\pi_u,\left(\frac{\pi_u^{'}}{\gamma}\right),{\left(\frac{\pi_u^{'}}{\gamma}\right)}^{'}\right).
\ee

The system of differential equations (\ref{D14}), (\ref{D15}) can be solved analytically:
$$\hat\omega^0=\int_{t,x}\gamma{\left[c(t)-\frac{1}{2n-1}{\left(\frac{2}{n}\right)}^{n/(n-1)}
\pi_u^{(2n-1)/(n-1)}-{\left(\frac{\pi_u^{'}}{\gamma}\right)}^2
\right]}^{1/2}+$$
$$+\int_{t,x}\pi_u^{'}\arcsin\left[\frac{\pi_u^{'}}{\gamma}{\left(
c(t)-\frac{1}{2n-1}{\left(\frac{2}{n}\right)}^{n/(n-1)}
\pi_u^{(2n-1)/(n-1)}\right)}^{-1/2}\right],$$
where
$\pi_u(\alpha,\dot\alpha,\alpha^{'},\alpha^{''};\beta,\beta^{'},\dot\beta^{'},\beta^{''};
\gamma,\dot\gamma,\ddot\gamma,\dot\gamma^{'})$
in the initial
variables is
\be\pi_u=\frac{1}{\alpha\gamma}\left[{\left(\frac{\beta^{'}-\dot\gamma}{\alpha}\right)}^{.}+
{\left(\frac{2\beta\dot\gamma-{(\alpha^2+\beta^2)}^{'}}{2\alpha\gamma}\right)}^{'}\right].\ee

We get a generalized De Rham variational
complex:
\be 0\stackrel{\delta}{\longrightarrow}\Lambda_*^0\stackrel{\delta}{\longrightarrow}
\Lambda_*^1\stackrel{\delta}{\longrightarrow}0
\ee
because the operator of the variational
differential $\delta$ is nilpotent: $\delta^2=0.$ So the generalized De Rham cohomology group is
trivial. Translating into a physical language we conclude that the functional of action (\ref{AS})
does not define any dynamical problem.


\chapter{Dynamics of the mixmaster model
\label{sect_AppE}}
\renewcommand{\theequation}{E.\arabic{equation}}
\setcounter{equation}{0}

\makeatletter
\renewcommand{\@evenhead}{\raisebox{0pt}[\headheight][0pt]
{\vbox{\hbox to\textwidth{\hfil \strut \thechapter . Appendix
\quad\rm\thepage}\hrule}}}
\renewcommand{\@oddhead}{\raisebox{0pt}[\headheight][0pt]
{\vbox{\hbox to\textwidth{\strut \thechapter .\quad
Dynamics of the mixmaster model
\hfil \rm\thepage}\hrule}}}
\makeatother
\section{Dynamics of the Misner model}

The metric of the mixmaster model \cite{MisnerMixmaster} is
\be
ds^2=N^2 dt^2- e^{2\alpha}\left(e^{2\beta}\right)_{ij}\omega^i\omega^j,
\ee
where the differential forms
$$\omega^1=\sin\psi \,d\theta-\cos\psi\sin\theta \,d\phi,$$
\be
\omega^2=\cos\psi \,d\theta+\sin\psi\sin\theta\, d\phi,
\ee
$$\omega^3=-(d\psi+cos\theta \, d\phi)$$
are expressed through the Euler angles $(\psi, \theta, \phi)$ on $SO(3)$ group. The structure constants
of the corresponding algebra $so(3)$ appear in the relations
$$d\omega^i=\frac{1}{2}\epsilon_{ijk}\omega^i\wedge\omega^j.$$
The symmetric traceless matrix $(\beta)_{ij}$
can be presented in the form
$$(\beta)_{ij}=diag (\beta_{+}+\beta_{-}\sqrt{3},\beta_{+}-\beta_{-}\sqrt{3}, -2\beta_{+}),$$
where $\beta_{+}, \beta_{-}$ are two field amplitudes as generalized coordinates.

The Misner's cosmological model
does not belong to the completely integrable systems
\cite{Levi}. It is an example of the pseudo--Euclidean generalized Toda chains at a
level of energy $H=0$ \cite{Bogo}. The Hamiltonian has the form
\be\label{MisnerHam} H=\frac{1}{2}(-p_\alpha^2+p_{+}^2+p_{-}^2)+\exp (4\alpha) V(\beta_{+},
\beta_{-}),\ee
 where the potential function
$V(\beta_{+}, \beta_{-})$
is an exponential polynomial:
$$V(\beta_{+},\beta_{-})=\exp(-8\beta_{+})+\exp(4\beta_{+}+4\sqrt{3}\beta_{-})+
\exp(4\beta_{+}-4\sqrt{3}\beta_{-})-
$$
$$-2\exp(4\beta_{+})-2\exp(-2\beta_{+}+2\sqrt{3}\beta_{-})-2\exp(-2\beta_{+}-2\sqrt{3}\beta_{-}).$$

The Hamiltonian of the generalized Toda chain has the form
\be H=\frac{1}{2}<{\bf p},{\bf p}>+\sum\limits_{i=1}^N c_i v_i,\ee
where $<,>$ is a scalar product in the Minkowski space ${\mathbb R}^{1,n-1}$, $c_i$ are some real
coefficients, $v_i\equiv \exp ({\bf a}_i,{\bf q})$, $(,)$ is a scalar product in the Euclidean space
${\mathbb R}^n$, and ${\bf a}_i$ are real vectors.
For the considered mixmaster model: $n=3, N=6.$
Pseudo-euclidity of a momentum space is a distinctive
peculiarity of gravitational problems so they cannot be referred to as analytical dynamics problems,
where the corresponding form quadratic in momenta is the kinetic energy.

\section{Kovalevski exponents}


On the other hand, the cosmological models can be considered as dynamical systems \cite{Pavlov6}.
So it is
possible to carry out strict methods of analysis traditionally used in the analytical mechanics,
and adopt them to systems like (\ref{MisnerHam}). Let us apply the Painlev$\acute{\rm e}$ test for
calculation of Kovalevski exponents \cite{Kovalevskaya}. The term ``Kovalevski exponents'' was
introduced in paper \cite{Yoshida}, thus marking an outstanding contribution of the Russian woman
to the solution of the important problem of integration of rigid body rotation.


Expanding the $2n$--dimensional phase space to the $2N$--dimensional one by homeomorphism $({\bf p},{\bf
q})\mapsto ({\bf v},{\bf u})$: \be v_i\equiv \exp ({\bf a}_i,{\bf q}),\qquad u_i\equiv<{\bf
a}_i,{\bf p}>,\qquad i=1,\ldots,N,\ee one gets a Hamiltonian system which equations of motion are
the autonomous homogeneous differential equations with polynomial right side:
\be\label{quasisystem} \dot v_i=u_i v_i,\quad \dot u_i=\sum\limits_{j=1}^N M_{ij} v_j,\qquad
i=1,\ldots, N.\ee The matrix $\hat M$ is constructed of scalar products of vectors ${\bf a}_i$ in the
Minkowski space ${\mathbb R}^{1,n-1}$:
$$M_{ij}\equiv -c_j<{\bf a}_i,{\bf a}_j>.$$


The system of equations (\ref{quasisystem}) is quasi-homogeneous. The power of quasi-homogeneity on
variables $u_i$ is one, and on $v_i$ is two. The property of integrability of a dynamical system
appears in a character of singularities of solutions so only singular points represent particular
interest for investigation. The differential equations have the following partial meromorphic
solutions:
\be u_i=\frac{U_i}{t},\qquad v_i=\frac{V_i}{t^2},\qquad i=1,\ldots, N,
\ee
the coefficients
$U_i, V_i$ satisfy the system of algebraic equations
$$2V_i=U_i V_i,\qquad -U_i=\sum\limits_{j=1}^N M_{ij}V_j.$$


Now let us analyze the special types of solutions. Let
$V_1\neq 0,$ the rest $V_2, V_3,\ldots,
V_N=0,$ then we get a solution: if $M_{11}\neq 0,$ then
$$V_1=\frac{2}{M_{11}},\quad U_1=-2,\quad
U_2=-\frac{2M_{21}}{V_{11}},\ldots ,U_N=-\frac{2M_{N1}}{M_{11}}.$$
Analogously the last solutions will be obtained. If for some $i$:
$V_i\neq 0$, and $V_j=0$ for all $j\neq i,$ then at $M_{ii}\neq 0$
we get
$$U_i=-2,\qquad U_j=-2M_{ji}/M_{ii}\qquad \mbox{for all} \qquad i\neq j.$$
It follows
from the obtained solutions that the significant point of analysis
is a nonequality of the corresponding diagonal element of the matrix
$\hat M$ to zero that it is possible in the case of isotropy of the
vector ${\bf a}_i$. It is a principal distinctive feature of
pseudo-Euclidean chains.


For investigating single--valuedness of the obtained solutions we use the Lyapunov method
\cite{Kovalevskaya} based on studying the behavior of their variations:
$$\frac{d}{dt}(\delta u_i)=\sum\limits_{j=1}^N M_{ij}\delta v_j,$$
$$\frac{d}{dt}(\delta v_i)=\frac{U_i\delta v_i}{t}+\frac{V_i\delta
u_i}{t^2},\qquad i=1,\ldots, N.$$ We seek their solutions in the
form of
$$\delta u_i=\xi_i t^{\rho-1},\quad \delta v_i=\eta_i
t^{\rho-2},\quad i=1,\ldots, N.$$
Then for searching the coefficients
$\xi_i, \eta_i$ one gets a linear homogeneous system of equations
with the parameter $\rho$:
\be\label{rho1}(\rho-2-U_i)\eta_i=V_i\xi_i,\ee
\be\label{rho2}(\rho-1)\xi_i=\sum_{j=1}^N M_{ij}\eta_j,\qquad
i=1,\ldots, N.
\ee
Values of the parameter $\rho$ are called the {\it Kovalevski exponents}.


Let us consider solutions when $V_i\neq 0.$ If $\eta_1\neq 0$ and the rest $\eta_i=0,$ then from
the first system of equations (\ref{rho1}) one gets $\xi_1=M_{11}\rho\eta_1/2$, a substitution of it
into the second system (\ref{rho2}) gives a condition of values of the parameter $\rho$:
$$\rho (\rho-1)-2=0,$$
{\it id est} $\rho_1=-1, \rho_2=2.$
The remaining equations (\ref{rho2}) give us solutions $\xi_i=\xi_i (\eta_1, \rho).$


Let $\eta_2\neq 0$, then $\eta_3, \eta_4,\ldots, \eta_N=0,$
$$\rho=2-2\frac{M_{21}}{M_{11}},\quad
\rho\eta_1=2\frac{\xi_1}{M_{11}}.$$
The second system gives functions $\xi_i=\xi_i (\eta_2),
i=1,2,\ldots, N,$ and so on. As a result, having looked through all solutions of the first series for
the case $V_1\neq 0,$ we obtain the formula for the spectrum $\rho$:
$$\rho=2-2\frac{<{\bf a}_i,{\bf a}_1>}{<{\bf a}_1,{\bf a}_1>},\ldots,
i=2,3,\ldots,N.$$


As a final result, having considered the rest solutions, we obtain a formula for the Kovalevski
exponents $\rho$ that generalizes the Adler---van Moerbeke formula \cite{AvM} for the case of
indefinite spaces: \be\label{AvMeq} \rho=2-2\frac{<{\bf a}_i,{\bf a}_k>}{<{\bf a}_k,{\bf
a}_k>},\qquad i\neq k,\qquad {<{\bf a}_k,{\bf a}_k>}\neq 0.\ee


The requirement $\rho\in {\mathbb Z}$ is a necessary condition for meromorphy of solutions on a
complex plane of $t.$ It should be noticed that while obtaining formula (\ref{AvMeq}) no
restrictions on a metric signature were imposed. It is correct not only for spaces of the Minkowski
signature.


Now let us apply the elaborated method to analyze the integrability of the mixmaster model of
the Universe, ``root vectors'' of which have the form:
$${\bf a}_1 (4,-8,0),\quad {\bf a}_2 (4,4,4\sqrt{3}),\quad {\bf
a}_3(4,4,-4\sqrt{3}),$$
$${\bf a}_4 (4,4,0),\quad {\bf a}_5 (4,-2,2\sqrt{3}),\quad {\bf a}_6
(4,-2,-2\sqrt{3}).$$
``Cartan matrix'' composed of scalar products
of the ``root vectors'' in the Minkowski space has the form:
$$<{\bf a}_i,{\bf a}_j>
=48\left(
\begin{array}{rrrccc}
1&-1&-1&-1&0&0\\
-1&1&-1&0&0&-1\\
-1&-1&1&0&-1&0\\
-1&0&0&0&-1/2&-1/2\\
0&0&-1&-1/2&0&-1/2\\
0&-1&0&-1/2&-1/2&0
\end{array}
\right).$$


One gets three ``root vectors'' disposed out of a light cone (space--like vectors), the rest three
are isotropic on the light cone. Using the generalized Adler---van Moerbeke formula (\ref{AvMeq}),
taking account of zero norm of three vectors, we get integer  $\rho_1=2, \rho_2=4$. As the Killing
metric in the generalized Adler---van Moerbeke formula (\ref{AvMeq}) is indefinite, it should be
pointed out in the classification scheme of the noncompact Lie algebras for getting exact solutions
of Toda lattices, as it was done in \cite{Bogo}.

Due to isotropic character of three vectors, we transit from the Misner phase variables to some
other \cite{Pavlov6}
$$(\alpha, \beta_{+}, \beta_{-}; p_\alpha, p_{+}, p_{-})\mapsto (X,Y,Z;
p_x,p_y,p_z).$$
Now the Hamiltonian has a more symmetric form:
$$X=\frac{1}{12}\exp (2(\alpha+\beta_{+}+\sqrt{3}\beta_{-})),\quad
Y=\frac{1}{12}\exp (2(\alpha+\beta_{+}-\sqrt{3}\beta_{-})),$$
$$Z=\frac{1}{12}\exp (2(\alpha-2\beta_{+}));$$
$$p_x=\frac{1}{12}(2p_\alpha+p_{+}+\sqrt{3}p_{-}),\quad
p_y=\frac{1}{12}(2p_\alpha+p_{+}-\sqrt{3}p_{-}),\quad
p_z=\frac{1}{6}(p_\alpha-p_{+}).$$

The equations of motion are represented as Hamiltonian equations on a direct sum of
two--dimensional solvable Lie algebras
$$g(6)=g(2)\oplus g(2)\oplus g(2):$$
\be\label{E9}
\{X, p_x\}=X,\qquad
\{Y, p_y\}=Y,\qquad \{Z,p_z\}=Z\ee with the Hamiltonian $H$:
$$H=-\frac{1}{2}(p_x^2+p_y^2+p_z^2)+\frac{1}{4}(p_x+p_y+p_z)^2-2(X^2+Y^2+Z^2)+(X+Y+Z)^2.$$


The Hamiltonian has a form of a kinetic energy of a top:
$$H=\frac{1}{2}\sum\limits_{i,j=1}^6 I_{ij}x^i x^j,$$
where the phase variables are enumerated as
$$x_1=X,\quad x_2=Y,\quad x_3=Z;\quad x_4=p_x,\quad x_5=p_y,\quad x_6=p_z,$$
and the energy tensor $I_{ij}$ has a block type. So the mixmaster
cosmological model can be considered as the Euler---Poincar$\acute{\rm e}$ top on
a Lie algebra (\ref{E9}). The Euler---Poincar$\acute{\rm e}$ equations are generalization
of the famous dynamical Euler equations describing a rotation of a
rigid body with the corresponding algebra of rotations $so(3).$


The partial meromorphic solution of the obtained system of differential equations is $x_i=C_i/t.$
Then the problem is reduced to investigation of a spectrum of Kovalevski's matrix
\be\label{matKov}
K_{ij}=(c_{jk}^i I_{kl}+c_{jk}^l I_{ki})C^l+\delta_{ij},
\ee
where $c_{ij}^k$ are the structure
constants of the algebra and $C_i$ are solutions of an algebraic system:
$$C_i+c_{ij}^k I_{jl} C_k C_l=0.$$
Calculations give an integer--valued spectrum of the matrix (\ref{matKov}): $$\rho=-1,1,1,2,2,2,$$
which point to the regular character of behavior of the considered dynamical system.

\end{document}